\documentclass[sn-basic]{sn-jnl}

\usepackage{graphicx}%
\usepackage{multirow}%
\usepackage{amsmath,amssymb,amsfonts}%
\usepackage{amsthm}%
\usepackage{mathrsfs}%
\usepackage[title]{appendix}%
\usepackage{xcolor}%
\usepackage{textcomp}%
\usepackage{manyfoot}%
\usepackage{booktabs}%
\usepackage{algorithm}%
\usepackage{algorithmicx}%
\usepackage{algpseudocode}%
\usepackage{listings}%

\usepackage{lmodern}
\usepackage{pifont}
\usepackage{enumitem}
\usepackage{array}

\newcolumntype{C}[1]{>{\centering\arraybackslash$}p{#1}<{$}} 
\newcolumntype{R}[1]{>{\raggedright\arraybackslash}p{#1}}    

\newcommand{\inquotes}[1]{`{#1}'}
\newcommand{\code}[1]{{\sc #1}}
\newcommand{\emdash}{\,---\,}

\newcommand{\cmark}{\ding{51}}
\newcommand{\xmark}{\ding{55}}

\newcommand{\msol}{\mathrm{M}_\odot}
\newcommand{\rsol}{\mathrm{R}_\odot}
\newcommand{\g}{\mathrm{g}}
\newcommand{\cm}{\mathrm{cm}}
\newcommand{\km}{\mathrm{km}}
\newcommand{\pc}{\mathrm{pc}}

\newcommand{\s}{\mathrm{s}}
\newcommand{\yr}{\mathrm{yr}}

\newcommand{\ddt}[1]{\frac{\partial{#1}}{\partial t}}
\newcommand{\abs}[1]{\vert{#1}\vert}
\newcommand{\vel}{\mathbf{v}}
\newcommand{\Bvec}{\mathbf{B}}
\newcommand{\pth}{p_\mathrm{th}}
\newcommand{\eint}{e_\mathrm{int}}
\newcommand{\cs}{c_\mathrm{s}}
\newcommand{\kB}{k_\mathrm{B}}
\newcommand{\mH}{m_\mathrm{H}}

\newcommand{\mach}{\mathcal{M}}
\newcommand{\ldriv}{L_\mathrm{driv}}
\newcommand{\vturb}{v_\mathrm{turb}}
\newcommand{\tturb}{t_\mathrm{turb}}
\newcommand{\rek}{\mathrm{Re}}
\newcommand{\rem}{\mathrm{Rm}}
\newcommand{\nushear}{\nu_\mathrm{shear}}
\newcommand{\nubulk}{\nu_\mathrm{bulk}}

\newcommand{\rsoft}{r_\mathrm{soft}}
\newcommand{\rbyrsoft}{\tilde{r}}
\newcommand{\ggas}{\mathbf{g}_\mathrm{gas}}
\newcommand{\gsinks}{\mathbf{g}_\mathrm{sinks}}
\newcommand{\phigas}{\Phi_\mathrm{gas}}
\newcommand{\phisinks}{\Phi_\mathrm{sinks}}
\newcommand{\lJ}{\lambda_\mathrm{J}}
\newcommand{\rhosink}{\rho_\mathrm{sink}}
\newcommand{\rsink}{r_\mathrm{sink}}
\newcommand{\Rsink}{\mathbf{R}_\mathrm{sink}}
\newcommand{\msink}{M_\mathrm{sink}}
\newcommand{\psink}{\mathbf{P}_\mathrm{sink}}
\newcommand{\lsink}{\mathbf{L}_\mathrm{sink}}
\newcommand{\ssink}{\mathbf{S}_\mathrm{sink}}
\newcommand{\rout}{r_\mathrm{out}}
\newcommand{\thetaout}{\theta_\mathrm{out}}
\newcommand{\mout}{M_\mathrm{out}}
\newcommand{\vout}{\vel_\mathrm{out}}
\newcommand{\pout}{\mathbf{P}_\mathrm{out}}
\newcommand{\lout}{\mathbf{L}_\mathrm{out}}
\newcommand{\dt}{\Delta t}
\newcommand{\dx}{\Delta x}
\newcommand{\maccdot}{\dot{M}_\mathrm{acc}}
\newcommand{\macc}{M_\mathrm{acc}}
\newcommand{\racc}{\mathbf{R}_\mathrm{acc}}
\newcommand{\pacc}{\mathbf{P}_\mathrm{acc}}
\newcommand{\lacc}{\mathbf{L}_\mathrm{acc}}
\newcommand{\fmass}{f_\mathrm{m}}
\newcommand{\fang}{f_\mathrm{a}}

\newcommand{\nvec}{\mathbf{n}}
\newcommand{\Erad}{E_\mathrm{r}}
\newcommand{\Frad}{\mathbf{F}_\mathrm{r}}
\newcommand{\Prad}{\mathbb{P}_\mathrm{r}}
\newcommand{\TEdd}{\mathbb{T}}
\newcommand{\subnu}{{_{,\nu}}}
\newcommand{\kappaE}{\kappa_\mathrm{E}}
\newcommand{\kappaF}{\kappa_\mathrm{F}}
\newcommand{\kappaR}{\kappa_\mathrm{R}}
\newcommand{\kappaP}{\kappa_\mathrm{P}}
\newcommand{\sigmaSB}{\sigma_\mathrm{SB}}
\newcommand{\emis}{4\sigmaSB T^4}

\newcommand{\crploss}{b_\mathrm{loss}}
\newcommand{\Ecr}{E_\mathrm{CR}}
\newcommand{\pcr}{p_\mathrm{CR}}
\newcommand{\gamcr}{\gamma_\mathrm{CR}}

\begin{document}

\title[Simulating turbulence and star formation]{Computational advances and challenges in simulations of turbulence and star formation}

\author*[1]{\fnm{Christoph} \sur{Federrath}}\email{christoph.federrath@anu.edu.au}

\author[2]{\fnm{Stella} \sur{Offner}}\email{soffner@astro.as.utexas.edu}

\affil*[1]{\orgdiv{Research School of Astronomy and Astrophysics}, \orgname{Australian National University}, \orgaddress{\street{Cotter Road}, \city{Canberra}, \postcode{2611}, \state{ACT}, \country{Australia}}, \\ORCID: \href{https://orcid.org/0000-0002-0706-2306}{0000-0002-0706-2306}}

\affil[2]{\orgdiv{Astronomy Department}, \orgname{The University of Texas at Austin}, \orgaddress{\street{2515 Speedway, Stop C1400}, \city{Austin}, \postcode{78712-1205}, \state{TX}, \country{USA}}, \\ORCID: \href{https://orcid.org/0000-0003-1252-9916}{0000-0003-1252-9916}}

\abstract{
We review recent advances in the numerical modeling of turbulent flows and star formation. An overview of the most widely used simulation codes and their core capabilities is provided. We then examine methods for achieving the highest-resolution magnetohydrodynamical turbulence simulations to date, highlighting challenges related to numerical viscosity and resistivity. State-of-the-art approaches to modeling gravity and star formation are discussed in detail, including implementations of star particles and feedback from jets, winds, heating, ionization, and supernovae. We review the latest techniques for radiation hydrodynamics, including ray tracing, Monte Carlo, and moment methods, with comparisons between the flux-limited diffusion, moment-1, and variable Eddington tensor methods. The final chapter summarizes advances in cosmic-ray transport schemes, emphasizing their growing importance for connecting small-scale star formation physics with galaxy-scale evolution.
}



\maketitle

\newpage
\setcounter{tocdepth}{3}
\tableofcontents

\newpage
\section{Introduction} \label{sec:intro}

Star formation is a critical component of many astrophysical phenomena: the life cycle of stars drives the formation and evolution of galaxies, regulates the abundance and distribution of metals in the Universe, and sets the initial conditions for planet formation. Despite being governed by fundamentally classical physics processes{\emdash}magnetized gas dynamics, gravity, thermodynamics, and radiation{\emdash}the complex, nonlinear interactions defy analytic description. Star formation also spans vast spatial scales, from the kpc interstellar medium (ISM) down to stellar radii, and occurs across regimes ranging from ultra-low densities, exceeding achievable laboratory vacuum limits, to stellar densities. Observations are further limited by resolution, sensitivity, and optical depth effects. Consequently, numerical simulations powered by high-performance computing have become indispensable tools for studying star formation.

Over the past decade, numerical developments have greatly expanded the scope of star formation studies by extending dynamic range, improving computational efficiency, and introducing new methods. Modern codes now routinely resolve gas dynamics across many orders of magnitude in scale while incorporating key physical processes. This has yielded important insights into how these processes shape central metrics such as the star formation rate (SFR) and the initial mass function (IMF). Despite these advances, major challenges remain. Current simulations still fall short of achieving the physical parameters associated with magnetized turbulence, or of spanning the full range of scales required to capture both individual star formation and the impact of stellar feedback. The long timescales ($>$\,Myr) and stochastic nature of star formation demand large computational domains evolved over many timesteps. Moreover, the interplay between large and small scales calls for holistic approaches in which physics and chemistry across wide ranges are modeled simultaneously, with minimum impact from the choice of numerical methods and/or boundary conditions.

Numerical advances over the past decade have focused on developing magnetohydrodynamical (MHD) methods (Sect.~\ref{sec:hydro}) that robustly capture high-Mach number turbulence{\emdash}a cornerstone of the star-formation process (Sect.~\ref{sec:turbulence}). Other important advances include more accurate radiative transfer methods coupled to MHD, which can capture shadows and remain valid across the difficult transition between optically thin and thick regimes (Sect.~\ref{sec:rhd}). Another fundamental requirement for star formation studies is modeling the gravity of gas and stars (Sect.~\ref{sec:grav}). Despite significant progress, simulations still rely heavily on prescriptions for unresolved physics, so-called \inquotes{sub-resolution} models\footnote{Sometimes referred to as \inquotes{sub-grid} models; however, the term \inquotes{sub-resolution} is more general and includes particle-based methods.}. These encompass treatments of dissipation and unresolved turbulence, star formation itself (Sect.~\ref{sec:sf}), and a variety of stellar feedback processes{\emdash}stellar evolution, protostellar jets and outflows, heating, winds, ionization, and supernovae. Such models often depend on ad-hoc parameters and remain among the least benchmarked areas of development (Sect.~\ref{sec:feedback}). Finally, growing recognition of the role of cosmic rays has driven the development of cosmic-ray transport (CRT) methods (Sect.~\ref{sec:crhd}).

In this review we present new approaches to modeling turbulence, radiation, gravity, star formation, feedback, and CRs, with a particular focus on code capabilities, their numerical treatments, and parameterizations. We assess the strengths and weaknesses of these methods and highlight current limitations. The remainder of the Introduction lists the codes most widely applied in turbulence and star-formation studies (Sect.~\ref{sec:codes}) and highlights recent advances in code optimization and parallel scaling (Sect.~\ref{sec:parallel_scaling}).

\subsection{Numerical codes for turbulence and star formation} \label{sec:codes}

Over the past two decades, more than a dozen powerful codes for turbulence and star-formation applications have been developed and continuously refined. Table~\ref{tab:codes} summarizes the most widely used codes and their capabilities. These methods can be broadly divided into two categories: grid methods, which discretize flows by volume, and particle methods, which discretize flows by mass. Grid-based approaches typically model hydrodynamics using elements fixed in the reference frame of a stationary observer, while particle-based approaches adopt a Lagrangian framework, with mass elements moving with the fluid. Hybrid approaches, such as moving-mesh and meshless finite mass/volume methods, aim to combine the strengths of both.

Most of these codes have publicly available versions, promoting transparency and broad community use. Links to the available releases are provided in the Table notes. However, not all features listed in Table~\ref{tab:codes} are implemented in the public versions. Given the continuous development of these codes, the table cannot be guaranteed to be fully complete or up to date. We have nevertheless made every effort to ensure its accuracy by consulting the main developers of the respective codes and compiling the information based on their latest input.

\begin{table}[tp]
 \setlength{\tabcolsep}{2.0pt} 
  \centering
  \caption{Popular astro-simulation codes used for turbulence and star formation modeling.}
  \label{tab:codes}
  \begin{tabular}{ccccccc}
    \hline
    Name & Type & Hydro & Self-Gravity & Feedback & Radiation & CRT \\
    (1) & (2) & (3) & (4) & (5) & (6) & (7) \\
    \hline
    \code{arepo}$^1$ & MM & NI & TPM & R/S & M1/RT/MC & \cmark \\ 
    \code{art}$^2$ & AMR & H & FFT/MG & R/W/S & OT-VET & \cmark \\ 
    \code{athena}$^3$ & UG/SMR & NI & FFT/MG & R/W/S & M1/RT & \xmark \\ 
    \code{athena}$^{++}$$^4$ & AMR & NI & FFT/MG & \xmark & VET & \cmark \\ 
    \code{enzo}$^5$ & AMR & NI & FFT/MG & R/S & FLD/RT & \cmark  \\ 
    \code{flash}$^6$ & AMR & NI & FFT/MG/TPM & O/R/W/S & FLD/M1/VET/RT/MC & \cmark \\
    \code{gadget}$^7$ & SPH & I & TPM & R/S & OT-VET/RT & \cmark \\ 
    \code{gizmo}$^8$ & MLFMV & NI & TPM & O/R/W/S & FLD/M1/OT-VET/RT & \cmark \\ 
    \code{idefix}$^9$ & UG & NI & MG & \xmark & \xmark & \xmark \\ 
    \code{nirvana}$^{10}$  & AMR & NI & MG & W/S & FLD & \xmark \\ 
    \code{orion}$^{11}$ & AMR & I & MG & O/R/W & FLD/RT & \xmark \\
    \code{pencil}$^{12}$ & UG & NI & FFT & S & RT & \cmark  \\
    \code{phantom}$^{13}$ & SPH & NI & TPM & R/W/S & FLD/MC & \xmark \\ 
    \code{pluto}$^{14}$ & AMR & NI & MG & \xmark & M1/RT & \cmark \\ 
    \code{quokka}$^{15}$ & AMR & H & MG & S & M1 & \xmark \\
    \code{ramses}$^{16}$ & AMR & NI & MG & O/R/W/S & M1 & \cmark  \\
    \code{seren}$^{17}$ & SPH & H & TPM & R/S & FLD/RT & \xmark \\  
    \code{torus}$^{18}$ & AMR & H & MG & R/W/S & MC & \xmark \\  
    \code{zeus}$^{19}$ & UG & NI & FFT/MG & W/S & VET & \xmark \\ 
    \hline
  \end{tabular}
{\bf Notes:} Column~1: code name. Column~2: code type: UG~=~uniform grid, SMR~=~static mesh refinement, AMR~=~adaptive mesh refinement \citep{BergerColella1989} (includes UG capability), SPH~=~smoothed particle hydrodynamics \citep[e.g.,][]{Benz1988,Monaghan1988,PriceEtAl2018}, MM~=~moving mesh, MLFMV~=~meshless finite mass/volume \citep{Hopkins2013GIZMO}. Column~3: type of gas dynamics: H~=~hydrodynamics, I~=~ideal MHD (includes~H capability), NI~=~non-ideal MHD (Ohmic dissipation, ambipolar diffusion, Hall effect; or all three; includes I capability). Column~4: self-gravity treatment: FFT~=~fast Fourier transform (Sect.~\ref{sec:grav_fft}), MG~=~multi-grid (Sect.~\ref{sec:grav_mg}), TPM~=~tree particle-mesh (Sect.~\ref{sec:grav_tree}). Column~5: feedback processes: O~=~protostellar outflows (Sect.~\ref{sec:outflow}), R~=~radiative feedback in the form of pressure, ionization and/or heating (Sect.~\ref{sec:heating}), W~=~stellar winds (Sect.~\ref{sec:winds}, \ref{sec:ionization}), and S~=~supernova feedback (Sect.~\ref{sec:sn}). Column~6: radiation transfer methods: FLD~=~flux limited diffusion (Sect.~\ref{sec:rt_fld}), M1~=~moment-1 (Sect.~\ref{sec:rt_m1}), VET~=~variable Eddington tensor (Sect.~\ref{sec:rt_vet}), OT-VET~=~optically-thin VET, RT~=~ray tracing (Sect.~\ref{sec:rt_ray}), MC~=~Monte Carlo (Sect.~\ref{sec:rt_mcrt}). Column~7: cosmic ray transport (CRT; Sect.~\ref{sec:crhd}). A {\cmark} indicates availability of a certain functionality (without specifying further details), while {\xmark} indicates missing functionality. References/Links:
{\footnotesize
$^1$\url{https://gitlab.mpcdf.mpg.de/vrs/arepo} \citep{Springel2010},
$^2$\url{https://bitbucket.org/cartamr/cart/src/master}
\citep{KravtsovKlypinKhokhlov1997,KravtsovEtAl2002,RuddEtAl2008},
$^3$\url{https://princetonuniversity.github.io/Athena-Cversion},
$^4$\url{https://www.athena-astro.app} \citep{StoneEtAl2020},
$^5$ \url{https://github.com/enzo-project} \citep{BryanEtAl2014},
$^6$\url{https://flash.rochester.edu/site/flashcode} \citep{FryxellEtAl2000,DubeyEtAl2008},
$^7$\url{https://wwwmpa.mpa-garching.mpg.de/gadget4},
$^8$\url{https://github.com/pfhopkins/gizmo-public} \citep{Hopkins2015},
$^9$ \url{https://github.com/idefix-code/idefix} \citep{LesurEtAl2023}, 
$^{10}$\url{https://gitlab.aip.de/ziegler/NIRVANA} \citep{Ziegler2008},
$^{11}$\citet{LiEtAl2021},
$^{12}$\citet{BrandenburgEtAl2021},
$^{13}$\url{https://phantomsph.github.io} \citep{PriceEtAl2018},
$^{14}$\url{https://plutocode.ph.unito.it},
$^{15}$\url{https://github.com/quokka-astro/quokka} \citep{WibkingKrumholz2022},
$^{16}$\url{https://github.com/ramses-organisation/ramses} \citep{Teyssier2002},
$^{17}$\citet{HubberEtAl2011ascl,HubberEtAl2011},
$^{18}$\citet{HarriesEtAl2019},
$^{19}$\url{https://github.com/bwoshea/ZEUS-MP_2} \citep{StoneNorman1992a}.
}
\end{table}

\subsection{Parallel scalability and code optimization} \label{sec:parallel_scaling}

Although physical and numerical accuracy are paramount, efficient parallel scaling to large core counts is also essential to perform high-resolution simulations within practical computational budgets. Strong scaling tests evaluate the time-to-solution for fixed problem sizes: the total amount of calculation work remains constant as the number of cores is increased. Weak scaling tests increase the problem size and resolution proportionally with the number of compute cores, i.e., the amount of work per core is fixed. Good strong and weak scaling performance are both essential for advancing astrophysical applications, while also ensuring environmentally responsible use of supercomputing resources. Continuous code optimization and scaling improvements are therefore indispensable, for example, through the development of optimized techniques for load balancing \citep[e.g., \code{dispatch}; see][]{NordlundEtAl2018}.

\begin{figure}[t]
\centerline{\includegraphics[width=0.9\linewidth]{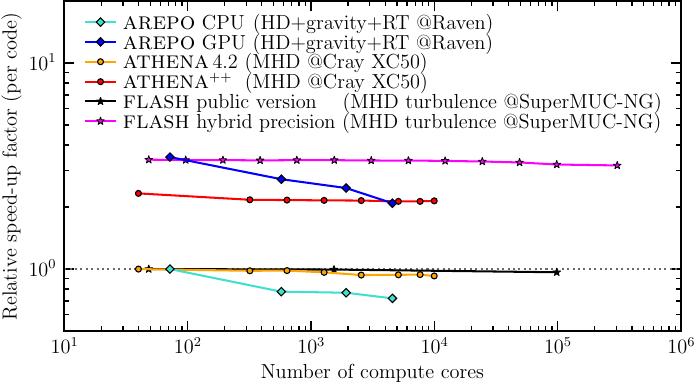}}
\caption{\label{fig:scaling}
Weak-scaling tests for three codes: \code{arepo} (diamonds), comparing CPU (turquoise) and GPU (blue) implementations of radiation transport (RT) in simulations including hydrodynamics and gravity, run on the \inquotes{Raven} system; \code{athena} (circles), comparing \code{athena}\,4.2 (orange) with the modernized \code{athena}$^{++}$ (red) for pure MHD on a Cray~XC50; and \code{flash} (stars), comparing the public version (black) with an optimized hybrid-precision version (magenta; Sect.~\ref{sec:hybrid-prec}) for MHD turbulence on \inquotes{SuperMUC-NG}. For each code, the two modes shown are directly comparable (but not across different codes). Substantial performance gains are evident: GPU acceleration (\code{arepo}), code modernization (\code{athena}), and hybrid precision (\code{flash}).}
\end{figure}

Figure~\ref{fig:scaling} highlights some of the recent developments in code optimization and weak scaling in three of the codes listed in Table~\ref{tab:codes}. 
Good weak scaling performance is particularly relevant for turbulence and star-formation simulations, where higher resolution, and thus larger problems, is always desired. Because the tests in Fig.~\ref{fig:scaling} involve different setups, physics, and machines, only the relative speed-up for a given code can be directly compared.

Overall, all three codes show excellent parallel efficiency up to thousands or even tens of thousands of cores. For pure MHD, both \code{athena}$^{++}$ and \code{flash} retain efficiencies above 90\% up to $10^4$--$10^5$~cores, i.e., very close to ideal scaling. Even the more challenging \code{arepo} tests, which combine hydrodynamics, self-gravity, and radiation transport (RT) \citep{ZierEtAl2024}, achieve 60--70\% efficiency up to nearly 5,000~cores{\emdash}impressive given the inherently non-local nature of gravity (Sect.~\ref{sec:grav}) and RT (Sect.~\ref{sec:rhd}).

The individual comparisons further illustrate substantial performance gains from recent developments. Accelerating RT in \code{arepo} with GPUs (see also Sect.~\ref{sec:gpu} below) yields an overall speed-up of 3--4$\times$, with the RT component itself running $\sim15\times$ faster on GPUs. Modernization of \code{athena} improves performance by $\sim2.3\times$ from version~4.2 to \code{athena}$^{++}$ \citep{StoneEtAl2020}. In \code{flash}, switching from the public version to an optimized hybrid-precision implementation (see details in Sect.~\ref{sec:hybrid-prec} below) produces a $\sim3.3\times$ speed-up for MHD turbulence, while retaining efficiency to more than $10^5$~cores.

These results demonstrate how continuous algorithmic and implementation improvements translate directly into higher scientific throughput. Such efforts are critical not only for enabling frontier simulations but also for maximizing the efficient and sustainable use of supercomputing resources. Future updates of this living review will include controlled scaling benchmarks for additional codes listed in Table~\ref{tab:codes}, extending earlier comparative studies \citep{KitsionasEtAl2009,PriceFederrath2010,KritsukEtAl2011Codes}.

\section{Hydrodynamics} \label{sec:hydro}

Here we introduce the basic set of MHD equations, which provide the foundation for the remainder of the review. We introduce the basics of different discretisation methods (mesh, particle, hybrid), magnetic field divergence control, and discuss challenges associated with the choice of initial and boundary conditions for star-formation simulations.

\subsection{Magnetohydrodynamics} \label{sec:mhd}

Astrophysical fluids generally have magnetic fields. While magnetic fields couple only to charged particles, even phases of the ISM such as molecular clouds (where stars form), composed almost entirely of neutral gas, retain a small degree of ionization (ionization fraction $\sim10^{-7}$). This is sufficient to generate electric currents, and through frequent collisions the ions strongly couple to the neutrals, making magnetic fields a critical component of the gas dynamics. Magnetic fields introduce additional physical processes, including magnetic pressure, tension, MHD waves, and resistive dissipation. Owing to this complexity, many early star-formation simulations neglected magnetic fields, but their crucial role is now widely recognized, and most current simulations include them.

A fully self-consistent description of magnetized systems requires kinetic theory and particle-in-cell (PIC) simulations \citep[e.g.,][]{StOngeKunz2018,RinconEtAl2016,AchikanathFederrathSeta2024}. This can, however, be reduced to a coupled set of fluid equations in which electrons, ions (protons, atomic and molecular ions), and neutrals (e.g., H, H$_2$, CO) are treated as separate fluids{\emdash}the framework usually referred to as \inquotes{non-ideal MHD}\footnote{Non-ideal MHD is commonly divided into three categories: (i) Ohmic resistivity, (ii) ambipolar diffusion, and (iii) the Hall effect. In this review, we include only Ohmic resistivity in the MHD equations. For a comprehensive review of non-ideal MHD methods relevant to star formation, see \citet{TeyssierCommercon2019}. We further note that non-ideal MHD effects are tied to magnetic reconnection events, where codes primarily rely on numerical diffusion to approximate physical reconnection \citep[see discussion in the review by][]{LazarianEtAl2020}}. Some studies follow this so-called \inquotes{multi-fluid} approach \citep{LiEtAl2008,LiMyersMcKee2012,BurgeEtAl2016,TritsisEtAl2022,TritsisBasuFederrath2025a,TritsisBasuFederrath2025b}. More commonly, however, strong ion-neutral coupling is assumed, yielding the ideal MHD limit, which is well justified on cloud and core scales, though it breaks down in dense accretion disks \citep[see e.g.,][]{XuKunz2021a,XuKunz2021b,MauxionLesurMaret2024}.

Unlike most turbulence studies, star-formation simulations, to the best of current knowledge, consistently neglect explicit dissipation. By contrast, MHD turbulence studies frequently include viscosity and at least Ohmic resistivity, as we do in the following equations. Even studies that omit explicit viscosity and resistivity inevitably introduce effective small-scale viscosity and resistivity through numerical diffusion, leading to partial ion-neutral decoupling. We will examine these effects in detail in Sect.~\ref{sec:turbulence}. With these foundations, we can now introduce the standard set of MHD equations.

\subsubsection{MHD equations}

The standard compressible MHD equations including viscous and resistive dissipation terms can be written as
\begin{eqnarray}
\ddt{\rho} & = & -\nabla\cdot(\rho\vel), \label{eq:mhd1} \\
\ddt{(\rho \vel)} & = & -\nabla\cdot\left(\rho\vel\vel - \frac{1}{4\pi}\Bvec\Bvec\right) - \nabla p + \nabla\cdot\left(2\rho\nu\mathbb{S}\right), \label{eq:mhd2} \\
\ddt{(\rho e)} & = & -\nabla\cdot\left[\left(\rho e+p\right)\vel - \frac{1}{4\pi}\left(\Bvec\cdot\vel\right)\Bvec\right] \nonumber \\
& & \quad\quad\quad + \nabla\cdot\left[2\rho\nu\mathbb{S}\vel+\frac{1}{4\pi}\Bvec\times\left(\eta\nabla\times\Bvec\right)\right], \label{eq:mhd3} \\
\ddt{\Bvec} & = & \nabla\times\left(\vel\times\Bvec\right) + \eta\nabla^2\Bvec, \label{eq:mhd4} \\
\phantom{\ddt{.}}\nabla\cdot\Bvec & = & 0. \label{eq:mhd5}
\end{eqnarray}
Here, $\rho$, $\vel$, $p=\pth+\abs{\Bvec}^2\!/(8\pi)$, $\Bvec$, and $\rho e=\rho\eint+\rho\abs{\vel}^2\!/2+\abs{\Bvec}^2\!/(8\pi)$ denote the gas density, velocity, pressure (thermal + magnetic), magnetic field, and total energy density (internal + kinetic + magnetic), respectively. Physical shear viscosity (kinematic viscosity $\nu$) in Eqs.~(\ref{eq:mhd2}) and~(\ref{eq:mhd3}) is included via the traceless rate-of-strain tensor,
\begin{equation} \label{eq:strain_tensor}
\mathbb{S} = \frac{1}{2}\left(\nabla\vel+\left(\nabla\vel\right)^\mathrm{T}\right)-\frac{1}{3}\left(\nabla\cdot\vel\right)\mathbb{I},
\end{equation}
where $\mathbb{I}$ is the identity matrix. In index notation, the components of $\mathbb{S}$ are $\mathbb{S}_{ij}=(1/2)(\partial_i v_j+\partial_j v_i)-(1/3)(\partial_k v_k)\delta_{ij}$. The product $\rho\nu$ appearing in Eqs.~(\ref{eq:mhd2}) and~(\ref{eq:mhd3}) is commonly referred to as the dynamic viscosity.

The evolution and diffusion of $\Bvec$ are governed by the induction equation (Eq.~\ref{eq:mhd4}), where the magnetic diffusivity is $\eta=1/(4\pi\sigma)$ with $\sigma$ the electric conductivity. Finally, we note that Eqs.~(\ref{eq:mhd2}) and~(\ref{eq:mhd3}) only include the shear viscosity $\nu$, while the bulk viscosity is usually assumed to vanish{\emdash}see Sect.~\ref{sec:shear_bulk_visc} for further discussion.

\subsubsection{Thermodynamics and equation of state} \label{sec:eos}

The MHD equations are closed with an equation of state (EOS). Most commonly this is the ideal gas EOS,
\begin{equation} \label{eq:ideal_gas_eos}
\pth = \rho\eint(\gamma-1) = \frac{\rho\kB T}{\mu\mH},
\end{equation}
where $T$ is the temperature, $\kB$ the Boltzmann constant, $\mH$ the mass of a hydrogen atom, $\mu$ the mean particle mass in units of $\mH$, and $\gamma$ the adiabatic index. The latter is given by $\gamma = 1+2/f$, where $f$ is the number of degrees of freedom excited for a given particle type (electrons, atoms/ions, or molecules), which always includes translational modes, and may include rotational or vibrational modes for molecules \citep[if they are excited; see e.g.,][]{ShardaKrumholzFederrath2019}. The sound speed is
\begin{equation}
\cs = \left(\frac{\partial\pth}{\partial\rho}\right)_s^{1/2}=\left(\frac{\gamma\kB T}{\mu\mH}\right)^{1/2},
\end{equation}
where the subscript $s$ denotes that the derivative is taken at constant entropy.

For simplicity, fundamental studies of MHD turbulence often assume an isothermal EOS, $\pth=\rho\cs^2$, with a constant $\cs$ fixed in space and time. This is a good approximation for molecular cloud turbulence and for the early stages of dense core formation \citep{WolfireEtAl1995,OmukaiEtAl2005,PavlovskiSmithMacLow2006,GloverMacLow2007a,GloverMacLow2007b,GloverFederrathMacLowKlessen2010}. More sophisticated treatments include heating and cooling via radiative transfer (RT; Sect.~\ref{sec:rhd}), which is crucial to capture the transition to optically thick gas at number densities $n=\rho/(\mu\mH)\gtrsim10^{10}\,\cm^{-3}$, where the gas heats up and fragmentation is suppressed \citep{JappsenEtAl2005,GuszejnovKrumholzHopkins2016,FederrathKrumholzHopkins2017,GuszejnovEtAl2018,MathewFederrath2020,HennebelleEtAl2020,HennebelleEtAl2022}.

Intermediate approaches lie between assuming an isothermal EOS and solving full RT. One common choice is a polytropic EOS of the form $\pth=K\rho^\Gamma$ \citep{PassotVazquez1998,LiKlessenMacLow2003}, with polytropic index $\Gamma$ and coefficient $K$. These can be specified as piecewise functions to approximate the heating of gas as it becomes optically thick \citep[see Eqs.~3 and~4 in][]{FederrathEtAl2014}. In this case the energy equation (Eq.~\ref{eq:mhd3}) can be omitted, since the MHD system is closed without it. Other strategies include incorporating approximations for stellar heating (Sect.~\ref{sec:heating}) or adding heating/cooling terms to Eq.~(\ref{eq:mhd3}) from pre-computed tables \citep{KoyamaInutsuka2002,VazquezSemadeniEtAl2007,MandalFederrathKoertgen2020}.


\subsection{Mesh vs.~particle, and hybrid methods for hydrodynamics}

Numerical methods for solving the equations of hydrodynamics (HD) and MHD can be broadly divided into mesh-based and particle-based approaches (see list of codes in Tab.~\ref{tab:codes}). Mesh-based (typically Eulerian) methods discretize the fluid variables on a fixed or adaptively refined grid and solve the conservation equations using finite-volume or finite-difference schemes. These methods are particularly well suited for capturing shocks and fluid instabilities, because modern Godunov-type schemes solve Riemann problems at cell interfaces and ensure accurate shock capturing with minimal numerical diffusion \citep{Godunov1959,vanLeer1979,Toro1997,LeVeque2002}. Adaptive mesh refinement (AMR) further enhances resolution dynamically in regions of interest while maintaining computational efficiency \citep{BergerColella1989}. Grid-based methods have been widely used in astrophysics codes, and are particularly effective for modeling turbulence, shocks, and magnetized flows \citep{KitsionasEtAl2009,PriceFederrath2010,KritsukEtAl2011Codes}.

Particle-based (typically Lagrangian) methods, most notably smoothed particle hydrodynamics (SPH), represent the fluid using discrete particles that move with the flow \citep{Lucy1977,GingoldMonaghan1977,Monaghan1992,Price2012SPH}. Hydrodynamic quantities are computed via kernel interpolation over neighboring particles, which naturally provides adaptive resolution in high-density regions and exact conservation of mass, momentum, and energy by construction. SPH methods are especially advantageous for problems involving self-gravity, large dynamic range in density, and complex geometries, such as star formation and galaxy evolution \citep[e.g.,][]{Springel2005}. However, traditional SPH formulations have historically exhibited challenges in capturing fluid instabilities and mixing due to artificial surface tension effects and zeroth-order consistency errors \citep{AgertzEtAl2007}, although modern formulations incorporating improved kernels, dissipation switches, and pressure-based formulations have significantly alleviated these issues \citep{Price2008,Hopkins2013GIZMO}.

Hybrid approaches aim to combine the advantages of both Eulerian and Lagrangian methods. Moving-mesh schemes, such as implemented in \code{arepo}, use a finite-volume Godunov method on a mesh defined by a Voronoi tessellation whose generating points move with the flow, providing Galilean invariance and reduced advection errors while maintaining accurate shock capturing \citep{Springel2010}. Other hybrid approaches, such as meshless finite-mass (MFM) and meshless finite-volume (MFV) methods implemented in \code{gizmo}, employ Godunov-type solvers in a mesh-free framework using Lagrangian resolution elements, combining the adaptive resolution and conservation properties of particle methods with the accuracy of modern finite-volume schemes \citep{Hopkins2015}. These hybrid methods have demonstrated improved performance in capturing shocks, turbulence, and multiphase flows compared to traditional SPH, while retaining many of its advantages, and are increasingly used in astrophysical simulations ranging from star formation to cosmological structure formation.

\subsection{MHD divergence control}

A key challenge in numerical MHD is maintaining the solenoidal constraint, $\nabla \cdot \mathbf{B} = 0$ (Eq.~\ref{eq:mhd5}), which reflects the absence of magnetic monopoles and is essential for physical consistency. If this condition is violated, numerical errors can generate unphysical forces parallel to the magnetic field, leading to incorrect dynamics and potential numerical instability \citep{BrackbillBarnes1980}. Broadly speaking, numerical approaches fall into three categories: constrained transport (CT), divergence cleaning methods, and formulations that control divergence errors through the discretisation itself. Among these, CT methods \citep{EvansHawley1988} are widely regarded as the most accurate, because they preserve $\nabla \cdot \mathbf{B} = 0$ to machine precision by evolving magnetic fluxes on staggered grids. This is achieved by discretising Faraday's law in a manner consistent with Stokes' theorem, ensuring that the divergence-free property is maintained exactly by construction. CT has become the standard in many grid-based MHD codes.

Alternative approaches include divergence cleaning and divergence-controlled formulations, which are particularly useful in mesh-free or particle-based methods. Hyperbolic or parabolic divergence cleaning schemes \citep{DednerEtAl2002} introduce an auxiliary scalar field that propagates and damps divergence errors away from their source, reducing violations of the constraint without enforcing it exactly. These methods are widely used in grid-based \citep[e.g.,][]{WaaganFederrathKlingenberg2011} and particle-based codes \citep{TriccoPrice2012,Hopkins2015}. Another class of approaches, such as the eight-wave formulation \citep{PowellEtAl1999}, modifies the MHD equations to transport divergence errors passively with the flow, preventing their accumulation but not eliminating them. While cleaning and transport methods are generally more flexible and easier to implement in complex or unstructured geometries, they do not guarantee exact preservation of the divergence-free condition, and their effectiveness depends on resolution and numerical parameters. As a result, CT remains the preferred approach where feasible, while cleaning methods provide a practical and robust alternative for mesh-free and hybrid numerical schemes. For example, \citet{WaaganFederrathKlingenberg2011} have quantified the impact of divergence cleaning in their method and find that the fractional energy associated with $\nabla \cdot \mathbf{B} = 0$ is always $<10^{-6}$, even in highly complex three-dimensional problems involving supersonic MHD turbulence.

\subsection{Initial and boundary conditions} \label{sec:ics}

A recurring issue in simulations of star formation is the interconnection of physical processes across vastly different scales. This coupling operates in both directions: large-scale dynamics can strongly influence the evolution of star formation on small scales, while feedback from stars can in turn affect the large-scale environment. This poses a severe challenge, as we would ideally like to include all relevant length and time scales in order to solve the star formation problem self-consistently{\emdash}that is, starting from a much larger scale (e.g., the scale of an entire galaxy, or at least a significant patch thereof) than the one we ultimately wish to resolve, such as a forming star cluster, while simultaneously capturing the effects of the galactic environment and feedback from nearby star-forming regions. Achieving this at the maximum possible resolution to ensure numerical convergence and robustness of the results remains elusive with current numerical methods and supercomputing capabilities. Therefore, all calculations must adopt approximations, both in terms of the initial conditions and the boundary conditions.

Popular techniques for initializing a star formation simulation include placing a uniform sphere of gas with a chosen radial density profile \citep[e.g.,][]{BateBonnellBromm2003,FederrathBanerjeeClarkKlessen2010,GirichidisEtAl2011,GirichidisEtAl2012a,GirichidisEtAl2012b} in a box with isolated boundary conditions for hydrodynamics and gravity, and adding a turbulent velocity field at the start to promote the formation of structure reminiscent of molecular clouds. Another common approach models a sub-volume of a molecular cloud with periodic boundary conditions and drives turbulence with an idealized energy injection mechanism \citep[e.g.,][]{KlessenHeitschMacLow2000,JappsenEtAl2005,FederrathKlessen2012,KrumholzKleinMcKee2012,GuszejnovEtAl2022,MathewFederrathSeta2023}. Both techniques have advantages and disadvantages: although real molecular clouds are neither truly periodic nor completely isolated from the surrounding ISM, some approximation in boundary conditions is unavoidable. Moreover, since turbulence decays on roughly a crossing time \citep{StoneOstrikerGammie1998,MacLow1999}, some form of continuous driving is required to sustain realistic turbulent motions.

A recent method combines the advantage of continuous turbulence driving with the ability to use isolated gas and gravity boundaries by imposing an external gravitational potential that confines the cloud during the driving phase \citep{LaneEtAl2022}. This and the aforementioned idealized approaches allow for controlled parameter studies, as properties such as the sonic Mach number, mean density, and size of the target cloud or region can be set explicitly via the initial conditions. However, these methods still suffer from artificial boundary and initial conditions\footnote{Periodic boundary conditions may appear particularly unrealistic, but given the large spatial extent of the ISM in a galaxy, a cloud patch of a few parsecs in size can be regarded as a specific realization of a much larger turbulent medium. In this sense, periodic boundaries can be viewed as mimicking the tidal effects and gas flows induced by surrounding, similar cloud regions.}.

Finally, a method that avoids the use of artificial turbulence driving altogether involves initializing the simulation on a much larger scale than the ultimate scale of interest, and then \inquotes{zooming in} on that region by selectively adding higher resolution elements or by re-simulating an extracted subdomain. This technique has been employed for both large-scale cloud simulations of star formation \citep[e.g.,][]{ZamoraAvilesVazquezSemadeni2014,SeifriedEtAl2017,IbanezMejiaEtAl2017,KimOstriker2017,GrisdaleEtAl2018,VazquezSemadeniEtAl2019,HaidEtAl2019,ZhaoEtAl2024,HopkinsEtAl2024,PillsworthEtAl2025}, as well as for smaller-scale simulations of individual accretion disks \citep[e.g.,][]{OffnerEtAl2009,KuffmeierHaugbolleNordlund2017,Bate2018,LebreuillyEtAl2021,HeRicotti2023,YangFederrath2025,MayerEtAl2025}. Ideally, once the zoom-in region is sufficiently refined, the scales of interest are far enough removed from the numerical boundaries of the parent simulation that boundary effects are negligible within the region of interest, allowing for re-simulation of this subregion. In the case of zoom-ins, the high-resolution region is no longer treated as isolated and it continues to exchange mass and energy with the larger environment via accretion and feedback effects as it evolves. Likewise, the initial conditions are more \inquotes{natural} in this approach, as they emerge self-consistently from the larger-scale flows in the parent simulation, such as from the formation of the dense, collapsing part of the cloud itself. While this is certainly a more realistic approach to initializing a star-formation simulation, it comes at the cost of making parameter studies considerably harder, as there is little direct control over the cloud parameters, apart from selecting different zoom-in regions from the larger parent simulation.

\subsection{Numerical advances and challenges}

\subsubsection{Positivity-preserving schemes} \label{sec:robust_mhd}

A major advance in the development of robust numerical schemes for MHD has been the introduction of positivity-preserving solvers. Building on the theory of nonlinear stability for finite-volume methods, \citet{BouchutKlingenbergWaagan2007} formulated a relaxation-based approximate Riemann solver that guarantees non-negative density and pressure while remaining consistent with entropy stability. The relaxation framework replaces the MHD fluxes with a multi-wave system that is easier to control numerically, yet faithfully represents the full wave structure of MHD. Practical three- and five-wave implementations \citep{WaaganFederrathKlingenberg2011} demonstrated that this approach provides a computationally efficient alternative to standard approximate solvers, with the added benefit of strict positivity preservation{\emdash}an essential feature for simulations involving strong shocks, rarefactions, and vacuum-like states, such as those encountered in supersonic turbulence and star-formation applications. More recent adaptations and extensions of this include works by \citet{DerigsEtAl2016} and \citet{BirkeEtAl2021}, with the latter allowing accurate solutions with significantly reduced dissipation in flows with low Mach numbers \citep[see also][]{LeidiEtAl2022,WattEtAl2025}.

For example, in simulations of supersonic turbulence with $\mach\sim10$ \citep{LeeDeaneFederrath2009}, even robust solvers such as HLLD (Harten--Lax--van~Leer Discontinuities) \citep{MiyoshiKusano2005} require careful tuning and the use of relatively low CFL factors \citep{CourantFriedrichsLewy1928} to maintain stability. By contrast, turbulence simulations exceeding $\mach\sim20$ can be evolved stably at high resolution with the positivity-preserving HLLxR scheme \citep[Harten--Lax--van~Leer Relaxation, where the \inquotes{x} denotes the 3-wave or 5-wave implementation; see][]{WaaganFederrathKlingenberg2011}, even though increasing resolution steepens gradients and amplifies density contrasts in isothermal shocks, making stability more challenging \citep{Federrath2013}. A similar advantage has been demonstrated in simulations of jet launching and propagation \citep[e.g.,][]{SeifriedEtAl2012}, where the HLLxR scheme has proven highly effective. We note that the original \citet{WaaganFederrathKlingenberg2011} scheme uses $\nabla\cdot\Bvec$ cleaning instead of constrained transport \citep[CT; ][]{EvansHawley1988} to approximate $\nabla\cdot\Bvec=0$, however, the extension of this scheme in \citet{WattEtAl2025} makes direct use of CT.

\subsubsection{Hybrid precision} \label{sec:hybrid-prec}

A promising way to accelerate future simulations while simultaneously increasing numerical resolution is to employ a \inquotes{hybrid-precision} approach. In such schemes, all hydrodynamical variables are stored in single precision (4~bytes per floating-point number), while critical operations are performed in double precision (8~bytes per floating-point number). This design retains the accuracy of full double-precision calculations while reducing memory usage and compute time. Operations that require double precision{\emdash}such as summations over large numbers of resolution elements (e.g., for volume or mass averages){\emdash}are explicitly forced to run in double precision.

\begin{figure}[t]
\centerline{\includegraphics[width=1.0\linewidth]{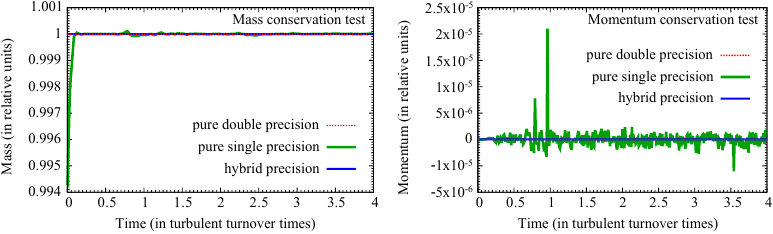}}
\caption{\label{fig:hb}
Comparison of double-precision, single-precision, and hybrid-precision schemes for supersonic MHD turbulence. The hybrid scheme (blue line) matches the accuracy of double precision (red dotted line), conserving gas mass (left) and momentum (right), while single precision (green line) exhibits significant errors. At the same time, the hybrid scheme provides a computing speed-up and reduces memory use by a factor of $\sim2$ relative to double precision.}
\end{figure}

For example, \citet{FederrathEtAl2021} implemented a hybrid-precision algorithm for MHD turbulence simulations, achieving performance gains of \mbox{$\sim3$--$4\times$} over the standard (public) version of the \code{flash} code (Fig.~\ref{fig:scaling}). Independent of other optimizations, switching to hybrid precision alone typically yields a \mbox{$\lesssim2\times$} speed-up, provided communication is bandwidth-limited rather than latency-limited{\emdash}a condition that can be met by packing MPI messages and thereby reducing communication frequency. Additional optimizations, such as removing unused 3D fields and inlining the EOS into the MHD solver, further improved runtime and memory usage, leading to the overall factor of 3--4 reported in \citet{FederrathEtAl2021}. Figure~\ref{fig:hb} compares pure double-precision and single-precision runs with the hybrid scheme for a standard turbulence setup. Gas mass (left panel) and momentum (right panel) remain well conserved in the hybrid scheme, while significant errors appear in pure single-precision mode.

Hybrid-precision methods reproduce high-order statistical moments and power spectra of turbulence without significant loss of accuracy. They therefore hold strong promise for enabling very high-resolution simulations while reducing computational cost and carbon footprint. The method was recently extended to a particle-in-cell (PIC) code \citep{AchikanathFederrathSeta2024}, demonstrating its potential beyond grid-based hydrodynamics. Future extensions to other physics modules, such as gravity and radiation, may provide further benefits. Hybrid-precision schemes are becoming increasingly relevant in HPC, particularly with the rise of GPU computing, where limited working memory and performance constraints often necessitate single-precision operations\footnote{Even FP64 GPUs typically run at $2\times$ speed in single precision compared to double precision.}.

\subsubsection{GPU acceleration} \label{sec:gpu}

GPU acceleration has become central to HPC, and more codes are adapting to GPUs for MHD and other physics capabilities. However, using GPUs well hinges on performance portability across rapidly changing vendor ecosystems. Two community portability layers are often used: Kokkos\footnote{\url{https://github.com/kokkos}} (templated C$^{++}$ abstractions targeting CUDA/HIP/SYCL/HPX/OpenMP) and AMReX\footnote{\url{https://amrex-codes.github.io/amrex}} (GPU-aware, block-structured AMR framework). Both address common hotspots: finite-volume loops, Riemann solves, reconstruction, and divergence control, while exposing GPU-friendly iteration and memory models. Persistent challenges include bandwidth-bound kernels, AMR data motion and kernel launch overheads, multi-GPU scaling with overlap of communication/compute, and maintaining numerical reproducibility and stability when using mixed precision (cf.~Fig.~\ref{fig:hb}) or fast-math options.

The \code{athenak}, \code{flash-x}, and \code{quokka} astrophysics codes are a few example codes supporting GPUs. \code{athenak}\footnote{\url{https://github.com/IAS-Astrophysics/athenak}} \citep{StoneEtAl2026} is a performance-portable implementation of the \code{athena}$^{++}$ AMR framework that uses Kokkos to run MHD on CPUs and GPUs, providing a single code base for heterogeneous systems and multi-GPU capability via Kokkos backends. Under significant development, \code{flash-x}\footnote{\url{https://github.com/Flash-X}} \citep{DubeyEtAl2022,DhruvEtAl2023} aims to adopt a code-generation–driven performance-portability strategy and options for Fortran–Kokkos as well as Fortran-AMReX interoperability layers to offload selected kernels to GPUs while preserving the long-standing multiphysics architecture provided in \code{flash}. \code{quokka}\footnote{\url{https://github.com/quokka-astro}} \citep{WibkingKrumholz2022} targets GPU-resident MHD as well as radiation hydrodynamics with AMR using AMReX, with strong single-GPU throughput and a unified CPU/GPU code base.

\subsection{Summary of challenges and future directions}

The MHD equations provide the foundation for modeling magnetized astrophysical fluids, yet their accurate and efficient numerical solution remains challenging. While ideal MHD is often justified on cloud and core scales, non-ideal effects can become important in dense regions, and consistently capturing these processes across the vast range of relevant scales remains difficult. Even in ideal MHD, numerical diffusion, resolution limits, and thermodynamic simplifications introduce uncertainties that must be carefully evaluated. Maintaining the solenoidal constraint of the magnetic field, $\nabla\cdot\Bvec=0$, is another central requirement; constrained transport provides the most accurate solution on structured meshes, while divergence-cleaning approaches remain essential for mesh-free and hybrid methods.

Mesh-based, particle-based, and hybrid discretization strategies each offer distinct strengths. Grid-based Godunov schemes excel at shock capturing and turbulence modeling, particle methods naturally adapt to large density contrasts, and hybrid approaches increasingly combine these advantages. Advances such as positivity-preserving solvers have improved robustness in highly supersonic and magnetized flows, yet stability and accuracy at extreme Mach numbers and density contrasts remain demanding.

Initial and boundary conditions pose additional limitations, as simulations cannot yet self-consistently cover the full range of scales from galactic environments to protostellar disks. Idealized setups enable controlled studies but introduce artificial constraints, whereas zoom-in techniques offer greater realism at higher computational cost with less flexibility in exploring the influence of different physical parameters.

Hybrid-precision algorithms and GPU acceleration are enabling higher resolution and improved physical fidelity, although performance portability and numerical robustness across heterogeneous architectures remain active areas of development. Continued progress in numerical methods for MHD and high-performance computing will be essential for advancing turbulence and star-formation simulations.

\section{Turbulence} \label{sec:turbulence}

Turbulence is a fundamental ingredient of star formation, with its properties shaping the structure and dynamics of molecular clouds, the birthplaces of stars \citep{ScaloElmegreen2004,ElmegreenScalo2004,MacLowKlessen2004,McKeeOstriker2007,HennebelleFalgarone2012,PadoanEtAl2014}. The statistical nature of turbulent gas, including MHD \citep[e.g.,][]{BeresnyakLazarian2019}, its density probability distribution function and power spectra, provides essential input for theories of the SFR \citep{KrumholzMcKee2005,PadoanNordlund2011,HennebelleChabrier2011,FederrathKlessen2012,Burkhart2018,HennebelleBrucyColman2024} and the stellar IMF \citep{PadoanNordlund2002,HennebelleChabrier2008,HennebelleChabrier2009,HennebelleChabrier2011,HennebelleChabrier2013,Hopkins2012a,Hopkins2013IMF,GuszejnovHopkinsKrumholz2017,LeeEtAl2020}. Capturing turbulence is therefore critical for numerical simulations of star formation, and we begin this review by examining the main challenges and current state of the art in turbulence modeling.

\subsection{Challenges in modeling turbulent flows} \label{sec:challenges_turbulence}

The primary challenge in modeling supersonic turbulence lies in capturing the dynamic range required to achieve fully-developed turbulent flows{\emdash}that is, to reproduce the self-similar characteristics and statistics of turbulence across the relevant density, velocity, and spatial scales. For molecular clouds, the turbulence spans more than six orders of magnitude in scale, from the cloud size of $\sim10$--$100\,\mathrm{pc}$ down to the dissipation scale, below AU scales. Importantly, this may not be a single universal \inquotes{cascade} (or spectrum){\emdash}although often approximated as such{\emdash}because different energy injection mechanisms can drive turbulence on different scales within molecular clouds. Turbulence continues to play a crucial role in fragmentation and in shaping the dynamics of accretion disks, where turbulence may be driven by shearing and magnetic instabilities.

The difficulty can be quantified by the Reynolds number, a dimensionless parameter measuring the ratio of advective transport to viscous diffusion. It is defined as
\begin{equation}
\rek = \frac{\ldriv\,\vturb}{\nu},
\end{equation}
where $\vturb$ is the characteristic turbulent velocity at the largest scale (here, the cloud scale $\ldriv$), and $\nu$ is the kinematic viscosity (cf.~Sect.~\ref{sec:mhd}). Different Reynolds numbers correspond to different flow behavior. For \mbox{$\rek\lesssim10$}, flows remain laminar. At $\rek\sim100$, a K\'arm\'an vortex street appears, marking the transition toward turbulence, which becomes fully developed for $\rek\gtrsim1,\!000$ \citep{Frisch1995}.

In the ISM, $\rek$ typically ranges from $\sim10^5$ to $10^{10}$, confirming that it is in a state of ubiquitous, fully developed turbulence. However, even the most powerful current supercomputers only achieve simulations with effective $\rek\sim10^5$ (see below). The challenge is the enormous dynamic range required, which demands either extremely high grid resolution or, in particle-based methods, very large numbers of particles.

As discussed in detail in Sect.~\ref{sec:num_visc} below, for grid-based codes, $\rek$ depends on the linear grid resolution $N$, as $\rek=(N/N_\rek)^{p_\rek}$, where $N_\rek$ is the number of grid cells resolving the viscous scale (where $\rek=1$), and $p_\rek=1+\zeta$ is determined by the scaling of the turbulent velocity dispersion with size $\ell$. In general, $\vturb(\ell)\propto\ell^\zeta$, with $\zeta=1/3$ and $p_{\rek}=4/3$ for Kolmogorov (incompressible) turbulence \citep{Kolmogorov1941c}, and $\zeta=1/2$ and $p_{\rek}=3/2$ for Burgers (shock-dominated) turbulence \citep{Burgers1948}. Intuitively, one expects $N_\rek\sim1$, since eddies smaller than a grid cell cannot be resolved. Early numerical studies suggested $N_\rek\sim0.5$ \citep{BenziEtAl2008}, but more recent, higher-resolution simulations indicate $N_{\rek}\sim1.5$ for subsonic and $\sim3$ for supersonic turbulence \citep{ShivakumarFederrath2025}. Using $N_\rek=1.5$ and $p_\rek=4/3$--$3/2$, we find $\rek\sim(2.7$--$5.4)\times10^2$, $(8.5$--$17)\times10^3$, and $(1.2$--$5.4)\times10^5$ for $N=100$, $1,\!000$, and $10,\!000$, respectively. Thus, fully turbulent flows with $\rek\gtrsim1,\!000$ require $N\gtrsim270$. Much higher resolution is necessary, however, to establish a turbulent cascade with sufficient dynamic range to measure inertial-range statistics such as scaling exponents and correlation functions.

\begin{figure}[t]
\centerline{\includegraphics[width=0.89\linewidth]{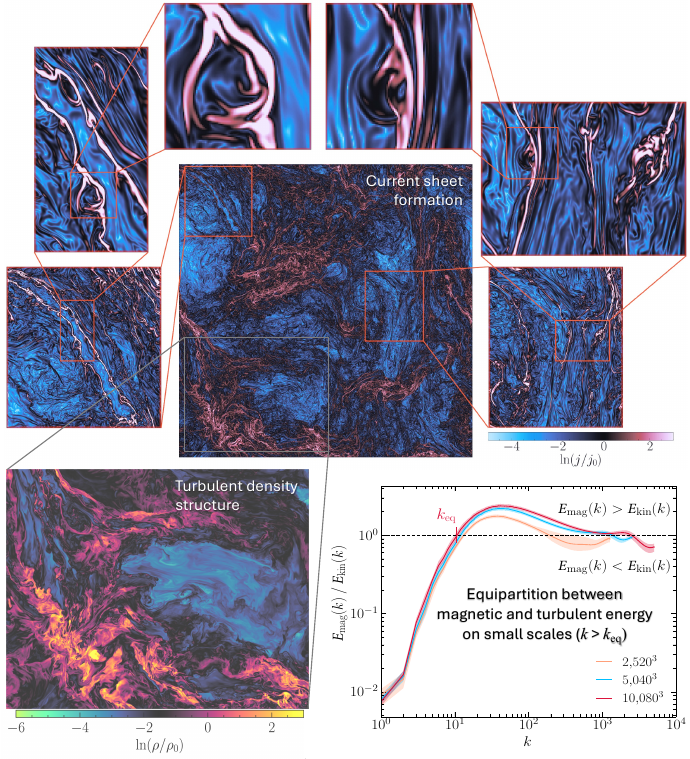}}
\caption{
Zoom into the magnetic current structures (top panels) and into the gas density (bottom left), together with the power spectrum of the magnetic-to-kinetic energy ratio $E_\mathrm{mag}/E_\mathrm{kin}$ (bottom right) in a supersonic MHD turbulence simulation with $10,\!080^3$ grid cells from \citet{BeattieEtAl2025}. Supersonic turbulence produces density contrasts spanning several orders of magnitude and intricate current structures across scales, ultimately forming strongly magnetized plasma structures on small scales. Accurately modeling these requires powerful numerical schemes and extreme resolution. A key finding is that while the magnetic field has only a modest impact on large scales (low $k$), magnetic energy always approaches equipartition with kinetic energy on small scales ($k>k_\mathrm{eq}$). Numerical convergence of the energy content below $k_\mathrm{eq}$ requires resolutions $\gtrsim5,\!000^3$. Adapted from Figs.~1 and 2, and Suppl.~Fig.~3 in \citet{BeattieEtAl2025}.}
\label{fig:beattie}
\end{figure}

The current state of the art has reached resolutions of $\sim10,\!000^3$. For example, \citet{IshiharaEtAl2020} achieved $12,\!288^3$ grid cells in incompressible hydrodynamic turbulence. \citet{DongEtAl2022} ran subsonic MHD turbulence with $10,\!000\times10,\!000\times5,\!000$ grid cells to study magnetic reconnection in the cascade, while \citet{KempskiEtAl2025} reached $10,\!240^3$ for subsonic MHD turbulence. Such extreme resolutions are required to capture the geometry of magnetic current sheets at the smallest scales. In the supersonic regime relevant for molecular clouds and star formation, \citet{FederrathEtAl2021} and \citet{BeattieEtAl2025} performed turbulence simulations with $\gtrsim10,\!000^3$ grid cells to resolve the transition from supersonic to subsonic turbulence. This \inquotes{sonic scale} may be a critical ingredient for star formation \citep{VazquezBallesterosKlessen2003,KrumholzMcKee2005,FederrathKlessen2012} and filament formation \citep{ArzoumanianEtAl2013,SmithGloverKlessen2014,SmithEtAl2016,Federrath2016,FederrathEtAl2016,ArzoumanianEtAl2018}. A visualization of the \citet{BeattieEtAl2025} simulation is shown in Fig.~\ref{fig:beattie}. A critical outcome of this is that the magnetic field always tends to be at least as strong as the kinetic field on sufficiently small scales, likely associated with the transition from supersonic to subsonic turbulence.

\subsection{Turbulence initialization and driving} \label{sec:turb_driv}

Since turbulence decays quickly on about a turnover timescale (defined below) \citep{StoneOstrikerGammie1998,MacLowEtAl1998,MacLow1999}, but is ubiquitously observed in the ISM, it must be maintained by some form of energy injection, so-called \inquotes{turbulence driving}. Since there is no closed-form description of turbulence, simulations must begin with an initialization phase to establish a fully-developed turbulent state with the correct statistical properties of density, velocity, and magnetic fields, and their correlations. This is achieved via turbulence driving, which injects energy and momentum until the statistical properties (velocity dispersion, level of density fluctuations, etc.) converge to a quasi-equilibrium state, to mimic the combined effect of energy injection from physical sources such as stellar feedback (supernova explosions, jets/outflows, winds, radiation), galactic or local shear, magneto-rotational instability (MRI), or accretion \citep{Elmegreen2009,FederrathEtAl2017iaus}.

To drive turbulence one can apply an external acceleration field ${\bf F}$ to the MHD system, as source terms $\rho\mathbf{F}$ and $\rho\mathbf{F}\cdot\vel$ on the RHS of Eqs.~(\ref{eq:mhd2}) and~(\ref{eq:mhd3}). In order to have control over the properties of the driving (amplitude, mode mixture, correlation time), the field $\mathbf{F}$ is often generated via a stochastic Ornstein-Uhlenbeck (OU) process \citep{EswaranPope1988,SchmidtEtAl2009,FederrathDuvalKlessenSchmidtMacLow2010,PriceFederrath2010}, which evolves smoothly in space and time. Its auto-correlation time is usually set to the turbulent (eddy) turnover time, $\tturb=\ldriv/(\mach\cs)$, on the largest (integral) scales, where $\ldriv=L/2$ is the driving scale, $L$ the domain size, and $\mach$ the sonic Mach number defined by $v_\mathrm{turb}=\mach\cs$. These choices provide consistency between time and length scales in a periodic computational domain where the driving is often confined to large scales, providing fully-developed turbulence on smaller scales. However, the influence of different choices for the correlation time have also been explored recently \citep{GreteEtAl2025}.

In this case, the driving is constructed in Fourier space, usually injecting kinetic energy at the lowest wave numbers, $1<\abs{\mathbf{k}}L/2\pi<3$, with the peak at $k=2$ (the $L/2$ scale). While different driving spectra can be used, a good choice is a driving amplitude that declines parabolically toward $k=1$ and $k=3$, confining direct driving to a narrow band of scales and allowing turbulence to develop naturally on $k\geq3$. Some works have explored alternative injection scales, with spectra peaking at $k=3$--$4$ or $7$--$8$ \citep{Klessen2001,WalchWhitworthGirichidis2012,KochEtAl2019}, or even across an entire power-law spectrum \citep{NamFederrathKrumholz2021}, to study the role of the driving scale in turbulence statistics, the emergence of inverse cascades \citep{BrandenburgSharmaVachaspati2023}, and implications for star formation and the IMF \citep{Klessen2001,Bate2009init,GirichidisEtAl2011,NamFederrathKrumholz2021}.

By applying a Fourier-space projection, $\mathbf{F}$ can be decomposed into solenoidal ($\nabla\cdot\mathbf{F}=0$) and compressive ($\nabla\times\mathbf{F}=0$) parts via Helmholtz decomposition. In index notation, the projection operator is
\begin{equation}
\mathbb{P}_{ij}^\zeta(\mathbf{k}) = \zeta\,\delta_{ij}+(1-2\zeta)\,k_i k_j/\abs{\mathbf{k}}^2.
\end{equation}
This yields a ratio of compressive-to-total driving power of
\begin{equation}
\frac{F_\mathrm{comp}}{F_\mathrm{tot}} = \frac{(1-\zeta)^2}{1-2\zeta+3\zeta^2}.
\end{equation}
Thus, $\zeta=1$ produces purely solenoidal driving, while $\zeta=0$ yields purely compressive driving. Intermediate $\zeta$ values give mixtures, with $\zeta=1/2$ producing $F_\mathrm{comp}/F_\mathrm{tot}=1/3$ (\inquotes{natural mixture}), equivalent to randomly selecting modes in three dimensions, where on average one axis is compressive and two are solenoidal \citep{FederrathKlessenSchmidt2008}.

The effects of the driving mode on structure formation have been widely studied \citep[e.g.,][]{KochEtAl2017,MohapatraFederrathSharma2022}. In particular, \citet{FederrathKlessen2012} and \citet{MathewFederrathSeta2023} show that compressive driving yields $\sim10\times$ higher SFR and reduces the IMF characteristic mass by a factor of $\sim2$ compared to solenoidal driving. While the resolution may not fully resolve the low-mass IMF and binary formation, they find that solenoidal driving results in a measurable difference in the IMF compared to compressive driving.

Many codes have implemented turbulence driving via the OU process, however, only few support the Helmholtz mode decomposition. A turbulence-driving module that implements the concepts above is publicly available as a maintained C$^{++}$ code on GitHub \citep[][\url{https://github.com/chfeder/turbulence_generator}]{FederrathEtAl2022ascl}, with MHD code plugins that currently support \code{flash} and \code{quokka} directly as Git sub-modules. It also provides more general-purpose Python interfaces. This basic code was also implemented in \code{arepo}, \code{gadget}, \code{phantom}, and \code{pluto}.

In addition to, or as an alternative approach to driving via the OU process, simulations can include direct physical drivers such as supernovae \citep[e.g.,][]{KorpiEtAl1999,JoungMacLow2006,MeeBrandenburg2006,GresselEtAl2008,GentEtAl2013,GentEtAl2020,GentEtAl2021}, protostellar jets \citep[e.g.,][]{NakamuraLi2007,CunninghamEtAl2009,FederrathEtAl2014,MurrayGoyalChang2018,GuszejnovEtAl2021}, or MRI \citep[e.g.,][]{BalbusHawleyStone1996,PessahChanPsaltis2007,FromangStone2009}. While these provide a more natural and realistic mechanism for energy injection and turbulence driving, they make it more difficult to perform controlled, systematic parameter studies and to characterize the driving in terms of general properties, such as the mode mixture, compared to OU driving.

\subsection{Physical shear and bulk viscosity} \label{sec:shear_bulk_visc}

The traceless rate-of-strain tensor $\mathbb{S}$ (Eq.~\ref{eq:strain_tensor}), used in the MHD momentum and energy equations (Eqs.~\ref{eq:mhd2} and~\ref{eq:mhd3}), accounts only for shear viscosity. More generally, however, viscous stresses are proportional to all components of the velocity gradient tensor $\nabla\vel$, and compressible gases therefore exhibit both shear and bulk (volume) viscosity. Accordingly, the generalized Navier-Stokes equation includes not only the shear viscosity $\nushear$, but also the bulk viscosity coefficient $\nubulk$, which becomes relevant for compressible flows where $\nabla\cdot\vel\neq0$ \citep{BeattieEtAl2025visc}. Therefore, writing Eq.~(\ref{eq:mhd2}) (for simplicity omitting magnetic terms), including $\nushear$ and $\nubulk$ yields
\begin{eqnarray}
\ddt{(\rho\vel)} & = & -\nabla\cdot\left(\rho\vel\vel\right) - \nabla\pth \nonumber \\
 & & + \nabla\cdot\left(2\rho\nushear\mathbb{S}\right) + \nabla\cdot\left[\rho\nubulk(\nabla\cdot\vel)\mathbb{I}\right].
\end{eqnarray}

For monatomic ideal gases, the bulk viscosity is identically zero, as shown by kinetic theory \citep{Mihalas1984}. In contrast, for polyatomic molecules, $\nubulk$ can be non-zero, but only if relaxation processes occur on timescales comparable to or longer than typical fluid timescales \citep{Mihalas1984}. The magnitude of $\nubulk$ therefore depends strongly on the composition of the polyatomic gas, and direct measurements remain uncertain \citep{Tisza1942}. Because turbulence studies have traditionally focused on incompressible gases, bulk viscosity is often omitted from the governing equations. However, in compressible gases and plasmas{\emdash}particularly relevant for the ISM and star formation{\emdash}bulk viscosity may play a significant role and thus warrants further investigation.

A recent parameter study provides new insights into the role of bulk viscosity. While the relative importance of shear and bulk viscosity depends on gas composition and remains uncertain due to laboratory measurement challenges, \citet{BeattieEtAl2025visc} estimate ratios as high as $\nubulk/\nushear\sim1$--$100$ for the molecular phase of the ISM, where stars form. This implies measurably enhanced kinetic energy dissipation on small scales, with the strongest impact on the compressible component of the velocity field. However, \citet{BeattieEtAl2025visc} also find that bulk viscosity has relatively little influence on magnetic field amplification by the turbulent dynamo. This is because the dynamo is primarily powered by vorticity, even in highly compressible plasmas \citep{FederrathEtAl2011,FederrathEtAl2014,AchikanathEtAl2021}, and the vortical component of the velocity field is not directly affected by bulk viscosity.

\subsection{Sub-resolution models for turbulence}

Given the challenges of resolving turbulent flows (see Sect.~\ref{sec:challenges_turbulence}), it is essential to recognize the limitations imposed by finite resolution and to develop methods for quantifying numerical dissipation. These methods can then be used to model turbulence on unresolved scales and to estimate key physical parameters on the grid scale in cosmological or galaxy-scale simulations. Such parameters include the turbulent Mach number, the virial parameter, the effective driving mode of turbulence, and the plasma beta (in the presence of magnetic fields). These sub-resolution properties provide crucial inputs for turbulence-based star formation models, for example by regulating the amount of local stellar feedback to be injected in large-scale simulations. In this way, sub-resolution turbulence models serve as a critical bridge between small-scale physics, which cannot be directly resolved, and the global dynamics of galaxies and star-forming regions, thereby enabling simulations to incorporate the essential impact of turbulence across vastly different scales.

\subsubsection{Numerical viscosity and resistivity} \label{sec:num_visc}

The first step in sub-resolution modeling is to quantify and predict the levels of numerical viscosity and resistivity{\emdash}specifically, the effective kinetic and magnetic Reynolds numbers that can be achieved for a given grid or particle resolution. To address this, \citet{KrielEtAl2022,KrielEtAl2025} conducted an extensive suite of turbulence simulations in both the subsonic and supersonic regimes, explicitly including physical viscosity and resistivity terms. By systematically varying $\nu$ and $\eta$ while changing the grid resolution, they demonstrated that converged results are obtained only above a threshold resolution $N_\mathrm{thresh}$ (where $N$ is the number of grid cells per dimension). If $N<N_\mathrm{thresh}$ for a target $\rek$ and/or $\rem$, then numerical viscosity and/or resistivity dominate over the explicit terms, such that the effective $\rek$ and $\rem$ fall below the intended values.

\begin{figure}[t]
\centerline{\includegraphics[width=1.0\linewidth]{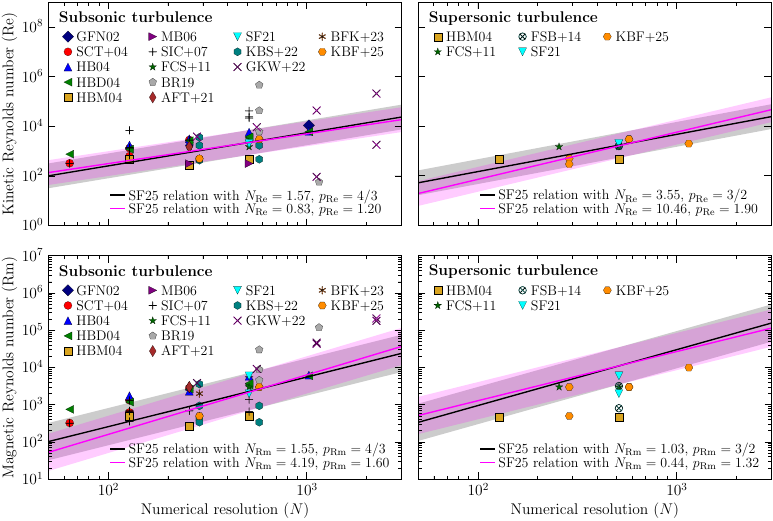}}
\caption{
Kinetic Reynolds number ($\rek$, top) and magnetic Reynolds number ($\rem$, bottom) as functions of the number of resolution elements ($N$) along one side of a cubic domain, shown for the subsonic (left) and supersonic (right) turbulence regimes. Data points represent simulations with explicit dissipation, i.e., $\rek$ and $\rem$ values set by the authors of the respective works (see legend): GFN02 \citep{GotohFukayamaNakano2002}, SCT+04 \citep{SchekochihinEtAl2004}, HB04 \citep{HaugenBrandenburg2004dyn}, HBD04 \citep{HaugenBrandenburgDobler2004}, HBM04 \citep{HaugenBrandenburgMee2004}, MB06 \citep{MeeBrandenburg2006}, SIC+07 \citep{SchekochihinEtAl2007}, FCS+11 \citep{FederrathEtAl2011}, BR19 \citep{BrandenburgRempel2019}, AFT+21 \citep{AchikanathEtAl2021}, SF21 \citep{SetaFederrath2021}, KBS+22 \citep{KrielEtAl2022}, GKW+22 \citep{GalishnikovaEtAl2022}, BFK+23 \citep{BeattieEtAl2023}, and KBF+25 \citep{KrielEtAl2025} for the subsonic regime, and HBM04, FCS+11, FSB+14 \citep{FederrathSchoberBovinoSchleicher2014}, SF21, and KBF+25 for the supersonic regime. The lines show $\rek$-$N$ and $\rem$–$N$ relations derived by \citet{ShivakumarFederrath2025}, with parameters listed in each panel. These relations provide estimates of the maximum $\rek$ and $\rem$ achievable for a given $N$ due solely to numerical dissipation. Shaded regions indicate the associated uncertainties, accounting for methodological variations as well as differences across 14~numerical schemes, including different grid-based methods and SPH.}
\label{fig:sf25}
\end{figure}

Using the relations in \citet{KrielEtAl2022,KrielEtAl2025} as a foundation, \citet{ShivakumarFederrath2025} conducted a large parameter study with varying $N$ and purely numerical dissipation, deriving relations between the resolution $N$ and the achievable $\rek$ and $\rem$,
\begin{align}
\rek &= \left(N/N_\rek\right)^{p_\rek}, \\
\rem &= \left(N/N_\rem\right)^{p_\rem},
\end{align}
where the driving scale of turbulence is taken as $L/2$, with $L$ the domain side length covered by $N$~resolution elements in each direction. The parameters $N_\rek$, $p_\rek$, $N_\rem$, and $p_\rem$ were determined from high-resolution implicit large-eddy simulations (ILES) spanning $N^3=144^3$ to $2,\!576^3$, using spectral fitting. Figure~\ref{fig:sf25} summarizes the achievable Re and Rm as functions of $N$ based on these relations. Solid lines show the fits from \citet{ShivakumarFederrath2025}, with shaded regions spanning one order of magnitude to represent maximum uncertainty. Two sets of relations are displayed: black lines fix $p_\rek=p_\rem=4/3$ for subsonic turbulence and $p_\rek=p_\rem=3/2$ for supersonic turbulence, consistent with theoretical scaling expectations \citep{FederrathEtAl2021,ShivakumarFederrath2025}. Magenta lines, in contrast, allow $p_\rek$ and $p_\rem$ to vary freely. Parameter values are reported in the figure legends. The shaded bands capture uncertainties from fitting, methodological variations, and code-to-code differences across a wide range of numerical schemes compiled in \citet{ShivakumarFederrath2025}, including grid-based solvers (with/without Riemann solvers, flux limiters, etc.) and SPH \citep{PriceEtAl2018}.

The data points in Figure~\ref{fig:sf25} represent simulations from the literature reporting target $\rek$ and $\rem$ values obtained with explicit viscosity and resistivity in the MHD equations (Eqs.~\ref{eq:mhd1}--\ref{eq:mhd5}). However, even with explicit dissipation included, numerical dissipation remains present and may dominate depending on $N$. Simulations lying significantly above the SF25 relations are dominated by numerical dissipation, while those within or below the relations are converged with respect to their chosen $\rek$ and/or $\rem$.  

In practice, this means that simulations cannot reach high Reynolds numbers simply by lowering the explicit viscosity or resistivity; sufficient resolution is also essential. For example, the SF25 relations show that at grid resolutions routinely achievable today ($N^3\sim1,\!024^3$), the effective $\rek\sim10^3-10^4$ for both subsonic and supersonic turbulence. While this is nominally turbulent, i.e., $\rek\gtrsim1000$ \citep{Frisch1995}, it remains far below the ISM values of $\rek\sim10^9$, which would require $N\sim10^6-10^7$ depending on turbulence regime and the uncertainties in the SF25 relations. The largest current simulations ($N\sim10,\!000$) reach $\rek\sim10^5-10^6$ \citep{IshiharaEtAl2020,FederrathEtAl2021,BeattieEtAl2025,KempskiEtAl2025}. Further progress will depend on improved algorithms, continued code optimization, and advances in supercomputing technology (cf.~Sect.~\ref{sec:parallel_scaling}).

\subsubsection{Local kinetic energy dissipation rate} \label{sec:local_diss}

To quantify the local viscosity (numerical and, if present, explicit), we need to determine what the kinetic energy of the system would have been in the absence of dissipation. Although this may appear elusive, it can be obtained by deriving an evolution equation for the kinetic energy itself\footnote{By \inquotes{perfectly evolved kinetic energy} we mean dissipation-free within the order of accuracy and limitations of the numerical scheme.}, analogous to the total energy equation. Applying $\vel\cdot(\dots)$ to Eq.~(\ref{eq:mhd2}) yields \citep[see][]{TroccoliFederrath2026},
\begin{equation} \label{eq:ekin}
\ddt{E_\mathrm{kin}} = -\nabla\cdot\left[\left(E_\mathrm{kin}+p\right)\vel - \frac{1}{4\pi}\left(\Bvec\cdot\vel\right)\Bvec\right] + \pth(\nabla\cdot\vel),
\end{equation}
where $E_\mathrm{kin}=\rho\abs{\vel}^2/2$ is the kinetic energy density, and $p=\pth+\abs{\Bvec}^2/(8\pi)$ the total (thermal + magnetic) pressure. Note that the last term only involves $\pth$. Unlike Eq.~(\ref{eq:mhd3}), explicit viscosity terms are omitted here, since the aim is to track numerical dissipation and obtain a local, time-dependent measure of it.

Evolving Eq.~(\ref{eq:ekin}) alongside Eqs.~(\ref{eq:mhd1})--(\ref{eq:mhd5}) offers a direct measure of local dissipation. Denoting post-timestep quantities with a prime, the dissipation-free kinetic energy from Eq.~(\ref{eq:ekin}) is $E_\mathrm{kin}'$. In contrast, the kinetic energy from the standard MHD update is $\rho'\abs{\vel'}^2/2$, which is dissipative\footnote{In practice, MHD codes compute the new thermal energy by subtracting $\rho'\abs{\vel'}^2/2$ from the updated total energy, effectively adding dissipated kinetic energy to the thermal reservoir.}. The key point is that Eq.~(\ref{eq:mhd2}) updates the velocity in a momentum-conserving, rather than energy-conserving, manner. The local kinetic energy dissipation rate is therefore
\begin{equation} \label{eq:epskin}
\varepsilon_\mathrm{kin} = \frac{E_\mathrm{kin}' - \rho'\abs{\vel'}^2/2}{\dt},
\end{equation}
providing a direct measure of the kinetic energy lost through numerical discretization and solver-specific diffusion (e.g., artificial viscosity or slope limiting).

This framework also enables estimates of the sub-resolution turbulent velocity,
\begin{equation}
v_\mathrm{turb,unresolved}=\left(2E_\mathrm{kin}'/\rho' - \abs{\vel'}^2\right)^{1/2},
\end{equation}
from which one can compute the sub-resolution turbulent Mach number or virial parameter. These are central inputs to turbulence-regulated star formation models that determine the SFR \citep{KrumholzMcKee2005,PadoanNordlund2011,HennebelleChabrier2011,FederrathKlessen2012,BurkhartMocz2019,HennebelleBrucyColman2024} and the IMF \citep{PadoanNordlund2002,HennebelleChabrier2008,HennebelleChabrier2009,HennebelleChabrier2013,Hopkins2013IMF,GuszejnovEtAl2022,MathewFederrathSeta2025}. Coupling such dissipation diagnostics with turbulence-based star formation prescriptions offers a promising pathway for sub-resolution star formation models in galaxy and cosmological simulations, where cloud scales are marginally resolved \citep[e.g.,][]{SemenovKravtsovGnedin2016,KangEtAl2025}. A similar approach has recently been implemented in \citet{Semenov2025}.


\subsection{Summary of challenges and future directions}

Turbulence remains one of the fundamental unsolved problems in physics and mathematics. Numerical methods are indispensable for advancing our understanding, but modeling truly turbulent flows, such as those in the ISM, requires Reynolds numbers of order $\sim10^9${\emdash}well beyond the reach of current supercomputers. This gap between physical reality and numerical feasibility remains a central challenge for the field.

Despite these limitations, major progress has been made in recent years. Advances in code optimization, such as the hybrid-precision method (Sect.~\ref{sec:hybrid-prec}), now allow simulations at resolutions up to $10,\!000^3$~grid cells, corresponding to effective Reynolds numbers of order $10^5$, based on improved understanding of how numerical viscosity scales with grid resolution (Sect.~\ref{sec:num_visc}). At the same time, many codes are being modernized for GPU acceleration, opening additional opportunities to push toward higher effective Reynolds numbers in the near future. Another critical advance has been the development of positivity-preserving MHD schemes (Sect.~\ref{sec:robust_mhd}), which remain robust in the presence of strong shocks and discontinuities. These solvers are particularly important for modeling the highly compressible, supersonic turbulence that dominates the star-forming ISM, enabling both fundamental turbulence studies and more reliable simulations of star formation.

Further progress has also been made in capturing local dissipation rates. This includes both numerical and physical dissipation, with extensions beyond the standard shear viscosity to include bulk viscosity effects (Sect.~\ref{sec:shear_bulk_visc}). New methods evolve the kinetic energy equation directly (Sect.~\ref{sec:local_diss}) alongside the standard MHD equations (Sect.~\ref{sec:mhd}), providing novel ways of estimating sub-resolution turbulent velocity dispersions and energies. These quantities are crucial for subgrid star-formation models and can be applied in galaxy-scale simulations. A promising direction for future work is to extend these approaches to quantify the local magnetic energy dissipation rate, providing a more complete view of turbulent energy transfer and its impact on star formation and ISM dynamics.

\section{Radiation hydrodynamics} \label{sec:rhd}

A defining characteristic of stars is that they shine. Forming stars heat their natal gas, and young massive stars ($\gtrsim8\,\msol$) further shape their environment through ionization and radiation pressure \citep{RosenEtAl2020}. In many regimes, the energy and momentum carried by radiation are comparable to those of the gas, making radiation-matter interactions critical for modeling star formation.

Radiation hydrodynamics (RHD), however, remains one of the most difficult areas of computational astrophysics, both numerically and computationally. All current simulations rely on approximations of the full equations. Many studies omit radiation transfer (RT) altogether, replacing it with simplified heating and cooling prescriptions such as modified equations of state or pre-computed tables. These can be adequate in some regimes but fail in others{\emdash}notably in massive star formation, where RT is essential.

It is useful to distinguish RT from radiation feedback. Radiation transport describes how photons propagate and interact with matter, while radiation feedback refers to emission from stars or clusters, which may or may not be coupled to detailed RHD. In simulations that include RHD, feedback typically enters as a source term for radiation energy and flux.

In this section, we review RHD methods\footnote{A recent review on RT methods for star formation is provided by \citet{Wuensch2024}.} used in star-formation modeling. We outline the relevant physical regimes, opacity treatments, and the role of laboratory, co-moving, and mixed frames. We then present the basic RT equations, discuss ray-tracing, and derive the moment equations with three closures: flux-limited diffusion (FLD), moment-1 (M1), and variable Eddington tensor (VET). Section~\ref{sec:rt_mcrt} summarizes advances in Monte Carlo radiation transport (MCRT), and we conclude with a direct comparison of MCRT, FLD, M1, and VET in Sect.~\ref{sec:rt_comp}.

\subsection{Optically-thin vs.~optically-thick regimes} \label{sec:rt_regimes}

Consider a system of size $\ell$ with velocity $v$ and photon mean free path $\lambda_\mathrm{mf}$. The hydrodynamic flow time is $\ell/v$, while the mean time between photon-matter interactions is $\lambda_\mathrm{mf}/c$. The optical depth is $\tau=\ell/\lambda_\mathrm{mf}$. In the optically-thin limit, $\tau\ll1$ (streaming limit), radiation propagates freely at $\sim c$, decouples from the gas, and escapes, allowing efficient cooling. Here isothermality is often a good approximation, since local heat (e.g., produced by shocks or dissipation) is quickly radiated away.

In contrast, for $\tau\gg1$ (diffusion limit), gas and radiation are tightly coupled, photons undergo many scatterings, and radiation propagates diffusively, much slower than $c$. Radiation becomes effectively \inquotes{trapped} in the material, leading to local heating.

The diffusion limit can be subdivided depending on the importance of gas motion \citep{Mihalas1984}. In the static diffusion limit ($\beta\tau\ll1$, with $\beta=v/c$), radiation and gas are strongly coupled and flows are non-relativistic. In the dynamic diffusion limit ($\beta\tau\gg1$), radiation is advected with the gas, and terms describing advection and radiation work dominate over emission and absorption. Even non-relativistic systems with extreme $\tau$, such as stellar interiors ($\tau\sim10^{11}$), can enter this regime. For most star-formation problems{\emdash}dense cores and accretion disks{\emdash}conditions lie in the static diffusion limit \citep{KrumholzEtAl2007}. Jets and outflows, however, carve low-density polar cavities (cf.~Sect.~\ref{sec:outflow}), making the diffusion approximation inaccurate in those directions.

In present-day star-forming regions (unlike in the primordial Universe), dust dominates the opacity, which depends on frequency and temperature \citep{SemenovEtAl2003}. As a result, mean opacities vary by orders of magnitude: dense cores are typically optically thick, while diffuse cloud regions remain optically thin. Accurate RT methods must therefore treat both regimes and, crucially, the transition between them.

\subsubsection{Opacities} \label{sec:rt_opacities}

A central ingredient in radiation transfer is the opacity, $\kappa_\nu$, which determines how radiation interacts with matter. Opacities are intrinsically frequency-dependent, reflecting the microphysical absorption and scattering processes relevant at different photon energies. For practical applications, these frequency-dependent opacities are often averaged to produce a single effective value. Two of the most common averages are the Planck mean opacity,
\begin{equation} \label{eq:kappaP}
\kappaP(T) = \frac{\int_0^\infty \kappa_\nu B_\nu(T)\,d\nu}{\int_0^\infty B_\nu(T)\,d\nu},
\end{equation}
and the Rosseland mean opacity,
\begin{equation} \label{eq:kappaR}
\kappaR(T) = \frac{\int_0^\infty \frac{\partial B_\nu(T)}{\partial T}\,d\nu}{\int_0^\infty \kappa_\nu^{-1}\,\frac{\partial B_\nu(T)}{\partial T}\,d\nu}.
\end{equation}
Here $B_\nu(T)$ is the Planck function. The Planck mean $\kappaP$ weights the opacity by the local emissivity spectrum and is most relevant when describing absorption of stellar or thermal emission. The Rosseland mean $\kappaR$, by contrast, weights low-opacity windows more strongly and is appropriate in the diffusion limit, where radiative flux is carried preferentially through the most transparent frequency channels. Both averages are widely used in RHD simulations, depending on the physical regime.

In present-day star-forming regions, dust provides the dominant source of opacity. The effective dust opacity depends on grain size, composition, and shape, and varies strongly with frequency \citep{SemenovEtAl2003}. In practice, many studies employ pre-computed, tabulated dust opacities as functions of temperature and density \citep[e.g.,][]{SemenovEtAl2003}, which implicitly capture frequency-averaged effects, i.e., they provide $\kappaP(T)$ and $\kappaR(T)$.

Different physical regimes require different opacity prescriptions. For example, in primordial (Population~III) star formation, where metals and dust are absent, the relevant opacities arise from bound-free and free-free transitions and electron scattering. The main contributors in this case are atomic H, He, and molecular H$_2$ (via ro-vibrational lines), with Thomson scattering becoming important at high temperatures \citep{MayerDuschl2005}.

At photon energies above the Lyman limit ($h\nu>13.6\,\mathrm{eV}$), ionizing radiation dominates. Here, the key opacity sources are the photoionization cross sections of neutral hydrogen and helium \citep{VernerEtAl1996,OsterbrockFerland2006}. These frequency-dependent cross sections set the structure of ionization fronts and determine the coupling between radiation and gas in H\,{\sc ii} regions around massive stars. Simulations of reionization and stellar feedback therefore typically adopt tabulated H and He opacities, augmented by Thomson scattering for the highest photon energies \citep{ShardaMenon2025}.

Thus, while dust dominates most present-day star-forming environments, opacity physics is highly problem-dependent. Accurate modeling requires not only capturing the correct frequency dependence, but also adopting the relevant opacity sources for the environment under consideration. Moreover, the frequency dependence of the opacity introduces challenges related to the choice of reference frame in which to formulate the relevant equations.

\subsubsection{Laboratory, co-moving, and mixed reference frames} \label{sec:rt_frames}

The difficulty of solving the RT can be reduced both by approximations and by choosing a suitable frame. In the laboratory frame the observer is at rest, while in the co-moving frame we move with the fluid. Ideally, RT would be formulated in the lab frame to remain consistent with the MHD equations (Eqs.~\ref{eq:mhd1}--\ref{eq:mhd5}), where conservation laws are easiest to implement, especially in Eulerian schemes. However, RT involves frequency-dependent opacities, which in the lab frame require Doppler corrections due to gas-photon relative motion. In the co-moving frame, opacities can be evaluated directly.

To balance these advantages, most RHD methods use a mixed-frame formulation: radiation quantities (energy density, flux, pressure tensor) are defined in the lab frame, while opacities are computed in the co-moving frame. This simplifies microphysics but introduces additional source terms that expand in orders of $v/c$ \citep{Mihalas1984,MihalasAuer2001}. In practice, it is sufficient to retain the leading terms in $v/c$, with the required order depending on whether the system is in the streaming, static, or dynamic diffusion limit. The most general formulation, valid across all regimes, includes terms up to $\mathcal{O}(v^2/c^2)$ \citep[see tab.~1 in][]{KrumholzEtAl2007}, which is adequate for the non-relativistic flows relevant to the ISM and star formation, where $v/c\lesssim10^{-3}$.

\subsection{Basic equations of radiation transfer} \label{sec:rt_eqs}

\subsubsection{Time-dependent radiation transfer}

The basic RT equation describes how the specific intensity ($I_\nu$) changes along a ray. Consider an element of material of length $ds$ along a ray with propagation direction $\nvec$, such that \(d\mathbf{x} = \nvec\, ds\). The infinitesimal difference in the radiant energy carried by the ray segment entering at $(\mathbf{x},t)$ and emerging at \mbox{$(\mathbf{x}+d\mathbf{x},t+dt)$}, is
\begin{equation}
\left[I_\nu(\mathbf{x}+d\mathbf{x},t+dt,\nvec)-I_\nu(\mathbf{x},t,\nvec)\right]dA\,d\Omega\,d\nu\,dt. \label{eq:rt_lhs}
\end{equation}
Here $I_\nu$ ($\mathrm{erg\,\cm^{-2}\,sr^{-1}\,Hz^{-1}\,s^{-1}}$) is the specific intensity at frequency $\nu$, $dA$ is the surface area perpendicular to the direction of propagation $\nvec$, $d\Omega$ is the solid angle, and $dt$ is the time interval. Neglecting scattering\footnote{Neglecting direct scattering is often a reasonable approximation in star-formation studies, where true absorption usually dominates, or scattering can be folded into an effective opacity.} this change in energy is given by the competition between emission and absorption along the path element $ds$ \citep[e.g.,][]{MihalasAuer2001},
\begin{equation}
\left[j_\nu(\mathbf{x},t,\nvec)-\rho\kappa_\nu(\mathbf{x},t,\nvec)I_\nu(\mathbf{x},t,\nvec)\right]ds\,dA\,d\Omega\,d\nu\,dt, \label{eq:rt_rhs}
\end{equation}
where $j_\nu$ ($\mathrm{erg\,\cm^{-3}\,sr^{-1}\,Hz^{-1}\,s^{-1}}$) and $\rho\kappa_\nu$ ($\cm^{-1}$) are the emissivity and extinction coefficient, respectively, with the opacity $\kappa_\nu$ ($\cm^2\,\g^{-1}$). 
Combining Eqs.~(\ref{eq:rt_lhs}) and~(\ref{eq:rt_rhs}), we have
\begin{equation}
I_\nu(\mathbf{x}+d\mathbf{x},t+dt,\nvec)-I_\nu(\mathbf{x},t,\nvec) = \left[j_\nu(\mathbf{x},t,\nvec)-\rho\kappa_\nu(\mathbf{x},t,\nvec)I_\nu(\mathbf{x},t,\nvec)\right]ds.
\end{equation}
Expanding the left-hand side to first order,
\begin{equation}
I_\nu(\mathbf{x}+d\mathbf{x},t+dt,\nvec)-I_\nu(\mathbf{x},t,\nvec) = \frac{\partial I_\nu}{\partial t}\,dt+\nabla I_\nu\cdot d\mathbf{x},
\end{equation}
and using $d\mathbf{x}=\nvec\,ds=\nvec\,c\,dt$ along the ray, we obtain a PDE for the intensity,
\begin{equation} \label{eq:rt}
\left(\frac{1}{c}\ddt{}+\nvec\cdot\nabla\right)I_\nu = j_\nu-\rho\kappa_\nu I_\nu,
\end{equation}
where we have omitted the explicit dependencies $(\mathbf{x},t,\nvec)$ for compactness. This equation depends on seven variables: time, three spatial dimensions, two angular directions (via $\nvec$), and photon frequency. Time-dependent RHD calculations must solve the RT equation at every time step and then couple the result to the MHD equations \mbox{(Eqs.~\ref{eq:mhd1}--\ref{eq:mhd5})}. Altogether this defines a complex, coupled set of PDEs that is extremely challenging to solve.

\subsubsection{Time-independent radiation transfer} \label{sec:rt_eqs_t_indep}

A common simplification is the time-independent RT equation, obtained by neglecting the $\partial/\partial t$ term in Eq.~(\ref{eq:rt}), valid when the radiation field is in steady state compared to the hydrodynamic evolution. In this case, the equation reduces to
\begin{equation}
\nvec\cdot\nabla I_\nu = j_\nu - \rho\kappa_\nu I_\nu.
\end{equation}
Along a ray parametrized by path length $s$, this becomes an ordinary differential equation,
\begin{equation} \label{eq:rt_t_indep}
\frac{dI_\nu}{ds} = j_\nu - \rho\kappa_\nu I_\nu = \rho\kappa_\nu(S_\nu -  I_\nu),
\end{equation}
with the source function $S_\nu=j_\nu/(\rho\kappa_\nu)$. This has the well-known exponential solution
\begin{equation} \label{eq:rt_t_indep_sol}
\begin{aligned}
I_\nu(s) & = I_\nu(0)\,e^{-\tau_\nu(s)} + \int_0^s j_\nu(s')\,e^{-[\tau_\nu(s)-\tau_\nu(s')]}\,ds' \\
         & = I_\nu(0)\,e^{-\tau_\nu(s)} + \int_0^s \rho(s')\,\kappa_\nu(s')\,S_\nu(s')\,e^{-[\tau_\nu(s)-\tau_\nu(s')]}\,ds' \\
         & = I_\nu(0)\,e^{-\tau_\nu(s)} + \int_0^{\tau_\nu(s)} S_\nu(\tau_\nu')\,e^{-[\tau_\nu(s)-\tau_\nu']}\,d\tau_\nu',
\end{aligned}
\end{equation}
where $\tau_\nu(s)=\int_0^s \rho(s')\kappa_\nu(s')\,ds'$ is the optical depth along the path. The first term describes attenuation of the incident radiation, while the second term accounts for local emission along the ray.

\subsection{Ray-based methods} \label{sec:rt_ray}

Ray-based methods solve the RT equation (Eq.~\ref{eq:rt}) along characteristics (rays). Their accuracy and efficiency depend on ray length and angular sampling. Most star-formation applications use either long characteristics or hybrid long–short characteristics. Long characteristics trace rays with multiple crossings of resolution elements (grid cells or SPH particles), while short characteristics solve the RT equation by interpolating between neighboring cells \citep{AbelWandelt2002,DavisStoneJiang2012}. The former is more accurate but harder to parallelize, since rays cross multiple processor domains, requiring intensive communication. Short characteristics are easier to parallelize but are also more diffusive. Hybrid schemes combine both: long characteristics near sources and short characteristics for long-range propagation, as in the \code{flash} implementation \citep{RijkhorstEtAl2006,BuntemeyerEtAl2016}.

Relative to moment methods (Sect.~\ref{sec:rt_moment_methods}), ray-based approaches excel at modeling radiation from point sources but perform poorly in the diffusion limit unless many rays are used. Hybrid ray–moment methods address this by splitting the flux into a direct stellar component and a diffuse (reprocessed) component,
\begin{equation} \label{eq:flux_split_stellar_reprocessed}
\Frad = \mathbf{F}_\star + \mathbf{F}_{\rm reprocessed},
\end{equation}
with $\mathbf{F}_\star$ computed by ray tracing and $\mathbf{F}_{\rm reprocessed}$ solved via a moment method \citep[e.g.,][]{MurrayEtAl1994}.

For $\mathbf{F}_\star(r)$, as a function of distance $r$ from the star, neglecting time dependence, scattering, and in-situ emission, the time-independent RT equation (Eq.~\ref{eq:rt_t_indep_sol}) with $j_\nu=0$ yields
\begin{equation}
I(r) = I_0\,e^{-\tau(r)},
\end{equation}
where the optical depth is
\begin{equation}
\tau(r) = \int_{R_\star}^{r} \kappaP(T_\star)\,\rho(r')\,dr',
\end{equation}
with $R_\star$ the stellar radius and $\kappaP(T_\star)$ the Planck-mean opacity (Eq.~\ref{eq:kappaP}) at $T_\star$. The stellar flux then follows
\begin{equation}
\mathbf{F}_\star(r) = F_\star(R_\star)\,\left(\frac{R_\star}{r}\right)^2 e^{-\tau(r)}\,\frac{\mathbf{r}}{r},
\end{equation}
where $F_\star(R_\star)=L_\star/(4\pi R_\star^2)=\sigmaSB T_\star^4$ is the stellar surface flux for a star of luminosity $L_\star$, and $\sigmaSB$ is the Stefan--Boltzmann constant.

The accuracy of ray tracing depends on angular sampling. Long characteristics oversample near the source but undersample at large radii. HEALPix \citep{GorskiEtAl2005} addresses this by dividing the sphere into equal-area elements and enabling adaptive ray splitting, maintaining uniform angular resolution and efficient parallelization in AMR and SPH codes \citep{AbelWandelt2002,WiseAbel2011,BaczynskiGloverKlessen2015,BuntemeyerEtAl2016,RosenEtAl2017,KimEtAl2017}.

Hybrid methods following the approach in Eq.~(\ref{eq:flux_split_stellar_reprocessed}) are implemented in \code{flash} \citep{RijkhorstEtAl2006,BuntemeyerEtAl2016,PetersEtAl2010,MenonEtAl2022}, \code{orion2} \citep{RosenEtAl2017}, and \code{pluto} \citep{KuiperEtAl2010RT,KlassenEtAl2014}. These methods are well suited for massive star formation, where they capture both direct stellar radiation pressure and dust-reprocessed flux. They regulate accretion, drive instabilities and outflows, and influence fragmentation \citep{RosenEtAl2016,RosenEtAl2019,MenonEtAl2023}, potentially setting the final stellar mass \citep{KuiperHosokawa2018}. Adaptive UV ray-tracing methods have also been applied to photoionization \citep{KimEtAl2017,KimKimOstriker2019,KimOstrikerFilippova2021,MenonEtAl2023}.


\subsection{Moment methods} \label{sec:rt_moment_methods}

One of the most widely adopted classes of RHD approaches are moment methods, which reduce the dimensionality of the RT problem by integrating over angles. This procedure is analogous to deriving the MHD equations from the Boltzmann equation via the Chapman--Enskog method. The accuracy of the resulting equations depends critically on the adopted closure relation. In the following, we derive the moments of the RT equation and then discuss the three most common closures: the Eddington approximation (FLD), the M1 closure, and the VET closure.

By averaging over all angles, low-order closures such as FLD perform well in the optically-thick, diffusive limit. Higher-order closures like M1 and VET also capture the transition between optically-thick and optically-thin gas. While all moment methods average over angles to obtain a mean intensity, M1 and VET additionally retain information about the radiation flux, thus preserving directional information. As a class, moment methods are computationally efficient, with cost scaling as $N\log N$, where $N$ is the number of resolution elements. Implementation can be challenging, however, particularly for VET, which requires a preliminary ray-tracing step (cf.~Sect.~\ref{sec:rt_ray}) to compute the Eddington tensor before solving the RHD moment equations. The latter is typically done with solver libraries such as the Portable, Extensible Toolkit for Scientific Computation \citep[PETSc;][]{BalayEtAl1997}, the High Performance Preconditioners package \citep[Hypre;][]{FalgoutYang2002}, or the Adaptive Mesh Refinement Exascale library \citep[AMReX;][]{ZhangEtAl2019}.

\subsubsection{Moments of the radiation transport equation} \label{sec:rt_moments}

Equation~(\ref{eq:rt}) describes the directional propagation of radiation at frequency $\nu$, subject to emission and absorption. Solving it in full is prohibitively complex, since every point in space and time emits and absorbs at multiple frequencies and angles. To simplify, we adopt the \inquotes{gray} approximation, integrating over all frequencies (an assumption that can later be relaxed with multi-frequency bins). We then take angular averages of the radiation intensity, defining the moments of the radiation field:
\begin{eqnarray}
\hspace{-0.5cm}
{\rm Radiation~energy~density:}\quad \Erad & = & \int\Erad\subnu\,d\nu = \int\left(\frac{1}{c}\int I_\nu\,d\Omega\right)\!d\nu, \label{eq:Erad} \\
{\rm Radiation~flux:}\quad \Frad & = & \int\Frad\subnu\,d\nu = \int\left(\int I_\nu\nvec\,d\Omega\right)\!d\nu, \label{eq:Frad} \\
{\rm Radiation~pressure:}\quad \Prad & = & \int\Prad\subnu\,d\nu = \int\left(\frac{1}{c}\int I_\nu\nvec\nvec\,d\Omega\right)\!d\nu. \label{eq:Prad}
\end{eqnarray}

Using these definitions, we can derive more tractable equations for the radiation field by taking successive angular moments of Eq.~(\ref{eq:rt}), i.e., $\iint(\cdots)\nvec^m\,d\Omega\,d\nu$ for $m=0,1,\dots$. The $m=0$ and $m=1$ moments describe the evolution of radiation energy and flux, respectively. As an example, integrating Eq.~(\ref{eq:rt}) gives
\begin{eqnarray} \label{eq:Erad_moment}
\iint\left(\frac{1}{c}\ddt{} + \nvec\cdot\nabla\right) I_\nu\,d\Omega\,d\nu & = & \iint \left(j_\nu-\rho\kappa_\nu I_\nu\right)d\Omega\,d\nu \nonumber \\ \implies
\ddt{\Erad} + \nabla\cdot\Frad & = & \iint \left(j_\nu-\rho\kappa_\nu I_\nu\right)d\Omega\,d\nu.
\end{eqnarray}
This relation is analogous to the hydrodynamic continuity equation: the local change in $\Erad$ is set by the net flux divergence, plus emission and absorption terms on the RHS.

If the gas were at rest, the RHS could be written as
\begin{equation}
\iint \left(j_\nu-\rho\kappa_\nu I_\nu\right)d\Omega\,d\nu = J - \rho\kappaE c\Erad,
\end{equation}
where $J=\rho\kappaP\emis$ is the emission rate in LTE (via Kirchhoff’s law), with $\kappaP$ the Planck mean opacity (Eq.~\ref{eq:kappaP}). The absorption term involves the energy mean opacity,
\begin{equation} \label{eq:kappaE}
\kappaE = \frac{\int_0^\infty\kappa_\nu\Erad\subnu\,d\nu}{\int_0^\infty\Erad\subnu\,d\nu}.
\end{equation}
Similarly, the $m=1$ moment introduces the flux mean opacity,
\begin{equation} \label{eq:kappaF}
\kappaF = \frac{\int_0^\infty\kappa_\nu\abs{\Frad\subnu}\,d\nu}{\int_0^\infty\abs{\Frad\subnu}\,d\nu}.
\end{equation}

In reality, the gas moves. The LHS operators of Eq.~(\ref{eq:Erad_moment}) are exact in the lab frame, but evaluating the RHS is complicated because photons undergo Doppler shifts, aberration, and advection relative to the comoving frame where opacities are most naturally defined. Thus, absorption and emission couple differently in lab vs.~comoving frames. To treat this consistently, most RHD approaches adopt a mixed-frame formalism, valid to leading order in $v/c$ (cf.~Sect.~\ref{sec:rt_frames}), which introduces additional terms involving $\Erad$, $\Frad$, and $\Prad$ \citep{MihalasAuer2001,KrumholzEtAl2007,MenonEtAl2022}.

A fundamental issue is that the hierarchy of moment equations never closes: the $m$th moment depends on the $(m+1)$th. For example, Eq.~(\ref{eq:Erad_moment}) for $\Erad$ depends on $\Frad$, while $\Frad$ depends on $\Prad$, and so forth. Closure is obtained by assuming a relation of the form
\begin{equation}
\Prad = \TEdd\,\Erad, \label{eq:TEdd}
\end{equation}
where the Eddington tensor $\TEdd$ encodes the angular structure of the radiation field. The choice of $\TEdd$ determines the accuracy of the method. Common closures include:
\begin{enumerate}
\item Diffusion limit (FLD): $\TEdd=\mathbb{I}/3$.
\item Free-streaming limit: $\TEdd=\nvec\nvec$, aligning $\Prad$ with $\Frad$.
\item M1 closure: $\TEdd$ as a function of the reduced flux $\abs{\Frad}/(c\Erad)$.
\item VET closure: $\TEdd$ computed directly from Eqs.~(\ref{eq:Erad}) and~(\ref{eq:Prad}) using $I_\nu$ from the time-independent RT solution (Eq.~\ref{eq:rt_t_indep_sol}), typically via ray tracing.
\end{enumerate}
We will discuss and compare the FLD, M1, and VET closures in detail below.

\subsubsection{Complete set of RHD moment equations} \label{sec:rt_moment_eqs}

Combining the above, the mixed-frame formulation of the RHD moment equations is \citep{MihalasAuer2001,MenonEtAl2022}
\begin{gather}
\ddt{(\rho\vel)} = -\nabla\cdot(\rho\vel\vel) - \nabla\pth + \mathbf{G}, \label{eq:rhd_gas_mom} \\
\ddt{(\rho e)} = -\nabla\cdot[(\rho e + \pth)\vel] + cG^0, \label{eq:rhd_gas_ener} \\
\ddt{\Erad} = -\nabla\cdot\Frad - cG^0, \label{eq:Erad_eq} \\
\ddt{\Frad} = -\nabla\cdot(c^2\Erad\TEdd) - c^2\mathbf{G}, \label{eq:Frad_eq}
\end{gather}
with $\pth$ and $e$ as defined in Sect.~\ref{sec:mhd}. The source terms $G^0$ and $\mathbf{G}$ represent absorption and emission of radiation energy and momentum. They appear with opposite sign in the MHD momentum and energy equations, ensuring self-consistent gas–radiation coupling.

To leading order in $v/c$, and assuming a direction-independent flux spectrum \citep{KrumholzEtAl2007}, these terms are \citep{MihalasAuer2001,MenonEtAl2022}
\begin{eqnarray} \label{eq:G}
G^0 & = & \rho\kappaE\Erad - \frac{1}{c}\rho\kappaP\emis
+ \rho\left(\kappaF-2\kappaE\right)\frac{\Frad}{c}\cdot\frac{\vel}{c} \nonumber \\
& & +\,\rho\left(\kappaE-\kappaF\right)\Erad\left(\frac{\abs{\vel}^2}{c^2}+\frac{\vel\vel}{c^2}:\TEdd\right), \label{eq:G0} \\
\mathbf{G} & = & \rho\kappaF\left[\frac{\Frad}{c}-\Erad(\mathbb{I}+\TEdd)\frac{\vel}{c}\right]. \label{eq:Gvec}
\end{eqnarray}
The mean opacities $\kappaP$, $\kappaE$, and $\kappaF$ used here are defined in Eqs.~(\ref{eq:kappaP}), (\ref{eq:kappaE}), and (\ref{eq:kappaF}), and are evaluated in the co-moving frame. This choice makes them straightforward to compute, while the remaining quantities are handled in the lab frame—facilitating mixed-frame implementation. In general, however, solving Eqs.~(\ref{eq:Erad_eq}) and~(\ref{eq:Frad_eq}) requires implicit evaluation of $\kappaE$ and $\kappaF$ through frequency integration over $\Erad$ and $\Frad$. In practice, useful approximations exist: $\kappaE\approx\kappaP$ near LTE, while for optically-thick gas (diffusion limit) $\kappaF\approx\kappaR$, with $\kappaR$ defined in Eq.~(\ref{eq:kappaR}). These substitutions are widely used and may remain reasonable even before reaching the strict diffusion limit. This system of equations is closed once a choice for $\TEdd$ is specified, with the FLD, M1, and VET closures discussed in the following subsections.

\subsubsection{Flux-limited diffusion (FLD)} \label{sec:rt_fld}

FLD is the simplest moment closure, obtained by assuming an isotropic radiation field in the laboratory frame, so that the Eddington tensor is
\begin{equation} \label{eq:TEdd_fld}
\TEdd = \frac{1}{3}\,\mathbb{I}.
\end{equation}
In this limit, one eliminates the radiation momentum equation by prescribing a relation for the flux in terms of the radiation energy density (a diffusion law), and evolves only the $m=0$ moment (Eq.~\ref{eq:Erad_eq}) coupled to the gas via the source terms in Sect.~\ref{sec:rt_moment_eqs}. This approximation is often accurate in optically-thick regions where gas and radiation are strongly coupled (e.g., dense star-forming gas; \citealt{OffnerEtAl2012}), but it degrades in optically-thin environments (e.g., cloud outskirts, jet-cleared polar cavities), where diffusion is a poor description.

To prevent superluminal propagation of radiation fronts in optically-thin regions, FLD introduces a flux limiter that transitions smoothly between diffusion and free-streaming. The standard \citep{LevermorePomraning1981} form is
\begin{equation} \label{eq:fld_flux_limiter}
\Frad = -\frac{\zeta c}{\rho\kappaR}\,\nabla\Erad, \qquad
\zeta = \frac{1}{\xi}\left(\coth\xi - \frac{1}{\xi}\right), \qquad
\xi = \frac{\abs{\nabla\Erad}}{\rho\kappaR\Erad},
\end{equation}
where $\kappaR$ is the Rosseland mean opacity (Eq.~\ref{eq:kappaR}). In the diffusion limit \mbox{($\xi\to 0$)}, \mbox{$\zeta\to1/3$} and
\begin{equation}
\Frad \to -\frac{c}{3\rho\kappaR}\nabla\Erad,
\end{equation}
while in the streaming limit \mbox{($\xi\to \infty$)}, \mbox{$\zeta\to1/\xi$} and \mbox{$\abs{\Frad}\to c\,\Erad$} with $\Frad$ aligned antiparallel to $\nabla\Erad$.

Thus, the FLD system is obtained by inserting Eq.~(\ref{eq:fld_flux_limiter}) into the general RHD moment equations from Sect.~\ref{sec:rt_moment_eqs}. This includes radiation pressure through $\Prad=(\Erad/3)\,\mathbb{I}$ (e.g., relevant for massive star formation), through the mixed-frame source terms $G^0$ and $\mathbf{G}$ with the FLD flux substituted \citep[explicit forms are summarized in][]{KrumholzEtAl2007}. The incorporation of radiation pressure allows the treatment of massive star formation, where radiation regulates gas accretion \citep{RosenEtAl2020}. Simulations of massive star formation with non-axisymmetric geometries, however, find that radiation pressure is not sufficient to halt accretion onto the star \citep{KrumholzEtAl2009,CommerconEtAl2011}.

Gray and multi-group FLD schemes are widely used in star-formation simulations (cf.~Table~\ref{tab:codes}). Multi-group extensions, i.e., solving the FLD system for several frequency bins, improve temperature accuracy across a wide range of optical depths \citep{ShestakovOffner2008,KuiperKlessen2013,GonzalezEtAl2015}. Nonetheless, FLD has well-known limitations: by construction $\Frad \parallel -\nabla\Erad$, so it cannot cast shadows or represent crossing beams, and it can misdirect radiation forces in multi-source configurations \citep{HayesNorman2003,JiangStoneDavis2012,MenonEtAl2022}. These deficiencies motivate higher-order closures, such as M1 and VET (Sect.~\ref{sec:rt_m1} and~\ref{sec:rt_vet}), which retain angular information.


\subsubsection{Moment-1 (M1) closure} \label{sec:rt_m1}

Some limitations of FLD can be alleviated by retaining partial directional information of the radiation field. The M1 closure assumes rotational symmetry around the radiative flux direction. The Eddington tensor is then a linear combination of $\mathbb{I}$ (as in FLD) and the flux direction tensor $\hat{\mathbf{F}}_\mathrm{r}\hat{\mathbf{F}}_\mathrm{r}$, where $\hat{\mathbf{F}}_\mathrm{r} = \Frad/\abs{\Frad}$ \citep{Levermore1984,SkinnerOstriker2013},
\begin{equation} \label{eq:TEdd_m1}
\TEdd = \frac{1-\chi}{2}\mathbb{I} + \frac{3\chi-1}{2} \hat{\mathbf{F}}_\mathrm{r}\hat{\mathbf{F}}_\mathrm{r},
\end{equation}
with the Eddington factor $\chi$ chosen to ensure flux limiting \citep{Levermore1984},
\begin{equation} \label{eq:eddington_factor_m1}
\chi(f_r) = \frac{3+4f_r^2}{5+2\sqrt{4-3f_r^2}},
\end{equation}
where $f_r=\abs{\Frad}/(c\Erad)$ is the reduced flux. This closure smoothly interpolates between the diffusion \mbox{($f_r\to0\!\implies\!\chi\to1/3$)} and streaming \mbox{($f_r\to1\!\implies\!\chi\to1$)} limits, avoiding the need for an ad-hoc flux limiter.

The M1 system is thus the general RHD moment equations (Sect.~\ref{sec:rt_moment_eqs}) with $\TEdd$ given by Eqs.~(\ref{eq:TEdd_m1}) and~(\ref{eq:eddington_factor_m1}). While M1 outperforms FLD in optically-thin regions and can reproduce shadows from a single source, it fails in multi-source configurations where fluxes can spuriously cancel \citep{FrankHauckOlbrant2012,SkinnerOstriker2013,MenonEtAl2022}. This limitation arises because the M1 Eddington tensor is defined purely locally, depending only on $\Frad$ and $\Erad$.

An advantage is that the M1 system is hyperbolic, allowing radiation to be evolved like a fluid with wave speed $\lesssim c$. However, explicit timesteps are prohibitively small ($\Delta t\propto c^{-1}$) compared to the CFL timestep of the MHD system. A common remedy is the \inquotes{reduced speed of light approximation} ($c\to\tilde c \ll c$ in time-derivative terms only) \citep{GnedinAbel2001,SkinnerOstriker2013,HopkinsGrudic2019,MignonRisseEtAl2020}. This is accurate if $\tilde c \gg v_\mathrm{max}$, the maximum MHD signal speed \citep{RosdahlEtAl2013}, with tests showing proper ionization front propagation for $\tilde c\gtrsim30\,\mathrm{km\,s^{-1}}$ \citep{GeenEtAl2015b,GrudicEtAl2021}. Too small a $\tilde c$, however, underestimates radiation pressure in optically-thick gas, relevant to massive star formation. Additional cost reductions can be achieved via radiation sub-cycling \citep{SkinnerOstriker2013}.

\begin{figure}[t]
\centerline{\includegraphics[width=1.0\linewidth]{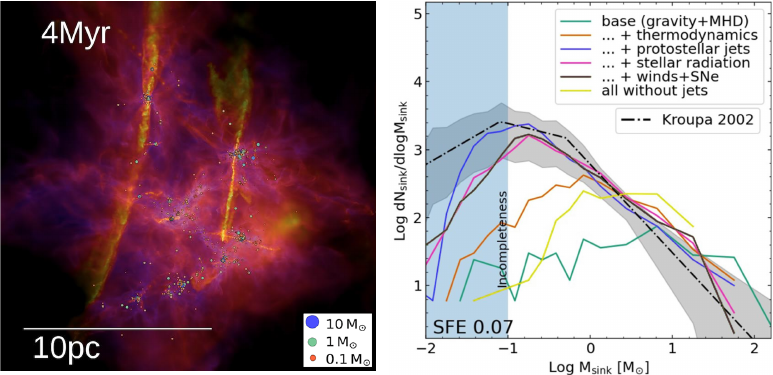}}
\caption{\label{fig:starforge}
STARFORGE simulation of a molecular cloud including an M1 multi-bin RHD method with ionization, radiation pressure, and other stellar feedback processes \citep{GuszejnovEtAl2022}.
{\bf Left panel:} Color map of the 1D line-of-sight velocity dispersion, increasing from purple ($0.1\,\mathrm{km\,s^{-1}}$) to orange ($10\,\mathrm{km\,s^{-1}}$). Surface density is indicated by brightness (lighter is denser). Kinematic maps highlight feedback features such as protostellar outflows. Circles mark the locations of sink particles, with size and color denoting stellar mass. {\bf Right panel:} Stellar IMF for simulations with different physics included, as indicated in the legend. Protostellar jets have the overall biggest effect (cf.~Sect.~\ref{sec:outflow}), with radiation primarily reducing the formation of very low-mass stars (cf.~Fig.~\ref{fig:heating}).
}
\end{figure}

M1 has been implemented in \code{athena} \citep{SkinnerOstriker2013}, \code{arepo} \citep{KannanEtAl2019}, \code{flash} \citep{MenonEtAl2022}, \code{gizmo} \citep{HopkinsGrudic2019}, \code{heracles} \citep{GonzalezAuditHuynh2007}, \code{pluto} \citep{MelonFuksmanMignone2019}, \code{quokka} \citep{WibkingKrumholz2022,HeWibkingKrumholz2024a,HeWibkingKrumholz2024b}, and \code{ramses} \citep{AubertTeyssier2008,RosdahlEtAl2013,RosdahlTeyssier2015,MignonRisseEtAl2020}. Its strictly local closure, similar to FLD, ensures that the M1 method is independent of the number of sources, and that it scales linearly with the number of resolution elements, as it is usually evolved with explicit time integration, making it well-suited for large-scale simulations. Applications to star-cluster formation find ionization and radiation pressure from massive stars significantly regulate efficiencies \citep{SkinnerOstriker2015,RaskuttiOstrikerSkinner2016,RaskuttiOstrikerSkinner2017,GrudicEtAl2021}. Multi-group M1 extensions following distinct radiation bands capture frequency-dependent radiation effects such as ionizing feedback more realistically \citep{AubertTeyssier2008,RosdahlEtAl2013,RosdahlTeyssier2015,GeenEtAl2015a,GeenEtAl2015b,GeenEtAl2016,GeenSolerHennebelle2017,GuszejnovEtAl2022,HeWibkingKrumholz2024a,HeWibkingKrumholz2024b}, with results showing cloud dispersal and reduced star formation efficiencies, and suppression of low-mass star formation through radiation heating (see Fig.~\ref{fig:starforge}).


\subsubsection{Variable Eddington tensor (VET) method} \label{sec:rt_vet}

The variable Eddington tensor (VET) method is the most advanced moment-based approach to RHD \citep[see e.g.,][]{HayesNorman2003,DavisStoneJiang2012,JiangStoneDavis2012,MenonEtAl2022}. Unlike the M1 closure, the Eddington tensor in VET is not determined locally but computed directly from angular quadratures of the frequency-averaged specific intensity $I_\mathrm{r}$ via Eqs.~(\ref{eq:Erad}) and~(\ref{eq:Prad}). The intensity $I_\mathrm{r}$ along path length $s$ follows the time-independent RT equation (cf.~Eqs.~\ref{eq:rt_t_indep}--\ref{eq:rt_t_indep_sol}) under the gray approximation with the Planck mean opacity $\kappaP$ (Eq.~\ref{eq:kappaP}):
\begin{equation} \label{eq:rt_vet}
\begin{aligned}
I_\mathrm{r}(s) & = \int_0^\infty I_\nu(s)\,d\nu \\
& \approx I_\mathrm{r}(0)\,e^{-\tau(s)} + \int_0^s \rho(s')\,\kappaP(s')\,S_\mathrm{r}(s')\,e^{-[\tau(s)-\tau(s')]}\,ds',
\end{aligned}
\end{equation}
with
\begin{equation}
\tau(s) = \int_0^s \rho(s')\kappaP(s')\,ds',
\end{equation}
and $S_\mathrm{r}$ the source function, typically the integrated Planck function for dust emission, $S_\mathrm{r}(s) = \sigmaSB T^4(s)/\pi$. This expression neglects scattering and $\mathcal{O}(v/c)$ terms from the mixed-frame formulation, which are expected to make only minor contributions to $\TEdd$.

In practice, Eq.~(\ref{eq:rt_vet}) is evaluated by ray tracing (Sect.~\ref{sec:rt_ray}) along discrete directions $\nvec$, producing $I_\mathrm{r}(\nvec)$ and thus $\TEdd$ via Eqs.~(\ref{eq:Erad}) and~(\ref{eq:Prad}). In \code{flash}, this is achieved using the parallel hybrid-characteristics ray tracer of \citet{BuntemeyerEtAl2016}, with AMR support, and directions discretized via HEALPix \citep{GorskiEtAl2005} into 12, 48, 192, 768, etc.~equal-area rays. The resulting $\TEdd$ is then inserted into the coupled RHD system (Eqs.~\ref{eq:rhd_gas_mom}--\ref{eq:Gvec}). Solving this implicit system is done efficiently in parallel using PETSc \citep{BalayEtAl1997}, with additional care at coarse-fine AMR boundaries \citep{MenonEtAl2022}.

The strength of VET is its accuracy across both optically-thin and optically-thick limits, as well as in the transition regime. It reproduces realistic shadows, radiation fields, and gas temperature distributions. Accuracy depends on the angular resolution \citep{DavisStoneJiang2012}: with HEALPix sampling, biases from grid alignment are minimized by rotating the base orientation between runs. \citet{MenonEtAl2022} showed that using 48~rays already yields very good results, with only marginal improvement at 192~rays; see their fig.~B1.

\paragraph{Extension to multi-frequency VET}

As mentioned at the end of Sect.~\ref{sec:rt_moment_eqs}, to avoid implicit evaluation of $\kappaE$ and $\kappaF$ in Eqs.~(\ref{eq:G0}) and~(\ref{eq:Gvec}), one usually approximates $\kappaE\approx\kappaP$ and $\kappaF\approx\kappaR$, which also ensures consistency with the steady-state equation for $\Erad$ \citep{MenonEtAl2022}. However, this choice is not simultaneously consistent with the steady-state equation for $\Frad$. In fact, no single gray opacity can make both moment equations exact{\emdash}only frequency-dependent opacities can achieve full consistency.

To improve upon this limitation, \citet{MenonEtAl2023} developed a multi-frequency RHD scheme combining the IR dust-radiation approach described above with an ionizing (UV) radiation treatment. The UV component computes the hydrogen ionization fraction by evaluating the optical depth of ionizing photons along rays. The ionization fraction then updates the opacity and iterates to convergence. A natural next step would be to include Doppler shifts of the ionizing lines, whereas the method of \citet{MenonEtAl2023} currently assumes hydrogen ionization in the rest frame.

\begin{figure}[t]
\centerline{\includegraphics[width=1.0\linewidth]{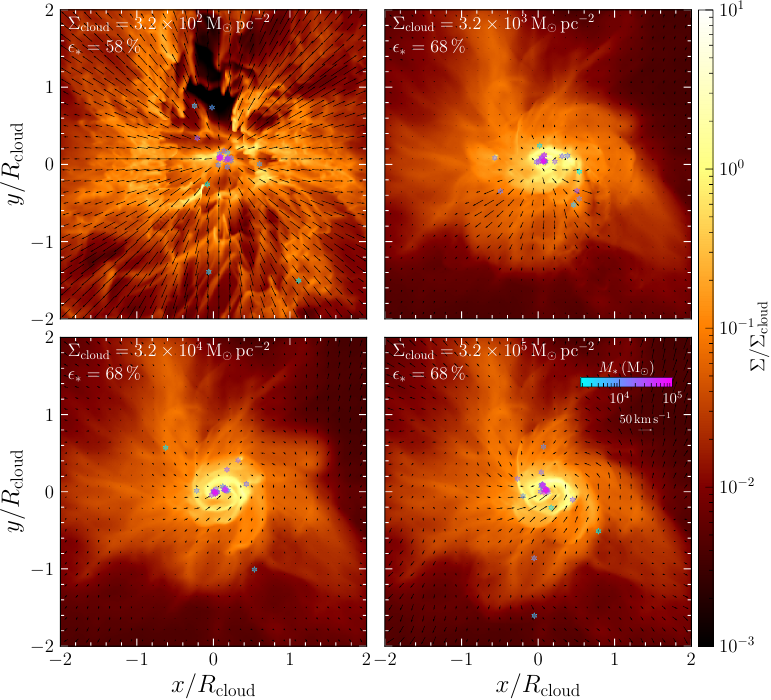}}
\caption{Massive star cluster formation and feedback using the combined VET IR and UV scheme from \citet{MenonEtAl2023}. Shown are simulations of clouds with increasing surface density. In Milky Way–like conditions (top left), the UV radiation component drives powerful outflows that disperse the cloud and limit the star formation efficiency $\epsilon_*$. In contrast, in extreme, dense environments (bottom right), neither UV nor IR feedback can disperse the cloud, and $\epsilon_*$ remains high. An animation is available at \url{https://shm-1996.github.io/movies/}.}
\label{fig:vet_ir_uv}
\end{figure}

Figure~\ref{fig:vet_ir_uv} demonstrates this IR+UV VET approach in simulations of massive star-cluster formation. Four clouds of increasing surface density are compared. In typical Galactic environments, UV feedback disperses clouds via radiation pressure-driven outflows, soon after cluster formation, significantly reducing the star formation efficiency, $\epsilon_*$. In very dense, extreme conditions, however, radiative feedback is largely ineffective, and the star formation efficiency remains high.

\subsection{Monte Carlo radiation transfer} \label{sec:rt_mcrt}

Monte Carlo radiation transfer (MCRT) is analogous to kinetic theory in gas dynamics: instead of explicitly solving the RT equation, MCRT uses a large number of test particles, or \inquotes{photon packets}, to model the radiation field. These packets sample the photon phase-space distribution, with each packet's behavior{\emdash}including direction of propagation and the probability of scattering or absorption{\emdash}determined probabilistically. As in kinetic theory, it is the ensemble behavior rather than individual packets that matters. With a sufficiently large number of packets, the ensemble accurately represents the RT process. For a detailed review of MCRT techniques, see \citet{NoebauerSim2019}. Here, we summarize the main characteristics of MCRT.

The accuracy of MCRT improves with the number of packets $N$, but the computational cost also scales proportionally with $N$. MCRT parallelizes efficiently, since each processor can follow an independent subset of packets with minimal communication \citep{Robitaille2011}. However, the method becomes inefficient in optically-thick regions, where packets undergo many scatterings, requiring numerous advancement steps to sample the radiation field across the domain.

MCRT methods naturally incorporate scattering and accurately follow the angular dependence of the radiation field, but they require a large number of packets to reach accuracies comparable to moment methods \citep[][see Fig.~\ref{fig:vet-comp} below]{DavisStoneJiang2012}. Like all Monte Carlo approaches, MCRT suffers from noise that scales as $N^{-1/2}$. For this reason, coupling MCRT directly to MHD simulations has long been considered prohibitively expensive, and the method has most often been used as a post-processing tool \citep[e.g.,][]{DullemondTurolla2000,ErcolanoEtAl2008,Robitaille2011,ReisslWolfBrauer2016,HaworthEtAl2018,HarriesEtAl2019}. Recent advances, however, have led to the development of in-situ MCRT-MHD methods that are competitive in speed and accuracy with moment methods. Implementations now exist in a number of hydrodynamics codes used for star-formation modeling, including \code{arepo} \citep{SmithEtAl2020}, \code{CMacIonize} \citep{VandenbrouckeWood2018}, \code{flash} \citep{TsangMilosavljevic2015}, \code{phantom} \citep{PetkovaEtAl2021}, and \code{torus} \citep{Harries2015}.

MCRT-MHD provides a powerful tool for problems with complex, high-intensity radiation fields or multiple scattering events. To make such calculations tractable, however, several numerical approximations and acceleration techniques are employed, including continuous absorption, energy deposition, and implicit thermal balance \citep{Lucy1999,BjorkmanWood2001}, photon weighting and luminosity boosting, and local packet merging and splitting \citep{SmithEtAl2020}.

\subsection{Comparison of RT methods} \label{sec:rt_comp}

\begin{figure}[t]
\centerline{\includegraphics[width=1.0\linewidth]{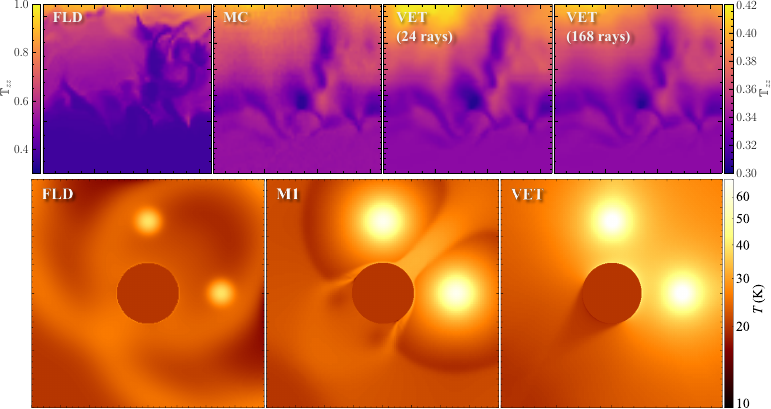}}
\caption{
Comparison of numerical methods for radiation transport (RT).
\textbf{Top panels:} Flux-limited diffusion (FLD, left), Monte Carlo (MC, 2nd panel), variable Eddington tensor (VET) with 24~rays (3rd panel), and VET with 168~rays (right) for a shearing-box patch of an accretion disc from \citet{HiroseKrolikStone2006} and \citet{DavisStoneJiang2012}. The disk gas density increases from top to bottom in each panel, with shearing flow along the horizontal direction. The panels show the $zz$-component of the Eddington tensor, where the left color map applies for FLD (for which $\TEdd_{zz}\to1$ in optically-thin gas, by construction), and the right color map applies for the MC and VET panels. FLD produces noticeably different structures and magnitudes of $\TEdd_{zz}$ compared to MC and VET, which are largely consistent. MC shows some small-scale noise but otherwise agrees well with VET using 168~rays. Even with only 24~rays, VET produces similar results, though a larger number of rays is preferable. \textbf{Bottom panels:} FLD (left), Moment-1 (M1, middle), and VET with 192~rays (right) in a shadow test by \citet{MenonEtAl2022}, where a dense dust cloud is placed at the center of the domain and two radiation sources are located at 12~and 3~o'clock. Panels show the gas temperature after one light-crossing time of the domain. VET is the only method that reproduces the correct temperature distribution, including the correct structure of the shadow behind the cloud. An animation of the bottom panels is available at \url{https://www.mso.anu.edu.au/~chfeder/pubs/vettam/vettam.html}.}
\label{fig:vet-comp}
\end{figure}

Several studies have carried out detailed comparisons between the FLD, M1, MC, and VET methods \citep{DavisStoneJiang2012,JiangStoneDavis2012,MenonEtAl2022}. Figure~\ref{fig:vet-comp} provides a direct comparison of all four methods. The top panels, taken from \citet{DavisStoneJiang2012}, show a shearing-box simulation of an accretion disk patch with physical properties described in \citet{HiroseKrolikStone2006} (density is highest at the bottom and decreases toward the disk atmosphere at the top). From left to right, the panels compare FLD, MC, VET with 24~rays, and VET with 168~rays. All panels display the $zz$-component of the Eddington tensor $\TEdd_{zz}$. FLD yields a markedly different radiation field distribution than MC or VET (note also the different color map range, as $\TEdd_{zz}$ in FLD approaches unity in the optically-thin limit, by construction). The MC result shows some small-scale noise despite using $\sim10^{10}$~photon packets and requiring $\sim100\times$ the compute time of VET. The close agreement between MC and VET strongly suggests both methods are accurate. Minor differences are visible between the 24~ray and 168~ray VET runs, with the latter more closely matching MC (aside from the MC noise).

The bottom panels of Fig.~\ref{fig:vet-comp} compare FLD, M1, and VET in a shadow test, where a dense, high-opacity gas cloud is placed at the center and two radiation sources are positioned at 12~and~3 o'clock in the computational domain \citep{MenonEtAl2022}. The surrounding medium is optically thin. In FLD, radiation diffuses around the cloud, producing no shadows. M1 creates directionally dependent solutions with a shadow behind the cloud, opposite the sources. However, the shadow is weaker along the mid-plane between the two sources and the radiation field shows spurious maxima and minima around the sources. By contrast, VET yields the correct radiation and gas temperature distribution, including well-defined shadows behind the cloud.

\subsection{Summary of challenges and future directions}

Although RT is fully understood in theory, solving it numerically is extremely challenging due to its high dimensionality (time, 3D space, 2~angles, and frequency; Sect.~\ref{sec:rt_regimes} and~\ref{sec:rt_eqs}). Approximations are therefore essential, including gray (frequency-integrated) treatments or reduced angular sampling in ray tracing (Sect.~\ref{sec:rt_ray}). Monte Carlo RT (MCRT; Sect.~\ref{sec:rt_mcrt}) is conceptually simple and highly parallelizable, but requires very large numbers of photon packets to reduce noise. Numerical tricks{\emdash}such as packet weighting, splitting/merging, continuous absorption, and luminosity boosting{\emdash}have made MCRT increasingly competitive for live RHD rather than just post-processing.

The most widely used methods remain RHD moment approaches (Sect.~\ref{sec:rt_moment_methods}), which are more efficient and easier to parallelize than full ray tracing. A mixed-frame treatment is required to handle frequency-integrated transport and mean opacities (Sect.~\ref{sec:rt_moments}), leading to the complete set of RHD moment equations (Sect.~\ref{sec:rt_moment_eqs}). Their accuracy hinges on the closure choice for the Eddington tensor: FLD (Sect.~\ref{sec:rt_fld}) and M1 (Sect.~\ref{sec:rt_m1}) rely on local quantities, while VET (Sect.~\ref{sec:rt_vet}) requires a formal RT solution (typically via ray tracing).

Comparisons of FLD, M1, VET, and MCRT (Sect.~\ref{sec:rt_comp}) highlight their respective strengths and weaknesses: FLD performs well only in optically-thick regions and cannot capture shadows. M1 reproduces single-source shadows but fails with multiple sources due to spurious flux cancellation. VET and MCRT both produce accurate radiation fields and shadowing; however, MCRT typically requires an order of magnitude more computation to match VET accuracy, though it stands to benefit from advances in GPU acceleration.

Another area of progress is multi-group RT, which distinguishes broad bands such as IR and UV. This enables more realistic treatments of feedback, especially from massive stars. However, incorporating full atomic and molecular line transfer with Doppler shifts (critical for, e.g., Population~III star formation) remains too computationally demanding for large-scale applications.

Finally, gas dissipation acts as a local radiation source (Sect.~\ref{sec:shear_bulk_visc}). Since numerical dissipation is resolution-dependent (Sect.~\ref{sec:num_visc}), coupling this heating consistently into the RHD system is non-trivial. Methods that explicitly track the local dissipation rate (Sect.~\ref{sec:local_diss}) may provide a promising way forward by linking dissipation-driven heating directly to the radiation source terms.

\section{Gravity} \label{sec:grav}

Modeling gravity is a fundamental requirement in star-formation simulations. Once the gravitational acceleration $\mathbf{g}(\mathbf{x},t)$ is computed, it is incorporated into the MHD system (Eqs.~\ref{eq:mhd1}--\ref{eq:mhd5}) by adding the terms $+\rho\mathbf{g}$ and $+\rho\vel\cdot\mathbf{g}$ to the RHS of Eqs.~(\ref{eq:mhd2}) and~(\ref{eq:mhd3}), respectively, as source terms, where $\mathbf{g} = -\nabla\Phi$ can be written as the negative gradient of the gravitational potential $\Phi$.

\subsection{Discrete mass distributions} \label{sec:grav_nbody}

To obtain $\mathbf{g}$, ideally, the gravitational acceleration $\mathbf{g}_i$ of a fluid element and/or mass particle $i$ at position $\mathbf{x}_i$ can be determined directly via a sum over the contributions from all $N$~mass elements $m_j$ at positions $\mathbf{x}_j$, as
\begin{equation} \label{eq:grav_sum}
\mathbf{g}_i = -G \sum_{j=1}^N m_j \frac{\mathbf{r}_j}{r_j^3}\,\mathcal{K}(r_j/\rsoft),
\end{equation}
where $G$ is the gravitational constant, $\mathbf{r}_j=\mathbf{x}_i-\mathbf{x}_j$ and $r_j=\abs{\mathbf{r}_j}$. Here the function $\mathcal{K}(r_j/\rsoft)$ is a cubic-spline softening kernel \citep{MonaghanLattanzio1985,PriceMonaghan2007},
\begin{equation} \label{eq:spline_softening}
\def\arraystretch{1.2}
\mathcal{K}(\rbyrsoft) = \left\{
\begin{array}{ll}
4\rbyrsoft^2\left(\frac{8}{3}\rbyrsoft-\frac{48}{5}\rbyrsoft^3+8\rbyrsoft^4\right) & \quad\mbox{for}\quad 0\leq\rbyrsoft<\frac{1}{2}, \\
4\rbyrsoft^2\left(\frac{16}{3}\rbyrsoft-12\rbyrsoft^2+\frac{48}{5}\rbyrsoft^3-\frac{8}{3}\rbyrsoft^4-\frac{1}{60}\rbyrsoft^{-2}\right) & \quad\mbox{for}\quad \frac{1}{2}\leq\rbyrsoft<1, \\
1 & \quad\mbox{for}\quad \rbyrsoft \geq 1\;, \\
\end{array} \right.
\end{equation}
with the dimensionless radius $\rbyrsoft=r/\rsoft$, where $\rsoft$ is a user-defined gravitational softening radius. Other softening functions are possible, but gravitational softening is required, because otherwise $\mathbf{g}\to\infty$ when particles approach one another. Using this particular kernel, we see that $\mathcal{K}\to0$ smoothly when $r\to0$, and $\mathcal{K}=1$ for $r\geq\rsoft$, such that we recover Newton's exact acceleration with Eqs.~(\ref{eq:grav_sum}) and~(\ref{eq:spline_softening}) outside the softening radius, which is preferred over simpler softening approaches, such as Plummer softening, where $\mathcal{K}(\rbyrsoft) = \rbyrsoft^3(\rbyrsoft^2+1)^{-3/2}$, which only approaches the exact solution for $r\gg\rsoft$.

Equation~(\ref{eq:grav_sum}) is essentially how a basic $N$-body method approaches the problem for discrete point masses and is the most accurate way to compute the gravitational acceleration. However, direct summation is by definition very slow, so most particle-based gas dynamics codes solve for the gravitational forces by hierarchically grouping close neighbors and constructing a tree (e.g., see Sect.~\ref{sec:grav_tree}).

\subsection{Continuous mass distributions}

For continuous mass distributions, modeling gravity
essentially requires the solution of the Poisson equation for the gravitational potential, $\Phi$,
\begin{equation} \label{eq:poisson}
\nabla^2 \Phi = 4\pi G \rho.
\end{equation}
There are three main methods to obtain a numerical solution of the Poisson equation, which are discussed in more detail below (FFT, multi-grid, tree-based), where all three can be adopted in grid-based codes, while only the third one is usually adopted in particle-based codes.

\subsection{Methods for solving the Poisson equation} \label{sec:poisson}

\subsubsection{FFT methods} \label{sec:grav_fft}

If the system has periodic boundary conditions and can be represented on a uniform grid, the solution to Poisson's equation can be obtained via Fourier transformation, i.e., in wavenumber ($\mathbf{k}$) space\footnote{In practice one would use a Fast Fourier Transform (FFT) method.}. In this case, $\nabla^2 \to -\abs{\mathbf{k}}^2$, and after FFT of $\rho\to\rho_k$, Eq.~(\ref{eq:poisson}) becomes
\begin{equation}
-\abs{\mathbf{k}}^2 \Phi_k = 4\pi G \rho_k,
\end{equation}
which is algebraically solved for $\Phi_k = -4\pi G \rho_k / \abs{\mathbf{k}}^2$. Ignoring the $k=0$ mode (mean density), because only density fluctuations matter for the potential in a periodic system, inverse FFT of $\Phi_k$ yields $\Phi$. While this method is fast (as FFT scales with $N\log N$), it has the obvious limitation that it only works on a uniform grid, and only for periodic boundary conditions \citep[although FFT-based methods that also work with free boundaries have been developed; see][]{GenoveseEtAl2006}, neither of which are usually the case for star-formation simulations.

\subsubsection{Multi-grid methods} \label{sec:grav_mg}

While FFT-based solvers are extremely efficient for periodic systems on a uniform grid, their applicability is restricted when dealing with non-periodic boundary conditions, and situations where high dynamic range in resolution is required. In astrophysical simulations, gravitational potentials often need to be resolved over many orders of magnitude in density and spatial scale. Multi-grid methods \citep{BriggsHensonMcCormick2000,TrottenbergOosterleeSchueller2001} are particularly suited for this as they employ a hierarchy of grids at different resolutions. The core idea is to smooth high-frequency error components on fine grids and correct low-frequency errors on coarser grids, accelerating convergence compared with simple relaxation methods. 

A key element of multi-grid is that the initial conditions for the iterative solution of Poisson's equation on the fine grids are obtained from interpolated solutions on the respective coarser grids. This hierarchical cycling between grid levels makes multi-grid solvers among the fastest iterative methods available for elliptic equations such as Poisson's equation. However, it should be noted that multi-grid methods therefore require a relatively low-resolution (coarse) grid on the lowest level of the grid hierarchy, because the lowest level is where initial guesses are required to initialize the iteration. This can cause problems if one wants to employ a large (high-resolution) base grid on the lowest-resolution level, because the initial iteration may take a very long time to converge. For star-formation applications that use a large molecular cloud as a basis, where turbulence needs to be resolved on a base grid of at least $256^3$~cells or more, this step in the multi-grid solve can become prohibitively expensive. The only solution is to add \inquotes{ghost} lower-level grids to artificially extend the multi-grid hierarchy to coarser-resolution grids. Alternatively, if the system represents a $\sim1-10\,\pc$-sized molecular cloud, which is always embedded in the larger-scale ISM of the galaxy, then periodic boundary conditions can be a reasonable approximation \citep{KlessenHeitschMacLow2000,JappsenEtAl2005,FederrathKlessen2012,KrumholzEtAl2016,MathewFederrathSeta2023,MathewFederrathSeta2025}, in which case the coarse-level multi-grid solve can be done via the FFT method above, followed by the standard multi-grid cycles to finer grids of the AMR hierarchy.

In practice, multi-grid approaches allow localized refinement while still solving the global problem consistently, making them well suited for self-gravitating astrophysical flows, adaptive mesh refinement (AMR) frameworks, and different types of boundary conditions. A few examples of multi-grid in astrophysical AMR codes are \code{flash} \citep{Ricker2008}, \code{enzo} \citep{BryanEtAl2014}, and \code{ramses} \citep{Teyssier2002}; see Table~\ref{tab:codes}. These methods provide the flexibility and scalability necessary for modern simulations of cosmic structure formation and star-forming systems.

\subsubsection{Tree-based methods}\label{sec:grav_tree}

Other widely used alternatives to FFT or multi-grid solvers are tree methods, which are particularly well suited for N-body simulations of self-gravitating systems. Instead of mapping the mass density onto a mesh, tree algorithms operate directly on particle distributions. The essential idea is to hierarchically subdivide the computational domain into nodes (or cells), forming an octree in 3D or a quadtree in 2D. At each level of the tree, the mass distribution of a node can be approximated by a multipole expansion when viewed from sufficiently large distances. This reduces the computational cost of force evaluation from the brute-force $\mathcal{O}(N^2)$ scaling to approximately $\mathcal{O}(N \log N)$, while retaining controllable accuracy. The Barnes–Hut (BH) method \citep{BarnesHut1986} is the classic example, introducing a simple opening-angle criterion to decide whether a distant group of particles can be treated as a single node or whether the tree must be further refined.

In the BH-tree algorithm, the decision to approximate a group of particles by a single pseudo-particle is governed by the opening-angle criterion,
\begin{equation}
\frac{l}{d} < \theta,
\end{equation}
where $l$ is the size of the node (typically the side length of the cubic cell enclosing the particles), $d$ is the distance from the particle under consideration to the node's center of mass (COM), and $\theta$ is a user-chosen tolerance parameter (commonly $\theta\sim0.5-0.7$, with lower $\theta$ resulting in higher accuracy). If this inequality is satisfied, the node subtends a sufficiently small angle and its contents are treated as a single body of mass $m_{\rm node}$ located at the COM of that node $\mathbf{x}_{\rm COM}$. The gravitational acceleration contribution then reduces to the standard point-mass expression, i.e., using Eq.~(\ref{eq:grav_sum}) with $m_j=m_{\rm node}$ and $\mathbf{r}_j=\mathbf{x}_i-\mathbf{x}_{\rm COM}$ for each node. If the criterion is not met, the node is opened and the test applied recursively to its children. This adaptive procedure ensures that distant groups of particles are efficiently approximated, while nearby or extended structures are resolved in detail, balancing speed and accuracy.

Tree-based methods have proven especially powerful in astrophysical contexts where mass distributions are highly clustered and the dynamic range is large. Unlike mesh-based solvers, they naturally adapt to particle positions without requiring a predefined grid, making them efficient for collisionless dynamics, galaxy formation, cosmological structure growth, and star-formation simulations. Hybrid approaches are also common, for instance the TreePM method \citep{Xu1995,BodeOstrikerXu2000,Bagla2002}, which combines a particle–mesh solver for long-range forces with a tree algorithm for short-range interactions, achieving both speed and flexibility. For a comprehensive introduction, the original \citet{BarnesHut1986} paper remains the canonical reference, complemented by later reviews such as \citet{Springel2005} for the \code{gadget} code family.

While tree-based methods have traditionally been applied in particle-based simulations, they can also be adapted for grid-based approaches by treating individual cells as effective point masses. An example is the BH-tree solver implemented in \code{flash} \citep{WuenschEtAl2018}. Moreover, the utility of tree algorithms extends well beyond solving the Poisson equation. Since the method hierarchically subdivides the domain into nodes characterized by angular size and distance relative to each evaluation point, it can be repurposed for ray-tracing applications. Prominent examples are TreeCol \citep{ClarkGloverKlessen2012} and TreeRay \citep{WuenschEtAl2022,KlepitkoEtAl2023,GachesEtAl2023}, which use the tree structure to estimate local column densities and shielding from external radiation fields. In these applications, the angular domain is typically discretized using the Hierarchical Equal Area isoLatitude Pixelization of the sphere \citep[HEALPix; see][]{GorskiEtAl2005}, which provides an efficient equal-area tessellation of the sphere and ensures uniform angular coverage.

\subsection{Combining the gravitational effects of gas and stars}

Most star-formation simulations contain both gas and stars, such that the total gravitational potential and acceleration are obtained by summing the contributions from both components,
\begin{eqnarray}
\Phi & = & \phigas + \phisinks, \\
\mathbf{g} & = & \ggas + \gsinks.
\end{eqnarray}
While this decomposition is straightforward, obtaining sufficiently accurate gravitational forces for both components is considerably more challenging.

A common approach in grid-based codes is to deposit the stellar masses onto the computational mesh, thus adding them to the gas density and solving Poisson's equation (Eq.~\ref{eq:poisson}) for the combined mass distribution. Although this provides a unified treatment of gravity, the resulting stellar potential is smoothed on the local resolution scale of the gravity solver, reducing the accuracy of stellar accelerations and orbital integration. This can become particularly problematic in dense star-forming regions, where close stellar encounters, binaries, and stellar dynamics play an important role.

A more accurate alternative is to treat the gas and stellar components with different gravity solvers. The gas self-gravity is computed with one of the Poisson methods described above (FFT, multi-grid, or tree-based methods), while the stellar component is evolved with a dedicated $N$-body solver. For moderate numbers of star particles, direct summation (Eq.~\ref{eq:grav_sum}) provides highly accurate stellar accelerations and is commonly used in conjunction with sink particles (Sect.~\ref{sec:sinks}). Because the computational cost scales as $N^2$, this approach remains efficient as long as the number of star particles is modest  (typically of the order of a few thousand) compared to the number of gas resolution elements, which is typically the case for present-day star-formation simulations.

More generally, hybrid gravity schemes couple different gravitational solvers for the gas and stellar components. A prominent example is the BRIDGE algorithm \citep{FujiiEtAl2007}, which combines direct $N$-body integration for collisional stellar systems with tree-based gravity (and, in subsequent applications, hydrodynamics) through a Hamiltonian operator-splitting scheme \citep[e.g.,][]{FujiiEtAl2012,PelupessyPortegiesZwart2012,WallEtAl2019,JoEtAl2024}. Such methods preserve accurate stellar dynamics while retaining the efficiency of Poisson solvers for the gas. Nevertheless, simultaneously resolving gas dynamics, stellar dynamics, and their mutual gravitational interaction over the full range of spatial and temporal scales relevant for star formation remains computationally challenging.

\subsection{Summary of challenges and future directions}

Gravitational interactions in simulations can broadly be divided into two categories: (i) interactions between point masses (e.g., stars) and (ii) interactions of continuous mass distributions (e.g., gas). For point-mass systems, direct $N$-body summation with gravitational softening (Sect.~\ref{sec:grav_nbody}) or tree-based methods (Sect.~\ref{sec:grav_tree}) are commonly employed. Continuous mass distributions, by contrast, require solving Poisson's equation. In particle-based approaches such as SPH or mesh-free methods, tree solvers are often used to approximate the continuous density field. In grid-based methods, the choice of solver depends on geometry and boundary conditions: FFT solvers (Sect.~\ref{sec:grav_fft}) are the most efficient for periodic domains and uniform grids, while multi-grid methods (Sect.~\ref{sec:grav_mg}) are best suited for hierarchically refined meshes with complex boundaries. Tree-based solvers can also be applied in grid codes, where they demonstrate both accuracy and efficiency. A further advantage of tree methods is their versatility, as they can be extended to compute column densities via ray-based techniques that exploit the tree structure.

Overall, gravitational solvers in modern star-formation simulations have become highly robust, yielding accurate solutions when convergence criteria are satisfied (e.g., the opening-angle parameter in tree codes or tolerance thresholds in iterative multi-grid solvers). The remaining challenges lie primarily in efficiently coupling different gravitational components and numerical methods. In particular, accurately modelling the mutual gravitational interaction between gas and stars requires hybrid approaches that preserve the accuracy of stellar dynamics while retaining the computational efficiency of Poisson solvers for the gas. As simulations continue to increase in dynamic range and the number of stars formed, developing scalable hybrid gravity algorithms that balance accuracy, efficiency, and conservation properties will remain an important area of future research.

\section{Star formation} \label{sec:sf}

As gravitational collapse proceeds, regions of runaway density enhancement form. Following collapse to the natural endpoint of star formation calculations{\emdash}stellar densities of $\sim1\,\mathrm{g\,cm^{-3}}${\emdash}requires a dynamic range of some 20~orders of magnitude in density. Meanwhile, numerical stability requires significantly reduced time-steps to treat the evolution of the accompanying small spatial scales at sufficient resolution (see Sect.~\ref{sec:jeans}). While calculations of the collapse of isolated dense cores have succeeded in modeling protostar formation self-consistently \citep[e.g.,][]{TomidaEtAl2013}, these are limited to relatively short evolutionary times of $\mathcal{O}(1)$~yr, though recent advances are pushing to later stages of disk formation and evolution \citep[e.g.,][]{XuKunz2021a,XuKunz2021b,MauxionLesurMaret2024}. The cost remains prohibitive for calculations of star cluster formation, which commence from parsec scales and aim to follow dynamical times of millions of years.

Consequently, most star formation calculations adopt significantly lower spatial resolution and instead follow processes in smaller, denser regions using a sub-resolution model attached to a particle that represents an individual star or stellar group termed a \inquotes{sink particle} \citep{BateBonnellPrice1995}. Sink particles accrete material from the gaseous domain (i.e., from the grid or from SPH particles), i.e., they are mass sinks, but have no internal structure, and thus require no further spatial or temporal resolution. Sophisticated treatments may attach detailed evolutionary models to the particles that follow time-dependent protostellar evolution and stellar feedback. Below we describe different numerical treatments for sink particles (Sect.~\ref{sec:sinks}), sub-resolution modeling of stellar evolution (Sect.~\ref{sec:stellar_evol}), and various forms of stellar feedback (Sect.~\ref{sec:outflow}--\ref{sec:sn}).

\subsection{Jeans resolution criterion} \label{sec:jeans}

Since stars form in dense, self-gravitating gas, the standard resolution criterion in simulations of star formation is based on resolving the Jeans length,
\begin{equation} \label{eq:lJ}
\lJ = \left(\frac{\pi\cs^2}{G\rho}\right)^{1/2},
\end{equation}
where $\cs$ is the sound speed and $G$ is the gravitational constant. \citet{TrueloveEtAl1997} demonstrated that $\lJ$ must be resolved by at least 4~grid cells to avoid artificial fragmentation. Subsequent studies have shown that much higher Jeans resolution ($\sim30$ cells) is needed to capture additional physical processes on the Jeans scale, including convergence of the turbulent solenoidal energy, capturing minimal dynamo amplification of magnetic fields, and resolving small-scale density substructure \citep{TurkAbelOShea2009,SurEtAl2010,FederrathEtAl2011,TurkEtAl2012,FederrathEtAl2014}. An equivalent resolution criterion can be formulated for particle-based methods such as SPH, moving-mesh, and meshless approaches (see Table~\ref{tab:codes}), by requiring adequate resolution of the Jeans mass,
$M_\mathrm{J}=(4\pi/3)(\lJ/2)^3\rho$,
as introduced by \citet{BateBurkert1997}.

\begin{figure}[t]
\centerline{\includegraphics[width=1.0\linewidth]{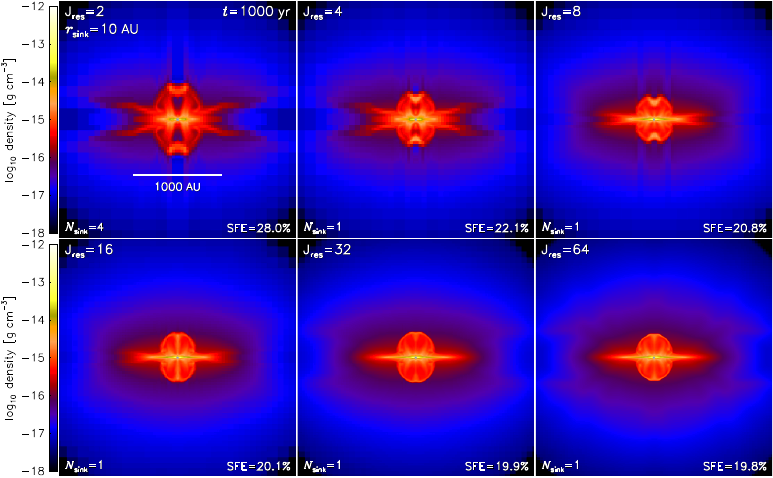}}
\caption{\label{fig:jeans}
Jeans resolution study of accretion disk and outflow formation, $1000\,\yr$ after the formation of the protostar, with edge-on slices through the disk. Panels show identical simulations but with Jeans resolutions of $\mathrm{J_{res}}=2$, 4, 8, 16, 32, and 64~cells per $\lJ$. For $\mathrm{J_{res}}=2$, artificial fragmentation produces four sink particles (annotated as $N_\mathrm{sink}$), while only a single star forms for $\mathrm{J_{res}}\geq4$, confirming the \citet{TrueloveEtAl1997} criterion. However, disk and outflow structure, as well as the accretion rate (measured by the fraction of gas accreted, i.e., the star formation efficiency after $1000\,\yr$), only converge for $\mathrm{J_{res}}\gtrsim30$. Figure reproduced with permission from \citet{FederrathEtAl2014}, copyright by AAS.
}
\end{figure}

Most grid-based star formation studies now adopt a Jeans resolution of at least $\sim8$~cells per Jeans length. However, substantially higher resolution may be required to properly capture turbulence, dynamo action, and core/disk structure on the Jeans scale. Several studies suggest that at least $\sim30$~cells are needed \citep{SurEtAl2010,FederrathEtAl2011,TurkEtAl2012}, which poses a major computational challenge. Compared to a resolution of 8~cells per $\lJ$, increasing to 30~cells requires $\sim(30/8)^3\gtrsim50$ times more computational resources in 3D. While this is an enormous cost, the investment may be essential for accurately resolving turbulence and magnetic-field amplification. Moreover, systematic tests of core and disk formation, such as those shown in Fig.~\ref{fig:jeans}, reveal substantial structural differences with increasing Jeans resolution{\emdash}for example, significant disk flaring persists when $\lJ$ is resolved with $\lesssim20$~cells and only disappears at $\sim30$~cells per $\lJ$.

\subsection{Star particles} \label{sec:sinks}

Since the Jeans resolution criterion (Eq.~\ref{eq:lJ}) cannot be upheld indefinitely with rising density during collapse, star particles have become a crucial tool for modeling collapse, accretion, and star formation. This method is often referred to as \inquotes{star particle} or \inquotes{sink particle} method \citep{BateBonnellPrice1995,KrumholzMcKeeKlein2004,FederrathBanerjeeClarkKlessen2010,HubberWalchWhitworth2013,BleulerTeyssier2014}. Any code aiming to capture star-forming gas over a significant period of time during accretion and fragmentation will need a sink particle implementation\footnote{For example, the sink particle method implemented in \code{flash} is publicly available on GitHub: \url{https://github.com/chfeder/cfflash}}.

\subsubsection{Sink particle formation} \label{sec:sink_creation}

Sink particles are designed to represent star-forming gas undergoing gravitational collapse and accretion. Thus, only bound and collapsing gas should be eligible to form sink particles and be accreted. To enforce this, a control volume must be defined around each grid cell that exceeds a density threshold,
\begin{equation} \label{eq:rhosink}
\rhosink = \frac{\pi\cs^2}{4G\rsink^2},
\end{equation}
which is obtained from Eq.~(\ref{eq:lJ}) by setting the Jeans length $\lJ=2\,\rsink$ and solving for the density. Here, $2\rsink$ is the chosen sink particle diameter, which is typically set to $5$~grid-cell lengths (corresponding to $\rsink=2.5\,\dx$) at the maximum refinement level, in order to avoid artificial fragmentation (cf.~Sect.~\ref{sec:jeans}).

Grid cells that exceed the threshold in Eq.~(\ref{eq:rhosink}) must not immediately form sinks. Instead, a sequence of additional checks for collapse and gravitational instability is performed. First, a spherical control volume of radius $\rsink$ is defined, centered on the candidate cell with $\rho>\rhosink$. The total gravitational, thermal, kinetic, and magnetic energies ($E_\mathrm{grav}$, $E_\mathrm{th}$, $E_\mathrm{kin}$, $E_\mathrm{mag}$) are then determined by summation over all cells $i$ within the control volume, and using the gravitational potential provided by the Poisson solver (Sect.~\ref{sec:poisson}). We note that the energies are evaluated in the reference frame of the central cell{\emdash}in particular, the relevant kinetic energy in the following checks is the unordered, turbulent component of kinetic energy, corresponding to the motions that can counter the local gravitational collapse of the region.

Therefore, a sink particle should only be created if the gas within the control volume,
\begin{enumerate}
\item lies on the highest refinement level (for grid-based codes),
\item is not within $\rsink$ of an existing sink particle,
\item is converging from all directions, i.e., radial velocity $v_{r,i}<0$,
\item has a minimum of the gravitational potential at its center,
\item is gravitationally bound ($\abs{E_\mathrm{grav}} > E_\mathrm{th}+E_\mathrm{kin}+E_\mathrm{mag}$), and
\item is Jeans-unstable.
\end{enumerate}
\citet{GrudicEtAl2021} replace condition~4 with a tidal stability criterion, while \citet{HubberWalchWhitworth2013} instead employ a Hill-sphere criterion. Both serve the same purpose of ensuring that the candidate cell lies at the true center of a collapsing region. These checks ensure that only genuinely collapsing, star-forming gas is converted into sink particles. We further note that depending on the local physical conditions, some of these checks may be redundant{\emdash}nevertheless, in general it is safest to check for all of them.

The density threshold in Eq.~(\ref{eq:rhosink}) alone is insufficient, because local density enhancements can arise in shocks without leading to collapse. Therefore, the cell-by-cell Jeans criterion alone is inadequate, since it does not guarantee that sufficient mass is contained within the entire Jeans volume to become gravitationally bound. Ensuring collapse requires non-local consistency, as the Jeans length must be resolved by multiple grid cells (see Sect.~\ref{sec:jeans}).

The implementations by \citet{BateBonnellPrice1995}, \citet{FederrathBanerjeeClarkKlessen2010}, \citet{HubberWalchWhitworth2013}, and \citet{BleulerTeyssier2014} all adopt the set of creation checks outlined above (with slight variations in the details). By contrast, the sink method of \citet{KrumholzMcKeeKlein2004} relies solely on the local density criterion $\rho>\rhosink$ (Eq.~\ref{eq:rhosink}) for sink formation. As discussed in \citet{FederrathBanerjeeClarkKlessen2010}, this single-cell condition is insufficient to ensure that sink-forming gas is both bound and collapsing, and can therefore lead to spurious sink creation.

\begin{figure}[t]
\centerline{\includegraphics[width=1.0\linewidth]{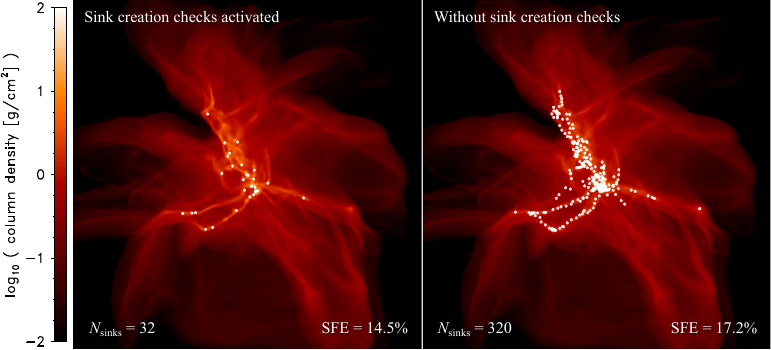}}
\caption{\label{fig:sink_checks}
The importance of sink particle creation checks for identifying truly collapsing regions. Shown are two identical turbulent cloud collapse simulations forming sink particles (shown as white circles): the left-hand panel uses the full set of sink creation checks (Sect.~\ref{sec:sink_creation}), while the right-hand panel uses only the density threshold criterion in Eq.~(\ref{eq:rhosink}). Without the additional checks, $10\times$ more sink particles are created (32 sinks vs.~320). The total stellar mass is also overestimated, with an SFE of 14.5\% compared to 17.2\%. Figure adapted from \citet{FederrathBanerjeeClarkKlessen2010}. An animation of these simulations is available at \url{https://www.mso.anu.edu.au/~chfeder/pubs/sinks/sinks.html}.}
\end{figure}

Figure~\ref{fig:sink_checks} illustrates this point with two otherwise identical simulations: one with the full set of sink creation checks enabled (left-hand panel), and one using only the density threshold criterion (right-hand panel). In the latter case, the number of sink particles is overproduced by an order of magnitude ($10\times$ more sinks), and the total accreted mass is also inflated, yielding a star formation efficiency (SFE) of 17.2\% instead of 14.5\%. Thus, when adopting the \citet{KrumholzMcKeeKlein2004} approach, it is critical to choose a sufficiently high density threshold for sink formation, so that collapse and boundedness can be assumed implicitly rather than explicitly tested.

When and where sink particle formation occurs is also sensitive to the assumed thermodynamics of the simulated gas. A gravitationally collapsing, isothermal cloud has no characteristic stopping scale and will undergo successive fragmentation to infinitely high density \citep{Hoyle1953}. In this case, convergence is not possible, since higher resolution will produce more fragmentation, regardless of the details of the sink implementation \citep[e.g.,][]{GuszejnovEtAl2018,LeeHennebelle2018}. Therefore, the sink creation checks are necessary but not sufficient for convergence. Numerical convergence of cloud fragmentation requires including either a barotropic equation of state or self-consistently solving the radiation transfer equations \citep{TrueloveEtAl1997,LeeHennebelle2018}.

\subsubsection{Sink particle accretion} \label{sec:sink_accretion}

In grid/AMR simulations, once a sink particle is created, it can accrete gas from the grid but only if the gas exceeds the density threshold, is inside the sink particle accretion radius, is bound to the particle and more bound to it than to any other, and is collapsing toward it. If all these criteria are fulfilled, the excess mass above the density threshold defined by Equation~(\ref{eq:rhosink}) is removed from the MHD system and added to the sink particle, such that mass, momentum, and angular momentum are conserved. The mass fraction $\Delta m_i$ to be accreted from cell $i$ with mass $m_i$ and volume $V_i$ is $\Delta m_i=m_i-\rhosink V_i$. Within the control volume $(4\pi/3)\rsink^3$ of a sink particle, we gather the mass, center of mass (COM), momentum, and angular momentum of the material to be accreted,
\begin{equation}
\def\arraystretch{1.2}
\setlength{\tabcolsep}{1.0pt}
\begin{tabular}{rrl}
\text{ Mass: } & $\macc$ & $= \sum_i \Delta m_i$ \\
\text{ Center of mass: } & $\racc$ & $= \sum_i \Delta m_i \mathbf{r}_i/\macc$ \\
\text{ Linear momentum: } & $\pacc$ & $= \sum_i \Delta m_i \mathbf{v}_i$ \\
\text{ Angular momentum: } & $\lacc$ & $= \sum_i \Delta m_i \mathbf{r}_i \times \mathbf{v}_i$,
\end{tabular}
\end{equation}
where $\mathbf{r}_i$ and $\mathbf{v}_i$ are the lab-frame position and velocity of the gas in cell~$i$. We then remove the mass $\Delta m_i$ from each affected cell and update the sink particle properties{\emdash}mass $\msink$, position $\Rsink$, linear momentum $\psink$, and angular momentum $\lsink${\emdash}such that the total mass, center of mass (COM), linear momentum, and angular momentum are conserved. The latter includes both the lab-frame angular momentum of the sink particle, $\lsink$, and its intrinsic spin, $\ssink$. Denoting post-accretion quantities with a prime, the updated sink properties are
\begin{equation}
\def\arraystretch{1.2}
\setlength{\tabcolsep}{1.0pt}
\begin{tabular}{rrl} \label{eq:sinkaccretion}
\text{Mass:} & $\msink'$ & $= \msink + \macc$ \\
\text{COM:} & $\Rsink'$ & $= (\msink \Rsink + \macc\racc) / \msink'$ \\
\text{Linear momentum:} & $\psink'$ & $= \psink + \pacc$ \\
\text{Angular momentum:} & $\lsink'$ & $= \Rsink' \times \psink'$ \\
\text{Spin:} & $\ssink'$ & $= \ssink + \lacc + \lsink - \lsink'$.
\end{tabular}
\end{equation}
Here, the spin $\ssink$ is required to absorb the excess angular momentum carried by the accreted material. Importantly, this excess is not simply equal to $\lacc$, because the sink position $\Rsink'$ and momentum $\psink'${\emdash}and thus its orbital angular momentum $\lsink'${\emdash}change during the accretion step. The corrective term $\lsink-\lsink'$ accounts for this shift, ensuring exact global angular momentum conservation \citep[see Appendix~B in][]{FederrathBanerjeeClarkKlessen2010}.

In SPH codes the accretion criteria are the same (within $\rsink$ and bound to the sink), and additionally satisfies an angular momentum criterion to ensure that it does not have an orbit that takes it outside $\rsink$ \citep{BateBonnellPrice1995}. The meshless code \code{gizmo} also requires that the volume element associated with accretion must be smaller than the spherical volume corresponding to $\rsink$ to ensure that it has sufficient spatial resolution to be accreted. The \code{gizmo} sink accretion prescription further introduces a sub-grid reservoir, where the mass to be accreted is first stored and subsequently accreted onto the sink according to the local freefall time \citep{GrudicEtAl2021}; this reduces the artifical variability of accreting entire particles of a fixed mass, while avoiding artificially over-smoothing. This problem does not exist in the grid-based method described above, where the fraction of accreted material always exactly matches that required to fulfill the Jeans criterion of the collapsing gas. We note that the grid-based accretion method by \citet{KrumholzMcKeeKlein2004} is different in that it combines Bondi-Hoyle accretion with accretion of all gas cells exceeding $\rhosink$. In practice, \citet{KrumholzMcKeeKlein2004} find that the Bondi-Hoyle pathway of accretion is usually subdominant compared to the direct accretion method by temporarily-created particles exceeding $\rhosink$, and subsequently merged into the main sink doing the accretion. Since all gas above $\rhosink$ is therefore accreted, without a check for boundedness and collapse, the accretion rate is somewhat overestimated in the \citet{KrumholzMcKeeKlein2004} method (cf.~Fig.~\ref{fig:sink_checks}).

Finally, we note that it has been argued that magnetic flux should also be accreted. Implementing this would require directly modifying the magnetic field whenever a sink particle is created or gains mass. However, none of the existing sink particle methods adopt this approach, in order to avoid complications associated with altering the local magnetic field and the risk of introducing $\nabla\cdot\Bvec$ errors. Instead, our method leaves $\Bvec$ unchanged and only accretes mass. The magnetic flux, and thus the associated magnetic pressure and tension, remain on the grid. This is in fact a desirable property: if the magnetic field were accreted, one would need to model its influence on the surrounding gas through sub-resolution prescriptions, which becomes unnecessary when the field is left intact.

\subsubsection{Sink particle dynamics} \label{sec:sink_dynamics}

With sink particles present there are three types of gravitational interactions that need to be considered:
\begin{enumerate}[label=(\roman*)]
\item the force of the gas on the sink particles (gas$\to$sink),
\item the force of the sink particles on the gas (sink$\to$gas),
\item the force of the sink particles on other sinks (sink$\leftrightarrow$sink).
\end{enumerate}
Each of these could be treated with distinct methods for calculating the gravitational acceleration. However, we note that (i) and (ii) should be done with the same method, which ensures strict symmetry between the two interactions that would otherwise violate linear and angular momentum conservation \citep{FederrathBanerjeeSeifriedClarkKlessen2011}.

Ideally, all three gravitational interactions are computed by direct $N$-body summation (see Sect.~\ref{sec:grav_nbody}) over all sink particles and grid cells or SPH particles, providing the most accurate gravitational acceleration. This is the case for example in the \code{flash} code\footnote{A detail worth noting is that while the sink$\leftrightarrow$sink interaction is softened with Eq.~(\ref{eq:spline_softening}) to capture tight sink-sink orbits as accurately as possible, interactions (i) and (ii) are softened with $\mathcal{K}=(r/\rsoft)^3$ for $r\leq\rsoft$ and $\mathcal{K}=1$ for $r>\rsoft$, resulting in a linearly rising gravitational acceleration (Eq.~\ref{eq:grav_sum}) \inquotes{inside} the sink particle, as for the purposes of gas$\to$sink and sink$\to$gas interactions, it is reasonable to approximate the sink as a gaseous object of radius $\rsink$.}. Many other codes only use direct summation for the sink$\leftrightarrow$sink interaction (iii), while they use the Poisson solver (usually in the form of multi-grid or tree-based; Sect.~\ref{sec:poisson}) to compute the gas$\to$sink and sink$\to$gas interactions. Using the Poisson solver first requires that the  sink mass is interpolated onto the grid \citep[or treating the sink mass as part of the tree-solve in SPH codes or other meshless codes like \code{gizmo}; see][]{GrudicEtAl2021}, and then solving the Poisson equation for the combined gas+sink potential \citep{KrumholzMcKeeKlein2004,BleulerTeyssier2014}. While this has the advantage of speed, i.e., $\mathcal{O}(N\log N)$ vs.~$\mathcal{O}(N^2)$, and simplicity in that one can use the existing Poisson solver, it comes at the price of a substantially smoothed acceleration field in the gas$\to$sink and sink$\to$gas interactions, which can cause inaccuracies in the gas and sink dynamics, especially in regions close to a sink particle, e.g., in the orbits of a sink particle around a dense gas accumulation \citep{FederrathBanerjeeClarkKlessen2010,FederrathBanerjeeSeifriedClarkKlessen2011}.

For time integration of sink particle positions and velocities, a second-order Leapfrog integrator is commonly employed. This is combined with both velocity- and acceleration-based timestep constraints, which allow close and highly eccentric orbits of sink particles to be resolved without introducing errors such as artificial perihelion shifts. These timestep constraints can, however, become very stringent{\emdash}particularly for closely approaching sink particles interacting via mechanism (iii){\emdash}and may force extremely small timesteps. To alleviate this, a sub-cycling method can be used: the sink$\leftrightarrow$sink interactions are updated in each sub-cycle, while gas$\to$sink and sink$\to$gas forces are held fixed until the sub-cycles catch up with the global timestep. The global timestep itself is limited by the CFL condition together with additional timestep constraints, including a gravity-based constraint that accounts only for interactions (i) and (ii) \citep[see implementation details and orbit tests in][]{FederrathBanerjeeClarkKlessen2010,FederrathBanerjeeSeifriedClarkKlessen2011}.

\subsubsection{Sink particle merging} \label{sec:sink_merging}

Sink particle merging is available in some implementations of sink particles. For example, in the \citet{KrumholzMcKeeKlein2004} implementation in \code{orion}, sink particle merging is a key step in modeling sink particle creation and accretion, as their implementation creates many temporary sink particle fragments that then need to undergo merging. In the \citet{FederrathBanerjeeClarkKlessen2010} implementation in \code{flash}, sink particle merging is optional, as accretion is handled directly with the gas on the grid (see above). If merging is activated, sink particles are only allowed to merge, if they are inside the accretion radii of one another, and if they are gravitationally bound and converging. The merged particle is moved to the center of mass of the merging particles, and their linear and angular momenta are assigned to the merged particle. In \code{gizmo}, sink merging only occurs if the pair has a binary semi-major axis less than $\rsink$ and the secondary is at least 10~times smaller than the primary \citep{GrudicEtAl2021}.

Merging can be used to model star formation in extremely dense environments, where stellar mergers have been considered a possibility to form massive stars \citep{ZinneckerYorke2007}. Under more normal conditions, however, merging is rather unlikely, considering that sink particles are often used to represent individual stars, for which the probability of merging is near zero due to the actual stellar size they represent and the velocity dispersion between forming stars. One possible exception occurs during the bloating phase of high-mass stars, during which their radii expand significantly \citep{BhandareEtAl2020}.

\subsection{Summary of challenges and future directions}

Achieving sufficient Jeans resolution is computationally demanding, but essential. At least $\sim30$~cells per Jeans length are required to accurately capture key physical processes such as the solenoidal turbulent energy content, the minimum amplification of magnetic fields, and the structure of disks on the Jeans scale (cf.~Sect.~\ref{sec:jeans}).

Rigorous sink particle creation checks are critical to ensure that only bound, collapsing gas forms sinks. Without these checks, spurious sinks can form and the accretion rate may be systematically overestimated (cf.~Sect.~\ref{sec:sink_creation}).

During gas accretion onto sink particles, strict conservation of mass, center of mass, linear momentum, and angular momentum must be maintained. Careful numerical implementation is required to guarantee this conservation (cf.~Sect.~\ref{sec:sink_accretion}).

Many implementations combine direct $N$-body summation for sink$\leftrightarrow$sink interactions with a grid-based Poisson solver (or a tree solver in particle-based methods) for sink--gas coupling. However, in dense gas systems, resolving accurate orbital dynamics requires direct summation of all three interaction channels: sink$\leftrightarrow$sink, gas$\to$sink, and sink$\to$gas (cf.~Sect.~\ref{sec:sink_dynamics}).

Together, these criteria define the minimum requirements for robust and physically reliable collapse calculations and sink particle implementations. 

Beyond individual star formation, sink (or star) particle methods are also widely used to represent entire star clusters in large-scale simulations of galaxy formation and evolution. In this context, sub-resolution prescriptions are employed to model the properties of the embedded stellar population{\emdash}including the IMF, chemical composition, and resulting feedback efficiencies{\emdash}at varying levels of detail. However, significant uncertainties remain in these approaches, owing to the many free parameters involved. We therefore anticipate substantial developments in the coming years, with increasingly sophisticated, multi-faceted models aimed at bridging the gap between small-scale star formation physics and galaxy-scale evolution.

\section{Stellar feedback} \label{sec:feedback}

Stars not only accrete gas, but also eject material, loaded with momentum and energy, and sometimes with species that can enrich and change the chemical composition of the ISM. This stellar feedback is a key process for many reasons. Jet feedback (Sect.~\ref{sec:outflow}) for instance transports angular momentum away from the accretion disk, allowing stars to accrete more efficiently through the protostellar disk. Radiation feedback in the form of heating (Sect.~\ref{sec:heating}) changes the thermodynamic conditions of the parental cloud, suppressing fragmentation of the gas. Stellar winds (Sect.~\ref{sec:winds}), ionization (Sect.~\ref{sec:ionization}), and supernova feedback (Sect.~\ref{sec:sn}) can drive powerful outflows and shocks and change the chemical composition of the gas in their surroundings.

On a per-star basis, a rough ordering of feedback mechanisms by characteristic energy budget is: protostellar jets and outflows, accretion heating, stellar winds, ionizing radiation, and supernovae. This sequence, however, should be interpreted with caution, as it depends on stellar mass, environment, and crucially the efficiency with which each feedback channel couples to the surrounding gas. In particular, ionizing radiation carries a large integrated energy budget, but only a fraction is converted into gas motions, whereas supernovae provide the dominant source of mechanical energy input into the ISM. Moreover, jets/outflows and accretion heating are associated with essentially all forming stars, including the numerous low-mass population, while stellar winds, ionizing radiation, and supernovae are dominated by massive stars. As a result, although protostellar feedback channels are individually less energetic, they collectively play a crucial role in regulating star formation, especially during the early stages.

Stellar feedback is generally modeled by adding source terms to the RHS of the mass, momentum, and energy equations (Eqs.~\ref{eq:mhd1}--\ref{eq:mhd5}). Generally, adding the entire feedback contribution to a single grid cell or particle produces numerical instability, especially when the equations are very stiff. Various prescriptions have been developed to distribute the stellar feedback to gas in the vicinity of the emitting star in a robust and efficient way. Grid-based codes typically add the mass, momentum, and energy to some number of neighboring cells according to a weighting function centered on the stellar location \citep{OffnerEtAl2009,CunninghamEtAl2011,FederrathEtAl2014,HopkinsGrudic2019,MathewFederrath2020}. Similarly, SPH methods distribute the radiated energy among the nearby particles that are exposed to the stellar radiation, i.e., have a direct sight-line to the source \citep{JonesBate2018} or insert new cells/particles \citep{GrudicEtAl2021}.


\subsection{Stellar and protostellar evolution} \label{sec:stellar_evol}

A key requirement for many of these feedback prescriptions is to track the evolution of the unresolved protostar, in order to estimate the stellar radius and stellar luminosity, which is required for modeling radiation feedback. Given the accretion rate (Sect.~\ref{sec:sink_accretion}), the protostar will grow in mass, producing accretion luminosity \citep{NakanoEtAl2000}, and start deuterium burning. The protostellar evolution phases can be summarized and implemented as \citep{OffnerEtAl2009}:
\begin{itemize}
\item {\bf Phase 0: Pre-collapse} -- The gas continues to collapse but the object has not yet reached stellar densities.
\item {\bf Phase I: No-burning Contraction} -- The object is adiabatically contracting but the central density has not yet reached a temperature of $1.5 \times 10^6$\,K, at which deuterium fusion begins.
\item {\bf Phase II: Core Deuterium Burning with Fixed $T_c$ } -- Deuterium fusion begins in the core and the protostar enters a convective phase during which the central core temperature, $T_c$, remains constant.
\item {\bf Phase III: Core Deuterium Burning with Variable $T_c$ } -- The deuterium in the core is depleted and the protostar contracts further, during which $T_c$ rises.
\item {\bf Phase IV: Shell Deuterium Burning} -- Deuterium burning begins in a shell outside the core, as the center continues to contract.
\item {\bf Zero Age Main Sequence (ZAMS)} -- The central temperature reaches $10^7$\,K, at which point hydrogen fusion begins and the protostar becomes a ZAMS star.
\end{itemize}
Thus, the protostellar evolution model primarily provides the stellar radius and luminosity during its accretion phase and its main sequence evolution. The accretion rate and resulting luminosity during the early stages, in particular, are crucial for determining the efficiency of various feedback processes that we describe in the following.

\subsection{Protostellar jet and outflow feedback} \label{sec:outflow}

Observations show that jets and outflows are launched from virtually all young protostar–disk systems, with ALMA now providing unprecedented detail and resolution. Jets and outflows play a central role in star formation: they help explain the low SFR and efficiency in turbulent molecular clouds, remove angular momentum from the rotating star-forming core/disk, and reduce the characteristic stellar mass in the IMF by a factor of $\sim3$ \citep{FederrathEtAl2014,MathewFederrath2021,GuszejnovEtAl2021}, where the exact reduction can depend on the assumed subgrid outflow parameters \citep{GuszejnovEtAl2021}. In addition, outflows can significantly contribute to driving and sustaining turbulence in star-forming regions \citep[e.g.,][]{LiNakamura2006,NakamuraLi2007,NakamuraEtAl2011,OffnerChaban2017,LebreuillyEtAl2024}, suggesting that star formation may in part regulate itself through outflow feedback. However, accurately incorporating jets and outflows into numerical simulations of star cluster formation remains a major challenge. To date, only a few codes include subgrid models for jet feedback{\emdash}e.g., \code{orion}, \citealt{CunninghamEtAl2011}; \code{flash}, \citealt{FederrathEtAl2014}; \code{gizmo}, \citealt{GuszejnovEtAl2021}; \code{Ramses}, \citealt{MurrayGoyalChang2018,VerliatEtAl2022}{\emdash}, each with varying levels of detail. Below we summarize how jet feedback can be implemented as a sub-resolution module, following the methods of \citet{FederrathEtAl2014}, and implemented in \code{flash}.

In this implementation, outflows are launched along the sink particle angular momentum vector once accretion has been calculated, consistent with the modular design of the code. The accretion and outflow modules are thus separated, with the outflow module directly depending on the accretion step but not vice versa. This ensures that in each timestep the system state after accretion is fully determined before outflow feedback is applied, allowing a prescribed fraction of the accreted material to be re-inserted and launched in bipolar outflows. Two loops over all sink particles and their surrounding grid cells (within the outflow radius; see details below) are required: the first gathers information about the cells that will host outflow injection, and the second updates the state variables to launch the outflows. In this way, the conservation of mass, momentum, and angular momentum is strictly maintained.

\subsubsection{Geometry of the outflow launching region}

\begin{figure}[t]
\centerline{\includegraphics[width=0.9\linewidth]{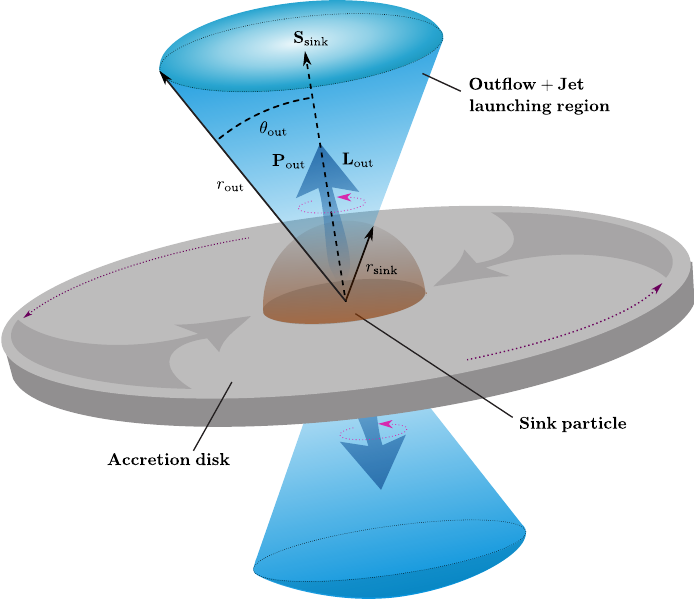}}
\caption{\label{fig:outflow_schematic}
Schematic (not to scale) of the sub-resolution outflow+jet model by \citet{FederrathEtAl2014}, showing
the basic geometry with the sink particle radius $\rsink$, sink spin $\ssink$, outflow radius $\rout$, outflow opening angle $\thetaout$, outflow linear momentum $\pout$, and outflow angular momentum $\lout$. The upper and lower components of the outflow region are shown as the two spherical sectors in blue, for the which the symmetry axis is defined by $\ssink$.
}
\end{figure}

Figure~\ref{fig:outflow_schematic} illustrates the sub-resolution outflow model of \citet{FederrathEtAl2014}, implemented in \code{flash} (see Sect.~\ref{sec:outflow_code_access}). The outflow is launched through two spherical sectors of radius $\rout$ and opening angle $\thetaout$, aligned with the spin axis of the sink particle, $\ssink$. We refer to these regions as the top and bottom \inquotes{outflow sectors}.

Observations and theoretical models indicate that outflows align with the rotation axis of the accretion disk \citep{AppenzellerMundt1989}. In our model, the sink spin{\emdash}acquired through the accretion of angular momentum (see Sect.~\ref{sec:sink_accretion}){\emdash}serves as a proxy for the unresolved disk spin axis. Although the opening angle is, in principle, a user-defined parameter, adopting $\thetaout=30^\circ$ is consistent with magneto-centrifugal acceleration of the jet component \citep{BlandfordPayne1982}, and this value is therefore used as the default in our sub-resolution model.

\subsubsection{Outflow mass transfer} \label{sec:outflow_mass}

The outflow model reinserts a fixed fraction of the accreted mass and launches it away from the sink particle to represent the outflow+jet component. The outflow mass $\mout$ inserted in each timestep $\dt$ is determined by the sink accretion rate $\maccdot$,
\begin{equation} \label{eq:mout}
\mout = \fmass\,\maccdot\,\dt,
\end{equation}
where $\fmass$ is a user-defined parameter. Theory, observations, and numerical simulations all suggest $\fmass\sim0.1-0.4$ \citep[][and references therein]{MatznerMcKee2000,CunninghamEtAl2011,FederrathEtAl2014}. In the \code{flash} implementation discussed here, we adopt $\fmass=0.3$ as a reasonable standard value. The results, however, are largely insensitive to the precise choice of $\fmass$, owing to the self-regulating nature of accretion–outflow coupling: if $\fmass$ is increased, $\maccdot$ is temporarily reduced, which in turn weakens the outflow until accretion resumes, thus establishing a self-regulated balance.

In practice, after each accretion step the outflow mass $\mout$ is computed, subtracted from the sink particle, and deposited into the gas within the outflow sectors, ensuring mass conservation. To achieve a smooth transition at the boundary of the launching region (defined by $\rout$ and $\thetaout$), we apply radial and angular smoothing functions,
\begin{eqnarray}
\mathcal{R}(r,\rout) & = &
\left\{
\def\arraystretch{1.2}
\begin{array}{C{3.3cm} l} 
\sin\!\left[\pi(r/\rout)\right] & \textnormal{for $r \leq \rout$}\\
0                               & \textnormal{for $r > \rout$}
\end{array}
\right., \label{eq:radsmooth} \\
\Theta(\theta,\thetaout) & = &
\left\{
\def\arraystretch{1.2}
\begin{array}{C{3.3cm} l} 
\cos^p\!\left[(\pi/2)(\theta/\thetaout)\right] & \textnormal{for $\abs{\theta}\leq \thetaout$}\\
0                                               & \textnormal{for $\abs{\theta} > \thetaout$}
\end{array}
\right., \label{eq:angsmooth}
\end{eqnarray}
with $p=1$ as the default smoothing power. These functions quickly approach zero at the edges of the outflow sectors, preventing sharp discontinuities. The resulting outflow morphology is insensitive to the exact choice of $p$ as long as $p \leq 4$. While the model would in principle function without smoothing, the application of these functions avoids numerical instabilities near the sector boundaries.

\subsubsection{Outflow momentum transfer} \label{sec:outflow_mom}

The momentum transferred to each of the two outflow sectors in the rest frame of the sink particle is $\pout=\pm\mout\vout/2$, where $\mout$ is the outflow mass (Eq.~\ref{eq:mout}). For the radial outflow velocity $\vout$, we adopt the Keplerian speed at the foot-point of a centrifugally-driven jet, close to the protostellar radius, as suggested by analytic models \citep{BlandfordPayne1982,ShibataUchida1985,ShibataUchida1986,PudritzNorman1986,WardleKoenigl1993,KoeniglPudritz2000}. The resulting outflow velocity is
\begin{equation} \label{eq:vout}
\abs{\vout} = \left(\frac{G\msink}{10\,\rsol}\right)^{1/2}\mathcal{V}(\theta/\thetaout) = 100\,\km\,\s^{-1}\left(\frac{\msink}{0.5\,\msol}\right)^{1/2}\mathcal{V}(\theta/\thetaout),
\end{equation}
which depends on the sink mass $\msink$ and on the two-component jet+outflow profile $\mathcal{V}(\theta/\thetaout)$ defined below.

Observations and jet-launching simulations \citep[e.g.,][]{Camenzind1990,MachidaEtAl2008,MachidaBasu2019} consistently show that outflows exhibit two components: a fast, collimated jet and a slower, wide-angle outflow. To capture both, we adopt a normalized velocity profile based on the angular smoothing function $\Theta(\theta/\thetaout)$ from Eq.~(\ref{eq:angsmooth}),
\begin{equation}
\mathcal{V}(\theta,\,\thetaout) = \frac{3}{4}\,\Theta(\theta,\,\thetaout/6) + \frac{1}{4}\,\Theta(\theta,\,\thetaout),
\end{equation}
where the first term represents the fast collimated jet and the second term the slower wide-angle outflow. Momentum transfer is implemented symmetrically between the two sectors to ensure exact global momentum conservation.

\subsubsection{Outflow angular momentum transfer} \label{sec:outflow_angmom}

Outflows and jets are observed to rotate \citep{BacciottiEtAl2002}, making them a key mechanism for transporting angular momentum away from the protostar and its disk, thereby enabling the star to grow in mass \citep{UchidaShibata1985,ShuAdamsLizano1987,KoeniglPudritz2000,PudritzEtAl2007,FrankEtAl2014}. Similar to the mass transfer described in Sect.~\ref{sec:outflow_mass}, we introduce a fraction $\fang$ of the accreted angular momentum, $\ssink' - \ssink$ (from Eq.~\ref{eq:sinkaccretion}), which is released along the sink particle's rotation axis $\ssink'$ after each accretion event. This fraction of angular momentum is transferred to the two outflow sectors as
\begin{equation}
\lout = \fang \, (\ssink' - \ssink)\,\frac{\ssink'}{\abs{\ssink'}}.
\end{equation}
Observationally, \citet{BacciottiEtAl2002} used Hubble Space Telescope data of the DG Tau flow to infer angular momentum fractions of $\fang=0.6-1.0$. This range is consistent with predictions from disk-wind models \citep{PelletierPudritz1992}, which yield $\fang\sim0.7-1.0$ for sub- to super-Alfv\'{e}nic accretion flows. Numerical simulations further support these values: \citet{BanerjeePudritz2006} and \citet{HennebelleFromang2008} measured $\fang\sim0.5-2$, with a time-averaged value of $\fang\sim0.9$. This agrees well with the observations of \citet{BacciottiEtAl2002}. Based on these findings, we adopt $\fang=0.9$ as the standard value in the jet+outflow model in \code{flash}, resulting in 90\% of the accreted angular momentum re-injected in the outflow sectors and removed from the disk-protostar system{\emdash}a physically reasonable choice for magnetically-driven jets and outflows \citep{PudritzEtAl2007}. This prescription ensures angular momentum is consistently removed from the accreting system, in agreement with both observations and theoretical expectations.

\subsubsection{Implementation, code validation, and access} \label{sec:outflow_code_access}

The practical implementation of the outflow module in \code{flash}, as described above, requires two loops over all grid cells, with each loop iterating over all sink particles. The first grid loop is used to gather information from the cells within the two outflow sectors and to test-insert the outflow mass, momentum, and angular momentum. In this stage, the state variables of the gas cells and the sink particles are not yet modified; instead, the collected information is used to ensure that all conservation laws can be satisfied exactly. The second grid loop then updates the gas cell properties to inject the outflowing mass, momentum, and angular momentum.

It is worth noting that no additional smoothing of the momentum and angular momentum injection is required, because both are proportional to the injected mass, which is already smoothed via Eqs.~(\ref{eq:radsmooth}) and~(\ref{eq:angsmooth}). Finally, the sink particle properties are updated to guarantee global conservation. As in the case of accretion, strict conservation requires slight repositioning of the sink particles driving outflows in order to satisfy two major constraints: (i) global conservation of the center of mass (COM), and (ii) ensuring that the linear momentum components injected into the two outflow sectors are parallel.

The sink particle outflow module implemented in \code{flash} is publicly available on GitHub\footnote{Sink particle jet/outflow module in the \code{flash} code: \url{https://github.com/chfeder/cfflash}}. The module has been calibrated, tested, and validated against dedicated resolved jet/outflow simulations in \citet{FederrathEtAl2014}. Key features of the method include:
\begin{itemize}
\item convergence of fundamental outflow properties (mass, linear and angular momentum, and jet speed) for a sufficiently resolved outflow radius, $\rout \gtrsim16\dx$, i.e., $\sim32$~cells in diameter,
\item a systematic quantification of the impact of the default parameter choices for $\fmass$ and $\fang$, demonstrating the self-regulation of the outflow–accretion system, and
\item built-in adaptivity of the sub-resolution model to changes in the absolute resolution scale of a simulation.
\end{itemize}
Adopting the open-source outflow module from \code{flash} in other codes is straightforward, owing to its design, which cleanly separates the accretion and outflow steps. As a result, the module only requires the sink particle properties after accretion, which then serve as input for the outflow injection.

\subsubsection{Comparison of jet/outflow implementations} \label{sec:outflow_comparison}

The first outflow model applied to star cluster formation was developed by \citet{LiNakamura2006} and \citet{NakamuraLi2007}. In their approach, a Lagrangian particle was created once the gas density exceeded 100~times the mean density. Twenty percent of the mass in a cubic region surrounding the particle was then transferred to it, without modeling subsequent accretion. Their feedback consisted of isotropic, radial point explosions, later extended by \citet{NakamuraLi2007} to include a collimated component with an opening angle of $30^\circ$. In this model, the outflow axis was aligned with the local magnetic field vector{\emdash}a choice that may inadvertently couple to the wound-up toroidal component of the disk field. Angular momentum transfer was not included.

\citet{WangEtAl2010} and \citet{NakamuraLi2011} improved on these early models by adding accretion onto sink particles. However, the outflow axis remained tied to the local magnetic field, and angular momentum transfer was still omitted. \citet{DaleBonnell2008} studied isotropic and collimated winds in massive star formation using SPH, but their prescription was not adaptive. In contrast, the model of \citet{CunninghamEtAl2011} launched outflows along the rotation axis of the sink particles, scaling the outflow mass and momentum with the sink accretion rate. Nevertheless, their model also neglected angular momentum transfer, likely underestimating the overall outflow impact.

The outflow model by \citet{FederrathEtAl2014} explicitly includes angular momentum transfer (see Sect.~\ref{sec:outflow_angmom}), a process that is highly efficient and likely the dominant mechanism for removing angular momentum from protostellar disks. It was tested against fully-resolved simulations of magnetized protostellar collapse and disk formation, with convergence studies demonstrating robustness. The model reproduces realistic mass, momentum, and energy injection rates, as well as jet velocities, even at resolutions $\sim1,\!000$ times lower than would otherwise be required in the absence of a sub-resolution prescription.

A more recent implementation by \citet{GrudicEtAl2021} in the \code{gizmo} code adopts methods broadly similar to \citet{FederrathEtAl2014}, but omits angular momentum injection. Their rationale is that angular momentum must already be shed by the disk before material accretes onto the star, and thus does not need to be explicitly injected back into the outflow. However, because the disk is generally unresolved in large-scale simulations, this argument overlooks part of the physics intended to be captured by a sub-resolution model. In particular, the \citet{FederrathEtAl2014} scheme returns angular momentum to larger scales through the jets, while the \citet{GrudicEtAl2021} implementation does not, potentially underestimating feedback effects on cluster-scale dynamics.

\subsection{Protostellar heating feedback} \label{sec:heating}

Accreting protostars radiate both through their intrinsic stellar luminosity, powered by nuclear burning, and through their accretion luminosity, which arises from the energy released by in-falling material. Most of the accretion luminosity is generated in shocks at the stellar surface and can be expressed as a function of the instantaneous stellar mass $M_\star$ and radius $R_\star$,
\begin{equation}
L_{\rm acc} = f_{\rm acc}\,\frac{G M_\star \dot M_\star}{R_\star},
\end{equation}
where $\dot M_\star$ is the accretion rate and $f_{\rm acc}$ is a coefficient of order unity that specifies the fraction of kinetic energy radiated away rather than expended to drive jets or retained by the star.

\citet{OffnerEtAl2009} introduced a detailed one-zone protostellar evolution model that follows the nuclear state of the star and tracks its radius as a function of mass and accretion history (cf.~Sect.~\ref{sec:stellar_evol}). When full radiation transfer (RT) is included in a simulation (Sect.~\ref{sec:rhd}), the resulting bolometric luminosity can be incorporated directly as a source term on the RHS of the RT equation \citep[e.g.,][]{OffnerEtAl2009,UrbanMartelEvans2010,KlassenPetersPudritz2012,KrumholzKleinMcKee2012,HennebelleEtAl2020,HennebelleEtAl2022,GrudicEtAl2021,MenonEtAl2022}.

A simpler way to model stellar heating, without solving the full RT problem, is to modify the gas temperature directly. Assuming the protostar radiates as a blackbody, the heating temperature profile can be written as
\begin{equation}
T_{\rm heat}^4(r) = \frac{L_{\rm acc}}{4\pi\sigmaSB r^2},
\end{equation}
with $\sigmaSB$ the Stefan--Boltzmann constant \citep{StamatellosWhitworthHubber2012}. \citet{MathewFederrath2020} extended this approach by relaxing the assumption of spherical symmetry and implementing a sub-resolution model that includes the accretion disk. This captures the effect of the asymmetric dust distribution on the circumstellar temperature. The orientation of the sub-resolution disk is aligned with the sink particle spin axis $\ssink$, as in the outflow module (cf.~Fig.~\ref{fig:outflow_schematic}). The resulting accretion- and protostar-induced luminosity yields an effective heating temperature $T_{\rm heat}(r,\theta)$ that depends on both distance from the star $r$ and polar angle $\theta$ \citep{MathewFederrath2020}\footnote{The protostellar heating method implemented in the \code{flash} code is publicly available on GitHub: \url{https://github.com/chfeder/cfflash}}.

Finally, this heating temperature is incorporated into the EOS (Sect.~\ref{sec:eos}) of the MHD system by modifying the thermal pressure \citep{GuszejnovKrumholzHopkins2016,FederrathKrumholzHopkins2017,MathewFederrath2020},
\begin{equation}
\pth \;\to\; \left[ \pth^4 + \left(\frac{\rho\kB T_{\rm heat}}{\mu\mH}\right)^4\right]^{1/4}.
\end{equation}

\begin{figure}[t]
\centerline{\includegraphics[width=1.0\linewidth]{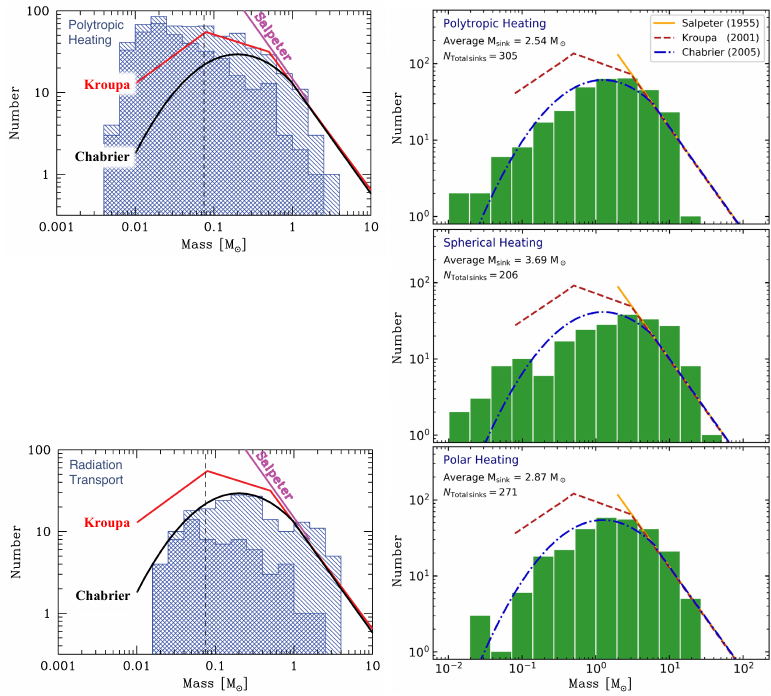}}
\caption{\label{fig:heating}
Impact of radiation on the IMF. Left: sink particle mass distributions from \citet{Bate2012}, comparing a polytropic EOS \citep[originally from][]{Bate2009cluster} (top) with RT in the FLD approximation (bottom). Right: simulations from \citet{MathewFederrath2020}, using the same polytropic EOS (top), spherical heating feedback (middle), and polar heating feedback (bottom). Without RT, or with only simple spherical heating, the number of low-mass stars is strongly overestimated. By contrast, RT and polar heating both suppress spurious low-mass stars. Polar heating provides a computationally efficient approximation to RT for low-mass star formation, but the IMF peak mass (here \mbox{$\sim1$--$2\,\msol$}) remains too high due to the neglect of additional feedback processes, especially jets and outflows. For clarity, the IMFs of \citet{Salpeter1955}, \citet{Kroupa2001}, and \citet{Chabrier2005} are shifted to highlight shape differences.}
\end{figure}

Figure~\ref{fig:heating} demonstrates how stellar heating alters the IMF\footnote{Note that the modified \citet{Kroupa2001} and \citet{Chabrier2005} IMFs shown in Fig.~\ref{fig:heating} are primarily intended as illustrative guides, not for direct comparison with the simulation data. The low-mass end of the IMF remains uncertain due to systematics related to binaries, incompleteness, and luminosity-to-mass conversions \citep{Chabrier2003,KroupaEtAl2013,Hopkins2018}. Ongoing debate continues regarding the most appropriate form of the IMF in different environments, with recent reviews and discussions provided by \citet{KroupaEtAl2026}, \citet{GjergoEtAl2026}, and \citet{JerabkovaEtAl2025}.}. In the simulations of \citet{Bate2012}, a simple polytropic EOS\footnote{We note that \citet{Bate2009cluster} used a polytropic EOS that switched from isothermal to $\Gamma=7/5$ at a density of $\rho_\mathrm{crit}=10^{-13}\,\g\,\cm^{-3}$, followed by sink particle creation at a density $1000\,\times$ higher than that. It is unclear what particular effect a switch to 7/5 vs.~5/3 for example, or the particular choice of $\rho_\mathrm{crit}$ and/or the separation between $\rho_\mathrm{crit}$ and the sink formation density might have. Regardless, a careful choice of polytropic EOS might be able to produce a reasonable IMF \citep{HennebelleEtAl2020}, however, the non-local effect of RT will ultimately matter.} produces an excess of low-mass objects, while including RT in the FLD approximation suppresses their formation\footnote{We note that the \citet{Bate2012} RT simulations do not explicitly include protostellar feedback. However, because the simulations reach very high densities, radiative trapping effectively mimics protostellar heating feedback.}. A similar trend is seen in the simulations of \citet{MathewFederrath2020}: spherical heating feedback reduces the overall fragmentation but still overproduces low-mass stars, whereas polar heating, which accounts for disk shielding, yields an IMF much closer to the RT case. Nonetheless, the IMF peak mass ($\sim1-2\,\msol$) is clearly overestimated, underscoring the importance of additional processes such as jet and outflow feedback \citep[cf.][]{MathewFederrath2021,GuszejnovEtAl2021,GuszejnovEtAl2022,MathewFederrathSeta2023,MathewFederrathSeta2025}.

In summary, the primary impact of stellar heating is the suppression of small-scale fragmentation, particularly within protostellar accretion disks \citep{KrumholzKleinMcKee2007,OffnerEtAl2010,JonesBate2018}. A similar effect can be partially reproduced by radiative trapping at high densities, even without explicitly modeling stellar feedback \citep{Bate2009rad,Bate2012}. Episodic accretion, during which intervals of efficient cooling occur, may mitigate radiative heating and allow disk fragmentation to proceed \citep{StamatellosWhitworthHubber2012}. Nevertheless, the microphysics of disk accretion remains poorly understood, and direct simulations of turbulent disk dynamics are currently beyond the resolution limits achievable in star-cluster calculations.

\subsection{Stellar wind feedback} \label{sec:winds}

Many simulation codes include sub-resolution models for radiatively driven mass loss from main-sequence (MS) and post-MS stars. Stellar winds from massive stars are a major source of nuclear-processed heavy elements and, after supernovae, represent the second most important channel of energy and momentum injection into the ISM \citep{Smith2014}. In contrast to protostellar outflows, stellar winds are relatively straightforward to implement: they are intrinsically spherical, contribute little angular momentum, and can be benchmarked against analytic models of wind-bubble evolution \citep{WeaverEtAl1977,KooMcKee1992,LancasterEtAl2021a}. While some uncertainty remains regarding the details of the mass-loss process{\emdash}particularly for late B- and A-type stars{\emdash}the general launching mechanisms and the dependence of wind properties on stellar mass and metallicity are well established \citep{CureAraya2023}.

Despite the central role of radiation pressure in driving stellar winds, most numerical prescriptions do not explicitly model radiation transport or radiation pressure. Instead, mass, momentum, and energy are injected directly into the computational domain, analogous to the treatment of SNe (see Sect.~\ref{sec:sn}). Mass-loss rates, wind velocities, and temperatures are typically adopted from stellar evolution models or empirical fits to observational data. In SPH and hybrid methods, winds are implemented by spawning new particles or cells at the source location \citep{DaleBonnell2008,PriceEtAl2018,GrudicEtAl2021}.

In mesh-based methods, the injected mass, momentum, and energy are deposited into a spherical region using a predefined stencil \citep{OffnerArce2015,GeenEtAl2015a,GattoEtAl2017,LancasterEtAl2021b}. The injection region must be sufficiently large to resolve spherical expansion. To model winds more realistically, \citet{LancasterEtAl2021b} introduced a hybrid thermal/kinetic energy scheme that interpolates between thermal energy{\emdash}dominant near the source{\emdash}and kinetic energy, which dominates at the shell edge. In addition, modifications to the CFL timestep criterion may be required to enhance stability and avoid numerical overshoot in the amount of injected feedback \citep{GrudicEtAl2021}.

Simulations that include stellar winds show that they generate a complex, fractal ISM structure, which in turn shapes the environment for subsequent SN explosions \citep{RogersPittard2013,WalchNaab2015,GeenEtAl2015a}. Feedback from winds of intermediate-mass stars contributes to driving and sustaining turbulence within molecular clouds \citep{OffnerLiu2018}. For massive stars, stellar winds play a key role in halting ongoing star formation and dispersing the natal gas reservoirs of their parent clouds \citep{GuszejnovEtAl2022}.


\subsection{Ionization feedback} \label{sec:ionization}

Ionization regulates accretion and strongly shapes the natal environments of massive stars \citep{KuiperHosokawa2018}. As with stellar winds, ionization feedback is primarily important for high-mass stars. Several methods implement ionization feedback from star particles by computing the instantaneous ionizing flux they emit \citep{RosenEtAl2020}. Because the Str\"omgren radius depends on the local gas number density, the effects of ionization can be approximated without solving the full RT problem. For example, \citet{DaleErcolanoClarke2007} compute the ionization front radius around a source, identify which nearby SPH particles lie within it, and reset their temperatures to $10^4$\,K, the approximate equilibrium temperature of ionized hydrogen.

More detailed approaches couple ionization feedback to stellar evolution models (Sect.~\ref{sec:stellar_evol}), estimating the ionizing photon flux from the stellar spectrum and computing the gas response via ray-tracing or other RT schemes \citep[][see Sect.~\ref{sec:rhd}]{PetersEtAl2010,GeenEtAl2015a,GrudicEtAl2021}. In clustered environments, photoionization may even trigger secondary star formation, although this remains difficult to confirm observationally \citep{DaleErcolanoBonnell2013}. On larger scales, photoionization plays a central role in dispersing molecular clouds and in reducing star formation efficiencies to $\lesssim10\%$ \citep{GeenEtAl2016,GeenSolerHennebelle2017,KimKimOstriker2018,KimOstrikerFilippova2021,GrudicEtAl2022,GuszejnovEtAl2022,MenonEtAl2023}. The combined influence of ionizing radiation and jets has also recently been studied in \citet{VerliatEtAl2022}, showing that both are crucial ingredients for massive cluster formation and evolution. Similar applies to the combination of ionization and wind feedback \citep{GeenDeKoter2022}.


\subsection{Supernova feedback} \label{sec:sn}

Supernova (SN) explosions are a critical component of star formation and galaxy evolution. They shape the multi-phase ISM, regulate the rate and efficiency of star formation \citep{McKeeOstriker1977}, and are among the dominant drivers of large-scale ISM turbulence \citep[e.g.,][]{MacLowKlessen2004,AvillezBreitschwerdt2005,JoungMacLow2006,TamburroEtAl2009,HennebelleIffrig2014,PadoanEtAl2016,PanEtAl2016,IffrigHennebelle2017,CollingEtAl2018,BeattieEtAl2025SN}. Consequently, most star formation codes include some prescription for SN energy, momentum, and mass injection (see Table~\ref{tab:codes}). In contrast to many other feedback processes, SN sub-resolution models can be benchmarked directly, since SNe evolve through a sequence of well-characterized phases, including the energy-conserving Sedov–Taylor phase \citep{Taylor1950,Sedov1959,Book1994}.

While the underlying physics of the blast wave is well understood, numerical implementations face several challenges and adopt varying approaches. Most models assume a fiducial explosion energy of $E_{\rm SN}=10^{51}$\,erg, with all stars above $M_\star>8\,M_\odot$ ending their lives as SNe. The relative fractions of thermal and kinetic energy that make up $E_{\rm SN}$ depend on the evolutionary phase being modeled and, critically, on how well the injected blast wave is resolved numerically. Energy, momentum, and ejecta mass are typically deposited within the local region surrounding the explosion, either by spawning new SPH particles or by distributing mass and energy across a group of grid cells with some weighting kernel \citep{WalchNaab2015,HaidEtAl2016,KimOstriker2017,HopkinsEtAl2018,GrudicEtAl2021}. Proper modelling of supernova feedback requires accounting for rapid radiative cooling losses, which can significantly reduce the effective energy and momentum coupling to the ISM if unresolved, motivating sub-resolution prescriptions that capture this behaviour \citep[e.g.,][]{KimOstriker2015,SimpsonEtAl2015}.

Special care is required to ensure that the injected feedback converges toward the analytic Sedov–Taylor solution. If mass and energy are deposited into too large a region, the explosion is artificially underpowered; if deposited into too small a region, the injected thermal energy may radiate away before driving expansion \citep{HuEtAl2016,HopkinsEtAl2018}. To maintain numerical stability and load-balancing, some methods distribute the injection over several timesteps \citep{GrudicEtAl2021}. It is also essential to enforce isotropy, ensuring that the feedback does not introduce spurious momentum anisotropies or directional biases in the expanding blast wave \citep{BrulsVollmoellerSchuessler1999,HopkinsEtAl2018}. Accurate SN feedback is therefore central to galaxy-scale simulations, where it sets the ISM structure and controls star formation efficiencies.


\subsection{Summary of challenges and future directions}

Star formation feedback relies heavily on sub-resolution modeling. Key ingredients include protostellar evolution models that track stellar radii and luminosities during accretion and onto the ZAMS (Sect.~\ref{sec:stellar_evol}). Outflows and jets must be included in simulations aiming to converge on the SFR and the IMF: without them, both the SFR and the characteristic stellar mass are overestimated by factors of $\sim2-3$. Remarkably, although angular momentum transport via protostellar jets is essential for enabling accretion and likely represents the primary solution to the angular momentum problem, only one of the three codes that currently support sub-resolution jet models coupled to dynamic star particles includes angular momentum injection (Sect.~\ref{sec:outflow}).

Additional advances have been made in simplified treatments of stellar heating feedback (Sect.~\ref{sec:heating}), which avoid the full radiation transport problem. Such approaches provide efficient approximations while still suppressing spurious fragmentation. Whereas jets and heating are relevant for stars of all masses, winds (Sect.~\ref{sec:winds}), ionization (Sect.~\ref{sec:ionization}), and supernova (SN) feedback (Sect.~\ref{sec:sn}) primarily affect massive stars. Wind models are comparatively straightforward due to their spherical symmetry and relatively well-constrained physical background. Ionization and SN blast-wave physics are also well understood, but ionization feedback ideally requires radiation transport (see Sect.~\ref{sec:rhd}), while SN feedback remains sensitive to the resolution scale and to how energy and momentum are partitioned between thermal and kinetic components.

Future developments aim to improve the robustness and physical fidelity of these feedback models. For SN feedback in particular, new approaches seek to initialize profiles based on the analytic Sedov–Taylor solution, while incorporating the correct timing, expansion phase, and the structure of the surrounding medium. Similarly, more comprehensive implementations of jet feedback, including angular momentum transport, and improved sub-resolution treatments of stellar heating will be critical. These advances are essential not only for realistic star cluster simulations, but also for bridging the gap to galaxy-scale studies where feedback regulates star formation and drives the multiphase ISM.

\section{Cosmic-ray hydrodynamics} \label{sec:crhd}

The importance of cosmic rays (CRs), i.e., charged particles traveling at relativistic velocities, to the structure of the ISM and galaxy evolution has long been recognized \citep{Beck2016}. CRs are thought to be primarily produced by diffusive shock acceleration (first-order Fermi acceleration) in supernova remnants \citep{Bell2004}. Within the ISM, the CR pressure is comparable to both thermal and magnetic pressures, making CRs a key regulator of the SFR \citep{BirnboimBalbergTeyssier2015,RuszkowskiYangZweibel2017}. Bursts of star formation and AGN activity can further accelerate CRs, driving large-scale galactic winds \citep{BoothEtAl2013,HanaszEtAl2013,GirichidisEtAl2016,RuszkowskiYangZweibel2017}.

Most cosmic-ray hydrodynamics (CRHD) studies to date have focused on the impact of CRs at galactic scales. However, their role at smaller scales is equally important. CRs are an important source of ionization in dense molecular gas, thereby altering chemical abundances and reaction rates \citep[e.g.,][]{BisbasEtAl2017,JoergensenBellocheGarrod2020}, gas temperatures \citep[e.g.,][]{PapadopoulosEtAl2011,GachesOffnerBisbas2019}, and the magnetic resistivity in molecular clouds, cores, and disks \citep[e.g.,][]{PadovaniEtAl2014,Wurster2021}, which in turn shapes the gas dynamics. In the extreme limit of CR-dominated star-forming environments, elevated gas temperatures could lead to a top-heavy IMF \citep{PapadopoulosEtAl2011}. Despite these implications, very few numerical studies have explicitly modeled CR propagation within molecular clouds; most star formation simulations either neglect CRs entirely or assume a constant ionization rate.

Recent theoretical and observational studies highlight that CR processes within star-forming regions cannot be ignored. Protostellar jets, accretion shocks, and expanding HII regions can accelerate low-energy ($\lesssim100\,\mathrm{GeV}$) CRs in situ \citep{PadovaniEtAl2020}, enhancing the local CR ionization rate with strong consequences for thermodynamics and chemistry \citep{GachesOffner2018,GachesOffnerBisbas2019}. Elevated CR ionization increases the coupling between the gas and magnetic fields and the efficacy of magnetic braking, which can either promote compact disk formation or suppress disks altogether \citep{ZhaoEtAl2020Hall}. A variety of numerical studies have shown that protostellar disks forming in weakly ionized, non-ideal MHD-dominated environments tend to be larger than those forming in higher-ionization conditions, satisfying the ideal MHD limit \citep{ZhaoEtAl2020Hall,ZhaoEtAl2021,Wurster2021,AhmadEtAl2025}.

Motivated by this growing recognition, CRT modules have now been implemented in several hydrodynamic codes (cf.~Table~\ref{tab:codes}). The associated numerical challenges closely parallel those of RHD (cf.~Sect.~\ref{sec:rhd}): high characteristic velocities ($v\sim c$), large effective particle numbers, and a broad dynamic range in energies. As a result, many RHD techniques—gray approximations, moment methods (e.g., diffusion), reduced-speed-of-light schemes, and flux limiters—are directly applicable to CRHD. In the following, we review the principal CRHD methods employed in modern star-formation codes. See also the recent comprehensive reviews of CRHD by \citet{RuszkowskiPfrommer2023} and \citet{Hopkins2026}.


\subsection{Uncertainties in CR propagation} \label{sec:cr_uncertainties}

In contrast to RHD, where the physics of radiation transfer and radiation–matter interactions is relatively well understood, CRHD suffers from significant uncertainties, foremost among them how CRs actually propagate in different physical regimes. These uncertainties can be divided into two main categories: those concerning the physics of CR transport and those concerning CR production mechanisms.

CRs primarily stream along magnetic field lines and change direction through scattering, produced by inhomogeneities in the field \citep[e.g., magnetic mirroring; see][]{CesarskyVolk1978}, ambient MHD turbulence \citep{YanLazarian2004}, or Alfv\'en waves resonantly excited by the CRs themselves \citep[i.e., via the streaming instability; see][]{Skilling1975}. The large uncertainties and complexities in the turbulent structure and dynamics of the magnetized ISM (cf.~Fig.~\ref{fig:beattie}), combined with the anisotropic nature of CR transport, imply that the degree of scattering and the extent to which CRs diffuse across field lines are uncertain by orders of magnitude. Transport is therefore often parameterized using diffusion coefficients parallel and perpendicular to the bulk magnetic field, typically treated as free parameters constrained empirically by observations \citep[e.g.,][]{OwenEtAl2021}.

As they propagate, CRs continuously lose energy through both collisional and non-collisional interactions with the thermal ISM gas. These energy losses are parameterized by a loss function that depends on the ionization state, density, magnetic field strength, and gas/dust composition. However, due to uncertain interaction cross sections and microphysics, even the dominant energy-loss mechanism in some regimes remains debated \citep[e.g.,][]{LazarianXu2022,GachesEtAl2024,Hopkins2026}.

While there is well-developed theory for the acceleration of CRs at supernova shocks and their energetics, which is supported by observations \citep{Caprioli2015}, the details depend on the local conditions \citep{HuEtAl2022,XuLazarian2022}. Meanwhile, the acceleration efficiency and resulting CR spectra from other sources have even larger uncertainties. Shock properties (e.g., magnetization, Mach number, and temperature) are poorly constrained in protostellar accretion, H\,{\sc ii}-region, and jet shocks, and these mechanisms appear to produce CR fluxes far lower than those from SNe \citep{KrumholzCrockerOffner2023}. Consequently, the limited observational constraints available for CR fluxes within molecular clouds{\emdash}such as non-thermal synchrotron emission, gamma-ray emission, and chemical abundances{\emdash}remain indirect \citep[e.g.,][]{RodriguezKamenetzkyEtAl2017,CabedoEtAl2023,PinedaEtAl2024,PandeyEtAl2026}. Additional sources of CRs, such as re-acceleration by ISM turbulence (second-order Fermi acceleration) or turbulent reconnection within molecular clouds, may also be important, but are highly uncertain \citep{DruryStrong2017,GachesWalchLazarian2021}.

Thus, while the numerical methods for CRHD described below are technically stable and robust, it is important to bear in mind that the fundamental uncertainties in CR transport physics may outweigh the approximations inherent in any given methodology.


\subsection{Moment methods for CRHD} \label{sec:cr_moment_methods}

Ideally, the transport of CRs would be modeled directly by solving for the CR phase space distribution, e.g., the Fokker--Planck equation derived by applying linear approximations \citep{Skilling1975,Schlickeiser1989}:
\begin{equation}
    \ddt{f} = - \vel\cdot\nabla f + \mathbf{\nabla}\cdot(\mathbb{D}_{xx}\mathbf{\nabla} f) + \frac{1}{3}(\mathbf{\nabla}\cdot\vel)p\frac{\partial f}{\partial p} + \frac{1}{p^2}\frac{\partial}{\partial p} \left[p^2\left(\crploss f + D_{pp}\frac{\partial f}{\partial p}\right)\right] + j, \label{eq:FP}
\end{equation}
where $f=f(\mathbf{x},\mathbf{p},t)$ is the isotropic part of the phase-space CR distribution function, assuming efficient pitch-angle scattering, and $\mathbb{D}_{xx}$ and $D_{pp}$ are the spatial diffusion tensor and the momentum diffusion coefficient, respectively. CR momentum losses are described by $\crploss(\mathbf{x},\mathbf{p},t)=dp/dt<0$, and $j = j(\mathbf{x},\mathbf{p},t)$ is a CR source term. The first term on the RHS in Eq.~(\ref{eq:FP}) represents CR advection with the gas flow ($\vel$), while the second term describes CR spatial diffusion. The third term models adiabatic behavior associated with gas expansion and compression. The fourth term describes both energy losses and second-order Fermi acceleration, i.e., re-acceleration due to small-scale magnetic turbulence. The last term, $j$, represents CRs accelerated due to first-order Fermi acceleration, i.e., acceleration produced by SNe and/or other shock processes. Cosmic-ray streaming adds an additional transport channel, in which CRs drift along magnetic field lines at approximately the Alfv\'en speed, which is effectively added to the fluid speed in the Fokker--Planck equation.

Solving the Fokker--Planck equation directly is the approach taken by dedicated CRT solvers such as \code{galprop} \citep{StrongMoskalenko1998}, \code{picard} \citep{Kissmann2014}, \code{dragon2} \citep{EvoliEtAl2017}, \code{crest} \citep{WinnerEtAl2019}, and \code{criptic} \citep{KrumholzCrockerSampson2022}. However, achieving sufficient accuracy requires resolving a large number of CR momentum bins, which becomes computationally prohibitive in full CRHD contexts \citep[e.g.,][]{WinnerEtAl2019}. Consequently, CRHD implementations in general-purpose hydrodynamic codes adopt a range of approximations, often analogous to those used in RHD (e.g., gray approximations, diffusion models and flux limiters).

\begin{figure}[t]
\centerline{\includegraphics[width=1.0\linewidth]{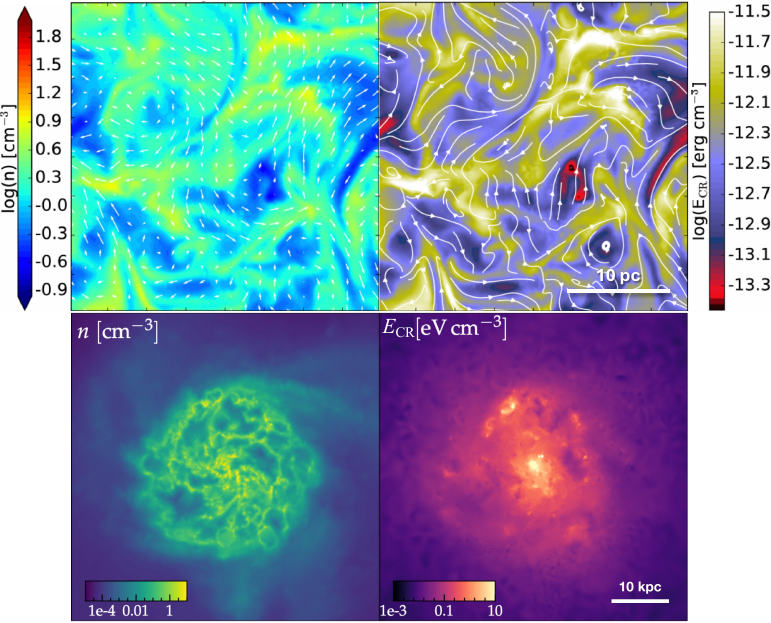}}
\caption{\label{fig:crt}
{\bf Top panels:} Gas number density (left) and CR energy density (right) in a simulation of magnetized, turbulent ISM gas using a single-moment CRHD scheme \citep{CommerconMarcowithDubois2019}. White arrows denote the velocity field, while streamlines show magnetic field lines. With a parallel diffusion coefficient $D=10^{22},\mathrm{cm^2,s^{-1}}$, well within the CR-trapping regime, strong CR energy inhomogeneities develop.
{\bf Bottom panels:} Gas number density (left) and CR energy density (right) in a galactic simulation using a two-moment CRHD scheme \citep{HopkinsEtAl2021b}. Here, the diffusion coefficient scales with the damping rate and the streaming velocity is tied to the Alfv\'en–ion speed. This model reproduces key observables, including the scattering-rate-weighted effective diffusivity, gamma-ray luminosity, and mean CR energy density of a Milky Way–like galaxy.}
\end{figure}

CRHD frameworks have been implemented in several AMR and mesh-free codes, including \code{arepo} \citep{ThomasPfrommerPakmor2021}, \code{athena} \citep{ZhaoXueNingOstriker2026}, \code{enzo} \citep{SalemBryan2014}, \code{flash} \citep{GirichidisEtAl2014CR,GirichidisEtAl2020}, \code{gizmo} \citep{HopkinsEtAl2021a}, \code{pluto} \citep{MignoneEtAl2018}, and \code{ramses} \citep{DuboisCommercon2016}. Owing to its similarity with RHD, approximate CRT is relatively straightforward to incorporate into codes that already support RHD solvers. Despite its reduced complexity, this treatment can effectively model CR-driven galactic winds, CR pressure support in the ISM, and the role of CRs in SN blast-wave energetics \citep{SalemBryan2014,DuboisCommercon2016,GirichidisEtAl2016}. It also enables parameter studies of the highly uncertain diffusion coefficients. For example, \citet{CommerconMarcowithDubois2019} identified a critical value of $D$, below which CRs become effectively trapped, producing CR pressure gradients that suppress thermal instability (see top panels of Fig.~\ref{fig:crt}).

\subsubsection{One-moment methods} \label{sec:cr_one_moment}

The simplest treatment of CRT assumes that CRs are a single fluid, distinct from the gas, that behaves diffusively, i.e., the CRs scatter and diffuse due to unresolved small-scale turbulence. The most basic implementation assumes frequent CR scattering isotropizes the pitch-angle distribution, leading to an effectively constant diffusion coefficient \citep{Zweibel2013}. In a slightly more nuanced approach transport is parameterized by a (usually constant) diffusion coefficient with components parallel and perpendicular to the magnetic-field direction to allow for anisotropic scattering. This is effectively a single-moment approach, similar to that adopted to treat gray FLD RT (cf.~Sect.~\ref{sec:rt_fld}). In this limit, the CR energy equation is
\begin{equation}
\ddt{\Ecr} = -\nabla\cdot(\Ecr\vel) - \pcr (\nabla\cdot\vel) + \nabla\cdot(\mathbb{D}\,\nabla\Ecr) + S_{\rm CR},
\end{equation}
where $\Ecr$ is the CR energy density, \mbox{$\pcr=(\gamcr-1)\Ecr$} is the CR pressure with $\gamcr=4/3$ for a relativistic gas, $\mathbb{D}$ is the CR diffusion tensor, and $S_{\rm CR}$ represents CR sources \citep[for a derivation, see][]{SchlickeiserLerche1985}. The first two terms on the RHS represent advection and adiabatic compression/expansion. The third and fourth terms describe diffusion and sources of CRs, respectively. CRs are coupled to the MHD system of equations (Sect.~\ref{sec:mhd}) by adding the diffusion term $\nabla\cdot(\mathbb{D}\nabla\Ecr)$ to gas energy equation (Eq.~\ref{eq:mhd3}) to account for thermal energy injected by CR diffusion. Likewise, the total gas energy density and pressure are updated to include the CR contributions, i.e., $\rho e \to \rho e + \Ecr$, and $p \to p + \pcr$.

\subsubsection{Two-moment methods} \label{sec:cr_two_moment}

The diffusion approximation for CRT neglects several key processes: the generation of Alfv\'en waves by CR streaming, damping of waves through ion–neutral collisions, and nonlinear Landau damping driven by anisotropic particle distributions \citep[see also][]{RinconEtAl2016,StOngeKunz2018,AchikanathEtAl2024,AchikanathFederrathSeta2024}. These processes are essential for accurately modeling CR transport in the multi-phase ISM \citep{SetaFederrath2022} and within molecular clouds. Moreover, the diffusion approach is effectively gray, since $\Ecr$ represents the spectrum-integrated CR energy density, providing no information on the spectral evolution or energy-dependent CR behavior.

A more sophisticated treatment is the two-moment (or flux-evolving) approach, which explicitly evolves the CR flux equation \citep{Zweibel2017,JiangOh2018,ThomasPfrommer2019,ChanEtAl2019,RuszkowskiPfrommer2023}. This method enables a more accurate treatment of CR streaming. The bottom panels of Figure~\ref{fig:crt} illustrate a two-moment CRHD simulation of a Milky Way-like galaxy, showing that this framework can reproduce several expected CR observables.

Despite methodological parallels to RHD, important differences arise. On galactic scales, one- and two-moment CRT approaches produce very similar bulk outcomes, such as SFRs and gamma-ray luminosities \citep{ChanEtAl2019}. \citet{ThomasPfrommer2022} demonstrate that, unlike RHD, the CR behavior is relatively insensitive to the choice of closure. This insensitivity stems from the fact that CR propagation is dominated by pitch-angle scattering, which tends to isotropize the distribution and ensures diffusion dominates over directional effects such as shadowing (cf.~Fig.~\ref{fig:vet-comp}).

For star-cluster formation, simulations that include all major physics and feedback processes along with CRT for a typical Galactic CR background recover a standard IMF \citep{FitzAxenEtAl2024}. In these models, ambient CRs do not have a significant effect inside dense clouds due to energy losses induced by the streaming instability \citep{BustardZweibel2021,FitzAxenEtAl2024}, creating CR pressure gradients between the cloud interior and exterior. This gradient enhances the final star formation efficiency. Furthermore, CRT simulations provide tentative evidence that environments with high CR energy densities{\emdash}whether from an elevated background or local CR acceleration{\emdash}can yield more top-heavy IMFs \citep{FitzAxenEtAl2024,FitzAxenEtAl2026}.

\subsection{Multi-bin spectral formalisms} \label{sec:cr_multi_bin}

Several recent approaches explicitly follow the evolution of the CR spectrum, including energy-dependent losses. These methods solve discretized forms of the Fokker--Planck equation for the CR number and energy densities, in close analogy to multi-frequency treatments of the RT equation. The first two moments of the Fokker--Planck equation are defined as
\begin{eqnarray}
n_{{\rm CR},i} & = & \int_{p_{i-1/2}}^{p_{i+1/2}} 4 \pi p^2 f(p)\,dp, \\
E_{{\rm CR},i} & = & \int_{p_{i-1/2}}^{p_{i+1/2}} 4 \pi p^2 f(p)\,T(p)\,dp,
\end{eqnarray}
where $p$ is the CR momentum, $n_{{\rm CR},i}$ and $E_{{\rm CR},i}$ are the CR number and energy densities in momentum bin $i$, and
\begin{equation}
T(p) = \sqrt{p^2c^2 + m_{\rm p}^2c^4} - m_{\rm p} c^2
\end{equation}
is the CR kinetic energy, with the proton mass $m_{\rm p}$.

In practice, these methods often assume a piecewise power-law representation of $f(p)$ within each bin,
\begin{equation}
f(p) = f_{i-1/2}\left(\frac{p}{p_{i-1/2}}\right)^{-q_i},
\end{equation}
with two degrees of freedom per bin: the normalization $f_{i-1/2}$ and the spectral slope $q_i$. Continuity of $f(p)$ enforces closure. Using these definitions, the two-moment equations for the CR number density and energy density in each momentum bin $i$ can be defined \citep[for details we refer to][]{GirichidisEtAl2020,HopkinsEtAl2022}. Cost savings in multi-bin methods are often achieved by using logarithmically-spaced momentum bins and adopting a reduced-speed-of-light approximation \citep[e.g.,][]{ChanEtAl2019,HopkinsEtAl2021b}.

Spectral CRT methods allow direct comparison with the observed CR spectrum in the Milky Way and other galaxies \citep{GirichidisEtAl2020,HopkinsEtAl2022}. They show that CR spectra and ionization rates can vary substantially across galactic environments. These calculations also indicate that turbulent re-acceleration plays only a minor role, while CRs produced in reverse supernova shocks dominate over other sources of acceleration \citep{HopkinsEtAl2022}.


\subsection{Monte Carlo CR transport} \label{sec:cr_mc}

In lieu of a continuous fluid approach, CRT may also be modeled discretely by random sampling from the CR distribution. This is analogous to the MC approach to RT, where the evolution of the radiation spectrum is modeled by following the interactions between individual photon packets and the gas (see Sect.~\ref{sec:rt_mcrt}). To date, no in situ MC~CRT methods have been implemented in star-formation hydrodynamic codes, although this is a promising future avenue due to advantages in scalability and spectral resolution, as well as potential for GPU acceleration.

A few examples of MC~CRT have been applied to star-formation problems in post processing, however. For example, \citet{FitzAxenEtAl2021a} developed a three-dimensional time-independent MC~CRT method that follows the propagation of CR protons and electrons through a grid of magnetic-field strengths and densities. The code solves the Boltzmann transport equation, rather than the quasi-linear Fokker--Planck equation, for each particle step, where the distance traveled is sampled from an exponential probability distribution \citep{HardingFryerMendel2016}. Scattering is modeled by sampling a random vector at every particle step, which emulates a turbulent magnetic field \citep[e.g.,][]{HardingFryerMendel2016}. In this way, the code models the sub-resolution turbulent magnetic field as an isotropic field, with a proton-scattering opacity. \citet{FraschettiEtAl2018} adopts a similar approach to follow the trajectories of CR particles accelerated in the stellar wind through the circumstellar disk. Likewise, they decompose the magnetic field into mean and turbulent components, where the mean component is given by MHD simulations and pitch-angle scattering produced by the turbulent component derived from an assumed \citet{Kolmogorov1941c} turbulent power spectrum \citep[see also][]{FraschettiGiacalone2012}.

Applications using such approaches suggest that the asymmetric magnetic field from core to stellar scales produces significant anisotropy in the CR flux that reaches the circumstellar disk \citep{FraschettiEtAl2018,FitzAxenEtAl2021b}. As in the radiation case, CRs can experience a flashlight effect, in which they preferentially leak out the outflow cavity due to a combination of lower-energy losses and magnetic focusing \citep{FitzAxenEtAl2021b}. However, as in other CR approaches, the effective diffusion coefficient and level of anisotropy is sensitive to assumptions about the sub-resolution magnetized turbulence \citep{LazarianXu2021,ZhangXu2023}.


\subsection{Summary of challenges and future directions}

Despite substantial progress in recent years, CRHD still faces major open challenges. Unlike RHD (cf.~Sect.~\ref{sec:rhd}), the basic transport physics of CRs remains highly uncertain (Sect.~\ref{sec:cr_uncertainties}). The degree to which CRs scatter along or across magnetic fields, the importance of streaming instabilities, and the impact of damping processes such as ion-neutral and Landau damping are all poorly constrained, particularly in the multi-phase ISM and in molecular clouds. Similarly, while supernovae are well-established CR accelerators, the efficiency and spectral properties of CRs from protostellar jets, accretion shocks, or expanding H\,{\sc ii} regions remain uncertain.

On the numerical side, most CRHD methods still rely on simplified one-moment diffusion or streaming treatments, often with gray approximations that integrate over the CR spectrum (Sect.~\ref{sec:cr_one_moment}). Multi-bin formalisms (Sect.~\ref{sec:cr_multi_bin}) now allow explicit spectral evolution, but at high computational cost. Two-moment methods (Sect.~\ref{sec:cr_two_moment}) improve the modeling of streaming and anisotropy, yet global galaxy simulations often find little difference in bulk outcomes such as the SFR compared to one-moment methods. Monte Carlo approaches (Sect.~\ref{sec:cr_mc}) promise higher accuracy and natural spectral resolution, but have so far only been applied in post-processing due to expense. The need for scalable implementations, possibly with GPU acceleration, is pressing.

Finally, observational constraints on CR transport are largely indirect, relying on gamma-ray emission, synchrotron radiation, or molecular abundances, which provide only limited leverage on diffusion coefficients and loss processes. Star-forming regions in particular show tentative evidence for enhanced CR ionization rates, but the magnitude and spatial structure of this enhancement remain debated.

Future progress will likely require multi-scale, multi-physics models that couple CR transport to MHD turbulence, chemistry, and radiative feedback, combined with hybrid fluid-particle methods that can capture both spectral evolution and anisotropy. Ultimately, predictive CRHD will depend on tighter synergy between theory, numerical models, and new observational diagnostics from facilities such as ALMA, JWST, and CTAO.

\section{Conclusions}

The physics of star formation spans a remarkable range of scales and processes, from MHD turbulence to radiation transfer, self-gravity, stellar feedback, and cosmic-ray dynamics. Each of these ingredients is now modeled with increasing realism in modern simulation codes (cf.~Table~\ref{tab:codes}), yet major challenges remain.

Despite significant advances, further development is needed in accurately solving the MHD equations under the extreme conditions relevant to star formation. Numerical schemes must simultaneously maintain stability, accuracy, $\nabla\cdot\Bvec=0$, and conservation properties across vast dynamic ranges in density, velocity, and magnetic field strength. Recent developments, including positivity-preserving solvers, hybrid-precision techniques, and GPU acceleration (Sect.~\ref{sec:hydro}), have substantially improved robustness and performance. Nevertheless, the treatment of boundary conditions remains an inherent limitation, as simulations cannot yet self-consistently resolve the full range of scales from galactic environments down to individual protostars.

For MHD turbulence, the fundamental difficulty lies in the enormous Reynolds numbers of the ISM, $\mathrm{Re}\sim10^9$, which remain far beyond current computational capabilities. While advances in numerical methods and analysis techniques, including improved dissipation tracking and scaling diagnostics (Sect.~\ref{sec:turbulence}), are enabling important progress, bridging the gap to fully realistic turbulence remains a long-term goal.

Radiation hydrodynamics (Sect.~\ref{sec:rhd}) presents perhaps the greatest numerical challenge due to its high dimensionality. Approximations remain unavoidable, whether gray moment methods with different closures (FLD, M1, VET) or Monte Carlo (MC) schemes. Each has strengths and weaknesses, with VET and MC providing the most accurate solutions but are either hard to implement (VET) or high in cost (MC). Multi-frequency extensions are now beginning to bridge the gap between idealized gray schemes and the complexity of realistic radiation from gas, dust, and stars, though full line transfer remains out of reach.

In gravity, solvers are generally robust (Sect.~\ref{sec:grav}), but accurately capturing the coupled evolution of gas and stars remains challenging, and may require more widespread use of direct $N$-body integration to reliably follow stellar dynamics in dense, gas-rich environments.

At the scale of star formation, robust sink particle treatments (Sect.~\ref{sec:sf}) demand strict checks for gravitational collapse and careful conservation of mass, momentum, and angular momentum. These criteria are essential to avoid spurious fragmentation and inaccurate accretion. Stellar feedback (Sect.~\ref{sec:feedback}) adds further complexity: jets and outflows are critical for angular momentum transport and IMF regulation, while heating, winds, ionization, and supernovae each operate in different mass regimes. Progress here requires more comprehensive implementations of jet feedback, improved thermal treatments, and resolution-independent SN feedback models.

CRHD (Sect.~\ref{sec:crhd}) shares some methodological parallels with RHD but faces even larger physical uncertainties. The propagation of CRs depends sensitively on poorly constrained scattering, streaming, and damping processes, while their sources outside supernovae remain uncertain. Numerically, one-moment diffusion models dominate current applications, though multi-bin spectral and two-moment approaches are emerging. Monte Carlo methods again offer conceptual simplicity and high accuracy but are not yet feasible in situ. Observational constraints are similarly limited, underscoring the need for closer ties between simulations and diagnostics such as gamma rays, synchrotron emission, and molecular abundances.

Looking forward, the most promising advances will come from methods that integrate across physics modules. Multi-scale, multi-physics approaches that couple turbulence, gravity, radiation, feedback, and CR transport will be required to create the most realistic models of star formation. Continued development of hybrid particle-fluid methods, combined with the increasing availability of GPU acceleration and exascale computing, will enable more accurate treatments of both microphysics and global ISM dynamics. 

Meanwhile, the revolution in artificial intelligence (AI) and scientific machine learning (ML) are providing novel architectures for solving systems of PDEs and ODEs \citep{KudelaMatousek2023}. These neural-network based approaches offer potential speed ups of many orders of magnitude compared to the traditional direct integrators described in this review. To date, these ``surrogates" have been applied to modeling astrochemistry, magnetized turbulence, radiative transfer, and supernova feedback sub-grid models \citep[e.g.,][]{HimesEtAl2022,SulzerBuck2023,PolettiOffnerWard2025,HirashimaEtAl2025,VermarienEtAl2025,RostBrancaBuck2025}.
Numerical simulations, generated at scale, provide essential input data for training and testing ML models.  Current work has barely scratched the surface of the possible astrophysical applications for these methods. Future developments in the robustness and accuracy of ML models for high-dimensional, multi-physics systems will enable these methods to become competitive with or even exceed the capabilities of present astrophysical codes, e.g., as suggested by the impressive performance of surrogates in the field of climate and weather modeling \citep{KurthEtAl2022,BoussiouxEtAl2022,AletEtAl2025}.

Ultimately, progress in theory and computation must go hand in hand with improved observational diagnostics, ensuring that simulations remain grounded in measurable reality while continuing to push toward ever higher fidelity.

\backmatter

\bmhead{Funding}
C.F.~acknowledges funding provided by the Australian Research Council (Discovery Projects DP230102280 and DP250101526), and the Australia-Germany Joint Research Cooperation Scheme (UA-DAAD). S.O.~acknowledges support from NSF~2107340, NSF~2107942, NASA~80NSSC23K047, NSF~2407522, a Peter O'Donnell Distinguished Researcher Fellowship, a Donald Harrington Fellowship, and the NSF under Cooperative Agreement 2421782 and the Simons Foundation grant MPS-AI-00010515 awarded to the NSF-Simons AI Institute for Cosmic Origins — CosmicAI.
We further acknowledge high-performance computing resources provided by the Leibniz Rechenzentrum and the Gauss Centre for Supercomputing (grants~pr32lo, pr48pi and GCS Large-scale project~10391), the Australian National Computational Infrastructure (grant~ek9) and the Pawsey Supercomputing Centre (project~pawsey0810) in the framework of the National Computational Merit Allocation Scheme and the ANU Merit Allocation Scheme.


\phantomsection
\addcontentsline{toc}{section}{References}
\bibliography{federrath}

\begin{thebibliography}{492}
\providecommand{\natexlab}[1]{#1}
\providecommand{\url}[1]{{#1}}
\providecommand{\urlprefix}{URL }
\providecommand{\doi}[1]{\url{https://doi.org/#1}}
\providecommand{\eprint}[2][]{\url{#2}}
 \bibcommenthead

\bibitem[{{Abel} and {Wandelt}(2002)}]{AbelWandelt2002}
{Abel} T, {Wandelt} BD (2002) {Adaptive ray tracing for radiative transfer
  around point sources}. \mnras 330(3):L53--L56.
  \doi{10.1046/j.1365-8711.2002.05206.x},
  {\href{https://arxiv.org/abs/astro-ph/0111033}{{arXiv:astro-ph/0111033}}}
  {[astro-ph]}

\bibitem[{{Achikanath Chirakkara} et~al.(2021){Achikanath Chirakkara},
  {Federrath}, {Trivedi}, and {Banerjee}}]{AchikanathEtAl2021}
{Achikanath Chirakkara} R, {Federrath} C, {Trivedi} P, et~al (2021) {Efficient
  Highly Subsonic Turbulent Dynamo and Growth of Primordial Magnetic Fields}.
  \prl 126(9):091103. \doi{10.1103/PhysRevLett.126.091103},
  {\href{https://arxiv.org/abs/2101.08256}{{arXiv:2101.08256}}} {[astro-ph.HE]}

\bibitem[{{Achikanath Chirakkara} et~al.(2024{\natexlab{a}}){Achikanath
  Chirakkara}, {Federrath}, and {Seta}}]{AchikanathFederrathSeta2024}
{Achikanath Chirakkara} R, {Federrath} C, {Seta} A (2024{\natexlab{a}})
  {AHKASH: a new Hybrid particle-in-cell code for simulations of astrophysical
  collisionless plasma}. \mnras 534(4):3761--3782.
  \doi{10.1093/mnras/stae2188},
  {\href{https://arxiv.org/abs/2409.12151}{{arXiv:2409.12151}}} {[astro-ph.IM]}

\bibitem[{{Achikanath Chirakkara} et~al.(2024{\natexlab{b}}){Achikanath
  Chirakkara}, {Seta}, {Federrath}, and {Kunz}}]{AchikanathEtAl2024}
{Achikanath Chirakkara} R, {Seta} A, {Federrath} C, et~al (2024{\natexlab{b}})
  {Critical magnetic Reynolds number of the turbulent dynamo in collisionless
  plasmas}. \mnras 528(1):937--953. \doi{10.1093/mnras/stad3967},
  {\href{https://arxiv.org/abs/2401.08499}{{arXiv:2401.08499}}} {[astro-ph.CO]}

\bibitem[{{Agertz} et~al.(2007){Agertz}, {Moore}, {Stadel}, {Potter},
  {Miniati}, {Read}, {Mayer}, {Gawryszczak}, {Kravtsov}, {Nordlund}, {Pearce},
  {Quilis}, {Rudd}, {Springel}, {Stone}, {Tasker}, {Teyssier}, {Wadsley}, and
  {Walder}}]{AgertzEtAl2007}
{Agertz} O, {Moore} B, {Stadel} J, et~al (2007) {Fundamental differences
  between SPH and grid methods}. \mnras 380:963--978.
  \doi{10.1111/j.1365-2966.2007.12183.x},
  {\href{https://arxiv.org/abs/arXiv:astro-ph/0610051}{{arXiv:astro-ph/0610051}}}

\bibitem[{{Ahmad} et~al.(2025){Ahmad}, {Gonz{\'a}lez}, {Hennebelle},
  {Lebreuilly}, and {Commer{\c{c}}on}}]{AhmadEtAl2025}
{Ahmad} A, {Gonz{\'a}lez} M, {Hennebelle} P, et~al (2025) {Birth of magnetized
  low-mass protostars and circumstellar disks}. \aap 696:A238.
  \doi{10.1051/0004-6361/202553663},
  {\href{https://arxiv.org/abs/2503.08637}{{arXiv:2503.08637}}} {[astro-ph.SR]}

\bibitem[{{Alet} et~al.(2025){Alet}, {Price}, {El-Kadi}, {Masters}, {Markou},
  {Andersson}, {Stott}, {Lam}, {Willson}, {Sanchez-Gonzalez}, and
  {Battaglia}}]{AletEtAl2025}
{Alet} F, {Price} I, {El-Kadi} A, et~al (2025) {Skillful joint probabilistic
  weather forecasting from marginals}. arXiv e-prints arXiv:2506.10772.
  \doi{10.48550/arXiv.2506.10772},
  {\href{https://arxiv.org/abs/2506.10772}{{arXiv:2506.10772}}} {[stat.ML]}

\bibitem[{{Appenzeller} and {Mundt}(1989)}]{AppenzellerMundt1989}
{Appenzeller} I, {Mundt} R (1989) {T Tauri stars}. \aapr 1:291--334.
  \doi{10.1007/BF00873081}

\bibitem[{{Arzoumanian} et~al.(2013){Arzoumanian}, {Andr{\'e}}, {Peretto}, and
  {K{\"o}nyves}}]{ArzoumanianEtAl2013}
{Arzoumanian} D, {Andr{\'e}} P, {Peretto} N, et~al (2013) {Formation and
  evolution of interstellar filaments. Hints from velocity dispersion
  measurements}. \aap 553:A119. \doi{10.1051/0004-6361/201220822},
  {\href{https://arxiv.org/abs/1303.3024}{{arXiv:1303.3024}}} {[astro-ph.SR]}

\bibitem[{{Arzoumanian} et~al.(2018){Arzoumanian}, {Shimajiri}, {Inutsuka},
  {Inoue}, and {Tachihara}}]{ArzoumanianEtAl2018}
{Arzoumanian} D, {Shimajiri} Y, {Inutsuka} Si, et~al (2018) {Molecular filament
  formation and filament-cloud interaction: Hints from Nobeyama 45 m telescope
  observations}. \pasj 70:96. \doi{10.1093/pasj/psy095},
  {\href{https://arxiv.org/abs/1807.08968}{{arXiv:1807.08968}}}

\bibitem[{{Aubert} and {Teyssier}(2008)}]{AubertTeyssier2008}
{Aubert} D, {Teyssier} R (2008) {A radiative transfer scheme for cosmological
  reionization based on a local Eddington tensor}. \mnras 387(1):295--307.
  \doi{10.1111/j.1365-2966.2008.13223.x},
  {\href{https://arxiv.org/abs/0709.1544}{{arXiv:0709.1544}}} {[astro-ph]}

\bibitem[{{Bacciotti} et~al.(2002){Bacciotti}, {Ray}, {Mundt}, {Eisl{\"o}ffel},
  and {Solf}}]{BacciottiEtAl2002}
{Bacciotti} F, {Ray} TP, {Mundt} R, et~al (2002) {Hubble Space Telescope/STIS
  Spectroscopy of the Optical Outflow from DG Tauri: Indications for Rotation
  in the Initial Jet Channel}. \apj 576:222--231. \doi{10.1086/341725},
  {\href{https://arxiv.org/abs/astro-ph/0206175}{{astro-ph/0206175}}}

\bibitem[{{Baczynski} et~al.(2015){Baczynski}, {Glover}, and
  {Klessen}}]{BaczynskiGloverKlessen2015}
{Baczynski} C, {Glover} SCO, {Klessen} RS (2015) {Fervent: chemistry-coupled,
  ionizing and non-ionizing radiative feedback in hydrodynamical simulations}.
  \mnras 454(1):380--411. \doi{10.1093/mnras/stv1906},
  {\href{https://arxiv.org/abs/1503.08987}{{arXiv:1503.08987}}} {[astro-ph.IM]}

\bibitem[{{Bagla}(2002)}]{Bagla2002}
{Bagla} JS (2002) {TreePM: A Code for Cosmological N-Body Simulations}. J
  Astrophys Astron 23(3-4):185--196. \doi{10.1007/BF02702282},
  {\href{https://arxiv.org/abs/astro-ph/9911025}{{arXiv:astro-ph/9911025}}}
  {[astro-ph]}

\bibitem[{Balay et~al.(1997)Balay, Gropp, McInnes, and Smith}]{BalayEtAl1997}
Balay S, Gropp WD, McInnes LC, et~al (1997) Efficient management of parallelism
  in object-oriented numerical software libraries. In: Arge E, Bruaset AM,
  Langtangen HP (eds) Modern Software Tools for Scientific Computing.
  Birkh{\"a}user Boston, Boston, MA, pp 163--202,
  \doi{10.1007/978-1-4612-1986-6_8}

\bibitem[{{Balbus} et~al.(1996){Balbus}, {Hawley}, and
  {Stone}}]{BalbusHawleyStone1996}
{Balbus} SA, {Hawley} JF, {Stone} JM (1996) {Nonlinear Stability,
  Hydrodynamical Turbulence, and Transport in Disks}. \apj 467:76.
  \doi{10.1086/177585}

\bibitem[{{Banerjee} and {Pudritz}(2006)}]{BanerjeePudritz2006}
{Banerjee} R, {Pudritz} RE (2006) {Outflows and Jets from Collapsing Magnetized
  Cloud Cores}. \apj 641:949--960. \doi{10.1086/500496},
  {\href{https://arxiv.org/abs/arXiv:astro-ph/0508374}{{arXiv:astro-ph/0508374}}}

\bibitem[{{Barnes} and {Hut}(1986)}]{BarnesHut1986}
{Barnes} J, {Hut} P (1986) {A hierarchical O(N log N) force-calculation
  algorithm}. \nat 324:446--449. \doi{10.1038/324446a0}

\bibitem[{{Bate}(2009{\natexlab{a}})}]{Bate2009cluster}
{Bate} MR (2009{\natexlab{a}}) {Stellar, brown dwarf and multiple star
  properties from hydrodynamical simulations of star cluster formation}. \mnras
  392:590--616. \doi{10.1111/j.1365-2966.2008.14106.x},
  {\href{https://arxiv.org/abs/0811.0163}{{arXiv:0811.0163}}}

\bibitem[{{Bate}(2009{\natexlab{b}})}]{Bate2009init}
{Bate} MR (2009{\natexlab{b}}) {The dependence of star formation on initial
  conditions and molecular cloud structure}. \mnras 397(1):232--248.
  \doi{10.1111/j.1365-2966.2009.14970.x},
  {\href{https://arxiv.org/abs/0905.3562}{{arXiv:0905.3562}}} {[astro-ph.SR]}

\bibitem[{{Bate}(2009{\natexlab{c}})}]{Bate2009rad}
{Bate} MR (2009{\natexlab{c}}) {The importance of radiative feedback for the
  stellar initial mass function}. \mnras 392:1363--1380.
  \doi{10.1111/j.1365-2966.2008.14165.x},
  {\href{https://arxiv.org/abs/0811.1035}{{arXiv:0811.1035}}}

\bibitem[{{Bate}(2012)}]{Bate2012}
{Bate} MR (2012) {Stellar, brown dwarf and multiple star properties from a
  radiation hydrodynamical simulation of star cluster formation}. \mnras
  419:3115--3146. \doi{10.1111/j.1365-2966.2011.19955.x},
  {\href{https://arxiv.org/abs/1110.1092}{{arXiv:1110.1092}}} {[astro-ph.SR]}

\bibitem[{{Bate}(2018)}]{Bate2018}
{Bate} MR (2018) {On the diversity and statistical properties of protostellar
  discs}. \mnras 475(4):5618--5658. \doi{10.1093/mnras/sty169},
  {\href{https://arxiv.org/abs/1801.07721}{{arXiv:1801.07721}}} {[astro-ph.SR]}

\bibitem[{{Bate} and {Burkert}(1997)}]{BateBurkert1997}
{Bate} MR, {Burkert} A (1997) {Resolution requirements for smoothed particle
  hydrodynamics calculations with self-gravity}. \mnras 288(4):1060--1072.
  \doi{10.1093/mnras/288.4.1060}

\bibitem[{{Bate} et~al.(1995){Bate}, {Bonnell}, and
  {Price}}]{BateBonnellPrice1995}
{Bate} MR, {Bonnell} IA, {Price} NM (1995) {Modelling accretion in protobinary
  systems}. \mnras 277(2):362--376. \doi{10.1093/mnras/277.2.362},
  {\href{https://arxiv.org/abs/astro-ph/9510149}{{arXiv:astro-ph/9510149}}}
  {[astro-ph]}

\bibitem[{{Bate} et~al.(2003){Bate}, {Bonnell}, and
  {Bromm}}]{BateBonnellBromm2003}
{Bate} MR, {Bonnell} IA, {Bromm} V (2003) {The formation of a star cluster:
  predicting the properties of stars and brown dwarfs}. \mnras 339:577--599.
  \doi{10.1046/j.1365-8711.2003.06210.x},
  {\href{https://arxiv.org/abs/arXiv:astro-ph/0212380}{{arXiv:astro-ph/0212380}}}

\bibitem[{{Beattie} et~al.(2023){Beattie}, {Federrath}, {Kriel}, {Mocz}, and
  {Seta}}]{BeattieEtAl2023}
{Beattie} JR, {Federrath} C, {Kriel} N, et~al (2023) {Growth or Decay - I:
  universality of the turbulent dynamo saturation}. \mnras 524(3):3201--3214.
  \doi{10.1093/mnras/stad1863},
  {\href{https://arxiv.org/abs/2209.10749}{{arXiv:2209.10749}}} {[astro-ph.GA]}

\bibitem[{{Beattie} et~al.(2025{\natexlab{a}}){Beattie}, {Federrath},
  {Klessen}, {Cielo}, and {Bhattacharjee}}]{BeattieEtAl2025}
{Beattie} JR, {Federrath} C, {Klessen} RS, et~al (2025{\natexlab{a}}) {The
  spectrum of magnetized turbulence in the interstellar medium}. Nature
  Astronomy 9:1195--1205. \doi{10.1038/s41550-025-02551-5},
  {\href{https://arxiv.org/abs/2504.07136}{{arXiv:2504.07136}}} {[astro-ph.GA]}

\bibitem[{{Beattie} et~al.(2025{\natexlab{b}}){Beattie}, {Federrath}, {Kriel},
  {Hew}, and {Bhattacharjee}}]{BeattieEtAl2025visc}
{Beattie} JR, {Federrath} C, {Kriel} N, et~al (2025{\natexlab{b}}) {Taking
  control of compressible modes: bulk viscosity and the turbulent dynamo}.
  \mnras 542(4):2669--2697. \doi{10.1093/mnras/staf1318},
  {\href{https://arxiv.org/abs/2312.03984}{{arXiv:2312.03984}}} {[astro-ph.GA]}

\bibitem[{{Beattie} et~al.(2025{\natexlab{c}}){Beattie}, {Kolborg},
  {Ramirez-Ruiz}, and {Federrath}}]{BeattieEtAl2025SN}
{Beattie} JR, {Kolborg} AN, {Ramirez-Ruiz} E, et~al (2025{\natexlab{c}}) {So
  Long Kolmogorov: The Forward and Backward Turbulence Cascades in a
  Supernovae-driven, Multiphase Interstellar Medium}. \apj 994(2):193.
  \doi{10.3847/1538-4357/ae07cd},
  {\href{https://arxiv.org/abs/2501.09855}{{arXiv:2501.09855}}} {[astro-ph.GA]}

\bibitem[{{Beck}(2016)}]{Beck2016}
{Beck} R (2016) {Magnetic fields in spiral galaxies}. \aapr 24:4.
  \doi{10.1007/s00159-015-0084-4},
  {\href{https://arxiv.org/abs/1509.04522}{{arXiv:1509.04522}}}

\bibitem[{{Bell}(2004)}]{Bell2004}
{Bell} AR (2004) {Turbulent amplification of magnetic field and diffusive shock
  acceleration of cosmic rays}. \mnras 353(2):550--558.
  \doi{10.1111/j.1365-2966.2004.08097.x}

\bibitem[{{Benz}(1988)}]{Benz1988}
{Benz} W (1988) {Applications of smooth particle hydrodynamics (SPH) to
  astrophysical problems.} Comput Phys Commun 48:97--105.
  \doi{10.1016/0010-4655(88)90027-6}

\bibitem[{{Benzi} et~al.(2008){Benzi}, {Biferale}, {Fisher}, {Kadanoff},
  {Lamb}, and {Toschi}}]{BenziEtAl2008}
{Benzi} R, {Biferale} L, {Fisher} RT, et~al (2008) {Intermittency and
  Universality in Fully Developed Inviscid and Weakly Compressible Turbulent
  Flows}. \prl 100(23):234503. \doi{10.1103/PhysRevLett.100.234503}

\bibitem[{{Beresnyak} and {Lazarian}(2019)}]{BeresnyakLazarian2019}
{Beresnyak} A, {Lazarian} A (2019) {Turbulence in Magnetohydrodynamics}. De
  Gruyter, Berlin, Boston, \doi{doi:10.1515/9783110263282},
  \urlprefix\url{https://doi.org/10.1515/9783110263282}

\bibitem[{{Berger} and {Colella}(1989)}]{BergerColella1989}
{Berger} MJ, {Colella} P (1989) {Local adaptive mesh refinement for shock
  hydrodynamics}. \jcp 82:64--84. \doi{10.1016/0021-9991(89)90035-1}

\bibitem[{{Bhandare} et~al.(2020){Bhandare}, {Kuiper}, {Henning}, {Fendt},
  {Flock}, and {Marleau}}]{BhandareEtAl2020}
{Bhandare} A, {Kuiper} R, {Henning} T, et~al (2020) {Birth of convective
  low-mass to high-mass second Larson cores}. \aap 638:A86.
  \doi{10.1051/0004-6361/201937029},
  {\href{https://arxiv.org/abs/2004.12419}{{arXiv:2004.12419}}} {[astro-ph.SR]}

\bibitem[{{Birke} et~al.(2021){Birke}, {Chalons}, and
  {Klingenberg}}]{BirkeEtAl2021}
{Birke} C, {Chalons} C, {Klingenberg} C (2021) {A low Mach two-speed relaxation
  scheme for the compressible Euler equations with gravity}. arXiv e-prints
  arXiv:2112.02986. \doi{10.48550/arXiv.2112.02986},
  {\href{https://arxiv.org/abs/2112.02986}{{arXiv:2112.02986}}} {[math.NA]}

\bibitem[{{Birnboim} et~al.(2015){Birnboim}, {Balberg}, and
  {Teyssier}}]{BirnboimBalbergTeyssier2015}
{Birnboim} Y, {Balberg} S, {Teyssier} R (2015) {Galaxy evolution: modelling the
  role of non-thermal pressure in the interstellar medium}. \mnras
  447(4):3678--3692. \doi{10.1093/mnras/stu2717},
  {\href{https://arxiv.org/abs/1311.1206}{{arXiv:1311.1206}}} {[astro-ph.CO]}

\bibitem[{{Bisbas} et~al.(2017){Bisbas}, {van Dishoeck}, {Papadopoulos},
  {Sz{\H{u}}cs}, {Bialy}, and {Zhang}}]{BisbasEtAl2017}
{Bisbas} TG, {van Dishoeck} EF, {Papadopoulos} PP, et~al (2017) {Cosmic-ray
  Induced Destruction of CO in Star-forming Galaxies}. \apj 839(2):90.
  \doi{10.3847/1538-4357/aa696d},
  {\href{https://arxiv.org/abs/1703.08598}{{arXiv:1703.08598}}} {[astro-ph.GA]}

\bibitem[{{Bjorkman} and {Wood}(2001)}]{BjorkmanWood2001}
{Bjorkman} JE, {Wood} K (2001) {Radiative Equilibrium and Temperature
  Correction in Monte Carlo Radiation Transfer}. \apj 554(1):615--623.
  \doi{10.1086/321336},
  {\href{https://arxiv.org/abs/astro-ph/0103249}{{arXiv:astro-ph/0103249}}}
  {[astro-ph]}

\bibitem[{{Blandford} and {Payne}(1982)}]{BlandfordPayne1982}
{Blandford} RD, {Payne} DG (1982) {Hydromagnetic flows from accretion disks and
  the production of radio jets.} \mnras 199:883--903.
  \doi{10.1093/mnras/199.4.883}

\bibitem[{{Bleuler} and {Teyssier}(2014)}]{BleulerTeyssier2014}
{Bleuler} A, {Teyssier} R (2014) {Towards a more realistic sink particle
  algorithm for the RAMSES CODE}. \mnras 445(4):4015--4036.
  \doi{10.1093/mnras/stu2005},
  {\href{https://arxiv.org/abs/1409.6528}{{arXiv:1409.6528}}} {[astro-ph.SR]}

\bibitem[{{Bode} et~al.(2000){Bode}, {Ostriker}, and {Xu}}]{BodeOstrikerXu2000}
{Bode} P, {Ostriker} JP, {Xu} G (2000) {The Tree Particle-Mesh N-Body Gravity
  Solver}. \apjs 128(2):561--569. \doi{10.1086/313398},
  {\href{https://arxiv.org/abs/astro-ph/9912541}{{arXiv:astro-ph/9912541}}}
  {[astro-ph]}

\bibitem[{{Book}(1994)}]{Book1994}
{Book} DL (1994) {The Sedov self-similar point blast solutions in nonuniform
  media}. Shock Waves 4(1):1--10. \doi{10.1007/BF01414626}

\bibitem[{{Booth} et~al.(2013){Booth}, {Agertz}, {Kravtsov}, and
  {Gnedin}}]{BoothEtAl2013}
{Booth} CM, {Agertz} O, {Kravtsov} AV, et~al (2013) {Simulations of Disk
  Galaxies with Cosmic Ray Driven Galactic Winds}. \apjl 777(1):L16.
  \doi{10.1088/2041-8205/777/1/L16},
  {\href{https://arxiv.org/abs/1308.4974}{{arXiv:1308.4974}}} {[astro-ph.GA]}

\bibitem[{{Bouchut} et~al.(2007){Bouchut}, {Klingenberg}, and
  {Waagan}}]{BouchutKlingenbergWaagan2007}
{Bouchut} F, {Klingenberg} C, {Waagan} K (2007) {A multiwave approximate
  Riemann solver for ideal MHD based on relaxation. I: theoretical framework}.
  Numerische Mathematik 108:7--42

\bibitem[{{Boussioux} et~al.(2022){Boussioux}, {Zeng}, {Gu{\'e}nais}, and
  {Bertsimas}}]{BoussiouxEtAl2022}
{Boussioux} L, {Zeng} C, {Gu{\'e}nais} T, et~al (2022) {Hurricane Forecasting:
  A Novel Multimodal Machine Learning Framework}. Weather and Forecasting
  37(6):817--831. \doi{10.1175/WAF-D-21-0091.1},
  {\href{https://arxiv.org/abs/2011.06125}{{arXiv:2011.06125}}} {[cs.LG]}

\bibitem[{{Brackbill} and {Barnes}(1980)}]{BrackbillBarnes1980}
{Brackbill} JU, {Barnes} DC (1980) {The Effect of Nonzero {\ensuremath{\nabla}}
  {\textperiodcentered} B on the numerical solution of the magnetohydrodynamic
  equations}. Journal of Computational Physics 35(3):426--430.
  \doi{10.1016/0021-9991(80)90079-0}

\bibitem[{{Brandenburg} and {Rempel}(2019)}]{BrandenburgRempel2019}
{Brandenburg} A, {Rempel} M (2019) {Reversed Dynamo at Small Scales and Large
  Magnetic Prandtl Number}. \apj 879(1):57. \doi{10.3847/1538-4357/ab24bd},
  {\href{https://arxiv.org/abs/1903.11869}{{arXiv:1903.11869}}} {[astro-ph.SR]}

\bibitem[{{Brandenburg} et~al.(2023){Brandenburg}, {Sharma}, and
  {Vachaspati}}]{BrandenburgSharmaVachaspati2023}
{Brandenburg} A, {Sharma} R, {Vachaspati} T (2023) {Inverse cascading for
  initial magnetohydrodynamic turbulence spectra between Saffman and
  Batchelor}. J Plasma Phys 89(6):905890606. \doi{10.1017/S0022377823001253},
  {\href{https://arxiv.org/abs/2307.04602}{{arXiv:2307.04602}}}
  {[physics.plasm-ph]}

\bibitem[{Briggs et~al.(2000)Briggs, Henson, and
  McCormick}]{BriggsHensonMcCormick2000}
Briggs W, Henson V, McCormick S (2000) A Multigrid Tutorial, 2nd edn. Society
  for Industrial and Applied Mathematics

\bibitem[{{Bruls} et~al.(1999){Bruls}, {Vollm{\"o}ller}, and
  {Sch{\"u}ssler}}]{BrulsVollmoellerSchuessler1999}
{Bruls} JHMJ, {Vollm{\"o}ller} P, {Sch{\"u}ssler} M (1999) {Computing radiative
  heating on unstructured spatial grids}. \aap 348:233--248

\bibitem[{{Bryan} et~al.(2014){Bryan}, {Norman}, {O'Shea}, {Abel}, {Wise},
  {Turk}, {Reynolds}, {Collins}, {Wang}, {Skillman}, {Smith}, {Harkness},
  {Bordner}, {Kim}, {Kuhlen}, {Xu}, {Goldbaum}, {Hummels}, {Kritsuk}, {Tasker},
  {Skory}, {Simpson}, {Hahn}, {Oishi}, {So}, {Zhao}, {Cen}, {Li}, and {Enzo
  Collaboration}}]{BryanEtAl2014}
{Bryan} GL, {Norman} ML, {O'Shea} BW, et~al (2014) {ENZO: An Adaptive Mesh
  Refinement Code for Astrophysics}. \apjs 211:19.
  \doi{10.1088/0067-0049/211/2/19},
  {\href{https://arxiv.org/abs/1307.2265}{{arXiv:1307.2265}}} {[astro-ph.IM]}

\bibitem[{{Buntemeyer} et~al.(2016){Buntemeyer}, {Banerjee}, {Peters},
  {Klassen}, and {Pudritz}}]{BuntemeyerEtAl2016}
{Buntemeyer} L, {Banerjee} R, {Peters} T, et~al (2016) {Radiation hydrodynamics
  using characteristics on adaptive decomposed domains for massively parallel
  star formation simulations}. \na 43:49--69.
  \doi{10.1016/j.newast.2015.07.002},
  {\href{https://arxiv.org/abs/1501.04501}{{arXiv:1501.04501}}}

\bibitem[{{Burge} et~al.(2016){Burge}, {Van Loo}, {Falle}, and
  {Hartquist}}]{BurgeEtAl2016}
{Burge} CA, {Van Loo} S, {Falle} SAEG, et~al (2016) {Ambipolar diffusion
  regulated collapse of filaments threaded by perpendicular magnetic fields}.
  \aap 596:A28. \doi{10.1051/0004-6361/201629039},
  {\href{https://arxiv.org/abs/1609.06879}{{arXiv:1609.06879}}} {[astro-ph.GA]}

\bibitem[{{Burgers}(1948)}]{Burgers1948}
{Burgers} JM (1948) {A mathematical model illustrating the theory of
  turbulence}. Advances in Applied Mechanics 1:171--199

\bibitem[{{Burkhart}(2018)}]{Burkhart2018}
{Burkhart} B (2018) {The Star Formation Rate in the Gravoturbulent Interstellar
  Medium}. \apj 863(2):118. \doi{10.3847/1538-4357/aad002},
  {\href{https://arxiv.org/abs/1801.05428}{{arXiv:1801.05428}}} {[astro-ph.GA]}

\bibitem[{{Burkhart} and {Mocz}(2019)}]{BurkhartMocz2019}
{Burkhart} B, {Mocz} P (2019) {The Self-gravitating Gas Fraction and the
  Critical Density for Star Formation}. \apj 879(2):129.
  \doi{10.3847/1538-4357/ab25ed},
  {\href{https://arxiv.org/abs/1805.11104}{{arXiv:1805.11104}}} {[astro-ph.GA]}

\bibitem[{{Bustard} and {Zweibel}(2021)}]{BustardZweibel2021}
{Bustard} C, {Zweibel} EG (2021) {Cosmic-Ray Transport, Energy Loss, and
  Influence in the Multiphase Interstellar Medium}. \apj 913(2):106.
  \doi{10.3847/1538-4357/abf64c},
  {\href{https://arxiv.org/abs/2012.06585}{{arXiv:2012.06585}}} {[astro-ph.HE]}

\bibitem[{{Cabedo} et~al.(2023){Cabedo}, {Maury}, {Girart}, {Padovani},
  {Hennebelle}, {Houde}, and {Zhang}}]{CabedoEtAl2023}
{Cabedo} V, {Maury} A, {Girart} JM, et~al (2023) {Magnetically regulated
  collapse in the B335 protostar?. II. Observational constraints on gas
  ionization and magnetic field coupling}. \aap 669:A90.
  \doi{10.1051/0004-6361/202243813},
  {\href{https://arxiv.org/abs/2204.10043}{{arXiv:2204.10043}}} {[astro-ph.GA]}

\bibitem[{{Camenzind}(1990)}]{Camenzind1990}
{Camenzind} M (1990) {Magnetized Disk-Winds and the Origin of Bipolar
  Outflows.} Reviews in Modern Astronomy 3:234--265.
  \doi{10.1007/978-3-642-76238-3_17}

\bibitem[{{Caprioli}(2015)}]{Caprioli2015}
{Caprioli} D (2015) {Cosmic-ray acceleration and propagation}. In: 34th
  International Cosmic Ray Conference (ICRC2015), p~8,
  \doi{10.22323/1.236.0008},
  {\href{https://arxiv.org/abs/1510.07042}{{arXiv:1510.07042}}}

\bibitem[{{Cesarsky} and {Volk}(1978)}]{CesarskyVolk1978}
{Cesarsky} CJ, {Volk} HJ (1978) {Cosmic Ray Penetration into Molecular Clouds}.
  \aap 70:367

\bibitem[{{Chabrier}(2003)}]{Chabrier2003}
{Chabrier} G (2003) {Galactic Stellar and Substellar Initial Mass Function}.
  \pasp 115:763--795. \doi{10.1086/376392},
  {\href{https://arxiv.org/abs/arXiv:astro-ph/0304382}{{arXiv:astro-ph/0304382}}}

\bibitem[{{Chabrier}(2005)}]{Chabrier2005}
{Chabrier} G (2005) {The Initial Mass Function: From Salpeter 1955 to 2005}.
  In: {Corbelli} E, {Palla} F, {Zinnecker} H (eds) The Initial Mass Function 50
  Years Later, p~41, \doi{10.1007/978-1-4020-3407-7_5},
  {\href{https://arxiv.org/abs/astro-ph/0409465}{{arXiv:astro-ph/0409465}}}

\bibitem[{{Chan} et~al.(2019){Chan}, {Kere{\v{s}}}, {Hopkins}, {Quataert},
  {Su}, {Hayward}, and {Faucher-Gigu{\`e}re}}]{ChanEtAl2019}
{Chan} TK, {Kere{\v{s}}} D, {Hopkins} PF, et~al (2019) {Cosmic ray feedback in
  the FIRE simulations: constraining cosmic ray propagation with GeV
  {\ensuremath{\gamma}}-ray emission}. \mnras 488(3):3716--3744.
  \doi{10.1093/mnras/stz1895},
  {\href{https://arxiv.org/abs/1812.10496}{{arXiv:1812.10496}}} {[astro-ph.GA]}

\bibitem[{{Clark} et~al.(2012){Clark}, {Glover}, and
  {Klessen}}]{ClarkGloverKlessen2012}
{Clark} PC, {Glover} SCO, {Klessen} RS (2012) {TreeCol: a novel approach to
  estimating column densities in astrophysical simulations}. \mnras
  420(1):745--756. \doi{10.1111/j.1365-2966.2011.20087.x},
  {\href{https://arxiv.org/abs/1109.3861}{{arXiv:1109.3861}}} {[astro-ph.GA]}

\bibitem[{{Colling} et~al.(2018){Colling}, {Hennebelle}, {Geen}, {Iffrig}, and
  {Bournaud}}]{CollingEtAl2018}
{Colling} C, {Hennebelle} P, {Geen} S, et~al (2018) {Impact of galactic shear
  and stellar feedback on star formation}. \aap 620:A21.
  \doi{10.1051/0004-6361/201833161},
  {\href{https://arxiv.org/abs/1809.01037}{{arXiv:1809.01037}}} {[astro-ph.GA]}

\bibitem[{{Commer{\c c}on} et~al.(2011){Commer{\c c}on}, {Hennebelle}, and
  {Henning}}]{CommerconEtAl2011}
{Commer{\c c}on} B, {Hennebelle} P, {Henning} T (2011) {Collapse of Massive
  Magnetized Dense Cores Using Radiation Magnetohydrodynamics: Early
  Fragmentation Inhibition}. \apjl 742:L9. \doi{10.1088/2041-8205/742/1/L9},
  {\href{https://arxiv.org/abs/1110.2955}{{arXiv:1110.2955}}} {[astro-ph.SR]}

\bibitem[{{Commer{\c{c}}on} et~al.(2019){Commer{\c{c}}on}, {Marcowith}, and
  {Dubois}}]{CommerconMarcowithDubois2019}
{Commer{\c{c}}on} B, {Marcowith} A, {Dubois} Y (2019) {Cosmic-ray propagation
  in the bi-stable interstellar medium. I. Conditions for cosmic-ray trapping}.
  \aap 622:A143. \doi{10.1051/0004-6361/201833809},
  {\href{https://arxiv.org/abs/1811.11509}{{arXiv:1811.11509}}} {[astro-ph.GA]}

\bibitem[{{Courant} et~al.(1928){Courant}, {Friedrichs}, and
  {Lewy}}]{CourantFriedrichsLewy1928}
{Courant} R, {Friedrichs} K, {Lewy} H (1928) {\"Uber die partiellen
  Differenzengleichungen der mathematischen Physik}. Mathematische Annalen
  100:32--74

\bibitem[{{Cunningham} et~al.(2009){Cunningham}, {Frank}, {Carroll},
  {Blackman}, and {Quillen}}]{CunninghamEtAl2009}
{Cunningham} AJ, {Frank} A, {Carroll} J, et~al (2009) {Protostellar Outflow
  Evolution in Turbulent Environments}. \apj 692:816--826.
  \doi{10.1088/0004-637X/692/1/816},
  {\href{https://arxiv.org/abs/0804.4197}{{arXiv:0804.4197}}}

\bibitem[{{Cunningham} et~al.(2011){Cunningham}, {Klein}, {Krumholz}, and
  {McKee}}]{CunninghamEtAl2011}
{Cunningham} AJ, {Klein} RI, {Krumholz} MR, et~al (2011)
  {Radiation-hydrodynamic Simulations of Massive Star Formation with
  Protostellar Outflows}. \apj 740:107. \doi{10.1088/0004-637X/740/2/107},
  {\href{https://arxiv.org/abs/1104.1218}{{arXiv:1104.1218}}} {[astro-ph.SR]}

\bibitem[{{Cur{\'e}} and {Araya}(2023)}]{CureAraya2023}
{Cur{\'e}} M, {Araya} I (2023) {Radiation-Driven Wind Hydrodynamics of Massive
  Stars: A Review}. Galaxies 11(3):68. \doi{10.3390/galaxies11030068},
  {\href{https://arxiv.org/abs/2305.11666}{{arXiv:2305.11666}}} {[astro-ph.SR]}

\bibitem[{{Dale} and {Bonnell}(2008)}]{DaleBonnell2008}
{Dale} JE, {Bonnell} IA (2008) {The effect of stellar winds on the formation of
  a protocluster}. \mnras 391:2--13. \doi{10.1111/j.1365-2966.2008.13802.x},
  {\href{https://arxiv.org/abs/0808.1510}{{arXiv:0808.1510}}}

\bibitem[{{Dale} et~al.(2007){Dale}, {Ercolano}, and
  {Clarke}}]{DaleErcolanoClarke2007}
{Dale} JE, {Ercolano} B, {Clarke} CJ (2007) {A new algorithm for modelling
  photoionizing radiation in smoothed particle hydrodynamics}. \mnras
  382(4):1759--1767. \doi{10.1111/j.1365-2966.2007.12486.x},
  {\href{https://arxiv.org/abs/0705.3396}{{arXiv:0705.3396}}} {[astro-ph]}

\bibitem[{{Dale} et~al.(2013){Dale}, {Ercolano}, and
  {Bonnell}}]{DaleErcolanoBonnell2013}
{Dale} JE, {Ercolano} B, {Bonnell} IA (2013) {Ionization-induced star formation
  - V. Triggering in partially unbound clusters}. \mnras 431(2):1062--1076.
  \doi{10.1093/mnras/stt236},
  {\href{https://arxiv.org/abs/1302.1342}{{arXiv:1302.1342}}} {[astro-ph.GA]}

\bibitem[{{Davis} et~al.(2012){Davis}, {Stone}, and
  {Jiang}}]{DavisStoneJiang2012}
{Davis} SW, {Stone} JM, {Jiang} YF (2012) {A Radiation Transfer Solver for
  Athena Using Short Characteristics}. \apjs 199(1):9.
  \doi{10.1088/0067-0049/199/1/9},
  {\href{https://arxiv.org/abs/1201.2222}{{arXiv:1201.2222}}} {[astro-ph.IM]}

\bibitem[{{de Avillez} and {Breitschwerdt}(2005)}]{AvillezBreitschwerdt2005}
{de Avillez} MA, {Breitschwerdt} D (2005) {Global dynamical evolution of the
  ISM in star forming galaxies. I. High resolution 3D simulations: Effect of
  the magnetic field}. \aap 436:585--600. \doi{10.1051/0004-6361:20042146},
  {\href{https://arxiv.org/abs/arXiv:astro-ph/0502327}{{arXiv:astro-ph/0502327}}}

\bibitem[{{Dedner} et~al.(2002){Dedner}, {Kemm}, {Kr{\"o}ner}, {Munz},
  {Schnitzer}, and {Wesenberg}}]{DednerEtAl2002}
{Dedner} A, {Kemm} F, {Kr{\"o}ner} D, et~al (2002) {Hyperbolic Divergence
  Cleaning for the MHD Equations}. Journal of Computational Physics
  175(2):645--673. \doi{10.1006/jcph.2001.6961}

\bibitem[{{Derigs} et~al.(2016){Derigs}, {Winters}, {Gassner}, and
  {Walch}}]{DerigsEtAl2016}
{Derigs} D, {Winters} AR, {Gassner} GJ, et~al (2016) {A novel high-order,
  entropy stable, 3D AMR MHD solver with guaranteed positive pressure}. J
  Comput Phys 317:223--256. \doi{10.1016/j.jcp.2016.04.048},
  {\href{https://arxiv.org/abs/1605.03572}{{arXiv:1605.03572}}} {[astro-ph.GA]}

\bibitem[{{Dhruv} et~al.(2023){Dhruv}, {Jain}, {O'Neal}, {Weide}, and
  {Dubey}}]{DhruvEtAl2023}
{Dhruv} A, {Jain} R, {O'Neal} J, et~al (2023) {Framework and Methodology for
  Verification of a Complex Scientific Simulation Software, Flash-X}. World
  Congress in Computer Science, Computer Engineering, \& Applied Computing
  (CSCE'23) arXiv:2308.16180. \doi{10.48550/arXiv.2308.16180},
  {\href{https://arxiv.org/abs/2308.16180}{{arXiv:2308.16180}}} {[cs.SE]}

\bibitem[{{Dong} et~al.(2022){Dong}, {Wang}, {Huang}, {Comisso}, {Sandstrom},
  and {Bhattacharjee}}]{DongEtAl2022}
{Dong} C, {Wang} L, {Huang} YM, et~al (2022) {Reconnection-driven energy
  cascade in magnetohydrodynamic turbulence}. Science Advances 8(49):eabn7627.
  \doi{10.1126/sciadv.abn7627},
  {\href{https://arxiv.org/abs/2210.10736}{{arXiv:2210.10736}}} {[astro-ph.SR]}

\bibitem[{{Drury} and {Strong}(2017)}]{DruryStrong2017}
{Drury} LOC, {Strong} AW (2017) {Power requirements for cosmic ray propagation
  models involving diffusive reacceleration; estimates and implications for the
  damping of interstellar turbulence}. \aap 597:A117.
  \doi{10.1051/0004-6361/201629526},
  {\href{https://arxiv.org/abs/1608.04227}{{arXiv:1608.04227}}} {[astro-ph.HE]}

\bibitem[{{Dubey} et~al.(2008){Dubey}, {Fisher}, {Graziani}, {Jordan}, {Lamb},
  {Reid}, {Rich}, {Sheeler}, {Townsley}, and {Weide}}]{DubeyEtAl2008}
{Dubey} A, {Fisher} R, {Graziani} C, et~al (2008) {Challenges of Extreme
  Computing using the FLASH code}. In: {Pogorelov} NV, {Audit} E, {Zank} GP
  (eds) Numerical Modeling of Space Plasma Flows, p 145

\bibitem[{{Dubey} et~al.(2022){Dubey}, {Weide}, {O'Neal}, {Dhruv}, {Couch},
  {Harris}, {Klosterman}, {Jain}, {Rudi}, {Messer}, {Pajkos}, {Carlson}, {Chu},
  {Wahib}, {Chawdhary}, {Ricker}, {Lee}, {Antypas}, {Riley}, {Daley},
  {Ganapathy}, {Timmes}, {Townsley}, {Vanella}, {Bachan}, {Rich}, {Kumar},
  {Endeve}, {Hix}, {Mezzacappa}, and {Papatheodore}}]{DubeyEtAl2022}
{Dubey} A, {Weide} K, {O'Neal} J, et~al (2022) {Flash-X: A multiphysics
  simulation software instrument}. SoftwareX 19:101168.
  \doi{10.1016/j.softx.2022.101168},
  {\href{https://arxiv.org/abs/2208.11630}{{arXiv:2208.11630}}}
  {[physics.comp-ph]}

\bibitem[{{Dubois} and {Commer{\c{c}}on}(2016)}]{DuboisCommercon2016}
{Dubois} Y, {Commer{\c{c}}on} B (2016) {An implicit scheme for solving the
  anisotropic diffusion of heat and cosmic rays in the RAMSES code}. \aap
  585:A138. \doi{10.1051/0004-6361/201527126},
  {\href{https://arxiv.org/abs/1509.07037}{{arXiv:1509.07037}}} {[astro-ph.GA]}

\bibitem[{{Dullemond} and {Turolla}(2000)}]{DullemondTurolla2000}
{Dullemond} CP, {Turolla} R (2000) {An efficient algorithm for two-dimensional
  radiative transfer in axisymmetric circumstellar envelopes and disks}. \aap
  360:1187--1202.
  {\href{https://arxiv.org/abs/arXiv:astro-ph/0003456}{{arXiv:astro-ph/0003456}}}

\bibitem[{{Elmegreen}(2009)}]{Elmegreen2009}
{Elmegreen} BG (2009) {Star Formation in Disks: Spiral Arms, Turbulence, and
  Triggering Mechanisms}. In: {J.~Andersen, J.~Bland-Hawthorn, \&
  B.~Nordstr{\"o}m} (ed) IAU Symposium, p 289, \doi{10.1017/S1743921308027713}

\bibitem[{{Elmegreen} and {Scalo}(2004)}]{ElmegreenScalo2004}
{Elmegreen} BG, {Scalo} J (2004) {Interstellar Turbulence I: Observations and
  Processes}. \araa 42(1):211--273.
  \doi{10.1146/annurev.astro.41.011802.094859},
  {\href{https://arxiv.org/abs/astro-ph/0404451}{{arXiv:astro-ph/0404451}}}
  {[astro-ph]}

\bibitem[{{Ercolano} et~al.(2008){Ercolano}, {Young}, {Drake}, and
  {Raymond}}]{ErcolanoEtAl2008}
{Ercolano} B, {Young} PR, {Drake} JJ, et~al (2008) {X-Ray Enabled MOCASSIN: A
  Three-dimensional Code for Photoionized Media}. \apjs 175(2):534--542.
  \doi{10.1086/524378},
  {\href{https://arxiv.org/abs/0710.2103}{{arXiv:0710.2103}}} {[astro-ph]}

\bibitem[{{Eswaran} and {Pope}(1988)}]{EswaranPope1988}
{Eswaran} V, {Pope} SB (1988) {An examination of forcing in direct numerical
  simulations of turbulence}. Computers and Fluids 16:257--278

\bibitem[{{Evans} and {Hawley}(1988)}]{EvansHawley1988}
{Evans} CR, {Hawley} JF (1988) {Simulation of Magnetohydrodynamic Flows: A
  Constrained Transport Model}. \apj 332:659. \doi{10.1086/166684}

\bibitem[{{Evoli} et~al.(2017){Evoli}, {Gaggero}, {Vittino}, {Di Bernardo}, {Di
  Mauro}, {Ligorini}, {Ullio}, and {Grasso}}]{EvoliEtAl2017}
{Evoli} C, {Gaggero} D, {Vittino} A, et~al (2017) {Cosmic-ray propagation with
  DRAGON2: I. numerical solver and astrophysical ingredients}. \jcap
  2017(2):015. \doi{10.1088/1475-7516/2017/02/015},
  {\href{https://arxiv.org/abs/1607.07886}{{arXiv:1607.07886}}} {[astro-ph.HE]}

\bibitem[{Falgout and Yang(2002)}]{FalgoutYang2002}
Falgout RD, Yang UM (2002) hypre: A library of high performance
  preconditioners. In: Sloot PMA, Hoekstra AG, Tan CJK, et~al (eds)
  Computational Science --- ICCS 2002. Springer, Berlin, Heidelberg, pp
  632--641, \doi{https://doi.org/10.1007/3-540-47789-6_66}

\bibitem[{{Federrath}(2013)}]{Federrath2013}
{Federrath} C (2013) {On the universality of supersonic turbulence}. \mnras
  436:1245--1257. \doi{10.1093/mnras/stt1644},
  {\href{https://arxiv.org/abs/1306.3989}{{arXiv:1306.3989}}} {[astro-ph.SR]}

\bibitem[{{Federrath}(2016)}]{Federrath2016}
{Federrath} C (2016) {On the universality of interstellar filaments: theory
  meets simulations and observations}. \mnras 457:375--388.
  \doi{10.1093/mnras/stv2880},
  {\href{https://arxiv.org/abs/1510.05654}{{arXiv:1510.05654}}} {[astro-ph.SR]}

\bibitem[{{Federrath} and {Klessen}(2012)}]{FederrathKlessen2012}
{Federrath} C, {Klessen} RS (2012) {The Star Formation Rate of Turbulent
  Magnetized Clouds: Comparing Theory, Simulations, and Observations}. \apj
  761:156. \doi{10.1088/0004-637X/761/2/156},
  {\href{https://arxiv.org/abs/1209.2856}{{arXiv:1209.2856}}} {[astro-ph.SR]}

\bibitem[{{Federrath} et~al.(2008){Federrath}, {Klessen}, and
  {Schmidt}}]{FederrathKlessenSchmidt2008}
{Federrath} C, {Klessen} RS, {Schmidt} W (2008) {The Density Probability
  Distribution in Compressible Isothermal Turbulence: Solenoidal versus
  Compressive Forcing}. \apjl 688:L79--L82. \doi{10.1086/595280}

\bibitem[{{Federrath} et~al.(2010{\natexlab{a}}){Federrath}, {Banerjee},
  {Clark}, and {Klessen}}]{FederrathBanerjeeClarkKlessen2010}
{Federrath} C, {Banerjee} R, {Clark} PC, et~al (2010{\natexlab{a}}) {Modeling
  Collapse and Accretion in Turbulent Gas Clouds: Implementation and Comparison
  of Sink Particles in AMR and SPH}. \apj 713:269--290.
  \doi{10.1088/0004-637X/713/1/269},
  {\href{https://arxiv.org/abs/1001.4456}{{arXiv:1001.4456}}}

\bibitem[{{Federrath} et~al.(2010{\natexlab{b}}){Federrath}, {Roman-Duval},
  {Klessen}, {Schmidt}, and {Mac Low}}]{FederrathDuvalKlessenSchmidtMacLow2010}
{Federrath} C, {Roman-Duval} J, {Klessen} RS, et~al (2010{\natexlab{b}})
  {Comparing the statistics of interstellar turbulence in simulations and
  observations. Solenoidal versus compressive turbulence forcing}. \aap
  512:A81. \doi{10.1051/0004-6361/200912437}

\bibitem[{{Federrath} et~al.(2011{\natexlab{a}}){Federrath}, {Banerjee},
  {Seifried}, {Clark}, and
  {Klessen}}]{FederrathBanerjeeSeifriedClarkKlessen2011}
{Federrath} C, {Banerjee} R, {Seifried} D, et~al (2011{\natexlab{a}})
  {Implementing and comparing sink particles in AMR and SPH}. In: {J.~Alves,
  B.~G.~Elmegreen, J.~M.~Girart, \& V.~Trimble} (ed) Computational Star
  Formation, pp 425--428, \doi{10.1017/S1743921311000755},
  {\href{https://arxiv.org/abs/1007.2504}{{arXiv:1007.2504}}}

\bibitem[{{Federrath} et~al.(2011{\natexlab{b}}){Federrath}, {Chabrier},
  {Schober}, {Banerjee}, {Klessen}, and {Schleicher}}]{FederrathEtAl2011}
{Federrath} C, {Chabrier} G, {Schober} J, et~al (2011{\natexlab{b}}) {Mach
  Number Dependence of Turbulent Magnetic Field Amplification: Solenoidal
  versus Compressive Flows}. \prl 107(11):114504.
  \doi{10.1103/PhysRevLett.107.114504},
  {\href{https://arxiv.org/abs/1109.1760}{{arXiv:1109.1760}}}
  {[physics.flu-dyn]}

\bibitem[{{Federrath} et~al.(2014{\natexlab{a}}){Federrath}, {Schober},
  {Bovino}, and {Schleicher}}]{FederrathSchoberBovinoSchleicher2014}
{Federrath} C, {Schober} J, {Bovino} S, et~al (2014{\natexlab{a}}) {The
  Turbulent Dynamo in Highly Compressible Supersonic Plasmas}. \apjl
  797(2):L19. \doi{10.1088/2041-8205/797/2/L19},
  {\href{https://arxiv.org/abs/1411.4707}{{arXiv:1411.4707}}} {[astro-ph.GA]}

\bibitem[{{Federrath} et~al.(2014{\natexlab{b}}){Federrath}, {Schr{\"o}n},
  {Banerjee}, and {Klessen}}]{FederrathEtAl2014}
{Federrath} C, {Schr{\"o}n} M, {Banerjee} R, et~al (2014{\natexlab{b}})
  {Modeling Jet and Outflow Feedback during Star Cluster Formation}. \apj
  790:128. \doi{10.1088/0004-637X/790/2/128},
  {\href{https://arxiv.org/abs/1406.3625}{{arXiv:1406.3625}}} {[astro-ph.SR]}

\bibitem[{{Federrath} et~al.(2016){Federrath}, {Rathborne}, {Longmore},
  {Kruijssen}, {Bally}, {Contreras}, {Crocker}, {Garay}, {Jackson}, {Testi},
  and {Walsh}}]{FederrathEtAl2016}
{Federrath} C, {Rathborne} JM, {Longmore} SN, et~al (2016) {The Link between
  Turbulence, Magnetic Fields, Filaments, and Star Formation in the Central
  Molecular Zone Cloud G0.253+0.016}. \apj 832:143.
  \doi{10.3847/0004-637X/832/2/143},
  {\href{https://arxiv.org/abs/1609.05911}{{arXiv:1609.05911}}}

\bibitem[{{Federrath} et~al.(2017{\natexlab{a}}){Federrath}, {Krumholz}, and
  {Hopkins}}]{FederrathKrumholzHopkins2017}
{Federrath} C, {Krumholz} M, {Hopkins} PF (2017{\natexlab{a}}) {Converging on
  the Initial Mass Function of Stars}. In: Journal of Physics Conference
  Series, Journal of Physics Conference Series, vol 837. IOP, p 012007,
  \doi{10.1088/1742-6596/837/1/012007}

\bibitem[{{Federrath} et~al.(2017{\natexlab{b}}){Federrath}, {Rathborne},
  {Longmore}, {Kruijssen}, {Bally}, {Contreras}, {Crocker}, {Garay}, {Jackson},
  {Testi}, and {Walsh}}]{FederrathEtAl2017iaus}
{Federrath} C, {Rathborne} JM, {Longmore} SN, et~al (2017{\natexlab{b}}) {The
  link between solenoidal turbulence and slow star formation in G0.253+0.016}.
  In: {Crocker} RM, {Longmore} SN, {Bicknell} GV (eds) IAU Symposium, pp
  123--128, \doi{10.1017/S1743921316012357},
  {\href{https://arxiv.org/abs/1609.08726}{{arXiv:1609.08726}}}

\bibitem[{{Federrath} et~al.(2021){Federrath}, {Klessen}, {Iapichino}, and
  {Beattie}}]{FederrathEtAl2021}
{Federrath} C, {Klessen} RS, {Iapichino} L, et~al (2021) {The sonic scale of
  interstellar turbulence}. Nature Astronomy 5:365--371.
  \doi{10.1038/s41550-020-01282-z},
  {\href{https://arxiv.org/abs/2011.06238}{{arXiv:2011.06238}}} {[astro-ph.GA]}

\bibitem[{{Federrath} et~al.(2022){Federrath}, {Roman-Duval}, {Klessen},
  {Schmidt}, and {Mac Low}}]{FederrathEtAl2022ascl}
{Federrath} C, {Roman-Duval} J, {Klessen} RS, et~al (2022) {TG: Turbulence
  Generator}. Astrophysics Source Code Library, record ascl:2204.001,
  {\href{https://arxiv.org/abs/2204.001}{{ascl:2204.001}}}

\bibitem[{{Fitz Axen} et~al.(2021{\natexlab{a}}){Fitz Axen}, {Offner},
  {Gaches}, {Fryer}, {Hungerford}, and {Silsbee}}]{FitzAxenEtAl2021b}
{Fitz Axen} M, {Offner} SSS, {Gaches} BAL, et~al (2021{\natexlab{a}})
  {Transport of Protostellar Cosmic Rays in Turbulent Dense Cores}. \apj
  915(1):43. \doi{10.3847/1538-4357/abfc55},
  {\href{https://arxiv.org/abs/2105.00028}{{arXiv:2105.00028}}} {[astro-ph.GA]}

\bibitem[{{Fitz Axen} et~al.(2021{\natexlab{b}}){Fitz Axen}, {Speicher},
  {Hungerford}, and {Fryer}}]{FitzAxenEtAl2021a}
{Fitz Axen} M, {Speicher} J, {Hungerford} A, et~al (2021{\natexlab{b}}) {Cosmic
  ray transport in mixed magnetic fields and their role on the observed
  anisotropies}. \mnras 500(3):3497--3510. \doi{10.1093/mnras/staa3500},
  {\href{https://arxiv.org/abs/2101.07759}{{arXiv:2101.07759}}} {[astro-ph.HE]}

\bibitem[{{Fitz Axen} et~al.(2024){Fitz Axen}, {Offner}, {Hopkins}, {Krumholz},
  and {Grudi{\'c}}}]{FitzAxenEtAl2024}
{Fitz Axen} M, {Offner} S, {Hopkins} PF, et~al (2024) {Suppressed Cosmic-Ray
  Energy Densities in Molecular Clouds from Streaming Instability-regulated
  Transport}. \apj 973(1):16. \doi{10.3847/1538-4357/ad675a},
  {\href{https://arxiv.org/abs/2407.17597}{{arXiv:2407.17597}}} {[astro-ph.GA]}

\bibitem[{{Fitz Axen} et~al.(2026){Fitz Axen}, {Offner}, {Hopkins}, and
  {Grudi{\'c}}}]{FitzAxenEtAl2026}
{Fitz Axen} M, {Offner} S, {Hopkins} PF, et~al (2026) {Gauging the Impact of
  Cosmic-ray Feedback on the Stellar Initial Mass Function}. \apj 1001(2):152.
  \doi{10.3847/1538-4357/ae5624},
  {\href{https://arxiv.org/abs/2604.21754}{{arXiv:2604.21754}}} {[astro-ph.HE]}

\bibitem[{{Frank} et~al.(2014){Frank}, {Ray}, {Cabrit}, {Hartigan}, {Arce},
  {Bacciotti}, {Bally}, {Benisty}, {Eisl{\"o}ffel}, {G{\"u}del}, {Lebedev},
  {Nisini}, and {Raga}}]{FrankEtAl2014}
{Frank} A, {Ray} TP, {Cabrit} S, et~al (2014) {Jets and Outflows from Star to
  Cloud: Observations Confront Theory}. In: {Beuther} H, {Klessen} RS,
  {Dullemond} CP, et~al (eds) Protostars and Planets VI. University of Arizona
  Press, pp 451--474, \doi{10.2458/azu_uapress_9780816531240-ch020},
  {\href{https://arxiv.org/abs/1402.3553}{{arXiv:1402.3553}}}

\bibitem[{{Frank} et~al.(2012){Frank}, {Hauck}, and
  {Olbrant}}]{FrankHauckOlbrant2012}
{Frank} M, {Hauck} CD, {Olbrant} E (2012) {Perturbed, Entropy-Based Closure for
  Radiative Transfer}. arXiv e-prints arXiv:1208.0772.
  \doi{10.48550/arXiv.1208.0772},
  {\href{https://arxiv.org/abs/1208.0772}{{arXiv:1208.0772}}}
  {[physics.comp-ph]}

\bibitem[{{Fraschetti} and {Giacalone}(2012)}]{FraschettiGiacalone2012}
{Fraschetti} F, {Giacalone} J (2012) {Early-time Velocity Autocorrelation for
  Charged Particles Diffusion and Drift in Static Magnetic Turbulence}. \apj
  755(2):114. \doi{10.1088/0004-637X/755/2/114},
  {\href{https://arxiv.org/abs/1206.6494}{{arXiv:1206.6494}}} {[astro-ph.HE]}

\bibitem[{{Fraschetti} et~al.(2018){Fraschetti}, {Drake}, {Cohen}, and
  {Garraffo}}]{FraschettiEtAl2018}
{Fraschetti} F, {Drake} JJ, {Cohen} O, et~al (2018) {Mottled Protoplanetary
  Disk Ionization by Magnetically Channeled T Tauri Star Energetic Particles}.
  \apj 853(2):112. \doi{10.3847/1538-4357/aaa48b},
  {\href{https://arxiv.org/abs/1710.01323}{{arXiv:1710.01323}}} {[astro-ph.HE]}

\bibitem[{{Frisch}(1995)}]{Frisch1995}
{Frisch} U (1995) {Turbulence. The legacy of A.N. Kolmogorov}. {Cambridge
  University Press}, \doi{10.1017/CBO9781139170666}

\bibitem[{{Fromang} and {Stone}(2009)}]{FromangStone2009}
{Fromang} S, {Stone} JM (2009) {Turbulent resistivity driven by the
  magnetorotational instability}. \aap 507(1):19--28.
  \doi{10.1051/0004-6361/200912752},
  {\href{https://arxiv.org/abs/0906.4422}{{arXiv:0906.4422}}} {[astro-ph.SR]}

\bibitem[{{Fryxell} et~al.(2000){Fryxell}, {Olson}, {Ricker}, {Timmes},
  {Zingale}, {Lamb}, {MacNeice}, {Rosner}, {Truran}, and
  {Tufo}}]{FryxellEtAl2000}
{Fryxell} B, {Olson} K, {Ricker} P, et~al (2000) {FLASH: An Adaptive Mesh
  Hydrodynamics Code for Modeling Astrophysical Thermonuclear Flashes}. \apjs
  131:273--334. \doi{10.1086/317361}

\bibitem[{{Fujii} et~al.(2007){Fujii}, {Iwasawa}, {Funato}, and
  {Makino}}]{FujiiEtAl2007}
{Fujii} M, {Iwasawa} M, {Funato} Y, et~al (2007) {BRIDGE: A Direct-Tree Hybrid
  N-Body Algorithm for Fully Self-Consistent Simulations of Star Clusters and
  Their Parent Galaxies}. \pasj 59:1095. \doi{10.1093/pasj/59.6.1095},
  {\href{https://arxiv.org/abs/0706.2059}{{arXiv:0706.2059}}} {[astro-ph]}

\bibitem[{{Fujii} et~al.(2012){Fujii}, {Iwasawa}, {Saitoh}, and {Portegies
  Zwart}}]{FujiiEtAl2012}
{Fujii} MS, {Iwasawa} M, {Saitoh} TR, et~al (2012) {N-body Simulations of Star
  Clusters, Black Holes, and Their Host Systems using BRIDGE}. In:
  {Capuzzo-Dolcetta} R, {Limongi} M, {Tornamb{\`e}} A (eds) Advances in
  Computational Astrophysics: Methods, Tools, and Outcome, p 211

\bibitem[{{Gaches} and {Offner}(2018)}]{GachesOffner2018}
{Gaches} BAL, {Offner} SSR (2018) {Exploration of Cosmic-ray Acceleration in
  Protostellar Accretion Shocks and a Model for Ionization Rates in Embedded
  Protoclusters}. \apj 861(2):87. \doi{10.3847/1538-4357/aac94d},
  {\href{https://arxiv.org/abs/1805.03215}{{arXiv:1805.03215}}} {[astro-ph.GA]}

\bibitem[{{Gaches} et~al.(2019){Gaches}, {Offner}, and
  {Bisbas}}]{GachesOffnerBisbas2019}
{Gaches} BAL, {Offner} SSR, {Bisbas} TG (2019) {The Astrochemical Impact of
  Cosmic Rays in Protoclusters. I. Molecular Cloud Chemistry}. \apj 878(2):105.
  \doi{10.3847/1538-4357/ab20c7},
  {\href{https://arxiv.org/abs/1905.02232}{{arXiv:1905.02232}}} {[astro-ph.GA]}

\bibitem[{{Gaches} et~al.(2021){Gaches}, {Walch}, and
  {Lazarian}}]{GachesWalchLazarian2021}
{Gaches} BAL, {Walch} S, {Lazarian} A (2021) {Cosmic-Ray Acceleration from
  Turbulence in Molecular Clouds}. \apjl 917(2):L39.
  \doi{10.3847/2041-8213/ac1b2f},
  {\href{https://arxiv.org/abs/2108.03250}{{arXiv:2108.03250}}} {[astro-ph.HE]}

\bibitem[{{Gaches} et~al.(2023){Gaches}, {Walch}, {W{\"u}nsch}, and
  {Mackey}}]{GachesEtAl2023}
{Gaches} BAL, {Walch} S, {W{\"u}nsch} R, et~al (2023) {Tree-based solvers for
  adaptive mesh refinement code FLASH - IV. An X-ray radiation scheme to couple
  discrete and diffuse X-ray emission sources to the thermochemistry of the
  interstellar medium}. \mnras 522(3):4674--4690. \doi{10.1093/mnras/stad1206},
  {\href{https://arxiv.org/abs/2301.13237}{{arXiv:2301.13237}}} {[astro-ph.IM]}

\bibitem[{{Gaches} et~al.(2024){Gaches}, {Grassi}, {Vogt-Geisse}, {Bovolenta},
  {Vallance}, {Heathcote}, {Padovani}, {Bovino}, and {Gorai}}]{GachesEtAl2024}
{Gaches} BAL, {Grassi} T, {Vogt-Geisse} S, et~al (2024) {The Astrochemistry
  Low-energy Electron Cross-Section (ALeCS) database. I. Semi-empirical
  electron-impact ionization cross-section calculations and ionization rates}.
  \aap 684:A41. \doi{10.1051/0004-6361/202348293},
  {\href{https://arxiv.org/abs/2310.10739}{{arXiv:2310.10739}}} {[astro-ph.GA]}

\bibitem[{{Galishnikova} et~al.(2022){Galishnikova}, {Kunz}, and
  {Schekochihin}}]{GalishnikovaEtAl2022}
{Galishnikova} AK, {Kunz} MW, {Schekochihin} AA (2022) {Tearing Instability and
  Current-Sheet Disruption in the Turbulent Dynamo}. Physical Review X
  12(4):041027. \doi{10.1103/PhysRevX.12.041027},
  {\href{https://arxiv.org/abs/2201.07757}{{arXiv:2201.07757}}} {[astro-ph.HE]}

\bibitem[{{Gatto} et~al.(2017){Gatto}, {Walch}, {Naab}, {Girichidis},
  {W{\"u}nsch}, {Glover}, {Klessen}, {Clark}, {Peters}, {Derigs}, {Baczynski},
  and {Puls}}]{GattoEtAl2017}
{Gatto} A, {Walch} S, {Naab} T, et~al (2017) {The SILCC project - III.
  Regulation of star formation and outflows by stellar winds and supernovae}.
  \mnras 466(2):1903--1924. \doi{10.1093/mnras/stw3209},
  {\href{https://arxiv.org/abs/1606.05346}{{arXiv:1606.05346}}} {[astro-ph.GA]}

\bibitem[{{Geen} and {de Koter}(2022)}]{GeenDeKoter2022}
{Geen} S, {de Koter} A (2022) {Bottling the champagne: dynamics and radiation
  trapping of wind-driven bubbles around massive stars}. \mnras
  509(3):4498--4514. \doi{10.1093/mnras/stab3245},
  {\href{https://arxiv.org/abs/2111.03399}{{arXiv:2111.03399}}} {[astro-ph.GA]}

\bibitem[{{Geen} et~al.(2015{\natexlab{a}}){Geen}, {Hennebelle}, {Tremblin},
  and {Rosdahl}}]{GeenEtAl2015b}
{Geen} S, {Hennebelle} P, {Tremblin} P, et~al (2015{\natexlab{a}})
  {Photoionization feedback in a self-gravitating, magnetized, turbulent
  cloud}. \mnras 454(4):4484--4502. \doi{10.1093/mnras/stv2272},
  {\href{https://arxiv.org/abs/1507.02981}{{arXiv:1507.02981}}} {[astro-ph.GA]}

\bibitem[{{Geen} et~al.(2015{\natexlab{b}}){Geen}, {Rosdahl}, {Blaizot},
  {Devriendt}, and {Slyz}}]{GeenEtAl2015a}
{Geen} S, {Rosdahl} J, {Blaizot} J, et~al (2015{\natexlab{b}}) {A detailed
  study of feedback from a massive star}. \mnras 448(4):3248--3264.
  \doi{10.1093/mnras/stv251},
  {\href{https://arxiv.org/abs/1412.0484}{{arXiv:1412.0484}}} {[astro-ph.GA]}

\bibitem[{{Geen} et~al.(2016){Geen}, {Hennebelle}, {Tremblin}, and
  {Rosdahl}}]{GeenEtAl2016}
{Geen} S, {Hennebelle} P, {Tremblin} P, et~al (2016) {Feedback in Clouds II: UV
  photoionization and the first supernova in a massive cloud}. \mnras
  463(3):3129--3142. \doi{10.1093/mnras/stw2235},
  {\href{https://arxiv.org/abs/1607.05487}{{arXiv:1607.05487}}} {[astro-ph.GA]}

\bibitem[{{Geen} et~al.(2017){Geen}, {Soler}, and
  {Hennebelle}}]{GeenSolerHennebelle2017}
{Geen} S, {Soler} JD, {Hennebelle} P (2017) {Interpreting the star formation
  efficiency of nearby molecular clouds with ionizing radiation}. \mnras
  471(4):4844--4855. \doi{10.1093/mnras/stx1765},
  {\href{https://arxiv.org/abs/1703.10071}{{arXiv:1703.10071}}} {[astro-ph.GA]}

\bibitem[{Genovese et~al.(2006)Genovese, Deutsch, Neelov, Goedecker, and
  Beylkin}]{GenoveseEtAl2006}
Genovese L, Deutsch T, Neelov A, et~al (2006) Efficient solution of poisson’s
  equation with free boundary conditions. J Chem Phys 125(7):074105.
  \doi{10.1063/1.2335442}

\bibitem[{{Gent} et~al.(2013){Gent}, {Shukurov}, {Fletcher}, {Sarson}, and
  {Mantere}}]{GentEtAl2013}
{Gent} FA, {Shukurov} A, {Fletcher} A, et~al (2013) {The supernova-regulated
  ISM - I. The multiphase structure}. \mnras 432(2):1396--1423.
  \doi{10.1093/mnras/stt560},
  {\href{https://arxiv.org/abs/1204.3567}{{arXiv:1204.3567}}} {[astro-ph.GA]}

\bibitem[{{Gent} et~al.(2020){Gent}, {Mac Low}, {K{\"a}pyl{\"a}}, {Sarson}, and
  {Hollins}}]{GentEtAl2020}
{Gent} FA, {Mac Low} MM, {K{\"a}pyl{\"a}} MJ, et~al (2020) {Modelling
  supernova-driven turbulence}. Geophysical and Astrophysical Fluid Dynamics
  114(1-2):77--105. \doi{10.1080/03091929.2019.1634705},
  {\href{https://arxiv.org/abs/1806.01570}{{arXiv:1806.01570}}} {[astro-ph.GA]}

\bibitem[{{Gent} et~al.(2021){Gent}, {Mac Low}, {K{\"a}pyl{\"a}}, and
  {Singh}}]{GentEtAl2021}
{Gent} FA, {Mac Low} MM, {K{\"a}pyl{\"a}} MJ, et~al (2021) {Small-scale Dynamo
  in Supernova-driven Interstellar Turbulence}. \apjl 910(2):L15.
  \doi{10.3847/2041-8213/abed59},
  {\href{https://arxiv.org/abs/2010.01833}{{arXiv:2010.01833}}} {[astro-ph.GA]}

\bibitem[{{Gingold} and {Monaghan}(1977)}]{GingoldMonaghan1977}
{Gingold} RA, {Monaghan} JJ (1977) {Smoothed particle hydrodynamics - Theory
  and application to non-spherical stars}. \mnras 181:375--389

\bibitem[{{Girichidis} et~al.(2011){Girichidis}, {Federrath}, {Banerjee}, and
  {Klessen}}]{GirichidisEtAl2011}
{Girichidis} P, {Federrath} C, {Banerjee} R, et~al (2011) {Importance of the
  initial conditions for star formation - I. Cloud evolution and morphology}.
  \mnras 413:2741--2759. \doi{10.1111/j.1365-2966.2011.18348.x},
  {\href{https://arxiv.org/abs/1008.5255}{{arXiv:1008.5255}}} {[astro-ph.SR]}

\bibitem[{{Girichidis} et~al.(2012{\natexlab{a}}){Girichidis}, {Federrath},
  {Allison}, {Banerjee}, and {Klessen}}]{GirichidisEtAl2012b}
{Girichidis} P, {Federrath} C, {Allison} R, et~al (2012{\natexlab{a}})
  {Importance of the initial conditions for star formation - III. Statistical
  properties of embedded protostellar clusters}. \mnras 420:3264--3280.
  \doi{10.1111/j.1365-2966.2011.20250.x},
  {\href{https://arxiv.org/abs/1111.5440}{{arXiv:1111.5440}}} {[astro-ph.SR]}

\bibitem[{{Girichidis} et~al.(2012{\natexlab{b}}){Girichidis}, {Federrath},
  {Banerjee}, and {Klessen}}]{GirichidisEtAl2012a}
{Girichidis} P, {Federrath} C, {Banerjee} R, et~al (2012{\natexlab{b}})
  {Importance of the initial conditions for star formation - II.
  Fragmentation-induced starvation and accretion shielding}. \mnras
  420:613--626. \doi{10.1111/j.1365-2966.2011.20073.x},
  {\href{https://arxiv.org/abs/1110.1924}{{arXiv:1110.1924}}} {[astro-ph.SR]}

\bibitem[{{Girichidis} et~al.(2014){Girichidis}, {Naab}, {Walch}, and
  {Hanasz}}]{GirichidisEtAl2014CR}
{Girichidis} P, {Naab} T, {Walch} S, et~al (2014) {Anisotropic transport and
  early dynamical impact of Cosmic Rays around Supernova remnants}. arXiv
  e-prints arXiv:1406.4861. \doi{10.48550/arXiv.1406.4861},
  {\href{https://arxiv.org/abs/1406.4861}{{arXiv:1406.4861}}} {[astro-ph.HE]}

\bibitem[{{Girichidis} et~al.(2016){Girichidis}, {Naab}, {Walch}, {Hanasz},
  {Mac Low}, {Ostriker}, {Gatto}, {Peters}, {W{\"u}nsch}, {Glover}, {Klessen},
  {Clark}, and {Baczynski}}]{GirichidisEtAl2016}
{Girichidis} P, {Naab} T, {Walch} S, et~al (2016) {Launching Cosmic-Ray-driven
  Outflows from the Magnetized Interstellar Medium}. \apjl 816(2):L19.
  \doi{10.3847/2041-8205/816/2/L19},
  {\href{https://arxiv.org/abs/1509.07247}{{arXiv:1509.07247}}} {[astro-ph.GA]}

\bibitem[{{Girichidis} et~al.(2020){Girichidis}, {Pfrommer}, {Hanasz}, and
  {Naab}}]{GirichidisEtAl2020}
{Girichidis} P, {Pfrommer} C, {Hanasz} M, et~al (2020) {Spectrally resolved
  cosmic ray hydrodynamics - I. Spectral scheme}. \mnras 491(1):993--1007.
  \doi{10.1093/mnras/stz2961},
  {\href{https://arxiv.org/abs/1909.12840}{{arXiv:1909.12840}}} {[astro-ph.HE]}

\bibitem[{{Gjergo} et~al.(2026){Gjergo}, {Zhang}, {Kroupa}, {Sorokin}, {Yan},
  {Guo}, {Jerabkova}, {Zonoozi}, and {Haghi}}]{GjergoEtAl2026}
{Gjergo} E, {Zhang} ZY, {Kroupa} P, et~al (2026) {Massive Star Formation at
  Supersolar Metallicities: Constraints on the Initial Mass Function}. Research
  in Astronomy and Astrophysics 26(2):025003. \doi{10.1088/1674-4527/ae1f79},
  {\href{https://arxiv.org/abs/2509.20440}{{arXiv:2509.20440}}} {[astro-ph.GA]}

\bibitem[{{Glover} and {Mac Low}(2007{\natexlab{a}})}]{GloverMacLow2007a}
{Glover} SCO, {Mac Low} MM (2007{\natexlab{a}}) {Simulating the Formation of
  Molecular Clouds. I. Slow Formation by Gravitational Collapse from Static
  Initial Conditions}. \apjs 169:239--268. \doi{10.1086/512238},
  {\href{https://arxiv.org/abs/arXiv:astro-ph/0605120}{{arXiv:astro-ph/0605120}}}

\bibitem[{{Glover} and {Mac Low}(2007{\natexlab{b}})}]{GloverMacLow2007b}
{Glover} SCO, {Mac Low} MM (2007{\natexlab{b}}) {Simulating the Formation of
  Molecular Clouds. II. Rapid Formation from Turbulent Initial Conditions}.
  \apj 659:1317--1337. \doi{10.1086/512227},
  {\href{https://arxiv.org/abs/arXiv:astro-ph/0605121}{{arXiv:astro-ph/0605121}}}

\bibitem[{{Glover} et~al.(2010){Glover}, {Federrath}, {Mac Low}, and
  {Klessen}}]{GloverFederrathMacLowKlessen2010}
{Glover} SCO, {Federrath} C, {Mac Low} M, et~al (2010) {Modelling CO formation
  in the turbulent interstellar medium}. \mnras 404:2--29.
  \doi{10.1111/j.1365-2966.2009.15718.x},
  {\href{https://arxiv.org/abs/0907.4081}{{arXiv:0907.4081}}}

\bibitem[{{Gnedin} and {Abel}(2001)}]{GnedinAbel2001}
{Gnedin} NY, {Abel} T (2001) {Multi-dimensional cosmological radiative transfer
  with a Variable Eddington Tensor formalism}. \na 6(7):437--455.
  \doi{10.1016/S1384-1076(01)00068-9},
  {\href{https://arxiv.org/abs/astro-ph/0106278}{{arXiv:astro-ph/0106278}}}
  {[astro-ph]}

\bibitem[{{Godunov}(1959)}]{Godunov1959}
{Godunov} SK (1959) {A Difference Scheme for Numerical Solution of
  Discontinuous Solution of Hydrodynamic Equations}. Math Sbornik 47:271--306

\bibitem[{{Gonz{\'a}lez} et~al.(2007){Gonz{\'a}lez}, {Audit}, and
  {Huynh}}]{GonzalezAuditHuynh2007}
{Gonz{\'a}lez} M, {Audit} E, {Huynh} P (2007) {HERACLES: a three-dimensional
  radiation hydrodynamics code}. \aap 464(2):429--435.
  \doi{10.1051/0004-6361:20065486}

\bibitem[{{Gonz{\'a}lez} et~al.(2015){Gonz{\'a}lez}, {Vaytet},
  {Commer{\c{c}}on}, and {Masson}}]{GonzalezEtAl2015}
{Gonz{\'a}lez} M, {Vaytet} N, {Commer{\c{c}}on} B, et~al (2015) {Multigroup
  radiation hydrodynamics with flux-limited diffusion and adaptive mesh
  refinement}. \aap 578:A12. \doi{10.1051/0004-6361/201525971},
  {\href{https://arxiv.org/abs/1504.01894}{{arXiv:1504.01894}}} {[astro-ph.IM]}

\bibitem[{{G{\'o}rski} et~al.(2005){G{\'o}rski}, {Hivon}, {Banday}, {Wandelt},
  {Hansen}, {Reinecke}, and {Bartelmann}}]{GorskiEtAl2005}
{G{\'o}rski} KM, {Hivon} E, {Banday} AJ, et~al (2005) {HEALPix: A Framework for
  High-Resolution Discretization and Fast Analysis of Data Distributed on the
  Sphere}. \apj 622(2):759--771. \doi{10.1086/427976},
  {\href{https://arxiv.org/abs/astro-ph/0409513}{{arXiv:astro-ph/0409513}}}
  {[astro-ph]}

\bibitem[{{Gotoh} et~al.(2002){Gotoh}, {Fukayama}, and
  {Nakano}}]{GotohFukayamaNakano2002}
{Gotoh} T, {Fukayama} D, {Nakano} T (2002) {Velocity field statistics in
  homogeneous steady turbulence obtained using a high-resolution direct
  numerical simulation}. Physics of Fluids 14(3):1065--1081.
  \doi{10.1063/1.1448296}

\bibitem[{{Gressel} et~al.(2008){Gressel}, {Elstner}, {Ziegler}, and
  {R{\"u}diger}}]{GresselEtAl2008}
{Gressel} O, {Elstner} D, {Ziegler} U, et~al (2008) {Direct simulations of a
  supernova-driven galactic dynamo}. \aap 486(3):L35--L38.
  \doi{10.1051/0004-6361:200810195},
  {\href{https://arxiv.org/abs/0805.2616}{{arXiv:0805.2616}}} {[astro-ph]}

\bibitem[{{Grete} et~al.(2025){Grete}, {Scannapieco}, {Br{\"u}ggen}, and
  {Pan}}]{GreteEtAl2025}
{Grete} P, {Scannapieco} E, {Br{\"u}ggen} M, et~al (2025) {The Density
  Distribution of Compressively Forced, Supersonic Turbulence Depends on the
  Driving Correlation Time}. \apj 987(2):122. \doi{10.3847/1538-4357/add936},
  {\href{https://arxiv.org/abs/2505.23898}{{arXiv:2505.23898}}} {[astro-ph.GA]}

\bibitem[{{Grisdale} et~al.(2018){Grisdale}, {Agertz}, {Renaud}, and
  {Romeo}}]{GrisdaleEtAl2018}
{Grisdale} K, {Agertz} O, {Renaud} F, et~al (2018) {Physical properties and
  scaling relations of molecular clouds: the effect of stellar feedback}.
  \mnras 479(3):3167--3180. \doi{10.1093/mnras/sty1595},
  {\href{https://arxiv.org/abs/1801.03104}{{arXiv:1801.03104}}} {[astro-ph.GA]}

\bibitem[{{Grudi{\'c}} et~al.(2021){Grudi{\'c}}, {Guszejnov}, {Hopkins},
  {Offner}, and {Faucher-Gigu{\`e}re}}]{GrudicEtAl2021}
{Grudi{\'c}} MY, {Guszejnov} D, {Hopkins} PF, et~al (2021) {STARFORGE: Towards
  a comprehensive numerical model of star cluster formation and feedback}.
  \mnras 506(2):2199--2231. \doi{10.1093/mnras/stab1347},
  {\href{https://arxiv.org/abs/2010.11254}{{arXiv:2010.11254}}} {[astro-ph.IM]}

\bibitem[{{Grudi{\'c}} et~al.(2022){Grudi{\'c}}, {Guszejnov}, {Offner},
  {Rosen}, {Raju}, {Faucher-Gigu{\`e}re}, and {Hopkins}}]{GrudicEtAl2022}
{Grudi{\'c}} MY, {Guszejnov} D, {Offner} SSR, et~al (2022) {The dynamics and
  outcome of star formation with jets, radiation, winds, and supernovae in
  concert}. \mnras 512(1):216--232. \doi{10.1093/mnras/stac526},
  {\href{https://arxiv.org/abs/2201.00882}{{arXiv:2201.00882}}} {[astro-ph.GA]}

\bibitem[{{Guszejnov} et~al.(2016){Guszejnov}, {Krumholz}, and
  {Hopkins}}]{GuszejnovKrumholzHopkins2016}
{Guszejnov} D, {Krumholz} MR, {Hopkins} PF (2016) {The necessity of feedback
  physics in setting the peak of the initial mass function}. \mnras
  458:673--680. \doi{10.1093/mnras/stw315},
  {\href{https://arxiv.org/abs/1510.05040}{{arXiv:1510.05040}}} {[astro-ph.SR]}

\bibitem[{{Guszejnov} et~al.(2017){Guszejnov}, {Hopkins}, and
  {Krumholz}}]{GuszejnovHopkinsKrumholz2017}
{Guszejnov} D, {Hopkins} PF, {Krumholz} MR (2017) {Protostellar feedback in
  turbulent fragmentation: consequences for stellar clustering and
  multiplicity}. \mnras 468(4):4093--4106. \doi{10.1093/mnras/stx725},
  {\href{https://arxiv.org/abs/1610.00772}{{arXiv:1610.00772}}} {[astro-ph.GA]}

\bibitem[{{Guszejnov} et~al.(2018){Guszejnov}, {Hopkins}, {Grudi{\'c}},
  {Krumholz}, and {Federrath}}]{GuszejnovEtAl2018}
{Guszejnov} D, {Hopkins} PF, {Grudi{\'c}} MY, et~al (2018) {Isothermal
  Fragmentation: Is there a low-mass cut-off?} \mnras 480:182--191.
  \doi{10.1093/mnras/sty1847},
  {\href{https://arxiv.org/abs/1804.08574}{{arXiv:1804.08574}}}

\bibitem[{{Guszejnov} et~al.(2021){Guszejnov}, {Grudi{\'c}}, {Hopkins},
  {Offner}, and {Faucher-Gigu{\`e}re}}]{GuszejnovEtAl2021}
{Guszejnov} D, {Grudi{\'c}} MY, {Hopkins} PF, et~al (2021) {STARFORGE: the
  effects of protostellar outflows on the IMF}. \mnras 502(3):3646--3663.
  \doi{10.1093/mnras/stab278},
  {\href{https://arxiv.org/abs/2010.11249}{{arXiv:2010.11249}}} {[astro-ph.GA]}

\bibitem[{{Guszejnov} et~al.(2022){Guszejnov}, {Grudi{\'c}}, {Offner},
  {Faucher-Gigu{\`e}re}, {Hopkins}, and {Rosen}}]{GuszejnovEtAl2022}
{Guszejnov} D, {Grudi{\'c}} MY, {Offner} SSR, et~al (2022) {Effects of the
  environment and feedback physics on the initial mass function of stars in the
  STARFORGE simulations}. \mnras 515(4):4929--4952.
  \doi{10.1093/mnras/stac2060},
  {\href{https://arxiv.org/abs/2205.10413}{{arXiv:2205.10413}}} {[astro-ph.GA]}

\bibitem[{{Haid} et~al.(2016){Haid}, {Walch}, {Naab}, {Seifried}, {Mackey}, and
  {Gatto}}]{HaidEtAl2016}
{Haid} S, {Walch} S, {Naab} T, et~al (2016) {Supernova blast waves in
  wind-blown bubbles, turbulent, and power-law ambient media}. \mnras
  460(3):2962--2978. \doi{10.1093/mnras/stw1082},
  {\href{https://arxiv.org/abs/1604.04395}{{arXiv:1604.04395}}} {[astro-ph.GA]}

\bibitem[{{Haid} et~al.(2019){Haid}, {Walch}, {Seifried}, {W{\"u}nsch},
  {Dinnbier}, and {Naab}}]{HaidEtAl2019}
{Haid} S, {Walch} S, {Seifried} D, et~al (2019) {SILCC-Zoom: The early impact
  of ionizing radiation on forming molecular clouds}. \mnras 482(3):4062--4083.
  \doi{10.1093/mnras/sty2938},
  {\href{https://arxiv.org/abs/1810.08210}{{arXiv:1810.08210}}} {[astro-ph.GA]}

\bibitem[{{Hanasz} et~al.(2013){Hanasz}, {Lesch}, {Naab}, {Gawryszczak},
  {Kowalik}, and {W{\'o}lta{\'n}ski}}]{HanaszEtAl2013}
{Hanasz} M, {Lesch} H, {Naab} T, et~al (2013) {Cosmic Rays Can Drive Strong
  Outflows from Gas-rich High-redshift Disk Galaxies}. \apjl 777(2):L38.
  \doi{10.1088/2041-8205/777/2/L38},
  {\href{https://arxiv.org/abs/1310.3273}{{arXiv:1310.3273}}} {[astro-ph.GA]}

\bibitem[{{Harding} et~al.(2016){Harding}, {Fryer}, and
  {Mendel}}]{HardingFryerMendel2016}
{Harding} JP, {Fryer} CL, {Mendel} S (2016) {Explaining TeV Cosmic-Ray
  Anisotropies with Non-diffusive Cosmic-Ray Propagation}. \apj 822(2):102.
  \doi{10.3847/0004-637X/822/2/102},
  {\href{https://arxiv.org/abs/1510.02487}{{arXiv:1510.02487}}} {[astro-ph.HE]}

\bibitem[{{Harries}(2015)}]{Harries2015}
{Harries} TJ (2015) {Radiation-hydrodynamical simulations of massive star
  formation using Monte Carlo radiative transfer - I. Algorithms and numerical
  methods}. \mnras 448(4):3156--3166. \doi{10.1093/mnras/stv158},
  {\href{https://arxiv.org/abs/1501.05754}{{arXiv:1501.05754}}} {[astro-ph.SR]}

\bibitem[{{Harries} et~al.(2019){Harries}, {Haworth}, {Acreman}, {Ali}, and
  {Douglas}}]{HarriesEtAl2019}
{Harries} TJ, {Haworth} TJ, {Acreman} D, et~al (2019) {The TORUS radiation
  transfer code}. Astronomy and Computing 27:63.
  \doi{10.1016/j.ascom.2019.03.002},
  {\href{https://arxiv.org/abs/1903.06672}{{arXiv:1903.06672}}} {[astro-ph.SR]}

\bibitem[{{Haugen} et~al.(2004{\natexlab{a}}){Haugen}, {Brandenburg}, and
  {Dobler}}]{HaugenBrandenburgDobler2004}
{Haugen} NE, {Brandenburg} A, {Dobler} W (2004{\natexlab{a}}) {Simulations of
  nonhelical hydromagnetic turbulence}. \pre 70(1):016308.
  \doi{10.1103/PhysRevE.70.016308},
  {\href{https://arxiv.org/abs/astro-ph/0307059}{{arXiv:astro-ph/0307059}}}
  {[astro-ph]}

\bibitem[{{Haugen} and {Brandenburg}(2004)}]{HaugenBrandenburg2004dyn}
{Haugen} NEL, {Brandenburg} A (2004) {Suppression of small scale dynamo action
  by an imposed magnetic field}. \pre 70(3):036408.
  \doi{10.1103/PhysRevE.70.036408},
  {\href{https://arxiv.org/abs/astro-ph/0402281}{{arXiv:astro-ph/0402281}}}
  {[astro-ph]}

\bibitem[{{Haugen} et~al.(2004{\natexlab{b}}){Haugen}, {Brandenburg}, and
  {Mee}}]{HaugenBrandenburgMee2004}
{Haugen} NEL, {Brandenburg} A, {Mee} AJ (2004{\natexlab{b}}) {Mach number
  dependence of the onset of dynamo action}. \mnras 353:947--952.
  \doi{10.1111/j.1365-2966.2004.08127.x}

\bibitem[{{Haworth} et~al.(2018){Haworth}, {Glover}, {Koepferl}, {Bisbas}, and
  {Dale}}]{HaworthEtAl2018}
{Haworth} TJ, {Glover} SCO, {Koepferl} CM, et~al (2018) {Synthetic observations
  of star formation and the interstellar medium}. \nar 82:1--58.
  \doi{10.1016/j.newar.2018.06.001},
  {\href{https://arxiv.org/abs/1711.05275}{{arXiv:1711.05275}}} {[astro-ph.GA]}

\bibitem[{{Hayes} and {Norman}(2003)}]{HayesNorman2003}
{Hayes} JC, {Norman} ML (2003) {Beyond Flux-limited Diffusion: Parallel
  Algorithms for Multidimensional Radiation Hydrodynamics}. \apjs
  147(1):197--220. \doi{10.1086/374658},
  {\href{https://arxiv.org/abs/astro-ph/0207260}{{arXiv:astro-ph/0207260}}}
  {[astro-ph]}

\bibitem[{{He} and {Ricotti}(2023)}]{HeRicotti2023}
{He} CC, {Ricotti} M (2023) {Massive pre-stellar cores in
  radiation-magneto-turbulent simulations of molecular clouds}. \mnras
  522(4):5374--5392. \doi{10.1093/mnras/stad1289},
  {\href{https://arxiv.org/abs/2210.11629}{{arXiv:2210.11629}}} {[astro-ph.GA]}

\bibitem[{{He} et~al.(2024{\natexlab{a}}){He}, {Wibking}, and
  {Krumholz}}]{HeWibkingKrumholz2024b}
{He} CC, {Wibking} BD, {Krumholz} MR (2024{\natexlab{a}}) {A novel numerical
  method for mixed-frame multigroup radiation-hydrodynamics with GPU
  acceleration implemented in the QUOKKA code}. \mnras 535(4):3059--3076.
  \doi{10.1093/mnras/stae2580},
  {\href{https://arxiv.org/abs/2407.18304}{{arXiv:2407.18304}}} {[astro-ph.GA]}

\bibitem[{{He} et~al.(2024{\natexlab{b}}){He}, {Wibking}, and
  {Krumholz}}]{HeWibkingKrumholz2024a}
{He} CC, {Wibking} BD, {Krumholz} MR (2024{\natexlab{b}}) {An asymptotically
  correct implicit-explicit time integration scheme for finite volume
  radiation-hydrodynamics}. \mnras 531(1):1228--1242.
  \doi{10.1093/mnras/stae1244},
  {\href{https://arxiv.org/abs/2404.08247}{{arXiv:2404.08247}}} {[astro-ph.IM]}

\bibitem[{{Hennebelle} and {Chabrier}(2008)}]{HennebelleChabrier2008}
{Hennebelle} P, {Chabrier} G (2008) {Analytical Theory for the Initial Mass
  Function: CO Clumps and Prestellar Cores}. \apj 684:395--410.
  \doi{10.1086/589916},
  {\href{https://arxiv.org/abs/0805.0691}{{arXiv:0805.0691}}}

\bibitem[{{Hennebelle} and {Chabrier}(2009)}]{HennebelleChabrier2009}
{Hennebelle} P, {Chabrier} G (2009) {Analytical Theory for the Initial Mass
  Function. II. Properties of the Flow}. \apj 702:1428--1442.
  \doi{10.1088/0004-637X/702/2/1428},
  {\href{https://arxiv.org/abs/0907.2765}{{arXiv:0907.2765}}}

\bibitem[{{Hennebelle} and {Chabrier}(2011)}]{HennebelleChabrier2011}
{Hennebelle} P, {Chabrier} G (2011) {Analytical Star Formation Rate from
  Gravoturbulent Fragmentation}. \apjl 743:L29.
  \doi{10.1088/2041-8205/743/2/L29},
  {\href{https://arxiv.org/abs/1110.0033}{{arXiv:1110.0033}}} {[astro-ph.GA]}

\bibitem[{{Hennebelle} and {Chabrier}(2013)}]{HennebelleChabrier2013}
{Hennebelle} P, {Chabrier} G (2013) {Analytical Theory for the Initial Mass
  Function. III. Time Dependence and Star Formation Rate}. \apj 770:150.
  \doi{10.1088/0004-637X/770/2/150},
  {\href{https://arxiv.org/abs/1304.6637}{{arXiv:1304.6637}}} {[astro-ph.GA]}

\bibitem[{{Hennebelle} and {Falgarone}(2012)}]{HennebelleFalgarone2012}
{Hennebelle} P, {Falgarone} E (2012) {Turbulent molecular clouds}. \aapr 20:55.
  \doi{10.1007/s00159-012-0055-y},
  {\href{https://arxiv.org/abs/1211.0637}{{arXiv:1211.0637}}} {[astro-ph.GA]}

\bibitem[{{Hennebelle} and {Fromang}(2008)}]{HennebelleFromang2008}
{Hennebelle} P, {Fromang} S (2008) {Magnetic processes in a collapsing dense
  core. I. Accretion and ejection}. \aap 477:9--24.
  \doi{10.1051/0004-6361:20078309},
  {\href{https://arxiv.org/abs/0709.2886}{{arXiv:0709.2886}}}

\bibitem[{{Hennebelle} and {Iffrig}(2014)}]{HennebelleIffrig2014}
{Hennebelle} P, {Iffrig} O (2014) {Simulations of magnetized multiphase
  galactic disc regulated by supernovae explosions}. \aap 570:A81.
  \doi{10.1051/0004-6361/201423392},
  {\href{https://arxiv.org/abs/1405.7819}{{arXiv:1405.7819}}} {[astro-ph.GA]}

\bibitem[{{Hennebelle} et~al.(2020){Hennebelle}, {Commer{\c{c}}on}, {Lee}, and
  {Chabrier}}]{HennebelleEtAl2020}
{Hennebelle} P, {Commer{\c{c}}on} B, {Lee} YN, et~al (2020) {What Is the Role
  of Stellar Radiative Feedback in Setting the Stellar Mass Spectrum?} \apj
  904(2):194. \doi{10.3847/1538-4357/abbfab},
  {\href{https://arxiv.org/abs/2010.03539}{{arXiv:2010.03539}}} {[astro-ph.GA]}

\bibitem[{{Hennebelle} et~al.(2022){Hennebelle}, {Lebreuilly}, {Colman},
  {Elia}, {Fuller}, {Leurini}, {Nony}, {Schisano}, {Soler}, {Traficante},
  {Klessen}, {Molinari}, and {Testi}}]{HennebelleEtAl2022}
{Hennebelle} P, {Lebreuilly} U, {Colman} T, et~al (2022) {Influence of magnetic
  field and stellar radiative feedback on the collapse and the stellar mass
  spectrum of a massive star-forming clump}. \aap 668:A147.
  \doi{10.1051/0004-6361/202243803},
  {\href{https://arxiv.org/abs/2210.12475}{{arXiv:2210.12475}}} {[astro-ph.GA]}

\bibitem[{{Hennebelle} et~al.(2024){Hennebelle}, {Brucy}, and
  {Colman}}]{HennebelleBrucyColman2024}
{Hennebelle} P, {Brucy} N, {Colman} T (2024) {Inefficient star formation in
  high Mach number environments: I. The turbulent support analytical model}.
  \aap 690:A43. \doi{10.1051/0004-6361/202450524},
  {\href{https://arxiv.org/abs/2404.17368}{{arXiv:2404.17368}}} {[astro-ph.GA]}

\bibitem[{{Himes} et~al.(2022){Himes}, {Harrington}, {Cobb},
  {G{\"u}ne{\textcommabelow s} Baydin}, {Soboczenski}, {O'Beirne}, {Zorzan},
  {Wright}, {Scheffer}, {Domagal-Goldman}, and {Arney}}]{HimesEtAl2022}
{Himes} MD, {Harrington} J, {Cobb} AD, et~al (2022) {Accurate Machine-learning
  Atmospheric Retrieval via a Neural-network Surrogate Model for Radiative
  Transfer}. \psj 3(4):91. \doi{10.3847/PSJ/abe3fd},
  {\href{https://arxiv.org/abs/2003.02430}{{arXiv:2003.02430}}} {[astro-ph.IM]}

\bibitem[{{Hirashima} et~al.(2025){Hirashima}, {Moriwaki}, {Fujii}, {Hirai},
  {Saitoh}, {Makino}, {Steinwandel}, and {Ho}}]{HirashimaEtAl2025}
{Hirashima} K, {Moriwaki} K, {Fujii} MS, et~al (2025) {ASURA-FDPS-ML:
  Star-by-star Galaxy Simulations Accelerated by Surrogate Modeling for
  Supernova Feedback}. \apj 987(1):86. \doi{10.3847/1538-4357/add689},
  {\href{https://arxiv.org/abs/2410.23346}{{arXiv:2410.23346}}} {[astro-ph.GA]}

\bibitem[{{Hirose} et~al.(2006){Hirose}, {Krolik}, and
  {Stone}}]{HiroseKrolikStone2006}
{Hirose} S, {Krolik} JH, {Stone} JM (2006) {Vertical Structure of Gas
  Pressure-dominated Accretion Disks with Local Dissipation of Turbulence and
  Radiative Transport}. \apj 640(2):901--917. \doi{10.1086/499153},
  {\href{https://arxiv.org/abs/astro-ph/0510741}{{arXiv:astro-ph/0510741}}}
  {[astro-ph]}

\bibitem[{{Hopkins}(2018)}]{Hopkins2018}
{Hopkins} AM (2018) {The Dawes Review 8: Measuring the Stellar Initial Mass
  Function}. \pasa 35:e039. \doi{10.1017/pasa.2018.29},
  {\href{https://arxiv.org/abs/1807.09949}{{arXiv:1807.09949}}} {[astro-ph.GA]}

\bibitem[{{Hopkins}(2012)}]{Hopkins2012a}
{Hopkins} PF (2012) {An excursion-set model for the structure of giant
  molecular clouds and the interstellar medium}. \mnras 423:2016--2036.
  \doi{10.1111/j.1365-2966.2012.20730.x}

\bibitem[{{Hopkins}(2013{\natexlab{a}})}]{Hopkins2013GIZMO}
{Hopkins} PF (2013{\natexlab{a}}) {A general class of Lagrangian smoothed
  particle hydrodynamics methods and implications for fluid mixing problems}.
  \mnras 428:2840--2856. \doi{10.1093/mnras/sts210},
  {\href{https://arxiv.org/abs/1206.5006}{{arXiv:1206.5006}}} {[astro-ph.IM]}

\bibitem[{{Hopkins}(2013{\natexlab{b}})}]{Hopkins2013IMF}
{Hopkins} PF (2013{\natexlab{b}}) {A general theory of turbulent
  fragmentation}. \mnras 430:1653--1693. \doi{10.1093/mnras/sts704},
  {\href{https://arxiv.org/abs/1210.0903}{{arXiv:1210.0903}}} {[astro-ph.CO]}

\bibitem[{{Hopkins}(2015)}]{Hopkins2015}
{Hopkins} PF (2015) {A new class of accurate, mesh-free hydrodynamic simulation
  methods}. \mnras 450(1):53--110. \doi{10.1093/mnras/stv195},
  {\href{https://arxiv.org/abs/1409.7395}{{arXiv:1409.7395}}} {[astro-ph.CO]}

\bibitem[{{Hopkins}(2026)}]{Hopkins2026}
{Hopkins} PF (2026) {Cosmic Rays on Galaxy Scales: Progress and Pitfalls for
  CR-MHD Dynamical Models}. The Open Journal of Astrophysics 9:61465.
  \doi{10.33232/001c.161465},
  {\href{https://arxiv.org/abs/2509.07104}{{arXiv:2509.07104}}} {[astro-ph.GA]}

\bibitem[{{Hopkins} and {Grudi{\'c}}(2019)}]{HopkinsGrudic2019}
{Hopkins} PF, {Grudi{\'c}} MY (2019) {Numerical problems in coupling photon
  momentum (radiation pressure) to gas}. \mnras 483(3):4187--4196.
  \doi{10.1093/mnras/sty3089},
  {\href{https://arxiv.org/abs/1803.07573}{{arXiv:1803.07573}}} {[astro-ph.GA]}

\bibitem[{{Hopkins} et~al.(2018){Hopkins}, {Wetzel}, {Kere{\v{s}}},
  {Faucher-Gigu{\`e}re}, {Quataert}, {Boylan-Kolchin}, {Murray}, {Hayward}, and
  {El-Badry}}]{HopkinsEtAl2018}
{Hopkins} PF, {Wetzel} A, {Kere{\v{s}}} D, et~al (2018) {How to model
  supernovae in simulations of star and galaxy formation}. \mnras
  477(2):1578--1603. \doi{10.1093/mnras/sty674},
  {\href{https://arxiv.org/abs/1707.07010}{{arXiv:1707.07010}}} {[astro-ph.GA]}

\bibitem[{{Hopkins} et~al.(2021{\natexlab{a}}){Hopkins}, {Chan}, {Squire},
  {Quataert}, {Ji}, {Kere{\v{s}}}, and
  {Faucher-Gigu{\`e}re}}]{HopkinsEtAl2021a}
{Hopkins} PF, {Chan} TK, {Squire} J, et~al (2021{\natexlab{a}}) {Effects of
  different cosmic ray transport models on galaxy formation}. \mnras
  501(3):3663--3669. \doi{10.1093/mnras/staa3692},
  {\href{https://arxiv.org/abs/2004.02897}{{arXiv:2004.02897}}} {[astro-ph.GA]}

\bibitem[{{Hopkins} et~al.(2021{\natexlab{b}}){Hopkins}, {Squire}, {Chan},
  {Quataert}, {Ji}, {Kere{\v{s}}}, and
  {Faucher-Gigu{\`e}re}}]{HopkinsEtAl2021b}
{Hopkins} PF, {Squire} J, {Chan} TK, et~al (2021{\natexlab{b}}) {Testing
  physical models for cosmic ray transport coefficients on galactic scales:
  self-confinement and extrinsic turbulence at {\ensuremath{\sim}}GeV
  energies}. \mnras 501(3):4184--4213. \doi{10.1093/mnras/staa3691},
  {\href{https://arxiv.org/abs/2002.06211}{{arXiv:2002.06211}}} {[astro-ph.HE]}

\bibitem[{{Hopkins} et~al.(2022){Hopkins}, {Butsky}, {Panopoulou}, {Ji},
  {Quataert}, {Faucher-Gigu{\`e}re}, and {Kere{\v{s}}}}]{HopkinsEtAl2022}
{Hopkins} PF, {Butsky} IS, {Panopoulou} GV, et~al (2022) {First predicted
  cosmic ray spectra, primary-to-secondary ratios, and ionization rates from
  MHD galaxy formation simulations}. \mnras 516(3):3470--3514.
  \doi{10.1093/mnras/stac1791},
  {\href{https://arxiv.org/abs/2109.09762}{{arXiv:2109.09762}}} {[astro-ph.HE]}

\bibitem[{{Hopkins} et~al.(2024){Hopkins}, {Grudic}, {Kremer}, {Offner},
  {Guszejnov}, and {Rosen}}]{HopkinsEtAl2024}
{Hopkins} PF, {Grudic} MY, {Kremer} K, et~al (2024) {FORGE'd in FIRE III: The
  IMF in Quasar Accretion Disks from STARFORGE}. The Open Journal of
  Astrophysics 7:71. \doi{10.33232/001c.122857},
  {\href{https://arxiv.org/abs/2404.08046}{{arXiv:2404.08046}}} {[astro-ph.GA]}

\bibitem[{{Hoyle}(1953)}]{Hoyle1953}
{Hoyle} F (1953) {On the Fragmentation of Gas Clouds Into Galaxies and Stars.}
  \apj 118:513. \doi{10.1086/145780}

\bibitem[{{Hu} et~al.(2016){Hu}, {Naab}, {Walch}, {Glover}, and
  {Clark}}]{HuEtAl2016}
{Hu} CY, {Naab} T, {Walch} S, et~al (2016) {Star formation and molecular
  hydrogen in dwarf galaxies: a non-equilibrium view}. \mnras
  458(4):3528--3553. \doi{10.1093/mnras/stw544},
  {\href{https://arxiv.org/abs/1510.05644}{{arXiv:1510.05644}}} {[astro-ph.GA]}

\bibitem[{{Hu} et~al.(2022){Hu}, {Xu}, {Stone}, and {Lazarian}}]{HuEtAl2022}
{Hu} Y, {Xu} S, {Stone} JM, et~al (2022) {Turbulent Magnetic Field
  Amplification by the Interaction of a Shock Wave and Inhomogeneous Medium}.
  \apj 941(2):133. \doi{10.3847/1538-4357/ac9ebc},
  {\href{https://arxiv.org/abs/2207.06941}{{arXiv:2207.06941}}} {[astro-ph.GA]}

\bibitem[{{Hubber} et~al.(2011{\natexlab{a}}){Hubber}, {Batty}, {McLeod},
  {Whitworth}, {Bisbas}, {Stamatellos}, {Walch}, {Rawiraswattana}, and
  {Goodwin}}]{HubberEtAl2011ascl}
{Hubber} D, {Batty} C, {McLeod} A, et~al (2011{\natexlab{a}}) {SEREN: A SPH
  code for star and planet formation simulations}. Astrophysics Source Code
  Library, record ascl:1102.010

\bibitem[{{Hubber} et~al.(2011{\natexlab{b}}){Hubber}, {Batty}, {McLeod}, and
  {Whitworth}}]{HubberEtAl2011}
{Hubber} DA, {Batty} CP, {McLeod} A, et~al (2011{\natexlab{b}}) {SEREN - a new
  SPH code for star and planet formation simulations. Algorithms and tests}.
  \aap 529:A27. \doi{10.1051/0004-6361/201014949},
  {\href{https://arxiv.org/abs/1102.0721}{{arXiv:1102.0721}}} {[astro-ph.SR]}

\bibitem[{{Hubber} et~al.(2013){Hubber}, {Walch}, and
  {Whitworth}}]{HubberWalchWhitworth2013}
{Hubber} DA, {Walch} S, {Whitworth} AP (2013) {An improved sink particle
  algorithm for SPH simulations}. \mnras 430(4):3261--3275.
  \doi{10.1093/mnras/stt128},
  {\href{https://arxiv.org/abs/1301.4520}{{arXiv:1301.4520}}} {[astro-ph.IM]}

\bibitem[{{Ib{\'a}{\~n}ez-Mej{\'\i}a} et~al.(2017){Ib{\'a}{\~n}ez-Mej{\'\i}a},
  {Mac Low}, {Klessen}, and {Baczynski}}]{IbanezMejiaEtAl2017}
{Ib{\'a}{\~n}ez-Mej{\'\i}a} JC, {Mac Low} MM, {Klessen} RS, et~al (2017)
  {Feeding versus Falling: The Growth and Collapse of Molecular Clouds in a
  Turbulent Interstellar Medium}. \apj 850(1):62.
  \doi{10.3847/1538-4357/aa93fe},
  {\href{https://arxiv.org/abs/1705.01779}{{arXiv:1705.01779}}} {[astro-ph.GA]}

\bibitem[{{Iffrig} and {Hennebelle}(2017)}]{IffrigHennebelle2017}
{Iffrig} O, {Hennebelle} P (2017) {Structure distribution and turbulence in
  self-consistently supernova-driven ISM of multiphase magnetized galactic
  discs}. \aap 604:A70. \doi{10.1051/0004-6361/201630290},
  {\href{https://arxiv.org/abs/1703.10421}{{arXiv:1703.10421}}} {[astro-ph.GA]}

\bibitem[{{Ishihara} et~al.(2020){Ishihara}, {Kaneda}, {Morishita}, {Yokokawa},
  and {Uno}}]{IshiharaEtAl2020}
{Ishihara} T, {Kaneda} Y, {Morishita} K, et~al (2020) {Second-order velocity
  structure functions in direct numerical simulations of turbulence with
  R$_{{\ensuremath{\lambda}}}$ up to 2250}. Phys Rev Fluids 5(10):104608.
  \doi{10.1103/PhysRevFluids.5.104608}

\bibitem[{{Jappsen} et~al.(2005){Jappsen}, {Klessen}, {Larson}, {Li}, and {Mac
  Low}}]{JappsenEtAl2005}
{Jappsen} AK, {Klessen} RS, {Larson} RB, et~al (2005) {The stellar mass
  spectrum from non-isothermal gravoturbulent fragmentation}. \aap
  435:611--623. \doi{10.1051/0004-6361:20042178},
  {\href{https://arxiv.org/abs/arXiv:astro-ph/0410351}{{arXiv:astro-ph/0410351}}}

\bibitem[{{Jerabkova} et~al.(2025){Jerabkova}, {Romano}, {Kroupa}, {Andr{\'e}},
  {Chru{\'s}li{\'n}ska}, {Fontanot}, {Hopkins}, {Jadhav}, {Lah{\'e}n}, {Lee},
  {Mucciarelli}, {Salvadori}, {Wang}, {Yan}, {Andersen}, {Durrant}, {Louvet},
  {Lyubenova}, {Matteucci}, {Sharda}, {van de Ven}, and
  {Vazdekis}}]{JerabkovaEtAl2025}
{Jerabkova} T, {Romano} D, {Kroupa} P, et~al (2025) {Cosmic Threads:
  Interlinking the Stellar Initial Mass Function from Star-Birth to Galaxies}.
  arXiv e-prints arXiv:2509.06886. \doi{10.48550/arXiv.2509.06886},
  {\href{https://arxiv.org/abs/2509.06886}{{arXiv:2509.06886}}} {[astro-ph.GA]}

\bibitem[{{Jiang} and {Oh}(2018)}]{JiangOh2018}
{Jiang} YF, {Oh} SP (2018) {A New Numerical Scheme for Cosmic-Ray Transport}.
  \apj 854(1):5. \doi{10.3847/1538-4357/aaa6ce},
  {\href{https://arxiv.org/abs/1712.07117}{{arXiv:1712.07117}}} {[astro-ph.HE]}

\bibitem[{{Jiang} et~al.(2012){Jiang}, {Stone}, and
  {Davis}}]{JiangStoneDavis2012}
{Jiang} YF, {Stone} JM, {Davis} SW (2012) {A Godunov Method for
  Multidimensional Radiation Magnetohydrodynamics Based on a Variable Eddington
  Tensor}. \apjs 199(1):14. \doi{10.1088/0067-0049/199/1/14},
  {\href{https://arxiv.org/abs/1201.2223}{{arXiv:1201.2223}}} {[astro-ph.HE]}

\bibitem[{{Jo} et~al.(2024){Jo}, {Kim}, {Kim}, and {Bryan}}]{JoEtAl2024}
{Jo} Y, {Kim} S, {Kim} Jh, et~al (2024) {Evolution of Star Clusters within
  Galaxies Using Self-consistent Hybrid Hydro/N-body Simulations}. \apj
  974(2):193. \doi{10.3847/1538-4357/ad6b16},
  {\href{https://arxiv.org/abs/2408.03128}{{arXiv:2408.03128}}} {[astro-ph.GA]}

\bibitem[{{Jones} and {Bate}(2018)}]{JonesBate2018}
{Jones} MO, {Bate} MR (2018) {Sink particle radiative feedback in smoothed
  particle hydrodynamics models of star formation}. \mnras 480(2):2562--2577.
  \doi{10.1093/mnras/sty1969},
  {\href{https://arxiv.org/abs/1807.09849}{{arXiv:1807.09849}}} {[astro-ph.SR]}

\bibitem[{{J{\o}rgensen} et~al.(2020){J{\o}rgensen}, {Belloche}, and
  {Garrod}}]{JoergensenBellocheGarrod2020}
{J{\o}rgensen} JK, {Belloche} A, {Garrod} RT (2020) {Astrochemistry During the
  Formation of Stars}. \araa 58:727--778.
  \doi{10.1146/annurev-astro-032620-021927},
  {\href{https://arxiv.org/abs/2006.07071}{{arXiv:2006.07071}}} {[astro-ph.SR]}

\bibitem[{{Joung} and {Mac Low}(2006)}]{JoungMacLow2006}
{Joung} MKR, {Mac Low} MM (2006) {Turbulent Structure of a Stratified
  Supernova-driven Interstellar Medium}. \apj 653(2):1266--1279.
  \doi{10.1086/508795},
  {\href{https://arxiv.org/abs/astro-ph/0601005}{{arXiv:astro-ph/0601005}}}
  {[astro-ph]}

\bibitem[{{Kang} et~al.(2025){Kang}, {Kimm}, {Han}, {Katz}, {Devriendt},
  {Slyz}, and {Teyssier}}]{KangEtAl2025}
{Kang} C, {Kimm} T, {Han} D, et~al (2025) {Impact of star formation models on
  the growth of simulated galaxies at high redshifts}. \aap 693:A149.
  \doi{10.1051/0004-6361/202451502},
  {\href{https://arxiv.org/abs/2407.12090}{{arXiv:2407.12090}}} {[astro-ph.GA]}

\bibitem[{{Kannan} et~al.(2019){Kannan}, {Vogelsberger}, {Marinacci},
  {McKinnon}, {Pakmor}, and {Springel}}]{KannanEtAl2019}
{Kannan} R, {Vogelsberger} M, {Marinacci} F, et~al (2019) {AREPO-RT: radiation
  hydrodynamics on a moving mesh}. \mnras 485(1):117--149.
  \doi{10.1093/mnras/stz287},
  {\href{https://arxiv.org/abs/1804.01987}{{arXiv:1804.01987}}} {[astro-ph.IM]}

\bibitem[{{Kempski} et~al.(2025){Kempski}, {Fielding}, {Quataert}, {Ewart},
  {Grete}, {Kunz}, {Philippov}, and {Stone}}]{KempskiEtAl2025}
{Kempski} P, {Fielding} DB, {Quataert} E, et~al (2025) {Self-similar Cosmic-Ray
  Transport in High-resolution Magnetohydrodynamic Turbulence}. \apjl
  994(2):L49. \doi{10.3847/2041-8213/ae1ca3},
  {\href{https://arxiv.org/abs/2507.10651}{{arXiv:2507.10651}}} {[astro-ph.HE]}

\bibitem[{{Kim} and {Ostriker}(2015)}]{KimOstriker2015}
{Kim} CG, {Ostriker} EC (2015) {Momentum Injection by Supernovae in the
  Interstellar Medium}. \apj 802(2):99. \doi{10.1088/0004-637X/802/2/99},
  {\href{https://arxiv.org/abs/1410.1537}{{arXiv:1410.1537}}} {[astro-ph.GA]}

\bibitem[{{Kim} and {Ostriker}(2017)}]{KimOstriker2017}
{Kim} CG, {Ostriker} EC (2017) {Three-phase Interstellar Medium in Galaxies
  Resolving Evolution with Star Formation and Supernova Feedback (TIGRESS):
  Algorithms, Fiducial Model, and Convergence}. \apj 846(2):133.
  \doi{10.3847/1538-4357/aa8599},
  {\href{https://arxiv.org/abs/1612.03918}{{arXiv:1612.03918}}} {[astro-ph.GA]}

\bibitem[{{Kim} et~al.(2017){Kim}, {Kim}, {Ostriker}, and
  {Skinner}}]{KimEtAl2017}
{Kim} JG, {Kim} WT, {Ostriker} EC, et~al (2017) {Modeling UV Radiation Feedback
  from Massive Stars. I. Implementation of Adaptive Ray-tracing Method and
  Tests}. \apj 851(2):93. \doi{10.3847/1538-4357/aa9b80},
  {\href{https://arxiv.org/abs/1711.06277}{{arXiv:1711.06277}}} {[astro-ph.IM]}

\bibitem[{{Kim} et~al.(2018){Kim}, {Kim}, and {Ostriker}}]{KimKimOstriker2018}
{Kim} JG, {Kim} WT, {Ostriker} EC (2018) {Modeling UV Radiation Feedback from
  Massive Stars. II. Dispersal of Star-forming Giant Molecular Clouds by
  Photoionization and Radiation Pressure}. \apj 859(1):68.
  \doi{10.3847/1538-4357/aabe27},
  {\href{https://arxiv.org/abs/1804.04664}{{arXiv:1804.04664}}} {[astro-ph.GA]}

\bibitem[{{Kim} et~al.(2019){Kim}, {Kim}, and {Ostriker}}]{KimKimOstriker2019}
{Kim} JG, {Kim} WT, {Ostriker} EC (2019) {Modeling UV Radiation Feedback from
  Massive Stars. III. Escape of Radiation from Star-forming Giant Molecular
  Clouds}. \apj 883(1):102. \doi{10.3847/1538-4357/ab3d3d},
  {\href{https://arxiv.org/abs/1908.07549}{{arXiv:1908.07549}}} {[astro-ph.GA]}

\bibitem[{{Kim} et~al.(2021){Kim}, {Ostriker}, and
  {Filippova}}]{KimOstrikerFilippova2021}
{Kim} JG, {Ostriker} EC, {Filippova} N (2021) {Star Formation Efficiency and
  Dispersal of Giant Molecular Clouds with UV Radiation Feedback: Dependence on
  Gravitational Boundedness and Magnetic Fields}. \apj 911(2):128.
  \doi{10.3847/1538-4357/abe934},
  {\href{https://arxiv.org/abs/2011.07772}{{arXiv:2011.07772}}} {[astro-ph.GA]}

\bibitem[{{Kissmann}(2014)}]{Kissmann2014}
{Kissmann} R (2014) {PICARD: A novel code for the Galactic Cosmic Ray
  propagation problem}. Astroparticle Physics 55:37--50.
  \doi{10.1016/j.astropartphys.2014.02.002},
  {\href{https://arxiv.org/abs/1401.4035}{{arXiv:1401.4035}}} {[astro-ph.HE]}

\bibitem[{{Kitsionas} et~al.(2009){Kitsionas}, {Federrath}, {Klessen},
  {Schmidt}, {Price}, {Dursi}, {Gritschneder}, {Walch}, {Piontek}, {Kim},
  {Jappsen}, {Ciecielag}, and {Mac Low}}]{KitsionasEtAl2009}
{Kitsionas} S, {Federrath} C, {Klessen} RS, et~al (2009) {Algorithmic
  comparisons of decaying, isothermal, supersonic turbulence}. \aap
  508:541--560. \doi{10.1051/0004-6361/200811170},
  {\href{https://arxiv.org/abs/0810.4599}{{arXiv:0810.4599}}}

\bibitem[{{Klassen} et~al.(2012){Klassen}, {Peters}, and
  {Pudritz}}]{KlassenPetersPudritz2012}
{Klassen} M, {Peters} T, {Pudritz} RE (2012) {H II Region Variability and
  Pre-main-sequence Evolution}. \apj 758(2):137.
  \doi{10.1088/0004-637X/758/2/137},
  {\href{https://arxiv.org/abs/1208.6001}{{arXiv:1208.6001}}} {[astro-ph.GA]}

\bibitem[{{Klassen} et~al.(2014){Klassen}, {Kuiper}, {Pudritz}, {Peters},
  {Banerjee}, and {Buntemeyer}}]{KlassenEtAl2014}
{Klassen} M, {Kuiper} R, {Pudritz} RE, et~al (2014) {A General Hybrid Radiation
  Transport Scheme for Star Formation Simulations on an Adaptive Grid}. \apj
  797(1):4. \doi{10.1088/0004-637X/797/1/4},
  {\href{https://arxiv.org/abs/1410.4259}{{arXiv:1410.4259}}} {[astro-ph.GA]}

\bibitem[{{Klepitko} et~al.(2023){Klepitko}, {Walch}, {W{\"u}nsch}, {Seifried},
  {Dinnbier}, and {Haid}}]{KlepitkoEtAl2023}
{Klepitko} A, {Walch} S, {W{\"u}nsch} R, et~al (2023) {Tree-based solvers for
  adaptive mesh refinement code FLASH - III: a novel scheme for radiation
  pressure on dust and gas and radiative transfer from diffuse sources}. \mnras
  521(1):160--184. \doi{10.1093/mnras/stad385},
  {\href{https://arxiv.org/abs/2204.09072}{{arXiv:2204.09072}}} {[astro-ph.IM]}

\bibitem[{{Klessen}(2001)}]{Klessen2001}
{Klessen} RS (2001) {The Formation of Stellar Clusters: Mass Spectra from
  Turbulent Molecular Cloud Fragmentation}. \apj 556:837--846.
  \doi{10.1086/321626},
  {\href{https://arxiv.org/abs/arXiv:astro-ph/0104127}{{arXiv:astro-ph/0104127}}}

\bibitem[{{Klessen} et~al.(2000){Klessen}, {Heitsch}, and {Mac
  Low}}]{KlessenHeitschMacLow2000}
{Klessen} RS, {Heitsch} F, {Mac Low} MM (2000) {Gravitational Collapse in
  Turbulent Molecular Clouds. I. Gasdynamical Turbulence}. \apj 535:887--906.
  \doi{10.1086/308891},
  {\href{https://arxiv.org/abs/arXiv:astro-ph/9911068}{{arXiv:astro-ph/9911068}}}

\bibitem[{{Koch} et~al.(2017){Koch}, {Ward}, {Offner}, {Loeppky}, and
  {Rosolowsky}}]{KochEtAl2017}
{Koch} EW, {Ward} CG, {Offner} S, et~al (2017) {Identifying tools for comparing
  simulations and observations of spectral-line data cubes}. \mnras
  471(2):1506--1530. \doi{10.1093/mnras/stx1671},
  {\href{https://arxiv.org/abs/1707.05415}{{arXiv:1707.05415}}} {[astro-ph.GA]}

\bibitem[{{Koch} et~al.(2019){Koch}, {Rosolowsky}, {Boyden}, {Burkhart},
  {Ginsburg}, {Loeppky}, and {Offner}}]{KochEtAl2019}
{Koch} EW, {Rosolowsky} EW, {Boyden} RD, et~al (2019) {TURBUSTAT: Turbulence
  Statistics in Python}. \aj 158(1):1. \doi{10.3847/1538-3881/ab1cc0},
  {\href{https://arxiv.org/abs/1904.10484}{{arXiv:1904.10484}}} {[astro-ph.IM]}

\bibitem[{{Kolmogorov}(1941)}]{Kolmogorov1941c}
{Kolmogorov} AN (1941) {Dissipation of energy in locally isotropic turbulence}.
  Dokl Akad Nauk SSSR 32:16--18

\bibitem[{{K\"{o}nigl} and {Pudritz}(2000)}]{KoeniglPudritz2000}
{K\"{o}nigl} A, {Pudritz} RE (2000) {Disk Winds and the Accretion-Outflow
  Connection}. Protostars and Planets IV p 759.
  {\href{https://arxiv.org/abs/astro-ph/9903168}{{astro-ph/9903168}}}

\bibitem[{{Koo} and {McKee}(1992)}]{KooMcKee1992}
{Koo} BC, {McKee} CF (1992) {Dynamics of Wind Bubbles and Superbubbles. II.
  Analytic Theory}. \apj 388:103. \doi{10.1086/171133}

\bibitem[{{Korpi} et~al.(1999){Korpi}, {Brandenburg}, {Shukurov}, {Tuominen},
  and {Nordlund}}]{KorpiEtAl1999}
{Korpi} MJ, {Brandenburg} A, {Shukurov} A, et~al (1999) {A Supernova-regulated
  Interstellar Medium: Simulations of the Turbulent Multiphase Medium}. \apjl
  514(2):L99--L102. \doi{10.1086/311954}

\bibitem[{{Koyama} and {Inutsuka}(2002)}]{KoyamaInutsuka2002}
{Koyama} H, {Inutsuka} Si (2002) {An Origin of Supersonic Motions in
  Interstellar Clouds}. \apjl 564(2):L97--L100. \doi{10.1086/338978},
  {\href{https://arxiv.org/abs/astro-ph/0112420}{{arXiv:astro-ph/0112420}}}
  {[astro-ph]}

\bibitem[{{Kravtsov} et~al.(1997){Kravtsov}, {Klypin}, and
  {Khokhlov}}]{KravtsovKlypinKhokhlov1997}
{Kravtsov} AV, {Klypin} AA, {Khokhlov} AM (1997) {Adaptive Refinement Tree: A
  New High-Resolution N-Body Code for Cosmological Simulations}. \apjs
  111(1):73--94. \doi{10.1086/313015},
  {\href{https://arxiv.org/abs/astro-ph/9701195}{{arXiv:astro-ph/9701195}}}
  {[astro-ph]}

\bibitem[{{Kravtsov} et~al.(2002){Kravtsov}, {Klypin}, and
  {Hoffman}}]{KravtsovEtAl2002}
{Kravtsov} AV, {Klypin} A, {Hoffman} Y (2002) {Constrained Simulations of the
  Real Universe. II. Observational Signatures of Intergalactic Gas in the Local
  Supercluster Region}. \apj 571(2):563--575. \doi{10.1086/340046},
  {\href{https://arxiv.org/abs/astro-ph/0109077}{{arXiv:astro-ph/0109077}}}
  {[astro-ph]}

\bibitem[{{Kriel} et~al.(2022){Kriel}, {Beattie}, {Seta}, and
  {Federrath}}]{KrielEtAl2022}
{Kriel} N, {Beattie} JR, {Seta} A, et~al (2022) {Fundamental scales in the
  kinematic phase of the turbulent dynamo}. \mnras 513(2):2457--2470.
  \doi{10.1093/mnras/stac969},
  {\href{https://arxiv.org/abs/2204.00828}{{arXiv:2204.00828}}} {[astro-ph.SR]}

\bibitem[{{Kriel} et~al.(2025){Kriel}, {Beattie}, {Federrath}, {Krumholz}, and
  {Hew}}]{KrielEtAl2025}
{Kriel} N, {Beattie} JR, {Federrath} C, et~al (2025) {Fundamental MHD scales -
  II. The kinematic phase of the supersonic small-scale dynamo}. \mnras
  537(3):2602--2629. \doi{10.1093/mnras/staf188},
  {\href{https://arxiv.org/abs/2310.17036}{{arXiv:2310.17036}}} {[astro-ph.GA]}

\bibitem[{{Kritsuk} et~al.(2011){Kritsuk}, {Nordlund}, {Collins}, {Padoan},
  {Norman}, {Abel}, {Banerjee}, {Federrath}, {Flock}, {Lee}, {Li},
  {M{\"u}ller}, {Teyssier}, {Ustyugov}, {Vogel}, and
  {Xu}}]{KritsukEtAl2011Codes}
{Kritsuk} AG, {Nordlund} {\AA}, {Collins} D, et~al (2011) {Comparing Numerical
  Methods for Isothermal Magnetized Supersonic Turbulence}. \apj 737:13.
  \doi{10.1088/0004-637X/737/1/13},
  {\href{https://arxiv.org/abs/1103.5525}{{arXiv:1103.5525}}} {[astro-ph.SR]}

\bibitem[{{Kroupa}(2001)}]{Kroupa2001}
{Kroupa} P (2001) {On the variation of the initial mass function}. \mnras
  322:231--246. \doi{10.1046/j.1365-8711.2001.04022.x},
  {\href{https://arxiv.org/abs/arXiv:astro-ph/0009005}{{arXiv:astro-ph/0009005}}}

\bibitem[{{Kroupa} et~al.(2013){Kroupa}, {Weidner}, {Pflamm-Altenburg},
  {Thies}, {Dabringhausen}, {Marks}, and {Maschberger}}]{KroupaEtAl2013}
{Kroupa} P, {Weidner} C, {Pflamm-Altenburg} J, et~al (2013) {`The Stellar and
  Sub-Stellar Initial Mass Function of Simple and Composite Populations' in
  Planets, Stars and Stellar Systems.~Volume 5: Galactic Structure and Stellar
  Populations}. In: {Oswalt} TD, {Gilmore} G (eds) Planets, Stars and Stellar
  Systems.~Volume 5: Galactic Structure and Stellar Populations. Springer, p
  115, \doi{10.1007/978-94-007-5612-0_4}

\bibitem[{{Kroupa} et~al.(2026){Kroupa}, {Gjergo}, {Jerabkova}, and
  {Yan}}]{KroupaEtAl2026}
{Kroupa} P, {Gjergo} E, {Jerabkova} T, et~al (2026) {The initial mass function
  of stars}. In: Encyclopedia of Astrophysics, pp 173--210,
  \doi{10.1016/B978-0-443-21439-4.00035-3}

\bibitem[{{Krumholz} and {McKee}(2005)}]{KrumholzMcKee2005}
{Krumholz} MR, {McKee} CF (2005) {A General Theory of Turbulence-regulated Star
  Formation, from Spirals to Ultraluminous Infrared Galaxies}. \apj
  630:250--268. \doi{10.1086/431734},
  {\href{https://arxiv.org/abs/arXiv:astro-ph/0505177}{{arXiv:astro-ph/0505177}}}

\bibitem[{{Krumholz} et~al.(2004){Krumholz}, {McKee}, and
  {Klein}}]{KrumholzMcKeeKlein2004}
{Krumholz} MR, {McKee} CF, {Klein} RI (2004) {Embedding Lagrangian Sink
  Particles in Eulerian Grids}. \apj 611:399--412. \doi{10.1086/421935},
  {\href{https://arxiv.org/abs/arXiv:astro-ph/0312612}{{arXiv:astro-ph/0312612}}}

\bibitem[{{Krumholz} et~al.(2007{\natexlab{a}}){Krumholz}, {Klein}, and
  {McKee}}]{KrumholzKleinMcKee2007}
{Krumholz} MR, {Klein} RI, {McKee} CF (2007{\natexlab{a}})
  {Radiation-Hydrodynamic Simulations of Collapse and Fragmentation in Massive
  Protostellar Cores}. \apj 656:959--979. \doi{10.1086/510664},
  {\href{https://arxiv.org/abs/arXiv:astro-ph/0609798}{{arXiv:astro-ph/0609798}}}

\bibitem[{{Krumholz} et~al.(2007{\natexlab{b}}){Krumholz}, {Klein}, {McKee},
  and {Bolstad}}]{KrumholzEtAl2007}
{Krumholz} MR, {Klein} RI, {McKee} CF, et~al (2007{\natexlab{b}}) {Equations
  and Algorithms for Mixed-frame Flux-limited Diffusion Radiation
  Hydrodynamics}. \apj 667(1):626--643. \doi{10.1086/520791},
  {\href{https://arxiv.org/abs/astro-ph/0611003}{{arXiv:astro-ph/0611003}}}
  {[astro-ph]}

\bibitem[{{Krumholz} et~al.(2009){Krumholz}, {Klein}, {McKee}, {Offner}, and
  {Cunningham}}]{KrumholzEtAl2009}
{Krumholz} MR, {Klein} RI, {McKee} CF, et~al (2009) {The Formation of Massive
  Star Systems by Accretion}. Science 323(5915):754.
  \doi{10.1126/science.1165857},
  {\href{https://arxiv.org/abs/0901.3157}{{arXiv:0901.3157}}} {[astro-ph.SR]}

\bibitem[{{Krumholz} et~al.(2012){Krumholz}, {Klein}, and
  {McKee}}]{KrumholzKleinMcKee2012}
{Krumholz} MR, {Klein} RI, {McKee} CF (2012) {Radiation-hydrodynamic
  Simulations of the Formation of Orion-like Star Clusters. II. The Initial
  Mass Function from Winds, Turbulence, and Radiation}. \apj 754:71.
  \doi{10.1088/0004-637X/754/1/71},
  {\href{https://arxiv.org/abs/1203.2620}{{arXiv:1203.2620}}} {[astro-ph.SR]}

\bibitem[{{Krumholz} et~al.(2016){Krumholz}, {Myers}, {Klein}, and
  {McKee}}]{KrumholzEtAl2016}
{Krumholz} MR, {Myers} AT, {Klein} RI, et~al (2016) {What physics determines
  the peak of the IMF? Insights from the structure of cores in
  radiation-magnetohydrodynamic simulations}. \mnras 460:3272--3283.
  \doi{10.1093/mnras/stw1236},
  {\href{https://arxiv.org/abs/1603.04557}{{arXiv:1603.04557}}}

\bibitem[{{Krumholz} et~al.(2022){Krumholz}, {Crocker}, and
  {Sampson}}]{KrumholzCrockerSampson2022}
{Krumholz} MR, {Crocker} RM, {Sampson} ML (2022) {Cosmic ray interstellar
  propagation tool using It{\^o} Calculus (CRIPTIC): software for simultaneous
  calculation of cosmic ray transport and observational signatures}. \mnras
  517(1):1355--1380. \doi{10.1093/mnras/stac2712},
  {\href{https://arxiv.org/abs/2207.13838}{{arXiv:2207.13838}}} {[astro-ph.HE]}

\bibitem[{{Krumholz} et~al.(2023){Krumholz}, {Crocker}, and
  {Offner}}]{KrumholzCrockerOffner2023}
{Krumholz} MR, {Crocker} RM, {Offner} SSR (2023) {The cosmic ray ionization and
  {\ensuremath{\gamma}}-ray budgets of star-forming galaxies}. \mnras
  520(4):5126--5143. \doi{10.1093/mnras/stad459},
  {\href{https://arxiv.org/abs/2211.03488}{{arXiv:2211.03488}}} {[astro-ph.GA]}

\bibitem[{{Kudela} and {Matousek}(2023)}]{KudelaMatousek2023}
{Kudela} J, {Matousek} R (2023) {Combining Lipschitz and RBF surrogate models
  for high-dimensional computationally expensive problems}. Information
  Sciences 619:457--477. \doi{https://doi.org/10.1016/j.ins.2022.11.045},
  \urlprefix\url{https://www.sciencedirect.com/science/article/pii/S0020025522013342}

\bibitem[{{Kuffmeier} et~al.(2017){Kuffmeier}, {Haugb{\o}lle}, and
  {Nordlund}}]{KuffmeierHaugbolleNordlund2017}
{Kuffmeier} M, {Haugb{\o}lle} T, {Nordlund} {\r{A}} (2017) {Zoom-in Simulations
  of Protoplanetary Disks Starting from GMC Scales}. \apj 846(1):7.
  \doi{10.3847/1538-4357/aa7c64},
  {\href{https://arxiv.org/abs/1611.10360}{{arXiv:1611.10360}}} {[astro-ph.SR]}

\bibitem[{{Kuiper} and {Hosokawa}(2018)}]{KuiperHosokawa2018}
{Kuiper} R, {Hosokawa} T (2018) {First hydrodynamics simulations of radiation
  forces and photoionization feedback in massive star formation}. \aap
  616:A101. \doi{10.1051/0004-6361/201832638},
  {\href{https://arxiv.org/abs/1804.10211}{{arXiv:1804.10211}}} {[astro-ph.GA]}

\bibitem[{{Kuiper} and {Klessen}(2013)}]{KuiperKlessen2013}
{Kuiper} R, {Klessen} RS (2013) {The reliability of approximate radiation
  transport methods for irradiated disk studies}. \aap 555:A7.
  \doi{10.1051/0004-6361/201321404},
  {\href{https://arxiv.org/abs/1305.2197}{{arXiv:1305.2197}}} {[astro-ph.SR]}

\bibitem[{{Kuiper} et~al.(2010){Kuiper}, {Klahr}, {Dullemond}, {Kley}, and
  {Henning}}]{KuiperEtAl2010RT}
{Kuiper} R, {Klahr} H, {Dullemond} C, et~al (2010) {Fast and accurate
  frequency-dependent radiation transport for hydrodynamics simulations in
  massive star formation}. \aap 511:A81. \doi{10.1051/0004-6361/200912355},
  {\href{https://arxiv.org/abs/1001.3301}{{arXiv:1001.3301}}} {[astro-ph.SR]}

\bibitem[{{Kurth} et~al.(2022){Kurth}, {Subramanian}, {Harrington}, {Pathak},
  {Mardani}, {Hall}, {Miele}, {Kashinath}, and {Anandkumar}}]{KurthEtAl2022}
{Kurth} T, {Subramanian} S, {Harrington} P, et~al (2022) {FourCastNet:
  Accelerating Global High-Resolution Weather Forecasting using Adaptive
  Fourier Neural Operators}. arXiv e-prints arXiv:2208.05419.
  \doi{10.48550/arXiv.2208.05419},
  {\href{https://arxiv.org/abs/2208.05419}{{arXiv:2208.05419}}}
  {[physics.ao-ph]}

\bibitem[{{Lancaster} et~al.(2021{\natexlab{a}}){Lancaster}, {Ostriker}, {Kim},
  and {Kim}}]{LancasterEtAl2021a}
{Lancaster} L, {Ostriker} EC, {Kim} JG, et~al (2021{\natexlab{a}}) {Efficiently
  Cooled Stellar Wind Bubbles in Turbulent Clouds. I. Fractal Theory and
  Application to Star-forming Clouds}. \apj 914(2):89.
  \doi{10.3847/1538-4357/abf8ab},
  {\href{https://arxiv.org/abs/2104.07691}{{arXiv:2104.07691}}} {[astro-ph.GA]}

\bibitem[{{Lancaster} et~al.(2021{\natexlab{b}}){Lancaster}, {Ostriker}, {Kim},
  and {Kim}}]{LancasterEtAl2021b}
{Lancaster} L, {Ostriker} EC, {Kim} JG, et~al (2021{\natexlab{b}}) {Efficiently
  Cooled Stellar Wind Bubbles in Turbulent Clouds. II. Validation of Theory
  with Hydrodynamic Simulations}. \apj 914(2):90.
  \doi{10.3847/1538-4357/abf8ac},
  {\href{https://arxiv.org/abs/2104.07722}{{arXiv:2104.07722}}} {[astro-ph.GA]}

\bibitem[{{Lane} et~al.(2022){Lane}, {Grudi{\'c}}, {Guszejnov}, {Offner},
  {Faucher-Gigu{\`e}re}, and {Rosen}}]{LaneEtAl2022}
{Lane} HB, {Grudi{\'c}} MY, {Guszejnov} D, et~al (2022) {Less wrong: a more
  realistic initial condition for simulations of turbulent molecular clouds}.
  \mnras 510(4):4767--4778. \doi{10.1093/mnras/stab3739},
  {\href{https://arxiv.org/abs/2110.14816}{{arXiv:2110.14816}}} {[astro-ph.GA]}

\bibitem[{{Lazarian} and {Xu}(2021)}]{LazarianXu2021}
{Lazarian} A, {Xu} S (2021) {Diffusion of Cosmic Rays in MHD Turbulence with
  Magnetic Mirrors}. \apj 923(1):53. \doi{10.3847/1538-4357/ac2de9},
  {\href{https://arxiv.org/abs/2106.08362}{{arXiv:2106.08362}}} {[astro-ph.HE]}

\bibitem[{{Lazarian} and {Xu}(2022)}]{LazarianXu2022}
{Lazarian} A, {Xu} S (2022) {Damping of Alfv{\'e}n Waves in MHD Turbulence and
  Implications for Cosmic Ray Streaming Instability and Galactic Winds}.
  Frontiers in Physics 10:702799. \doi{10.3389/fphy.2022.702799},
  {\href{https://arxiv.org/abs/2201.05168}{{arXiv:2201.05168}}} {[astro-ph.GA]}

\bibitem[{{Lazarian} et~al.(2020){Lazarian}, {Eyink}, {Jafari}, {Kowal}, {Li},
  {Xu}, and {Vishniac}}]{LazarianEtAl2020}
{Lazarian} A, {Eyink} GL, {Jafari} A, et~al (2020) {3D turbulent reconnection:
  Theory, tests, and astrophysical implications}. Phys Plasmas 27(1):012305.
  \doi{10.1063/1.5110603},
  {\href{https://arxiv.org/abs/2001.00868}{{arXiv:2001.00868}}} {[astro-ph.HE]}

\bibitem[{{Lebreuilly} et~al.(2021){Lebreuilly}, {Hennebelle}, {Colman},
  {Commer{\c{c}}on}, {Klessen}, {Maury}, {Molinari}, and
  {Testi}}]{LebreuillyEtAl2021}
{Lebreuilly} U, {Hennebelle} P, {Colman} T, et~al (2021) {Protoplanetary Disk
  Birth in Massive Star-forming Clumps: The Essential Role of the Magnetic
  Field}. \apjl 917(1):L10. \doi{10.3847/2041-8213/ac158c},
  {\href{https://arxiv.org/abs/2107.08491}{{arXiv:2107.08491}}} {[astro-ph.SR]}

\bibitem[{{Lebreuilly} et~al.(2024){Lebreuilly}, {Hennebelle}, {Maury},
  {Gonz{\'a}lez}, {Traficante}, {Klessen}, {Testi}, and
  {Molinari}}]{LebreuillyEtAl2024}
{Lebreuilly} U, {Hennebelle} P, {Maury} A, et~al (2024) {Influence of
  protostellar outflows on star and protoplanetary disk formation in a massive
  star-forming clump}. \aap 683:A13. \doi{10.1051/0004-6361/202347913},
  {\href{https://arxiv.org/abs/2309.05397}{{arXiv:2309.05397}}} {[astro-ph.SR]}

\bibitem[{{Lee} et~al.(2009){Lee}, {Deane}, and
  {Federrath}}]{LeeDeaneFederrath2009}
{Lee} D, {Deane} AE, {Federrath} C (2009) {A New Multidimensional Unsplit MHD
  Solver in FLASH3}. In: {N.~V.~Pogorelov, E.~Audit, P.~Colella, \& G.~P.~Zank}
  (ed) Numerical Modeling of Space Plasma Flows: ASTRONUM-2008, p 243

\bibitem[{{Lee} and {Hennebelle}(2018)}]{LeeHennebelle2018}
{Lee} YN, {Hennebelle} P (2018) {Stellar mass spectrum within massive
  collapsing clumps. II. Thermodynamics and tidal forces of the first Larson
  core. A robust mechanism for the peak of the IMF}. \aap 611:A89.
  \doi{10.1051/0004-6361/201731523},
  {\href{https://arxiv.org/abs/1711.00319}{{arXiv:1711.00319}}} {[astro-ph.GA]}

\bibitem[{{Lee} et~al.(2020){Lee}, {Offner}, {Hennebelle}, {Andr{\'e}},
  {Zinnecker}, {Ballesteros-Paredes}, {Inutsuka}, and
  {Kruijssen}}]{LeeEtAl2020}
{Lee} YN, {Offner} SSR, {Hennebelle} P, et~al (2020) {The Origin of the Stellar
  Mass Distribution and Multiplicity}. \ssr 216(4):70.
  \doi{10.1007/s11214-020-00699-2},
  {\href{https://arxiv.org/abs/2006.05778}{{arXiv:2006.05778}}} {[astro-ph.GA]}

\bibitem[{{Leidi} et~al.(2022){Leidi}, {Birke}, {Andrassy}, {Higl}, {Edelmann},
  {Wiest}, {Klingenberg}, and {R{\"o}pke}}]{LeidiEtAl2022}
{Leidi} G, {Birke} C, {Andrassy} R, et~al (2022) {A finite-volume scheme for
  modeling compressible magnetohydrodynamic flows at low Mach numbers in
  stellar interiors}. \aap 668:A143. \doi{10.1051/0004-6361/202244665},
  {\href{https://arxiv.org/abs/2210.01641}{{arXiv:2210.01641}}} {[astro-ph.SR]}

\bibitem[{{Lesur} et~al.(2023){Lesur}, {Baghdadi}, {Wafflard-Fernandez},
  {Mauxion}, {Robert}, and {Van den Bossche}}]{LesurEtAl2023}
{Lesur} GRJ, {Baghdadi} S, {Wafflard-Fernandez} G, et~al (2023) {IDEFIX: A
  versatile performance-portable Godunov code for astrophysical flows}. \aap
  677:A9. \doi{10.1051/0004-6361/202346005},
  {\href{https://arxiv.org/abs/2304.13746}{{arXiv:2304.13746}}} {[astro-ph.IM]}

\bibitem[{{LeVeque}(2002)}]{LeVeque2002}
{LeVeque} RJ (2002) Finite Volume Methods for Hyperbolic Problems. Cambridge
  University Press

\bibitem[{{Levermore}(1984)}]{Levermore1984}
{Levermore} CD (1984) {Relating Eddington factors to flux limiters.} \jqsrt
  31(2):149--160. \doi{10.1016/0022-4073(84)90112-2}

\bibitem[{{Levermore} and {Pomraning}(1981)}]{LevermorePomraning1981}
{Levermore} CD, {Pomraning} GC (1981) {A flux-limited diffusion theory}. \apj
  248:321--334. \doi{10.1086/159157}

\bibitem[{{Li} et~al.(2021){Li}, {Cunningham}, {Gaches}, {Klein}, {Krumholz},
  {Lee}, {McKee}, {Offner}, {Rosen}, and {Skinner}}]{LiEtAl2021}
{Li} P, {Cunningham} A, {Gaches} B, et~al (2021) {ORION2: A
  magnetohydrodynamics code for star formation}. J Open Source Softw
  6(68):3771. \doi{10.21105/joss.03771}

\bibitem[{{Li} et~al.(2008){Li}, {McKee}, {Klein}, and {Fisher}}]{LiEtAl2008}
{Li} PS, {McKee} CF, {Klein} RI, et~al (2008) {Sub-Alfv{\'e}nic Nonideal MHD
  Turbulence Simulations with Ambipolar Diffusion. I. Turbulence Statistics}.
  \apj 684:380--394. \doi{10.1086/589874},
  {\href{https://arxiv.org/abs/0805.0597}{{arXiv:0805.0597}}}

\bibitem[{{Li} et~al.(2012){Li}, {Myers}, and {McKee}}]{LiMyersMcKee2012}
{Li} PS, {Myers} A, {McKee} CF (2012) {Ambipolar Diffusion Heating in Turbulent
  Systems}. \apj 760(1):33. \doi{10.1088/0004-637X/760/1/33},
  {\href{https://arxiv.org/abs/1210.0700}{{arXiv:1210.0700}}} {[astro-ph.SR]}

\bibitem[{{Li} et~al.(2003){Li}, {Klessen}, and {Mac
  Low}}]{LiKlessenMacLow2003}
{Li} Y, {Klessen} RS, {Mac Low} MM (2003) {The Formation of Stellar Clusters in
  Turbulent Molecular Clouds: Effects of the Equation of State}. \apj
  592:975--985. \doi{10.1086/375780},
  {\href{https://arxiv.org/abs/arXiv:astro-ph/0210479}{{arXiv:astro-ph/0210479}}}

\bibitem[{{Li} and {Nakamura}(2006)}]{LiNakamura2006}
{Li} ZY, {Nakamura} F (2006) {Cluster Formation in Protostellar Outflow-driven
  Turbulence}. \apjl 640:L187--L190. \doi{10.1086/503419},
  {\href{https://arxiv.org/abs/astro-ph/0512278}{{astro-ph/0512278}}}

\bibitem[{{Lucy}(1977)}]{Lucy1977}
{Lucy} LB (1977) {A numerical approach to the testing of the fission
  hypothesis}. \aj 82:1013--1024. \doi{10.1086/112164}

\bibitem[{{Lucy}(1999)}]{Lucy1999}
{Lucy} LB (1999) {Computing radiative equilibria with Monte Carlo techniques}.
  \aap 344:282--288

\bibitem[{{Mac Low}(1999)}]{MacLow1999}
{Mac Low} MM (1999) {The Energy Dissipation Rate of Supersonic,
  Magnetohydrodynamic Turbulence in Molecular Clouds}. \apj 524:169--178.
  \doi{10.1086/307784},
  {\href{https://arxiv.org/abs/arXiv:astro-ph/9809177}{{arXiv:astro-ph/9809177}}}

\bibitem[{{Mac Low} and {Klessen}(2004)}]{MacLowKlessen2004}
{Mac Low} MM, {Klessen} RS (2004) {Control of star formation by supersonic
  turbulence}. \rmp 76(1):125--194. \doi{10.1103/RevModPhys.76.125},
  {\href{https://arxiv.org/abs/astro-ph/0301093}{{arXiv:astro-ph/0301093}}}
  {[astro-ph]}

\bibitem[{{Mac Low} et~al.(1998){Mac Low}, {Klessen}, {Burkert}, and
  {Smith}}]{MacLowEtAl1998}
{Mac Low} MM, {Klessen} RS, {Burkert} A, et~al (1998) {Kinetic Energy Decay
  Rates of Supersonic and Super-Alfv{\'e}nic Turbulence in Star-Forming
  Clouds}. \prl 80:2754--2757.
  {\href{https://arxiv.org/abs/arXiv:astro-ph/9712013}{{arXiv:astro-ph/9712013}}}

\bibitem[{{Machida} and {Basu}(2019)}]{MachidaBasu2019}
{Machida} MN, {Basu} S (2019) {The First Two Thousand Years of Star Formation}.
  \apj 876(2):149. \doi{10.3847/1538-4357/ab18a7},
  {\href{https://arxiv.org/abs/1904.04424}{{arXiv:1904.04424}}} {[astro-ph.SR]}

\bibitem[{{Machida} et~al.(2008){Machida}, {Matsumoto}, and
  {Inutsuka}}]{MachidaEtAl2008}
{Machida} MN, {Matsumoto} T, {Inutsuka} S (2008) {Magnetohydrodynamics of
  Population III Star Formation}. \apj 685:690--704. \doi{10.1086/591074},
  {\href{https://arxiv.org/abs/0803.1224}{{arXiv:0803.1224}}}

\bibitem[{{Mandal} et~al.(2020){Mandal}, {Federrath}, and
  {K{\"o}rtgen}}]{MandalFederrathKoertgen2020}
{Mandal} A, {Federrath} C, {K{\"o}rtgen} B (2020) {Molecular cloud formation by
  compression of magnetized turbulent gas subjected to radiative cooling}.
  \mnras 493(3):3098--3113. \doi{10.1093/mnras/staa468},
  {\href{https://arxiv.org/abs/1910.13762}{{arXiv:1910.13762}}} {[astro-ph.GA]}

\bibitem[{{Mathew} and {Federrath}(2020)}]{MathewFederrath2020}
{Mathew} SS, {Federrath} C (2020) {Implementation of stellar heating feedback
  in simulations of star cluster formation: effects on the initial mass
  function}. \mnras 496(4):5201--5210. \doi{10.1093/mnras/staa1931},
  {\href{https://arxiv.org/abs/2007.01875}{{arXiv:2007.01875}}} {[astro-ph.GA]}

\bibitem[{{Mathew} and {Federrath}(2021)}]{MathewFederrath2021}
{Mathew} SS, {Federrath} C (2021) {The IMF and multiplicity of stars from
  gravity, turbulence, magnetic fields, radiation, and outflow feedback}.
  \mnras 507(2):2448--2467. \doi{10.1093/mnras/stab2338},
  {\href{https://arxiv.org/abs/2106.06521}{{arXiv:2106.06521}}} {[astro-ph.GA]}

\bibitem[{{Mathew} et~al.(2023){Mathew}, {Federrath}, and
  {Seta}}]{MathewFederrathSeta2023}
{Mathew} SS, {Federrath} C, {Seta} A (2023) {The role of the turbulence driving
  mode for the initial mass function}. \mnras 518(4):5190--5214.
  \doi{10.1093/mnras/stac3415},
  {\href{https://arxiv.org/abs/2208.08802}{{arXiv:2208.08802}}} {[astro-ph.GA]}

\bibitem[{{Mathew} et~al.(2025){Mathew}, {Federrath}, and
  {Seta}}]{MathewFederrathSeta2025}
{Mathew} SS, {Federrath} C, {Seta} A (2025) {The influence of the cloud virial
  parameter on the initial mass function}. \mnras 536(2):1932--1947.
  \doi{10.1093/mnras/stae2692},
  {\href{https://arxiv.org/abs/2410.13137}{{arXiv:2410.13137}}} {[astro-ph.GA]}

\bibitem[{{Matzner} and {McKee}(2000)}]{MatznerMcKee2000}
{Matzner} CD, {McKee} CF (2000) {Efficiencies of Low-Mass Star and Star Cluster
  Formation}. \apj 545:364--378. \doi{10.1086/317785},
  {\href{https://arxiv.org/abs/arXiv:astro-ph/0007383}{{arXiv:astro-ph/0007383}}}

\bibitem[{{Mauxion} et~al.(2024){Mauxion}, {Lesur}, and
  {Maret}}]{MauxionLesurMaret2024}
{Mauxion} J, {Lesur} G, {Maret} S (2024) {Modeling the secular evolution of
  embedded protoplanetary disks}. \aap 686:A253.
  \doi{10.1051/0004-6361/202348405},
  {\href{https://arxiv.org/abs/2403.16753}{{arXiv:2403.16753}}} {[astro-ph.SR]}

\bibitem[{{Mayer} et~al.(2025){Mayer}, {Naab}, {Caselli}, {Ivlev}, {Grassi},
  {Zier}, {Pakmor}, {Walch}, and {Springel}}]{MayerEtAl2025}
{Mayer} AC, {Naab} T, {Caselli} P, et~al (2025) {Protostellar discs in their
  natural habitat -- the formation of protostars and their accretion discs in
  the turbulent and magnetized interstellar medium}. \mnras 543(4):3321--3344.
  \doi{10.1093/mnras/staf1404},
  {\href{https://arxiv.org/abs/2506.14394}{{arXiv:2506.14394}}} {[astro-ph.GA]}

\bibitem[{{Mayer} and {Duschl}(2005)}]{MayerDuschl2005}
{Mayer} M, {Duschl} WJ (2005) {Rosseland and Planck mean opacities for
  primordial matter}. \mnras 358(2):614--631.
  \doi{10.1111/j.1365-2966.2005.08826.x},
  {\href{https://arxiv.org/abs/astro-ph/0411613}{{arXiv:astro-ph/0411613}}}
  {[astro-ph]}

\bibitem[{{McKee} and {Ostriker}(2007)}]{McKeeOstriker2007}
{McKee} CF, {Ostriker} EC (2007) {Theory of Star Formation}. \araa 45:565--687.
  \doi{10.1146/annurev.astro.45.051806.110602},
  {\href{https://arxiv.org/abs/0707.3514}{{arXiv:0707.3514}}}

\bibitem[{{McKee} and {Ostriker}(1977)}]{McKeeOstriker1977}
{McKee} CF, {Ostriker} JP (1977) {A theory of the interstellar medium: three
  components regulated by supernova explosions in an inhomogeneous substrate.}
  \apj 218:148--169. \doi{10.1086/155667}

\bibitem[{{Mee} and {Brandenburg}(2006)}]{MeeBrandenburg2006}
{Mee} AJ, {Brandenburg} A (2006) {Turbulence from localized random expansion
  waves}. \mnras 370:415--419. \doi{10.1111/j.1365-2966.2006.10476.x}

\bibitem[{{Melon Fuksman} and {Mignone}(2019)}]{MelonFuksmanMignone2019}
{Melon Fuksman} JD, {Mignone} A (2019) {A Radiative Transfer Module for
  Relativistic Magnetohydrodynamics in the PLUTO Code}. \apjs 242(2):20.
  \doi{10.3847/1538-4365/ab18ff},
  {\href{https://arxiv.org/abs/1903.10456}{{arXiv:1903.10456}}} {[astro-ph.IM]}

\bibitem[{{Menon} et~al.(2022){Menon}, {Federrath}, {Krumholz}, {Kuiper},
  {Wibking}, and {Jung}}]{MenonEtAl2022}
{Menon} SH, {Federrath} C, {Krumholz} MR, et~al (2022) {VETTAM: a scheme for
  radiation hydrodynamics with adaptive mesh refinement using the variable
  Eddington tensor method}. \mnras 512(1):401--423.
  \doi{10.1093/mnras/stac485},
  {\href{https://arxiv.org/abs/2202.08778}{{arXiv:2202.08778}}} {[astro-ph.IM]}

\bibitem[{{Menon} et~al.(2023){Menon}, {Federrath}, and
  {Krumholz}}]{MenonEtAl2023}
{Menon} SH, {Federrath} C, {Krumholz} MR (2023) {Outflows driven by direct and
  reprocessed radiation pressure in massive star clusters}. \mnras
  521(4):5160--5176. \doi{10.1093/mnras/stad856},
  {\href{https://arxiv.org/abs/2210.02818}{{arXiv:2210.02818}}} {[astro-ph.GA]}

\bibitem[{{Mignon-Risse} et~al.(2020){Mignon-Risse}, {Gonz{\'a}lez},
  {Commer{\c{c}}on}, and {Rosdahl}}]{MignonRisseEtAl2020}
{Mignon-Risse} R, {Gonz{\'a}lez} M, {Commer{\c{c}}on} B, et~al (2020) {A new
  hybrid radiative transfer method for massive star formation}. \aap 635:A42.
  \doi{10.1051/0004-6361/201936605},
  {\href{https://arxiv.org/abs/2001.10886}{{arXiv:2001.10886}}} {[astro-ph.IM]}

\bibitem[{{Mignone} et~al.(2018){Mignone}, {Bodo}, {Vaidya}, and
  {Mattia}}]{MignoneEtAl2018}
{Mignone} A, {Bodo} G, {Vaidya} B, et~al (2018) {A Particle Module for the
  PLUTO Code. I. An Implementation of the MHD-PIC Equations}. \apj 859(1):13.
  \doi{10.3847/1538-4357/aabccd},
  {\href{https://arxiv.org/abs/1804.01946}{{arXiv:1804.01946}}} {[astro-ph.HE]}

\bibitem[{{Mihalas} and {Auer}(2001)}]{MihalasAuer2001}
{Mihalas} D, {Auer} L (2001) {On laboratory-frame radiation hydrodynamics}.
  \jqsrt 71(1):61--97. \doi{10.1016/S0022-4073(01)00013-9}

\bibitem[{{Mihalas} and {Mihalas}(1984)}]{Mihalas1984}
{Mihalas} D, {Mihalas} BW (1984) {Foundations of radiation hydrodynamics}.
  Oxford University Press

\bibitem[{{Miyoshi} and {Kusano}(2005)}]{MiyoshiKusano2005}
{Miyoshi} T, {Kusano} K (2005) {A multi-state HLL approximate Riemann solver
  for ideal magnetohydrodynamics}. \jcp 208:315--344.
  \doi{10.1016/j.jcp.2005.02.017}

\bibitem[{{Mohapatra} et~al.(2022){Mohapatra}, {Federrath}, and
  {Sharma}}]{MohapatraFederrathSharma2022}
{Mohapatra} R, {Federrath} C, {Sharma} P (2022) {Multiphase turbulence in
  galactic haloes: effect of the driving}. \mnras 514(3):3139--3159.
  \doi{10.1093/mnras/stac1610},
  {\href{https://arxiv.org/abs/2206.03602}{{arXiv:2206.03602}}} {[astro-ph.GA]}

\bibitem[{{Monaghan}(1988)}]{Monaghan1988}
{Monaghan} JJ (1988) {An introduction to SPH}. Comput Phys Commun 48:89--96.
  \doi{10.1016/0010-4655(88)90026-4}

\bibitem[{{Monaghan}(1992)}]{Monaghan1992}
{Monaghan} JJ (1992) {Smoothed particle hydrodynamics}. \araa 30:543--574.
  \doi{10.1146/annurev.aa.30.090192.002551}

\bibitem[{{Monaghan} and {Lattanzio}(1985)}]{MonaghanLattanzio1985}
{Monaghan} JJ, {Lattanzio} JC (1985) {A refined particle method for
  astrophysical problems}. \aap 149(1):135--143

\bibitem[{{Murray} et~al.(2018){Murray}, {Goyal}, and
  {Chang}}]{MurrayGoyalChang2018}
{Murray} D, {Goyal} S, {Chang} P (2018) {The effects of protostellar jet
  feedback on turbulent collapse}. \mnras 475(1):1023--1035.
  \doi{10.1093/mnras/stx3153},
  {\href{https://arxiv.org/abs/1710.09415}{{arXiv:1710.09415}}} {[astro-ph.GA]}

\bibitem[{{Murray} et~al.(1994){Murray}, {Castor}, {Klein}, and
  {McKee}}]{MurrayEtAl1994}
{Murray} SD, {Castor} JI, {Klein} RI, et~al (1994) {Accretion Disk Coronae in
  High-Luminosity Systems}. \apj 435:631. \doi{10.1086/174842},
  {\href{https://arxiv.org/abs/astro-ph/9405016}{{arXiv:astro-ph/9405016}}}
  {[astro-ph]}

\bibitem[{{Nakamura} and {Li}(2007)}]{NakamuraLi2007}
{Nakamura} F, {Li} ZY (2007) {Protostellar Turbulence Driven by Collimated
  Outflows}. \apj 662:395--412. \doi{10.1086/517515},
  {\href{https://arxiv.org/abs/astro-ph/0703152}{{astro-ph/0703152}}}

\bibitem[{{Nakamura} and {Li}(2011)}]{NakamuraLi2011}
{Nakamura} F, {Li} ZY (2011) {Clustered Star Formation in Magnetic Clouds:
  Properties of Dense Cores Formed in Outflow-driven Turbulence}. \apj 740:36.
  \doi{10.1088/0004-637X/740/1/36},
  {\href{https://arxiv.org/abs/1107.3616}{{arXiv:1107.3616}}} {[astro-ph.SR]}

\bibitem[{{Nakamura} et~al.(2011){Nakamura}, {Kamada}, {Kamazaki}, {Kawabe},
  {Kitamura}, {Shimajiri}, {Tsukagoshi}, {Tachihara}, {Akashi}, {Azegami},
  {Ikeda}, {Kurono}, {Li}, {Miura}, {Nishi}, and {Umemoto}}]{NakamuraEtAl2011}
{Nakamura} F, {Kamada} Y, {Kamazaki} T, et~al (2011) {The Molecular Outflows in
  the {$\rho$} Ophiuchi Main Cloud: Implications for Turbulence Generation}.
  \apj 726:46. \doi{10.1088/0004-637X/726/1/46},
  {\href{https://arxiv.org/abs/1010.2290}{{arXiv:1010.2290}}} {[astro-ph.SR]}

\bibitem[{{Nakano} et~al.(2000){Nakano}, {Hasegawa}, {Morino}, and
  {Yamashita}}]{NakanoEtAl2000}
{Nakano} T, {Hasegawa} T, {Morino} JI, et~al (2000) {Evolution of Protostars
  Accreting Mass at Very High Rates: Is Orion IRc2 a Huge Protostar?} \apj
  534:976--983. \doi{10.1086/308765}

\bibitem[{{Nam} et~al.(2021){Nam}, {Federrath}, and
  {Krumholz}}]{NamFederrathKrumholz2021}
{Nam} DG, {Federrath} C, {Krumholz} MR (2021) {Testing the turbulent origin of
  the stellar initial mass function}. \mnras 503(1):1138--1148.
  \doi{10.1093/mnras/stab505},
  {\href{https://arxiv.org/abs/2102.08564}{{arXiv:2102.08564}}} {[astro-ph.GA]}

\bibitem[{{Noebauer} and {Sim}(2019)}]{NoebauerSim2019}
{Noebauer} UM, {Sim} SA (2019) {Monte Carlo radiative transfer}. Living Rev
  Comput Astrophys 5:1. \doi{10.1007/s41115-019-0004-9},
  {\href{https://arxiv.org/abs/1907.09840}{{arXiv:1907.09840}}} {[astro-ph.IM]}

\bibitem[{{Nordlund} et~al.(2018){Nordlund}, {Ramsey}, {Popovas}, and
  {K{\"u}ffmeier}}]{NordlundEtAl2018}
{Nordlund} {\r{A}}, {Ramsey} JP, {Popovas} A, et~al (2018) {DISPATCH: a
  numerical simulation framework for the exa-scale era - I. Fundamentals}.
  \mnras 477(1):624--638. \doi{10.1093/mnras/sty599},
  {\href{https://arxiv.org/abs/1705.10774}{{arXiv:1705.10774}}} {[astro-ph.IM]}

\bibitem[{{Offner} and {Arce}(2015)}]{OffnerArce2015}
{Offner} SSR, {Arce} HG (2015) {Impact of Winds from Intermediate-mass Stars on
  Molecular Cloud Structure and Turbulence}. \apj 811(2):146.
  \doi{10.1088/0004-637X/811/2/146},
  {\href{https://arxiv.org/abs/1508.07008}{{arXiv:1508.07008}}} {[astro-ph.GA]}

\bibitem[{{Offner} and {Chaban}(2017)}]{OffnerChaban2017}
{Offner} SSR, {Chaban} J (2017) {Impact of Protostellar Outflows on Turbulence
  and Star Formation Efficiency in Magnetized Dense Cores}. \apj 847(2):104.
  \doi{10.3847/1538-4357/aa8996},
  {\href{https://arxiv.org/abs/1709.01086}{{arXiv:1709.01086}}} {[astro-ph.GA]}

\bibitem[{{Offner} and {Liu}(2018)}]{OffnerLiu2018}
{Offner} SSR, {Liu} Y (2018) {Turbulent action at a distance due to stellar
  feedback in magnetized clouds}. Nature Astronomy 2:896--900.
  \doi{10.1038/s41550-018-0566-1},
  {\href{https://arxiv.org/abs/1809.03513}{{arXiv:1809.03513}}} {[astro-ph.SR]}

\bibitem[{{Offner} et~al.(2009){Offner}, {Klein}, {McKee}, and
  {Krumholz}}]{OffnerEtAl2009}
{Offner} SSR, {Klein} RI, {McKee} CF, et~al (2009) {The Effects of Radiative
  Transfer on Low-Mass Star Formation}. \apj 703:131--149.
  \doi{10.1088/0004-637X/703/1/131},
  {\href{https://arxiv.org/abs/0904.2004}{{arXiv:0904.2004}}} {[astro-ph.SR]}

\bibitem[{{Offner} et~al.(2010){Offner}, {Kratter}, {Matzner}, {Krumholz}, and
  {Klein}}]{OffnerEtAl2010}
{Offner} SSR, {Kratter} KM, {Matzner} CD, et~al (2010) {The Formation of
  Low-mass Binary Star Systems Via Turbulent Fragmentation}. \apj
  725(2):1485--1494. \doi{10.1088/0004-637X/725/2/1485},
  {\href{https://arxiv.org/abs/1010.3702}{{arXiv:1010.3702}}} {[astro-ph.SR]}

\bibitem[{{Offner} et~al.(2012){Offner}, {Robitaille}, {Hansen}, {McKee}, and
  {Klein}}]{OffnerEtAl2012}
{Offner} SSR, {Robitaille} TP, {Hansen} CE, et~al (2012) {Observing Simulated
  Protostars with Outflows: How Accurate Are Protostellar Properties Inferred
  from SEDs?} \apj 753:98. \doi{10.1088/0004-637X/753/2/98},
  {\href{https://arxiv.org/abs/1205.0246}{{arXiv:1205.0246}}} {[astro-ph.SR]}

\bibitem[{{Omukai} et~al.(2005){Omukai}, {Tsuribe}, {Schneider}, and
  {Ferrara}}]{OmukaiEtAl2005}
{Omukai} K, {Tsuribe} T, {Schneider} R, et~al (2005) {Thermal and Fragmentation
  Properties of Star-forming Clouds in Low-Metallicity Environments}. \apj
  626:627--643. \doi{10.1086/429955},
  {\href{https://arxiv.org/abs/arXiv:astro-ph/0503010}{{arXiv:astro-ph/0503010}}}

\bibitem[{{Osterbrock} and {Ferland}(2006)}]{OsterbrockFerland2006}
{Osterbrock} DE, {Ferland} GJ (2006) {Astrophysics of gaseous nebulae and
  active galactic nuclei}. University Science Books

\bibitem[{{Owen} et~al.(2021){Owen}, {On}, {Lai}, and {Wu}}]{OwenEtAl2021}
{Owen} ER, {On} AYL, {Lai} SP, et~al (2021) {Observational Signatures of
  Cosmic-Ray Interactions in Molecular Clouds}. \apj 913(1):52.
  \doi{10.3847/1538-4357/abee1a},
  {\href{https://arxiv.org/abs/2103.06542}{{arXiv:2103.06542}}} {[astro-ph.GA]}

\bibitem[{{Padoan} and {Nordlund}(2002)}]{PadoanNordlund2002}
{Padoan} P, {Nordlund} {\AA} (2002) {The Stellar Initial Mass Function from
  Turbulent Fragmentation}. \apj 576:870--879. \doi{10.1086/341790}

\bibitem[{{Padoan} and {Nordlund}(2011)}]{PadoanNordlund2011}
{Padoan} P, {Nordlund} {\r{A}} (2011) {The Star Formation Rate of Supersonic
  Magnetohydrodynamic Turbulence}. \apj 730(1):40.
  \doi{10.1088/0004-637X/730/1/40},
  {\href{https://arxiv.org/abs/0907.0248}{{arXiv:0907.0248}}} {[astro-ph.GA]}

\bibitem[{{Padoan} et~al.(2014){Padoan}, {Federrath}, {Chabrier}, {Evans},
  {Johnstone}, {J{\o}rgensen}, {McKee}, and {Nordlund}}]{PadoanEtAl2014}
{Padoan} P, {Federrath} C, {Chabrier} G, et~al (2014) {The Star Formation Rate
  of Molecular Clouds}. In: {Beuther} H, {Klessen} RS, {Dullemond} CP, et~al
  (eds) Protostars and Planets VI. University of Arizona Press, pp 77--100,
  \doi{10.2458/azu_uapress_9780816531240-ch004},
  {\href{https://arxiv.org/abs/1312.5365}{{arXiv:1312.5365}}}

\bibitem[{{Padoan} et~al.(2016){Padoan}, {Pan}, {Haugb{\o}lle}, and
  {Nordlund}}]{PadoanEtAl2016}
{Padoan} P, {Pan} L, {Haugb{\o}lle} T, et~al (2016) {Supernova Driving. I. The
  Origin of Molecular Cloud Turbulence}. \apj 822:11.
  \doi{10.3847/0004-637X/822/1/11},
  {\href{https://arxiv.org/abs/1509.04663}{{arXiv:1509.04663}}}

\bibitem[{{Padovani} et~al.(2014){Padovani}, {Galli}, {Hennebelle},
  {Commer{\c{c}}on}, and {Joos}}]{PadovaniEtAl2014}
{Padovani} M, {Galli} D, {Hennebelle} P, et~al (2014) {The role of cosmic rays
  on magnetic field diffusion and the formation of protostellar discs}. \aap
  571:A33. \doi{10.1051/0004-6361/201424035},
  {\href{https://arxiv.org/abs/1408.5901}{{arXiv:1408.5901}}} {[astro-ph.GA]}

\bibitem[{{Padovani} et~al.(2020){Padovani}, {Ivlev}, {Galli}, {Offner},
  {Indriolo}, {Rodgers-Lee}, {Marcowith}, {Girichidis}, {Bykov}, and
  {Kruijssen}}]{PadovaniEtAl2020}
{Padovani} M, {Ivlev} AV, {Galli} D, et~al (2020) {Impact of Low-Energy Cosmic
  Rays on Star Formation}. \ssr 216(2):29. \doi{10.1007/s11214-020-00654-1},
  {\href{https://arxiv.org/abs/2002.10282}{{arXiv:2002.10282}}} {[astro-ph.GA]}

\bibitem[{{Pan} et~al.(2016){Pan}, {Padoan}, {Haugb{\o}lle}, and
  {Nordlund}}]{PanEtAl2016}
{Pan} L, {Padoan} P, {Haugb{\o}lle} T, et~al (2016) {Supernova Driving. II.
  Compressive Ratio in Molecular-cloud Turbulence}. \apj 825:30.
  \doi{10.3847/0004-637X/825/1/30},
  {\href{https://arxiv.org/abs/1510.04742}{{arXiv:1510.04742}}}

\bibitem[{{Pandey} et~al.(2026){Pandey}, {Lenker}, {Lopez}, {Rosen}, {Linden},
  {Thompson}, {Offner}, {Auchettl}, and {Hirata}}]{PandeyEtAl2026}
{Pandey} P, {Lenker} SCII, {Lopez} LA, et~al (2026) {Investigating the
  {\ensuremath{\gamma}}-Ray Emission from Explosive Dispersal Outflows with
  Fermi-LAT}. \apj 1000(1):102. \doi{10.3847/1538-4357/ae4690},
  {\href{https://arxiv.org/abs/2509.02679}{{arXiv:2509.02679}}} {[astro-ph.HE]}

\bibitem[{{Papadopoulos} et~al.(2011){Papadopoulos}, {Thi}, {Miniati}, and
  {Viti}}]{PapadopoulosEtAl2011}
{Papadopoulos} PP, {Thi} WF, {Miniati} F, et~al (2011) {Extreme cosmic ray
  dominated regions: a new paradigm for high star formation density events in
  the Universe}. \mnras 414(2):1705--1714.
  \doi{10.1111/j.1365-2966.2011.18504.x},
  {\href{https://arxiv.org/abs/1009.2496}{{arXiv:1009.2496}}} {[astro-ph.CO]}

\bibitem[{{Passot} and {V{\'a}zquez-Semadeni}(1998)}]{PassotVazquez1998}
{Passot} T, {V{\'a}zquez-Semadeni} E (1998) {Density probability distribution
  in one-dimensional polytropic gas dynamics}. \pre 58(4):4501--4510.
  \doi{10.1103/PhysRevE.58.4501},
  {\href{https://arxiv.org/abs/physics/9802019}{{arXiv:physics/9802019}}}
  {[physics.flu-dyn]}

\bibitem[{{Pavlovski} et~al.(2006){Pavlovski}, {Smith}, and {Mac
  Low}}]{PavlovskiSmithMacLow2006}
{Pavlovski} G, {Smith} MD, {Mac Low} MM (2006) {Hydrodynamical simulations of
  the decay of high-speed molecular turbulence - II. Divergence from
  isothermality}. \mnras 368:943--958. \doi{10.1111/j.1365-2966.2006.10172.x},
  {\href{https://arxiv.org/abs/arXiv:astro-ph/0504504}{{arXiv:astro-ph/0504504}}}

\bibitem[{{Pelletier} and {Pudritz}(1992)}]{PelletierPudritz1992}
{Pelletier} G, {Pudritz} RE (1992) {Hydromagnetic disk winds in young stellar
  objects and active galactic nuclei}. \apj 394:117--138. \doi{10.1086/171565}

\bibitem[{{Pelupessy} and {Portegies
  Zwart}(2012)}]{PelupessyPortegiesZwart2012}
{Pelupessy} FI, {Portegies Zwart} S (2012) {The evolution of embedded star
  clusters}. \mnras 420(2):1503--1517. \doi{10.1111/j.1365-2966.2011.20137.x},
  {\href{https://arxiv.org/abs/1111.0992}{{arXiv:1111.0992}}} {[astro-ph.SR]}

\bibitem[{{Pencil Code Collaboration} et~al.(2021){Pencil Code Collaboration},
  {Brandenburg}, {Johansen}, {Bourdin}, {Dobler}, {Lyra}, {Rheinhardt},
  {Bingert}, {Haugen}, {Mee}, {Gent}, {Babkovskaia}, {Yang}, {Heinemann},
  {Dintrans}, {Mitra}, {Candelaresi}, {Warnecke}, {K{\"a}pyl{\"a}},
  {Schreiber}, {Chatterjee}, {K{\"a}pyl{\"a}}, {Li}, {Kr{\"u}ger}, {Aarnes},
  {Sarson}, {Oishi}, {Schober}, {Plasson}, {Sandin}, {Karchniwy}, {Rodrigues},
  {Hubbard}, {Guerrero}, {Snodin}, {Losada}, {Pekkil{\"a}}, and
  {Qian}}]{BrandenburgEtAl2021}
{Pencil Code Collaboration}, {Brandenburg} A, {Johansen} A, et~al (2021) {The
  Pencil Code, a modular MPI code for partial differential equations and
  particles: multipurpose and multiuser-maintained}. J Open Source Softw
  6(58):2807. \doi{10.21105/joss.02807},
  {\href{https://arxiv.org/abs/2009.08231}{{arXiv:2009.08231}}} {[astro-ph.IM]}

\bibitem[{{Pessah} et~al.(2007){Pessah}, {Chan}, and
  {Psaltis}}]{PessahChanPsaltis2007}
{Pessah} ME, {Chan} Ck, {Psaltis} D (2007) {Angular Momentum Transport in
  Accretion Disks: Scaling Laws in MRI-driven Turbulence}. \apjl
  668(1):L51--L54. \doi{10.1086/522585},
  {\href{https://arxiv.org/abs/0705.0352}{{arXiv:0705.0352}}} {[astro-ph]}

\bibitem[{{Peters} et~al.(2010){Peters}, {Banerjee}, {Klessen}, {Mac Low},
  {Galv{\'a}n-Madrid}, and {Keto}}]{PetersEtAl2010}
{Peters} T, {Banerjee} R, {Klessen} RS, et~al (2010) {H II Regions: Witnesses
  to Massive Star Formation}. \apj 711:1017--1028.
  \doi{10.1088/0004-637X/711/2/1017},
  {\href{https://arxiv.org/abs/1001.2470}{{arXiv:1001.2470}}}

\bibitem[{{Petkova} et~al.(2021){Petkova}, {Vandenbroucke}, {Bonnell}, and
  {Kruijssen}}]{PetkovaEtAl2021}
{Petkova} MA, {Vandenbroucke} B, {Bonnell} IA, et~al (2021) {Modelling of
  ionizing feedback with smoothed particle hydrodynamics and Monte Carlo
  radiative transfer on a Voronoi grid}. \mnras 507(1):858--878.
  \doi{10.1093/mnras/stab2178},
  {\href{https://arxiv.org/abs/2109.13953}{{arXiv:2109.13953}}} {[astro-ph.GA]}

\bibitem[{{Pillsworth} et~al.(2025){Pillsworth}, {Roscoe}, {Pudritz}, and
  {Koch}}]{PillsworthEtAl2025}
{Pillsworth} R, {Roscoe} E, {Pudritz} RE, et~al (2025) {Filamentary Hierarchies
  and Superbubbles. I. Characterizing Filament Properties across a Simulated
  Spiral Galaxy}. \apj 987(1):20. \doi{10.3847/1538-4357/add68a},
  {\href{https://arxiv.org/abs/2504.01099}{{arXiv:2504.01099}}} {[astro-ph.GA]}

\bibitem[{{Pineda} et~al.(2024){Pineda}, {Sipil{\"a}}, {Segura-Cox},
  {Valdivia-Mena}, {Neri}, {Kuffmeier}, {Ivlev}, {Offner}, {Maureira},
  {Caselli}, {Spezzano}, {Cunningham}, {Schmiedeke}, and
  {Chen}}]{PinedaEtAl2024}
{Pineda} JE, {Sipil{\"a}} O, {Segura-Cox} DM, et~al (2024) {Probing the physics
  of star formation (ProPStar). I. First resolved maps of the electron fraction
  and cosmic-ray ionization rate in NGC 1333}. \aap 686:A162.
  \doi{10.1051/0004-6361/202347997},
  {\href{https://arxiv.org/abs/2402.16202}{{arXiv:2402.16202}}} {[astro-ph.GA]}

\bibitem[{Poletti et~al.(2025)Poletti, Offner, and
  Ward}]{PolettiOffnerWard2025}
Poletti K, Offner SSR, Ward RA (2025) {Modeling turbulent and self-gravitating
  fluids with Fourier neural operators}. APL Machine Learning 3(2):026118.
  \doi{10.1063/5.0263607}, \urlprefix\url{https://doi.org/10.1063/5.0263607}

\bibitem[{{Powell} et~al.(1999){Powell}, {Roe}, {Linde}, {Gombosi}, and {De
  Zeeuw}}]{PowellEtAl1999}
{Powell} KG, {Roe} PL, {Linde} TJ, et~al (1999) {A Solution-Adaptive Upwind
  Scheme for Ideal Magnetohydrodynamics}. Journal of Computational Physics
  154(2):284--309. \doi{10.1006/jcph.1999.6299}

\bibitem[{{Price}(2008)}]{Price2008}
{Price} DJ (2008) {Modelling discontinuities and Kelvin Helmholtz instabilities
  in SPH}. Journal of Computational Physics 227(24):10040--10057.
  \doi{10.1016/j.jcp.2008.08.011},
  {\href{https://arxiv.org/abs/0709.2772}{{arXiv:0709.2772}}} {[astro-ph]}

\bibitem[{{Price}(2012)}]{Price2012SPH}
{Price} DJ (2012) {Smoothed particle hydrodynamics and magnetohydrodynamics}.
  Journal of Computational Physics 231(3):759--794.
  \doi{10.1016/j.jcp.2010.12.011},
  {\href{https://arxiv.org/abs/1012.1885}{{arXiv:1012.1885}}} {[astro-ph.IM]}

\bibitem[{{Price} and {Federrath}(2010)}]{PriceFederrath2010}
{Price} DJ, {Federrath} C (2010) {A comparison between grid and particle
  methods on the statistics of driven, supersonic, isothermal turbulence}.
  \mnras 406(3):1659--1674. \doi{10.1111/j.1365-2966.2010.16810.x},
  {\href{https://arxiv.org/abs/1004.1446}{{arXiv:1004.1446}}} {[astro-ph.GA]}

\bibitem[{{Price} and {Monaghan}(2007)}]{PriceMonaghan2007}
{Price} DJ, {Monaghan} JJ (2007) {An energy-conserving formalism for adaptive
  gravitational force softening in smoothed particle hydrodynamics and N-body
  codes}. \mnras 374:1347--1358. \doi{10.1111/j.1365-2966.2006.11241.x},
  {\href{https://arxiv.org/abs/arXiv:astro-ph/0610872}{{arXiv:astro-ph/0610872}}}

\bibitem[{{Price} et~al.(2018){Price}, {Wurster}, {Tricco}, {Nixon}, {Toupin},
  {Pettitt}, {Chan}, {Mentiplay}, {Laibe}, {Glover}, {Dobbs}, {Nealon},
  {Liptai}, {Worpel}, {Bonnerot}, {Dipierro}, {Ballabio}, {Ragusa},
  {Federrath}, {Iaconi}, {Reichardt}, {Forgan}, {Hutchison}, {Constantino},
  {Ayliffe}, {Hirsh}, and {Lodato}}]{PriceEtAl2018}
{Price} DJ, {Wurster} J, {Tricco} TS, et~al (2018) {Phantom: A Smoothed
  Particle Hydrodynamics and Magnetohydrodynamics Code for Astrophysics}. \pasa
  35:e031. \doi{10.1017/pasa.2018.25},
  {\href{https://arxiv.org/abs/1702.03930}{{arXiv:1702.03930}}} {[astro-ph.IM]}

\bibitem[{{Pudritz} and {Norman}(1986)}]{PudritzNorman1986}
{Pudritz} RE, {Norman} CA (1986) {Bipolar hydromagnetic winds from disks around
  protostellar objects}. \apj 301:571--586. \doi{10.1086/163924}

\bibitem[{{Pudritz} et~al.(2007){Pudritz}, {Ouyed}, {Fendt}, and
  {Brandenburg}}]{PudritzEtAl2007}
{Pudritz} RE, {Ouyed} R, {Fendt} C, et~al (2007) {Disk Winds, Jets, and
  Outflows: Theoretical and Computational Foundations}. Protostars and Planets
  V pp 277--294.
  {\href{https://arxiv.org/abs/arXiv:astro-ph/0603592}{{arXiv:astro-ph/0603592}}}

\bibitem[{{Raskutti} et~al.(2016){Raskutti}, {Ostriker}, and
  {Skinner}}]{RaskuttiOstrikerSkinner2016}
{Raskutti} S, {Ostriker} EC, {Skinner} MA (2016) {Numerical Simulations of
  Turbulent Molecular Clouds Regulated by Radiation Feedback Forces. I. Star
  Formation Rate and Efficiency}. \apj 829(2):130.
  \doi{10.3847/0004-637X/829/2/130},
  {\href{https://arxiv.org/abs/1608.04469}{{arXiv:1608.04469}}} {[astro-ph.GA]}

\bibitem[{{Raskutti} et~al.(2017){Raskutti}, {Ostriker}, and
  {Skinner}}]{RaskuttiOstrikerSkinner2017}
{Raskutti} S, {Ostriker} EC, {Skinner} MA (2017) {Numerical Simulations of
  Turbulent Molecular Clouds Regulated by Radiation Feedback Forces. II.
  Radiation-Gas Interactions and Outflows}. \apj 850(2):112.
  \doi{10.3847/1538-4357/aa965e},
  {\href{https://arxiv.org/abs/1711.06737}{{arXiv:1711.06737}}} {[astro-ph.GA]}

\bibitem[{{Reissl} et~al.(2016){Reissl}, {Wolf}, and
  {Brauer}}]{ReisslWolfBrauer2016}
{Reissl} S, {Wolf} S, {Brauer} R (2016) {Radiative transfer with POLARIS. I.
  Analysis of magnetic fields through synthetic dust continuum polarization
  measurements}. \aap 593:A87. \doi{10.1051/0004-6361/201424930},
  {\href{https://arxiv.org/abs/1604.05305}{{arXiv:1604.05305}}} {[astro-ph.IM]}

\bibitem[{{Ricker}(2008)}]{Ricker2008}
{Ricker} PM (2008) {A Direct Multigrid Poisson Solver for Oct-Tree Adaptive
  Meshes}. \apjs 176:293--300. \doi{10.1086/526425},
  {\href{https://arxiv.org/abs/0710.4397}{{arXiv:0710.4397}}}

\bibitem[{{Rijkhorst} et~al.(2006){Rijkhorst}, {Plewa}, {Dubey}, and
  {Mellema}}]{RijkhorstEtAl2006}
{Rijkhorst} EJ, {Plewa} T, {Dubey} A, et~al (2006) {Hybrid characteristics: 3D
  radiative transfer for parallel adaptive mesh refinement hydrodynamics}. \aap
  452(3):907--920. \doi{10.1051/0004-6361:20053401},
  {\href{https://arxiv.org/abs/astro-ph/0505213}{{arXiv:astro-ph/0505213}}}
  {[astro-ph]}

\bibitem[{{Rincon} et~al.(2016){Rincon}, {Califano}, {Schekochihin}, and
  {Valentini}}]{RinconEtAl2016}
{Rincon} F, {Califano} F, {Schekochihin} AA, et~al (2016) {Turbulent dynamo in
  a collisionless plasma}. Proc Natl Acad Sci 113(15):3950--3953.
  \doi{10.1073/pnas.1525194113},
  {\href{https://arxiv.org/abs/1512.06455}{{arXiv:1512.06455}}} {[astro-ph.CO]}

\bibitem[{{Robitaille}(2011)}]{Robitaille2011}
{Robitaille} TP (2011) {HYPERION: an open-source parallelized three-dimensional
  dust continuum radiative transfer code}. \aap 536:A79.
  \doi{10.1051/0004-6361/201117150},
  {\href{https://arxiv.org/abs/1112.1071}{{arXiv:1112.1071}}} {[astro-ph.IM]}

\bibitem[{{Rodr{\'\i}guez-Kamenetzky} et~al.(2017){Rodr{\'\i}guez-Kamenetzky},
  {Carrasco-Gonz{\'a}lez}, {Araudo}, {Romero}, {Torrelles}, {Rodr{\'\i}guez},
  {Anglada}, {Mart{\'\i}}, {Perucho}, and
  {Valotto}}]{RodriguezKamenetzkyEtAl2017}
{Rodr{\'\i}guez-Kamenetzky} A, {Carrasco-Gonz{\'a}lez} C, {Araudo} A, et~al
  (2017) {The Highly Collimated Radio Jet of HH 80-81: Structure and Nonthermal
  Emission}. \apj 851(1):16. \doi{10.3847/1538-4357/aa9895},
  {\href{https://arxiv.org/abs/1711.02554}{{arXiv:1711.02554}}} {[astro-ph.HE]}

\bibitem[{{Rogers} and {Pittard}(2013)}]{RogersPittard2013}
{Rogers} H, {Pittard} JM (2013) {Feedback from winds and supernovae in massive
  stellar clusters - I. Hydrodynamics}. \mnras 431(2):1337--1351.
  \doi{10.1093/mnras/stt255},
  {\href{https://arxiv.org/abs/1302.2443}{{arXiv:1302.2443}}} {[astro-ph.SR]}

\bibitem[{{Rosdahl} and {Teyssier}(2015)}]{RosdahlTeyssier2015}
{Rosdahl} J, {Teyssier} R (2015) {A scheme for radiation pressure and photon
  diffusion with the M1 closure in RAMSES-RT}. \mnras 449(4):4380--4403.
  \doi{10.1093/mnras/stv567},
  {\href{https://arxiv.org/abs/1411.6440}{{arXiv:1411.6440}}} {[astro-ph.IM]}

\bibitem[{{Rosdahl} et~al.(2013){Rosdahl}, {Blaizot}, {Aubert}, {Stranex}, and
  {Teyssier}}]{RosdahlEtAl2013}
{Rosdahl} J, {Blaizot} J, {Aubert} D, et~al (2013) {RAMSES-RT: radiation
  hydrodynamics in the cosmological context}. \mnras 436(3):2188--2231.
  \doi{10.1093/mnras/stt1722},
  {\href{https://arxiv.org/abs/1304.7126}{{arXiv:1304.7126}}} {[astro-ph.CO]}

\bibitem[{{Rosen} et~al.(2016){Rosen}, {Krumholz}, {McKee}, and
  {Klein}}]{RosenEtAl2016}
{Rosen} AL, {Krumholz} MR, {McKee} CF, et~al (2016) {An unstable truth: how
  massive stars get their mass}. \mnras 463(3):2553--2573.
  \doi{10.1093/mnras/stw2153},
  {\href{https://arxiv.org/abs/1607.03117}{{arXiv:1607.03117}}} {[astro-ph.SR]}

\bibitem[{{Rosen} et~al.(2017){Rosen}, {Krumholz}, {Oishi}, {Lee}, and
  {Klein}}]{RosenEtAl2017}
{Rosen} AL, {Krumholz} MR, {Oishi} JS, et~al (2017) {Hybrid Adaptive Ray-Moment
  Method (HARM$^{2}$): A highly parallel method for radiation hydrodynamics on
  adaptive grids}. J Comput Phys 330:924--942. \doi{10.1016/j.jcp.2016.10.048},
  {\href{https://arxiv.org/abs/1607.01802}{{arXiv:1607.01802}}} {[astro-ph.IM]}

\bibitem[{{Rosen} et~al.(2019){Rosen}, {Li}, {Zhang}, and
  {Burkhart}}]{RosenEtAl2019}
{Rosen} AL, {Li} PS, {Zhang} Q, et~al (2019) {Massive-star Formation via the
  Collapse of Subvirial and Virialized Turbulent Massive Cores}. \apj
  887(2):108. \doi{10.3847/1538-4357/ab54c6},
  {\href{https://arxiv.org/abs/1902.10153}{{arXiv:1902.10153}}} {[astro-ph.SR]}

\bibitem[{{Rosen} et~al.(2020){Rosen}, {Offner}, {Sadavoy}, {Bhandare},
  {V{\'a}zquez-Semadeni}, and {Ginsburg}}]{RosenEtAl2020}
{Rosen} AL, {Offner} SSR, {Sadavoy} SI, et~al (2020) {Zooming in on Individual
  Star Formation: Low- and High-Mass Stars}. \ssr 216(4):62.
  \doi{10.1007/s11214-020-00688-5},
  {\href{https://arxiv.org/abs/2005.07717}{{arXiv:2005.07717}}} {[astro-ph.SR]}

\bibitem[{{Rost} et~al.(2025){Rost}, {Branca}, and {Buck}}]{RostBrancaBuck2025}
{Rost} R, {Branca} L, {Buck} T (2025) {Emulating Radiative Transfer in
  Astrophysical Environments}. Differentiable Systems and Scientific Machine
  Learning workshop at EurIPS, arXiv:2511.08219.
  \doi{10.48550/arXiv.2511.08219},
  {\href{https://arxiv.org/abs/2511.08219}{{arXiv:2511.08219}}} {[astro-ph.IM]}

\bibitem[{{Rudd} et~al.(2008){Rudd}, {Zentner}, and {Kravtsov}}]{RuddEtAl2008}
{Rudd} DH, {Zentner} AR, {Kravtsov} AV (2008) {Effects of Baryons and
  Dissipation on the Matter Power Spectrum}. \apj 672(1):19--32.
  \doi{10.1086/523836},
  {\href{https://arxiv.org/abs/astro-ph/0703741}{{arXiv:astro-ph/0703741}}}
  {[astro-ph]}

\bibitem[{{Ruszkowski} and {Pfrommer}(2023)}]{RuszkowskiPfrommer2023}
{Ruszkowski} M, {Pfrommer} C (2023) {Cosmic ray feedback in galaxies and galaxy
  clusters}. \aapr 31(1):4. \doi{10.1007/s00159-023-00149-2},
  {\href{https://arxiv.org/abs/2306.03141}{{arXiv:2306.03141}}} {[astro-ph.HE]}

\bibitem[{{Ruszkowski} et~al.(2017){Ruszkowski}, {Yang}, and
  {Zweibel}}]{RuszkowskiYangZweibel2017}
{Ruszkowski} M, {Yang} HYK, {Zweibel} E (2017) {Global Simulations of Galactic
  Winds Including Cosmic-ray Streaming}. \apj 834(2):208.
  \doi{10.3847/1538-4357/834/2/208},
  {\href{https://arxiv.org/abs/1602.04856}{{arXiv:1602.04856}}} {[astro-ph.GA]}

\bibitem[{{Salem} and {Bryan}(2014)}]{SalemBryan2014}
{Salem} M, {Bryan} GL (2014) {Cosmic ray driven outflows in global galaxy disc
  models}. \mnras 437(4):3312--3330. \doi{10.1093/mnras/stt2121},
  {\href{https://arxiv.org/abs/1307.6215}{{arXiv:1307.6215}}} {[astro-ph.CO]}

\bibitem[{{Salpeter}(1955)}]{Salpeter1955}
{Salpeter} EE (1955) {The Luminosity Function and Stellar Evolution.} \apj
  121:161. \doi{10.1086/145971}

\bibitem[{{Scalo} and {Elmegreen}(2004)}]{ScaloElmegreen2004}
{Scalo} J, {Elmegreen} BG (2004) {Interstellar Turbulence II: Implications and
  Effects}. \araa 42(1):275--316. \doi{10.1146/annurev.astro.42.120403.143327},
  {\href{https://arxiv.org/abs/astro-ph/0404452}{{arXiv:astro-ph/0404452}}}
  {[astro-ph]}

\bibitem[{{Schekochihin} et~al.(2004){Schekochihin}, {Cowley}, {Taylor},
  {Maron}, and {McWilliams}}]{SchekochihinEtAl2004}
{Schekochihin} AA, {Cowley} SC, {Taylor} SF, et~al (2004) {Simulations of the
  Small-Scale Turbulent Dynamo}. \apj 612(1):276--307. \doi{10.1086/422547},
  {\href{https://arxiv.org/abs/astro-ph/0312046}{{arXiv:astro-ph/0312046}}}
  {[astro-ph]}

\bibitem[{{Schekochihin} et~al.(2007){Schekochihin}, {Iskakov}, {Cowley},
  {McWilliams}, {Proctor}, and {Yousef}}]{SchekochihinEtAl2007}
{Schekochihin} AA, {Iskakov} AB, {Cowley} SC, et~al (2007) {Fluctuation dynamo
  and turbulent induction at low magnetic Prandtl numbers}. New J Phys 9:300.
  \doi{10.1088/1367-2630/9/8/300} {[physics.flu-dyn]}

\bibitem[{{Schlickeiser}(1989)}]{Schlickeiser1989}
{Schlickeiser} R (1989) {Cosmic-Ray Transport and Acceleration. I. Derivation
  of the Kinetic Equation and Application to Cosmic Rays in Static Cold Media}.
  \apj 336:243. \doi{10.1086/167009}

\bibitem[{{Schlickeiser} and {Lerche}(1985)}]{SchlickeiserLerche1985}
{Schlickeiser} R, {Lerche} I (1985) {Cosmic gas dynamics. I - Basic equations
  and the dynamics of hot interstellar matter}. \aap 151(1):151--156

\bibitem[{{Schmidt} et~al.(2009){Schmidt}, {Federrath}, {Hupp}, {Kern}, and
  {Niemeyer}}]{SchmidtEtAl2009}
{Schmidt} W, {Federrath} C, {Hupp} M, et~al (2009) {Numerical simulations of
  compressively driven interstellar turbulence. I. Isothermal gas}. \aap
  494(1):127--145. \doi{10.1051/0004-6361:200809967},
  {\href{https://arxiv.org/abs/0809.1321}{{arXiv:0809.1321}}} {[astro-ph]}

\bibitem[{{Sedov}(1959)}]{Sedov1959}
{Sedov} LI (1959) {Similarity and Dimensional Methods in Mechanics}. New York:
  Academic Press

\bibitem[{{Seifried} et~al.(2012){Seifried}, {Pudritz}, {Banerjee}, {Duffin},
  and {Klessen}}]{SeifriedEtAl2012}
{Seifried} D, {Pudritz} RE, {Banerjee} R, et~al (2012) {Magnetic fields during
  the early stages of massive star formation - II. A generalized outflow
  criterion}. \mnras 422:347--366. \doi{10.1111/j.1365-2966.2012.20610.x},
  {\href{https://arxiv.org/abs/1109.4379}{{arXiv:1109.4379}}} {[astro-ph.SR]}

\bibitem[{{Seifried} et~al.(2017){Seifried}, {Walch}, {Girichidis}, {Naab},
  {W{\"u}nsch}, {Klessen}, {Glover}, {Peters}, and {Clark}}]{SeifriedEtAl2017}
{Seifried} D, {Walch} S, {Girichidis} P, et~al (2017) {SILCC-Zoom: the dynamic
  and chemical evolution of molecular clouds}. \mnras 472(4):4797--4818.
  \doi{10.1093/mnras/stx2343},
  {\href{https://arxiv.org/abs/1704.06487}{{arXiv:1704.06487}}} {[astro-ph.GA]}

\bibitem[{{Semenov} et~al.(2003){Semenov}, {Henning}, {Helling}, {Ilgner}, and
  {Sedlmayr}}]{SemenovEtAl2003}
{Semenov} D, {Henning} T, {Helling} C, et~al (2003) {Rosseland and Planck mean
  opacities for protoplanetary discs}. \aap 410:611--621.
  \doi{10.1051/0004-6361:20031279},
  {\href{https://arxiv.org/abs/astro-ph/0308344}{{arXiv:astro-ph/0308344}}}
  {[astro-ph]}

\bibitem[{{Semenov}(2025)}]{Semenov2025}
{Semenov} VA (2025) {Capturing Turbulence with Numerical Dissipation: A Simple
  Dynamical Model for Unresolved Turbulence in Hydrodynamic Simulations}. \apjs
  281(2):37. \doi{10.3847/1538-4365/ae0cc6},
  {\href{https://arxiv.org/abs/2410.23339}{{arXiv:2410.23339}}} {[astro-ph.GA]}

\bibitem[{{Semenov} et~al.(2016){Semenov}, {Kravtsov}, and
  {Gnedin}}]{SemenovKravtsovGnedin2016}
{Semenov} VA, {Kravtsov} AV, {Gnedin} NY (2016) {Nonuniversal Star Formation
  Efficiency in Turbulent ISM}. \apj 826(2):200.
  \doi{10.3847/0004-637X/826/2/200},
  {\href{https://arxiv.org/abs/1512.03101}{{arXiv:1512.03101}}} {[astro-ph.GA]}

\bibitem[{{Seta} and {Federrath}(2021)}]{SetaFederrath2021}
{Seta} A, {Federrath} C (2021) {Saturation mechanism of the fluctuation dynamo
  in supersonic turbulent plasmas}. Phys Rev Fluids 6(10):103701.
  \doi{10.1103/PhysRevFluids.6.103701},
  {\href{https://arxiv.org/abs/2109.11698}{{arXiv:2109.11698}}} {[astro-ph.GA]}

\bibitem[{{Seta} and {Federrath}(2022)}]{SetaFederrath2022}
{Seta} A, {Federrath} C (2022) {Turbulent dynamo in the two-phase interstellar
  medium}. \mnras 514(1):957--976. \doi{10.1093/mnras/stac1400},
  {\href{https://arxiv.org/abs/2202.08324}{{arXiv:2202.08324}}} {[astro-ph.GA]}

\bibitem[{{Sharda} and {Menon}(2025)}]{ShardaMenon2025}
{Sharda} P, {Menon} SH (2025) {Population III star formation in the presence of
  turbulence, magnetic fields, and ionizing radiation feedback}. \mnras
  540(2):1745--1764. \doi{10.1093/mnras/staf803},
  {\href{https://arxiv.org/abs/2405.18265}{{arXiv:2405.18265}}} {[astro-ph.GA]}

\bibitem[{{Sharda} et~al.(2019){Sharda}, {Krumholz}, and
  {Federrath}}]{ShardaKrumholzFederrath2019}
{Sharda} P, {Krumholz} MR, {Federrath} C (2019) {The role of the H$_{2}$
  adiabatic index in the formation of the first stars}. \mnras 490(1):513--526.
  \doi{10.1093/mnras/stz2618},
  {\href{https://arxiv.org/abs/1909.06269}{{arXiv:1909.06269}}} {[astro-ph.GA]}

\bibitem[{{Shestakov} and {Offner}(2008)}]{ShestakovOffner2008}
{Shestakov} AI, {Offner} SSR (2008) {A multigroup diffusion solver using pseudo
  transient continuation for a radiation-hydrodynamic code with patch-based
  AMR}. J Comput Phys 227(3):2154--2186. \doi{10.1016/j.jcp.2007.09.019},
  {\href{https://arxiv.org/abs/0710.4509}{{arXiv:0710.4509}}} {[astro-ph]}

\bibitem[{{Shibata} and {Uchida}(1985)}]{ShibataUchida1985}
{Shibata} K, {Uchida} Y (1985) {A magnetodynamic mechanism for the formation of
  astrophysical jets. I - Dynamical effects of the relaxation of nonlinear
  magnetic twists}. \pasj 37:31--46

\bibitem[{{Shibata} and {Uchida}(1986)}]{ShibataUchida1986}
{Shibata} K, {Uchida} Y (1986) {A magnetodynamic mechanism for the formation of
  astrophysical jets. II - Dynamical processes in the accretion of magnetized
  mass in rotation}. \pasj 38:631--660

\bibitem[{{Shivakumar} and {Federrath}(2025)}]{ShivakumarFederrath2025}
{Shivakumar} LM, {Federrath} C (2025) {Numerical viscosity and resistivity in
  MHD turbulence simulations}. \mnras 537(4):2961--2986.
  \doi{10.1093/mnras/staf160},
  {\href{https://arxiv.org/abs/2311.10350}{{arXiv:2311.10350}}} {[astro-ph.SR]}

\bibitem[{{Shu} et~al.(1987){Shu}, {Adams}, and {Lizano}}]{ShuAdamsLizano1987}
{Shu} FH, {Adams} FC, {Lizano} S (1987) {Star formation in molecular clouds -
  Observation and theory}. \araa 25:23--81.
  \doi{10.1146/annurev.aa.25.090187.000323}

\bibitem[{{Simpson} et~al.(2015){Simpson}, {Bryan}, {Hummels}, and
  {Ostriker}}]{SimpsonEtAl2015}
{Simpson} CM, {Bryan} GL, {Hummels} C, et~al (2015) {Kinetic Energy from
  Supernova Feedback in High-resolution Galaxy Simulations}. \apj 809(1):69.
  \doi{10.1088/0004-637X/809/1/69},
  {\href{https://arxiv.org/abs/1410.3822}{{arXiv:1410.3822}}} {[astro-ph.GA]}

\bibitem[{{Skilling}(1975)}]{Skilling1975}
{Skilling} J (1975) {Cosmic ray streaming - I. Effect of Alfv{\'e}n waves on
  particles.} \mnras 172:557--566. \doi{10.1093/mnras/172.3.557}

\bibitem[{{Skinner} and {Ostriker}(2013)}]{SkinnerOstriker2013}
{Skinner} MA, {Ostriker} EC (2013) {A Two-moment Radiation Hydrodynamics Module
  in Athena Using a Time-explicit Godunov Method}. \apjs 206(2):21.
  \doi{10.1088/0067-0049/206/2/21},
  {\href{https://arxiv.org/abs/1306.0010}{{arXiv:1306.0010}}} {[astro-ph.IM]}

\bibitem[{{Skinner} and {Ostriker}(2015)}]{SkinnerOstriker2015}
{Skinner} MA, {Ostriker} EC (2015) {Numerical Simulations of Turbulent
  Molecular Clouds Regulated by Reprocessed Radiation Feedback from Nascent
  Super Star Clusters}. \apj 809(2):187. \doi{10.1088/0004-637X/809/2/187},
  {\href{https://arxiv.org/abs/1507.06366}{{arXiv:1507.06366}}} {[astro-ph.GA]}

\bibitem[{{Smith} et~al.(2020){Smith}, {Kannan}, {Tsang}, {Vogelsberger}, and
  {Pakmor}}]{SmithEtAl2020}
{Smith} A, {Kannan} R, {Tsang} BTH, et~al (2020) {AREPO-MCRT: Monte Carlo
  Radiation Hydrodynamics on a Moving Mesh}. \apj 905(1):27.
  \doi{10.3847/1538-4357/abc47e},
  {\href{https://arxiv.org/abs/2008.01750}{{arXiv:2008.01750}}} {[astro-ph.GA]}

\bibitem[{{Smith}(2014)}]{Smith2014}
{Smith} N (2014) {Mass Loss: Its Effect on the Evolution and Fate of High-Mass
  Stars}. \araa 52:487--528. \doi{10.1146/annurev-astro-081913-040025},
  {\href{https://arxiv.org/abs/1402.1237}{{arXiv:1402.1237}}} {[astro-ph.SR]}

\bibitem[{{Smith} et~al.(2014){Smith}, {Glover}, and
  {Klessen}}]{SmithGloverKlessen2014}
{Smith} RJ, {Glover} SCO, {Klessen} RS (2014) {On the nature of star-forming
  filaments - I. Filament morphologies}. \mnras 445:2900--2917.
  \doi{10.1093/mnras/stu1915},
  {\href{https://arxiv.org/abs/1407.6716}{{arXiv:1407.6716}}}

\bibitem[{{Smith} et~al.(2016){Smith}, {Glover}, {Klessen}, and
  {Fuller}}]{SmithEtAl2016}
{Smith} RJ, {Glover} SCO, {Klessen} RS, et~al (2016) {On the nature of
  star-forming filaments - II. Subfilaments and velocities}. \mnras
  455:3640--3655. \doi{10.1093/mnras/stv2559},
  {\href{https://arxiv.org/abs/1509.03321}{{arXiv:1509.03321}}}

\bibitem[{{Springel}(2005)}]{Springel2005}
{Springel} V (2005) {The cosmological simulation code GADGET-2}. \mnras
  364(4):1105--1134. \doi{10.1111/j.1365-2966.2005.09655.x},
  {\href{https://arxiv.org/abs/astro-ph/0505010}{{arXiv:astro-ph/0505010}}}
  {[astro-ph]}

\bibitem[{{Springel}(2010)}]{Springel2010}
{Springel} V (2010) {E pur si muove: Galilean-invariant cosmological
  hydrodynamical simulations on a moving mesh}. \mnras 401:791--851.
  \doi{10.1111/j.1365-2966.2009.15715.x},
  {\href{https://arxiv.org/abs/0901.4107}{{arXiv:0901.4107}}} {[astro-ph.CO]}

\bibitem[{{St-Onge} and {Kunz}(2018)}]{StOngeKunz2018}
{St-Onge} DA, {Kunz} MW (2018) {Fluctuation Dynamo in a Collisionless, Weakly
  Magnetized Plasma}. \apjl 863(2):L25. \doi{10.3847/2041-8213/aad638},
  {\href{https://arxiv.org/abs/1806.11162}{{arXiv:1806.11162}}} {[astro-ph.HE]}

\bibitem[{{Stamatellos} et~al.(2012){Stamatellos}, {Whitworth}, and
  {Hubber}}]{StamatellosWhitworthHubber2012}
{Stamatellos} D, {Whitworth} AP, {Hubber} DA (2012) {Episodic accretion,
  protostellar radiative feedback, and their role in low-mass star formation}.
  \mnras 427(2):1182--1193. \doi{10.1111/j.1365-2966.2012.22038.x},
  {\href{https://arxiv.org/abs/1209.0765}{{arXiv:1209.0765}}} {[astro-ph.GA]}

\bibitem[{{Stone} and {Norman}(1992)}]{StoneNorman1992a}
{Stone} JM, {Norman} ML (1992) {ZEUS-2D: A radiation magnetohydrodynamics code
  for astrophysical flows in two space dimensions. I - The hydrodynamic
  algorithms and tests.} \apjs 80:753--790. \doi{10.1086/191680}

\bibitem[{{Stone} et~al.(1998){Stone}, {Ostriker}, and
  {Gammie}}]{StoneOstrikerGammie1998}
{Stone} JM, {Ostriker} EC, {Gammie} CF (1998) {Dissipation in Compressible
  Magnetohydrodynamic Turbulence}. \apjl 508:L99--L102. \doi{10.1086/311718},
  {\href{https://arxiv.org/abs/arXiv:astro-ph/9809357}{{arXiv:astro-ph/9809357}}}

\bibitem[{{Stone} et~al.(2020){Stone}, {Tomida}, {White}, and
  {Felker}}]{StoneEtAl2020}
{Stone} JM, {Tomida} K, {White} CJ, et~al (2020) {The Athena++ Adaptive Mesh
  Refinement Framework: Design and Magnetohydrodynamic Solvers}. \apjs
  249(1):4. \doi{10.3847/1538-4365/ab929b},
  {\href{https://arxiv.org/abs/2005.06651}{{arXiv:2005.06651}}} {[astro-ph.IM]}

\bibitem[{{Stone} et~al.(2026){Stone}, {Mullen}, {Fielding}, {Grete}, {Guo},
  {Kempski}, {Most}, {White}, and {Wong}}]{StoneEtAl2026}
{Stone} JM, {Mullen} PD, {Fielding} D, et~al (2026) {AthenaK: A
  Performance-portable Version of the Athena++ Adaptive Mesh Refinement
  Framework}. \apjs 283(1):27. \doi{10.3847/1538-4365/ae3717},
  {\href{https://arxiv.org/abs/2409.16053}{{arXiv:2409.16053}}} {[astro-ph.IM]}

\bibitem[{{Strong} and {Moskalenko}(1998)}]{StrongMoskalenko1998}
{Strong} AW, {Moskalenko} IV (1998) {Propagation of Cosmic-Ray Nucleons in the
  Galaxy}. \apj 509(1):212--228. \doi{10.1086/306470},
  {\href{https://arxiv.org/abs/astro-ph/9807150}{{arXiv:astro-ph/9807150}}}
  {[astro-ph]}

\bibitem[{{Sulzer} and {Buck}(2023)}]{SulzerBuck2023}
{Sulzer} I, {Buck} T (2023) {Speeding up astrochemical reaction networks with
  autoencoders and neural ODEs}. Machine Learning and the Physical Sciences
  Workshop at Neurips arXiv:2312.06015. \doi{10.48550/arXiv.2312.06015},
  {\href{https://arxiv.org/abs/2312.06015}{{arXiv:2312.06015}}} {[astro-ph.GA]}

\bibitem[{{Sur} et~al.(2010){Sur}, {Schlei\-cher}, {Banerjee}, {Federrath}, and
  {Klessen}}]{SurEtAl2010}
{Sur} S, {Schlei\-cher} DRG, {Banerjee} R, et~al (2010) {The Generation of
  Strong Magnetic Fields During the Formation of the First Stars}. \apjl
  721:L134--L138. \doi{10.1088/2041-8205/721/2/L134} {[astro-ph.CO]}

\bibitem[{{Tamburro} et~al.(2009){Tamburro}, {Rix}, {Leroy}, {Low}, {Walter},
  {Kennicutt}, {Brinks}, and {de Blok}}]{TamburroEtAl2009}
{Tamburro} D, {Rix} HW, {Leroy} AK, et~al (2009) {What is Driving the HI
  Velocity Dispersion?} \aj 137:4424--4435. \doi{10.1088/0004-6256/137/5/4424},
  {\href{https://arxiv.org/abs/0903.0183}{{arXiv:0903.0183}}}

\bibitem[{{Taylor}(1950)}]{Taylor1950}
{Taylor} G (1950) {The Formation of a Blast Wave by a Very Intense Explosion.
  I. Theoretical Discussion}. Proc R Soc London Ser A 201(1065):159--174.
  \doi{10.1098/rspa.1950.0049}

\bibitem[{{Teyssier}(2002)}]{Teyssier2002}
{Teyssier} R (2002) {Cosmological hydrodynamics with adaptive mesh refinement.
  A new high resolution code called RAMSES}. \aap 385:337--364.
  \doi{10.1051/0004-6361:20011817},
  {\href{https://arxiv.org/abs/astro-ph/0111367}{{arXiv:astro-ph/0111367}}}
  {[astro-ph]}

\bibitem[{{Teyssier} and {Commer{\c{c}}on}(2019)}]{TeyssierCommercon2019}
{Teyssier} R, {Commer{\c{c}}on} B (2019) {Numerical Methods for Simulating Star
  Formation}. Front Astron Space Sci 6:51. \doi{10.3389/fspas.2019.00051},
  {\href{https://arxiv.org/abs/1907.08542}{{arXiv:1907.08542}}} {[astro-ph.IM]}

\bibitem[{{Thomas} and {Pfrommer}(2019)}]{ThomasPfrommer2019}
{Thomas} T, {Pfrommer} C (2019) {Cosmic-ray hydrodynamics: Alfv{\'e}n-wave
  regulated transport of cosmic rays}. \mnras 485(3):2977--3008.
  \doi{10.1093/mnras/stz263},
  {\href{https://arxiv.org/abs/1805.11092}{{arXiv:1805.11092}}} {[astro-ph.HE]}

\bibitem[{{Thomas} and {Pfrommer}(2022)}]{ThomasPfrommer2022}
{Thomas} T, {Pfrommer} C (2022) {Comparing different closure relations for
  cosmic ray hydrodynamics}. \mnras 509(4):4803--4816.
  \doi{10.1093/mnras/stab3079},
  {\href{https://arxiv.org/abs/2105.08090}{{arXiv:2105.08090}}} {[astro-ph.HE]}

\bibitem[{{Thomas} et~al.(2021){Thomas}, {Pfrommer}, and
  {Pakmor}}]{ThomasPfrommerPakmor2021}
{Thomas} T, {Pfrommer} C, {Pakmor} R (2021) {A finite volume method for
  two-moment cosmic ray hydrodynamics on a moving mesh}. \mnras
  503(2):2242--2264. \doi{10.1093/mnras/stab397},
  {\href{https://arxiv.org/abs/2010.11960}{{arXiv:2010.11960}}} {[astro-ph.HE]}

\bibitem[{{Tisza}(1942)}]{Tisza1942}
{Tisza} L (1942) {Supersonic Absorption and Stokes' Viscosity Relation}. Phys
  Rev 61:531--536. \doi{10.1103/PhysRev.61.531}

\bibitem[{{Tomida} et~al.(2013){Tomida}, {Tomisaka}, {Matsumoto}, {Hori},
  {Okuzumi}, {Machida}, and {Saigo}}]{TomidaEtAl2013}
{Tomida} K, {Tomisaka} K, {Matsumoto} T, et~al (2013) {Radiation
  Magnetohydrodynamic Simulations of Protostellar Collapse: Protostellar Core
  Formation}. \apj 763(1):6. \doi{10.1088/0004-637X/763/1/6},
  {\href{https://arxiv.org/abs/1206.3567}{{arXiv:1206.3567}}} {[astro-ph.SR]}

\bibitem[{Toro(1997)}]{Toro1997}
Toro EF (1997) Riemann Solvers and Numerical Methods for Fluid Dynamics.
  Springer Berlin, Heidelberg, \doi{https://doi.org/10.1007/b79761}

\bibitem[{{Tricco} and {Price}(2012)}]{TriccoPrice2012}
{Tricco} TS, {Price} DJ (2012) {Constrained hyperbolic divergence cleaning for
  smoothed particle magnetohydrodynamics}. \jcp 231:7214--7236.
  \doi{10.1016/j.jcp.2012.06.039},
  {\href{https://arxiv.org/abs/1206.6159}{{arXiv:1206.6159}}} {[astro-ph.IM]}

\bibitem[{{Tritsis} et~al.(2022){Tritsis}, {Federrath}, {Willacy}, and
  {Tassis}}]{TritsisEtAl2022}
{Tritsis} A, {Federrath} C, {Willacy} K, et~al (2022) {Non-ideal
  magnetohydrodynamic simulations of subcritical pre-stellar cores with
  non-equilibrium chemistry}. \mnras 510(3):4420--4435.
  \doi{10.1093/mnras/stab3740},
  {\href{https://arxiv.org/abs/2112.11462}{{arXiv:2112.11462}}} {[astro-ph.GA]}

\bibitem[{{Tritsis} et~al.(2025{\natexlab{a}}){Tritsis}, {Basu}, and
  {Federrath}}]{TritsisBasuFederrath2025a}
{Tritsis} A, {Basu} S, {Federrath} C (2025{\natexlab{a}}) {Projection-angle
  effects when ``observing'' a turbulent magnetized collapsing molecular cloud:
  I. Chemistry and line transfer}. \aap 695:A18.
  \doi{10.1051/0004-6361/202452013},
  {\href{https://arxiv.org/abs/2503.01963}{{arXiv:2503.01963}}} {[astro-ph.GA]}

\bibitem[{{Tritsis} et~al.(2025{\natexlab{b}}){Tritsis}, {Basu}, and
  {Federrath}}]{TritsisBasuFederrath2025b}
{Tritsis} A, {Basu} S, {Federrath} C (2025{\natexlab{b}}) {Projection-angle
  effects when ``observing'' a turbulent magnetized collapsing molecular cloud:
  II. Magnetic field}. \aap 696:A35. \doi{10.1051/0004-6361/202453265}

\bibitem[{{Troccoli} and {Federrath}(2026)}]{TroccoliFederrath2026}
{Troccoli} E, {Federrath} C (2026) {The statistics and structure of dissipation
  in subsonic and supersonic turbulence}. \mnras 547(4):stag461.
  \doi{10.1093/mnras/stag461}

\bibitem[{Trottenberg et~al.(2001)Trottenberg, Oosterlee, and
  Schüller}]{TrottenbergOosterleeSchueller2001}
Trottenberg U, Oosterlee CW, Schüller A (2001) Multigrid, Texts in Applied
  Mathematics. Bd., vol~33. Academic Press, with contributions by A. Brandt, P.
  Oswald and K. Stüben

\bibitem[{{Truelove} et~al.(1997){Truelove}, {Klein}, {McKee}, {Holliman},
  {Howell}, and {Greenough}}]{TrueloveEtAl1997}
{Truelove} JK, {Klein} RI, {McKee} CF, et~al (1997) {The Jeans Condition: A New
  Constraint on Spatial Resolution in Simulations of Isothermal
  Self-gravitational Hydrodynamics}. \apjl 489:L179. \doi{10.1086/316779}

\bibitem[{{Tsang} and {Milosavljevi{\'c}}(2015)}]{TsangMilosavljevic2015}
{Tsang} BTH, {Milosavljevi{\'c}} M (2015) {Radiation pressure driving of a
  dusty atmosphere}. \mnras 453(1):1108--1120. \doi{10.1093/mnras/stv1707},
  {\href{https://arxiv.org/abs/1506.05121}{{arXiv:1506.05121}}} {[astro-ph.GA]}

\bibitem[{{Turk} et~al.(2009){Turk}, {Abel}, and {O'Shea}}]{TurkAbelOShea2009}
{Turk} MJ, {Abel} T, {O'Shea} B (2009) {The Formation of Population III
  Binaries from Cosmological Initial Conditions}. Science 325(5940):601.
  \doi{10.1126/science.1173540},
  {\href{https://arxiv.org/abs/0907.2919}{{arXiv:0907.2919}}} {[astro-ph.CO]}

\bibitem[{{Turk} et~al.(2012){Turk}, {Oishi}, {Abel}, and
  {Bryan}}]{TurkEtAl2012}
{Turk} MJ, {Oishi} JS, {Abel} T, et~al (2012) {Magnetic Fields in Population
  III Star Formation}. \apj 745:154. \doi{10.1088/0004-637X/745/2/154},
  {\href{https://arxiv.org/abs/1112.4479}{{arXiv:1112.4479}}} {[astro-ph.CO]}

\bibitem[{{Uchida} and {Shibata}(1985)}]{UchidaShibata1985}
{Uchida} Y, {Shibata} K (1985) {Magnetodynamical acceleration of CO and optical
  bipolar flows from the region of star formation}. \pasj 37:515--535

\bibitem[{{Urban} et~al.(2010){Urban}, {Martel}, and
  {Evans}}]{UrbanMartelEvans2010}
{Urban} A, {Martel} H, {Evans} NJII (2010) {Fragmentation and Evolution of
  Molecular Clouds. II. The Effect of Dust Heating}. \apj 710(2):1343--1364.
  \doi{10.1088/0004-637X/710/2/1343},
  {\href{https://arxiv.org/abs/0912.3819}{{arXiv:0912.3819}}} {[astro-ph.GA]}

\bibitem[{{van Leer}(1979)}]{vanLeer1979}
{van Leer} B (1979) {Towards the Ultimate Conservative Difference Scheme. V. A
  Second-Order Sequel to Godunov's Method}. Journal of Computational Physics
  32(1):101--136. \doi{10.1016/0021-9991(79)90145-1}

\bibitem[{{Vandenbroucke} and {Wood}(2018)}]{VandenbrouckeWood2018}
{Vandenbroucke} B, {Wood} K (2018) {The Monte Carlo photoionization and
  moving-mesh radiation hydrodynamics code CMacIonize}. Astronomy and Computing
  23:40. \doi{10.1016/j.ascom.2018.02.005}

\bibitem[{{V{\'a}zquez-Semadeni} et~al.(2003){V{\'a}zquez-Semadeni},
  {Ballesteros-Paredes}, and {Klessen}}]{VazquezBallesterosKlessen2003}
{V{\'a}zquez-Semadeni} E, {Ballesteros-Paredes} J, {Klessen} RS (2003) {A
  Holistic Scenario of Turbulent Molecular Cloud Evolution and Control of the
  Star Formation Efficiency: First Tests}. \apjl 585:L131--L134.
  \doi{10.1086/374325},
  {\href{https://arxiv.org/abs/arXiv:astro-ph/0301546}{{arXiv:astro-ph/0301546}}}

\bibitem[{{V{\'a}zquez-Semadeni} et~al.(2007){V{\'a}zquez-Semadeni},
  {G{\'o}mez}, {Jappsen}, {Ballesteros-Paredes}, {Gonz{\'a}lez}, and
  {Klessen}}]{VazquezSemadeniEtAl2007}
{V{\'a}zquez-Semadeni} E, {G{\'o}mez} GC, {Jappsen} AK, et~al (2007) {Molecular
  Cloud Evolution. II. From Cloud Formation to the Early Stages of Star
  Formation in Decaying Conditions}. \apj 657(2):870--883.
  \doi{10.1086/510771},
  {\href{https://arxiv.org/abs/astro-ph/0608375}{{arXiv:astro-ph/0608375}}}
  {[astro-ph]}

\bibitem[{{V{\'a}zquez-Semadeni} et~al.(2019){V{\'a}zquez-Semadeni}, {Palau},
  {Ballesteros-Paredes}, {G{\'o}mez}, and
  {Zamora-Avil{\'e}s}}]{VazquezSemadeniEtAl2019}
{V{\'a}zquez-Semadeni} E, {Palau} A, {Ballesteros-Paredes} J, et~al (2019)
  {Global hierarchical collapse in molecular clouds. Towards a comprehensive
  scenario}. \mnras 490(3):3061--3097. \doi{10.1093/mnras/stz2736},
  {\href{https://arxiv.org/abs/1903.11247}{{arXiv:1903.11247}}} {[astro-ph.GA]}

\bibitem[{{Verliat} et~al.(2022){Verliat}, {Hennebelle}, {Gonz{\'a}lez}, {Lee},
  and {Geen}}]{VerliatEtAl2022}
{Verliat} A, {Hennebelle} P, {Gonz{\'a}lez} M, et~al (2022) {Influence of
  protostellar jets and HII regions on the formation and evolution of stellar
  clusters}. \aap 663:A6. \doi{10.1051/0004-6361/202141765},
  {\href{https://arxiv.org/abs/2202.02237}{{arXiv:2202.02237}}} {[astro-ph.GA]}

\bibitem[{{Vermari{\"e}n} et~al.(2025){Vermari{\"e}n}, {Bisbas}, {Viti},
  {Zhao}, {Tang}, and {Ravichandran}}]{VermarienEtAl2025}
{Vermari{\"e}n} G, {Bisbas} TG, {Viti} S, et~al (2025) {NeuralPDR: neural
  differential equations as surrogate models for photodissociation regions}.
  Machine Learning: Science and Technology 6(2):025069.
  \doi{10.1088/2632-2153/ade4ee},
  {\href{https://arxiv.org/abs/2506.14270}{{arXiv:2506.14270}}} {[astro-ph.GA]}

\bibitem[{{Verner} et~al.(1996){Verner}, {Ferland}, {Korista}, and
  {Yakovlev}}]{VernerEtAl1996}
{Verner} DA, {Ferland} GJ, {Korista} KT, et~al (1996) {Atomic Data for
  Astrophysics. II. New Analytic Fits for Photoionization Cross Sections of
  Atoms and Ions}. \apj 465:487. \doi{10.1086/177435},
  {\href{https://arxiv.org/abs/astro-ph/9601009}{{arXiv:astro-ph/9601009}}}
  {[astro-ph]}

\bibitem[{{Waagan} et~al.(2011){Waagan}, {Federrath}, and
  {Klingenberg}}]{WaaganFederrathKlingenberg2011}
{Waagan} K, {Federrath} C, {Klingenberg} C (2011) {A robust numerical scheme
  for highly compressible magnetohydrodynamics: Nonlinear stability,
  implementation and tests}. Journal of Computational Physics
  230(9):3331--3351. \doi{10.1016/j.jcp.2011.01.026},
  {\href{https://arxiv.org/abs/1101.3007}{{arXiv:1101.3007}}} {[astro-ph.IM]}

\bibitem[{{Walch} and {Naab}(2015)}]{WalchNaab2015}
{Walch} S, {Naab} T (2015) {The energy and momentum input of supernova
  explosions in structured and ionized molecular clouds}. \mnras
  451(3):2757--2771. \doi{10.1093/mnras/stv1155},
  {\href{https://arxiv.org/abs/1410.0011}{{arXiv:1410.0011}}} {[astro-ph.GA]}

\bibitem[{{Walch} et~al.(2012){Walch}, {Whitworth}, and
  {Girichidis}}]{WalchWhitworthGirichidis2012}
{Walch} S, {Whitworth} AP, {Girichidis} P (2012) {The influence of the
  turbulent perturbation scale on pre-stellar core fragmentation and disc
  formation}. \mnras 419:760--770. \doi{10.1111/j.1365-2966.2011.19741.x},
  {\href{https://arxiv.org/abs/1109.0280}{{arXiv:1109.0280}}} {[astro-ph.GA]}

\bibitem[{{Wall} et~al.(2019){Wall}, {McMillan}, {Mac Low}, {Klessen}, and
  {Portegies Zwart}}]{WallEtAl2019}
{Wall} JE, {McMillan} SLW, {Mac Low} MM, et~al (2019) {Collisional N-body
  Dynamics Coupled to Self-gravitating Magnetohydrodynamics Reveals Dynamical
  Binary Formation}. \apj 887(1):62. \doi{10.3847/1538-4357/ab4db1},
  {\href{https://arxiv.org/abs/1901.01132}{{arXiv:1901.01132}}} {[astro-ph.GA]}

\bibitem[{{Wang} et~al.(2010){Wang}, {Li}, {Abel}, and
  {Nakamura}}]{WangEtAl2010}
{Wang} P, {Li} ZY, {Abel} T, et~al (2010) {Outflow Feedback Regulated Massive
  Star Formation in Parsec-Scale Cluster-Forming Clumps}. \apj 709:27--41.
  \doi{10.1088/0004-637X/709/1/27},
  {\href{https://arxiv.org/abs/0908.4129}{{arXiv:0908.4129}}} {[astro-ph.SR]}

\bibitem[{{Wardle} and {K\"{o}nigl}(1993)}]{WardleKoenigl1993}
{Wardle} M, {K\"{o}nigl} A (1993) {The structure of protostellar accretion
  disks and the origin of bipolar flows}. \apj 410:218--238.
  \doi{10.1086/172739}

\bibitem[{{Watt} et~al.(2025){Watt}, {Federrath}, {Birke}, and
  {Klingenberg}}]{WattEtAl2025}
{Watt} J, {Federrath} C, {Birke} C, et~al (2025) {Mitigating numerical
  dissipation in simulations of subsonic turbulent flows}. \mnras
  544(4):4256--4270. \doi{10.1093/mnras/staf1973},
  {\href{https://arxiv.org/abs/2511.09806}{{arXiv:2511.09806}}}
  {[physics.flu-dyn]}

\bibitem[{{Weaver} et~al.(1977){Weaver}, {McCray}, {Castor}, {Shapiro}, and
  {Moore}}]{WeaverEtAl1977}
{Weaver} R, {McCray} R, {Castor} J, et~al (1977) {Interstellar bubbles. II.
  Structure and evolution.} \apj 218:377--395. \doi{10.1086/155692}

\bibitem[{{Wibking} and {Krumholz}(2022)}]{WibkingKrumholz2022}
{Wibking} BD, {Krumholz} MR (2022) {QUOKKA: a code for two-moment AMR radiation
  hydrodynamics on GPUs}. \mnras 512(1):1430--1449.
  \doi{10.1093/mnras/stac439},
  {\href{https://arxiv.org/abs/2110.01792}{{arXiv:2110.01792}}} {[astro-ph.IM]}

\bibitem[{{Winner} et~al.(2019){Winner}, {Pfrommer}, {Girichidis}, and
  {Pakmor}}]{WinnerEtAl2019}
{Winner} G, {Pfrommer} C, {Girichidis} P, et~al (2019) {Evolution of cosmic ray
  electron spectra in magnetohydrodynamical simulations}. \mnras
  488(2):2235--2252. \doi{10.1093/mnras/stz1792},
  {\href{https://arxiv.org/abs/1903.01467}{{arXiv:1903.01467}}} {[astro-ph.HE]}

\bibitem[{{Wise} and {Abel}(2011)}]{WiseAbel2011}
{Wise} JH, {Abel} T (2011) {ENZO+MORAY: radiation hydrodynamics adaptive mesh
  refinement simulations with adaptive ray tracing}. \mnras 414(4):3458--3491.
  \doi{10.1111/j.1365-2966.2011.18646.x},
  {\href{https://arxiv.org/abs/1012.2865}{{arXiv:1012.2865}}} {[astro-ph.IM]}

\bibitem[{{Wolfire} et~al.(1995){Wolfire}, {Hollenbach}, {McKee}, {Tielens},
  and {Bakes}}]{WolfireEtAl1995}
{Wolfire} MG, {Hollenbach} D, {McKee} CF, et~al (1995) {The neutral atomic
  phases of the interstellar medium}. \apj 443:152--168. \doi{10.1086/175510}

\bibitem[{{W{\"u}nsch}(2024)}]{Wuensch2024}
{W{\"u}nsch} R (2024) {Radiation transport methods in star formation
  simulations}. Frontiers in Astronomy and Space Sciences 11:1346812.
  \doi{10.3389/fspas.2024.1346812},
  {\href{https://arxiv.org/abs/2403.05410}{{arXiv:2403.05410}}} {[astro-ph.IM]}

\bibitem[{{W{\"u}nsch} et~al.(2018){W{\"u}nsch}, {Walch}, {Dinnbier}, and
  {Whitworth}}]{WuenschEtAl2018}
{W{\"u}nsch} R, {Walch} S, {Dinnbier} F, et~al (2018) {Tree-based solvers for
  adaptive mesh refinement code FLASH - I: gravity and optical depths}. \mnras
  475(3):3393--3418. \doi{10.1093/mnras/sty015},
  {\href{https://arxiv.org/abs/1708.06142}{{arXiv:1708.06142}}} {[astro-ph.IM]}

\bibitem[{{W{\"u}nsch} et~al.(2021){W{\"u}nsch}, {Walch}, {Dinnbier},
  {Seifried}, {Haid}, {Klepitko}, {Whitworth}, and
  {Palou{\v{s}}}}]{WuenschEtAl2022}
{W{\"u}nsch} R, {Walch} S, {Dinnbier} F, et~al (2021) {Tree-based solvers for
  adaptive mesh refinement code FLASH - II: radiation transport module
  TreeRay}. \mnras 505(3):3730--3754. \doi{10.1093/mnras/stab1482},
  {\href{https://arxiv.org/abs/2105.09644}{{arXiv:2105.09644}}} {[astro-ph.IM]}

\bibitem[{{Wurster}(2021)}]{Wurster2021}
{Wurster} J (2021) {Do we need non-ideal magnetohydrodynamic to model
  protostellar discs?} \mnras 501(4):5873--5891. \doi{10.1093/mnras/staa3943},
  {\href{https://arxiv.org/abs/2101.04129}{{arXiv:2101.04129}}} {[astro-ph.SR]}

\bibitem[{{Xu}(1995)}]{Xu1995}
{Xu} G (1995) {A New Parallel N-Body Gravity Solver: TPM}. \apjs 98:355.
  \doi{10.1086/192166},
  {\href{https://arxiv.org/abs/astro-ph/9409021}{{arXiv:astro-ph/9409021}}}
  {[astro-ph]}

\bibitem[{{Xu} and {Lazarian}(2022)}]{XuLazarian2022}
{Xu} S, {Lazarian} A (2022) {Shock Acceleration with Oblique and Turbulent
  Magnetic Fields}. \apj 925(1):48. \doi{10.3847/1538-4357/ac3824},
  {\href{https://arxiv.org/abs/2111.04759}{{arXiv:2111.04759}}} {[astro-ph.HE]}

\bibitem[{{Xu} and {Kunz}(2021{\natexlab{a}})}]{XuKunz2021a}
{Xu} W, {Kunz} MW (2021{\natexlab{a}}) {Formation and evolution of protostellar
  accretion discs - I. Angular-momentum budget, gravitational self-regulation,
  and numerical convergence}. \mnras 502(4):4911--4929.
  \doi{10.1093/mnras/stab314},
  {\href{https://arxiv.org/abs/2101.00131}{{arXiv:2101.00131}}} {[astro-ph.SR]}

\bibitem[{{Xu} and {Kunz}(2021{\natexlab{b}})}]{XuKunz2021b}
{Xu} W, {Kunz} MW (2021{\natexlab{b}}) {Formation and evolution of protostellar
  accretion discs - II. From 3D simulation to a simple semi-analytic model of
  Class 0/I discs}. \mnras 508(2):2142--2168. \doi{10.1093/mnras/stab2715},
  {\href{https://arxiv.org/abs/2109.07535}{{arXiv:2109.07535}}} {[astro-ph.SR]}

\bibitem[{{Yan} and {Lazarian}(2004)}]{YanLazarian2004}
{Yan} H, {Lazarian} A (2004) {Cosmic-Ray Scattering and Streaming in
  Compressible Magnetohydrodynamic Turbulence}. \apj 614(2):757--769.
  \doi{10.1086/423733},
  {\href{https://arxiv.org/abs/astro-ph/0408172}{{arXiv:astro-ph/0408172}}}
  {[astro-ph]}

\bibitem[{{Yang} and {Federrath}(2025)}]{YangFederrath2025}
{Yang} TQ, {Federrath} C (2025) {Protostellar disc structure and dynamics
  during star formation from cloud-scale initial conditions}. \mnras
  541(2):1969--1987. \doi{10.1093/mnras/staf1088},
  {\href{https://arxiv.org/abs/2501.07626}{{arXiv:2501.07626}}} {[astro-ph.SR]}

\bibitem[{{Zamora-Avil{\'e}s} and
  {V{\'a}zquez-Semadeni}(2014)}]{ZamoraAvilesVazquezSemadeni2014}
{Zamora-Avil{\'e}s} M, {V{\'a}zquez-Semadeni} E (2014) {An Evolutionary Model
  for Collapsing Molecular Clouds and their Star Formation Activity. II. Mass
  Dependence of the Star Formation Rate}. \apj 793(2):84.
  \doi{10.1088/0004-637X/793/2/84},
  {\href{https://arxiv.org/abs/1308.4918}{{arXiv:1308.4918}}} {[astro-ph.SR]}

\bibitem[{{Zhang} and {Xu}(2023)}]{ZhangXu2023}
{Zhang} C, {Xu} S (2023) {Numerical Testing of Mirror Diffusion of Cosmic
  Rays}. \apjl 959(1):L8. \doi{10.3847/2041-8213/ad0fe5},
  {\href{https://arxiv.org/abs/2311.18001}{{arXiv:2311.18001}}} {[astro-ph.HE]}

\bibitem[{{Zhang} et~al.(2019){Zhang}, {Almgren}, {Beckner}, {Bell},
  {Blaschke}, {Chan}, {Day}, {Friesen}, {Gott}, {Graves}, {Katz}, {Myers},
  {Nguyen}, {Nonaka}, {Rosso}, {Williams}, and {Zingale}}]{ZhangEtAl2019}
{Zhang} W, {Almgren} A, {Beckner} V, et~al (2019) {AMReX: a framework for
  block-structured adaptive mesh refinement}. J Open Source Softw 4(37):1370.
  \doi{10.21105/joss.01370}

\bibitem[{{Zhao} et~al.(2020){Zhao}, {Caselli}, {Li}, {Krasnopolsky}, {Shang},
  and {Lam}}]{ZhaoEtAl2020Hall}
{Zhao} B, {Caselli} P, {Li} ZY, et~al (2020) {Hall effect in protostellar disc
  formation and evolution}. \mnras 492(3):3375--3395.
  \doi{10.1093/mnras/staa041},
  {\href{https://arxiv.org/abs/2009.07796}{{arXiv:2009.07796}}} {[astro-ph.SR]}

\bibitem[{{Zhao} et~al.(2021){Zhao}, {Caselli}, {Li}, {Krasnopolsky}, {Shang},
  and {Lam}}]{ZhaoEtAl2021}
{Zhao} B, {Caselli} P, {Li} ZY, et~al (2021) {The interplay between ambipolar
  diffusion and Hall effect on magnetic field decoupling and protostellar disc
  formation}. \mnras 505(4):5142--5163. \doi{10.1093/mnras/stab1295},
  {\href{https://arxiv.org/abs/2009.07820}{{arXiv:2009.07820}}} {[astro-ph.SR]}

\bibitem[{{Zhao} et~al.(2024){Zhao}, {Pudritz}, {Pillsworth}, {Robinson}, and
  {Wadsley}}]{ZhaoEtAl2024}
{Zhao} B, {Pudritz} RE, {Pillsworth} R, et~al (2024) {Filamentary Hierarchies
  and Superbubbles: Galactic Multiscale Magnetohydrodynamic Simulations of
  Giant Molecular Cloud to Star Cluster Formation}. \apj 974(2):240.
  \doi{10.3847/1538-4357/ad67e2},
  {\href{https://arxiv.org/abs/2405.18474}{{arXiv:2405.18474}}} {[astro-ph.GA]}

\bibitem[{{Zhao} et~al.(2026){Zhao}, {Bai}, and
  {Ostriker}}]{ZhaoXueNingOstriker2026}
{Zhao} X, {Bai} XN, {Ostriker} EC (2026) {Cosmic-ray Magnetohydrodynamics: A
  New Two-moment Framework with Numerical Implementation}. \apjs 283(1):37.
  \doi{10.3847/1538-4365/ae3e7f},
  {\href{https://arxiv.org/abs/2509.04387}{{arXiv:2509.04387}}} {[astro-ph.HE]}

\bibitem[{{Ziegler}(2008)}]{Ziegler2008}
{Ziegler} U (2008) {The NIRVANA code: Parallel computational MHD with adaptive
  mesh refinement}. Comput Phys Commun 179(4):227--244.
  \doi{10.1016/j.cpc.2008.02.017}

\bibitem[{{Zier} et~al.(2024){Zier}, {Kannan}, {Smith}, {Vogelsberger}, and
  {Verbeek}}]{ZierEtAl2024}
{Zier} O, {Kannan} R, {Smith} A, et~al (2024) {Adapting AREPO-RT for exascale
  computing: GPU acceleration and efficient communication}. \mnras
  533(1):268--286. \doi{10.1093/mnras/stae1837},
  {\href{https://arxiv.org/abs/2404.17630}{{arXiv:2404.17630}}} {[astro-ph.IM]}

\bibitem[{{Zinnecker} and {Yorke}(2007)}]{ZinneckerYorke2007}
{Zinnecker} H, {Yorke} HW (2007) {Toward Understanding Massive Star Formation}.
  \araa 45:481--563. \doi{10.1146/annurev.astro.44.051905.092549},
  {\href{https://arxiv.org/abs/0707.1279}{{arXiv:0707.1279}}}

\bibitem[{{Zweibel}(2013)}]{Zweibel2013}
{Zweibel} EG (2013) {The microphysics and macrophysics of cosmic rays}. Phys
  Plasmas 20(5):055501. \doi{10.1063/1.4807033}

\bibitem[{{Zweibel}(2017)}]{Zweibel2017}
{Zweibel} EG (2017) {The basis for cosmic ray feedback: Written on the wind}.
  Phys Plasmas 24(5):055402. \doi{10.1063/1.4984017}

\end{thebibliography}

\end{document}